\documentclass[12pt]{ociamthesis}  % default square logo
\usepackage{nicematrix}
 \usepackage{multirow}
 \usepackage[normalem]{ulem}
 \useunder{\uline}{\ul}{}
\usepackage{multicol}
\usepackage{caption}
\usepackage[edges]{forest}
\usepackage{graphicx,wrapfig,lipsum}
\usepackage{amsmath,empheq}
\usepackage{dirtree}
\usepackage{chemmacros}
\usepackage{titlesec}
\usepackage{breqn}
\usepackage{amsthm}
\usepackage[style=numeric,minbibnames=3,maxbibnames=3,autopunct=true,babel=hyphen,hyperref=true,abbreviate=false,backend=biber]{biblatex}
\usepackage{tikz}

\usepackage[utf8]{inputenc}
\usepackage[english]{babel}
\usepackage{hyperref}
\usepackage[ruled]{algorithm2e}
\usepackage{amssymb}
\usepackage{graphicx}
\usepackage{cancel}
\usepackage{epstopdf}
\usepackage{mathtools}
\usepackage{float}
\usepackage{subcaption}
\usepackage{setspace}
\usepackage{xcolor}
\usepackage{bm}
\usepackage[export]{adjustbox}
\usepackage{relsize}
\usepackage{listings}
\usepackage[toc,page]{appendix}
\title{Gene Expression Time Delays in Reaction-Diffusion Systems}     %your thesis title}   %note \\[1ex] is a line break in the title

\author{Alec Sargood}             %your name
\college{St Anne's College}  %your college

\degree{MSc Mathematical Modelling and Scientific Computing}     %the degree
\degreedate{Trinity Term, September 2021}         %the degree date

\begin{document}

%this baselineskip gives sufficient line spacing for an examiner to easily
%markup the thesis with comments
\baselineskip=18pt plus1pt

%set the number of sectioning levels that get number and appear in the contents
\setcounter{secnumdepth}{3}
\setcounter{tocdepth}{3}

\maketitle                  % create a title page from the preamble info
\section*{\centering{Acknowledgements}}

I would like to thank my supervisors Dr. Andrew Krause and Professor Eamonn Gaffney for all of their advice and guidance throughout my dissertation. I would like to especially thank Andrew for his continued and tireless effort to support me, not only throughout the dissertation process, but also throughout the MMSC. From meeting Andrew at my MMSC interview and having him as a departmental supervisor in Michaelmas term, to being able to work on modelling case studies and the dissertation under his supervision in Hilary and Trinity terms, I am extremely grateful for all of his hard work and mentorship that has undoubtedly shaped my academic experience at Oxford.\\\\
Completing and coping with the arduous demands of the MMSC course, whilst juggling a global pandemic, would not have been possible had it not been for my fellow MMSC cohort. This has been my most academically challenging year, yet also been made one of my most enjoyable by the fantastic people around me. In this vein, I would also like to thank Dr. Kathryn Gillow for her attentiveness and advice, both academically and pastorally, throughout the year.\\\\
I would most like to thank my family for their unconditional support and belief in me. I am certain that I would not be where I am today without them.\\\\
Finally, I want to dedicate part of this work to my closest friend, Rodion Matveev, who tragically lost his life on the 14th of August 2021.
  % include an acknowledgements.tex file
\begin{abstract}

Gene expression time delays, modelling the complex biological processes of gene transcription and translation, have been shown to play an important role in cellular dynamics. Time delays, motivated by the gene expression process, can also greatly affect the behaviour of reaction-diffusion systems. In this dissertation, we explore their effects on Turing pattern mechanisms. By incorporating time delays, modelled as both a fixed parameter and as a continuous distribution, into classical reaction-diffusion systems that exhibit Turing instabilities, we investigate the changing behaviour of these systems. We find that an introduction of increasing time delay increases the time taken for spatially inhomogeneous patterns to stabilise, and the two are related linearly. We also present results to show, through a linear stability analysis, that an increasing time delay can act both to expand or shrink the Turing space of a certain reaction-diffusion mechanism, depending on the placement of time-delayed terms. Significantly, we find that modelling time delays as a continuous distribution has a negligible impact on qualitative or quantitative aspects of the results seen compared with a fixed time delay of the mean of the distribution. These findings serve to highlight the importance of considering gene expression time delays when modelling biological patterning events, as well as requiring a complete understanding of the cellular dynamics before attempting to apply Turing mechanisms to explain biological phenomena. The results also suggest, at least for the distributions considered in this dissertation, that fixed delay and distributed delay models have almost identical dynamics. This allows one to use simpler fixed delay models rather than the more complicated distributed delay variants.

\end{abstract}
          % include the abstract

\begin{romanpages}          % start roman page numbering
\tableofcontents            % generate and include a table of contents
\end{romanpages}            % end roman page numbering

%now include the files of latex for each of the chapters etc
\chapter{Introduction}

\section{Background}\label{section:background}

The self-organisation of cells into an apparent order appears across many different fields within biology. For example, the distribution of cells during the developmental process of an embryo, the growth of cancerous tissue \cite{morph}, vertebrate limb development \cite{miura1,glimm,miura2}, and pattern formation on animal coats (e.g. spots on a jaguar \cite{painter}, feathers on birds \cite{bailleul}). Wolpert \cite{wolpert} presented the idea that, underpinning the development of shape and form (morphogenesis) is a cell's ability to differentiate according to its position in space and time. Furthermore, the concentration of certain chemicals (morphogens), or the concentration gradients of certain morphogens across a spatial domain of cells, affects the cell differentiation mechanism, and thus cells adopt a state relative to the concentration of a specific morphogen that they are exposed to.

The mechanism allowing cells to adopt an appropriate state is known as differential gene expression, and depends crucially on the communication between cells, achieved through cellular signalling \cite{gaffmonk}. Typical reaction-diffusion systems assume a negligible timescale on which the cellular signalling and gene expression processes occur. The gene expression process is however extremely complex and proceeds through several stages \cite{gaffmonk}, including gene transcription and gene translation. These sub-processes can take large amounts of time, and it has been experimentally shown that these time delays are typically on the order of minutes, but in some cases can be as large as a few hours \cite{gaffmonk,tennyson}. These time delays can therefore be on the same order of magnitude as the pattern formation process itself. For example, the basic body plan of a zebrafish is established in less than 24 hours \cite{gaffmonk,kimmel}. It is therefore important to consider these delays when studying pattern formation in Turing mechanisms.

In 1952 \cite{turing}, Alan Turing proposed that the morphogenesis process could be mathematically modelled on a purely chemical basis via the interaction of morphogens, whose evolutions are described by a system of coupled reaction-diffusion equations. Turing showed that a stable steady state, robust to small perturbations in the spatially homogeneous setting (no diffusion), could become unstable and sensitive to small perturbations with the introduction of diffusion, leading to spatially inhomogeneous patterns. Cell fate decisions are then based on these morphogen concentrations, where regions of high morphogen or low morphogen concentration can lead to different cell fate decisions. Turing's model is therefore one of pre-patterning, where the morphogen pattern concentrations across a spatial domain are modelled, which in turn lead to cell fate decisions at a later stage. Typical reaction-diffusion systems in the context of Turing pattern formation consist of two partial differential equations, describing the interaction and evolution of two morphogens, the \textit{activator} and \textit{inhibitor}. Empirical evidence suggests that Turing instabilities are present in real biological systems, and can be used to explain complex biological phenomena \cite{yigaffneyli,molecular,miura,miura2,sick}. However, whether Turing patterns can be found experimentally in biological systems with simple two-species systems is still very much an active field of research \cite{bespoke}.

Time delays have been investigated in the context of Turing patterns, both numerically and analytically, through the incorporation of constant fixed time delays. One of the canonical reaction-diffusion mechanisms that exhibits Turing instabilities is the Schnakenberg model \cite{schnakenberg}. This model has been extensively studied in the context of Turing pattern formation with incorporated gene expression time delays. Two biologically motivated variants, the ligand-internalisation (LI) and reverse ligand-binding (RLB) models, have been considered in the literature. We briefly outline the biological motivation for these variants in \ref{section:fixedeq}. The Schnakenberg reaction kinetics can be described as \textit{cross} kinetics \cite{leegaffney}, where the inhibitor upregulates the activator, which in turn downregulates the inhibitor. The LI model places gene expression time delay in purely the activator's dynamics, whereas the RLB model contains time delay in both the activator's and inhibitor's dynamics.

The numerical results in \cite{gaffmonk}, which studied the LI variant of the Schnakenberg model, showed that the time taken until pattern formation occurs drastically increases as the gene expression time delay in the model is increased, and that small delays, on the order of minutes, can cause a large increase in time-to-pattern, on the order of several hours, compared to a model with no time delay. This highlights the importance of studying gene expression time delays, especially when considering patterning events that occur on a fast timescale. The two papers \cite{jiang, yigaffneyli} consider both the LI and RLB variants of the Schnakenberg model. Using linear analysis and dynamical systems theory, the results in both suggest that the RLB model can exhibit spatially inhomogeneous temporal oscillations, as well as de-stabilisation of spatially inhomogeneous steady states, inhibiting pattern formation via Turing instabilities. The results in \cite{yigaffneyli}  specifically suggest that for the LI model in particular, extensive ligand internalisation, i.e increasing the time delay in the activator's dynamics, can antagonise the formation of patterns from Turing instabilities, shrinking the parameter space where Turing instabilities may occur. We explore these observations in more detail by conducting bifurcation analysis neglected in the investigation carried out in \cite{yigaffneyli}.

Another typical reaction-diffusion system studied in the literature is the Gierer-Meinhard (GM) model \cite{gm}. General results in both \cite{leegaffney,leegaffmonk} suggest that time delay causes a significant effect on the time taken until pattern formation occurs. Both papers also suggest that linear theory is insufficient in determining the presence of oscillations, and \cite{leegaffmonk} suggests that the severity of the effect that time delays will have on the timescales on which patterning events will occur cannot be accurately predicted from linear theory. A final observation from \cite{gaffmonk,leegaffmonk}, for both LI and RLB models, is that an increasing time delay may also increase the sensitivity of the final pattern formation to variation in initial conditions.

More recently analysis of one-dimensional spike solutions of the GM model in \cite{fadai1,fadai2}\footnote{We note that the spike solution analysis considered in \cite{fadai1,fadai2} differs from the linear stability of homogeneous steady states considered in \cite{leegaffmonk,gaffmonk,leegaffney}.}, show that the biological interpretation, and thus the placement of delay terms in the model, can affect the size of the parameter regimes for which the spike solution is linearly stable. It was found that, depending on the positioning of time-delayed terms, an increasing time delay can act as a stabilising or de-stabilising agent, enlarging or shrinking the stable parameter region of the spike. Further details of spike solutions of the GM model and their stability analysis can also be found in \cite{spike}. This analysis highlighted the importance of time delay positioning in the GM model for the stability of spike solutions. We aim to show an analogous result for the Turing space of the GM model, namely that altering the time-delayed terms within the model can change the effect that an increasing time delay has on the Turing space (the parameter regions such that Turing instabilities can occur).

Paradoxically, we see that although Turing's models can be used to explain and reproduce complex biological phenomena \cite{leegaffney}, the results seem to be dependent on gene expression dynamics, and as most of the current literature shows, gene expression time delays can provide difficulty in applying Turing's models to real systems. In summary, most of the literature shows that gene expression time delays increase the time-to-pattern, and that depending on the positioning of the time-delayed terms, an increasing time-delay can increase sensitivity of the final pattern formation to initial and boundary conditions, as well as shrink Turing spaces and antagonise pattern formation. The former effect in particular is an obstacle to using Turing mechanisms due to the timescales involved. Time delays can induce much larger delays in pattern onset, compared to the otherwise fast pattern onset that would occur without time delay, calling into question how relevant and applicable Turing's models are for patterning events on fast timescales.

We note that the current literature on Turing pattern formation in development is only concerned with fixed time delays. On a cellular level however, the biological processes responsible for gene expression are inherently stochastic \cite{raj,elowitz,mcadams,paulsson}. Time delays, and more specifically distributed delays, have been motivated throughout mathematical biology. Distributed time delays have been incorporated to model biological phenomena such as hematopoiesis, and lactose operon dynamics \cite{newdist}, Wnt/$\beta$-catenin signalling pathways \cite{signal}, and Oncolytic virotherapy treatment for cancer \cite{cancer}. A distributed delay can be thought of as a more `general and realistic' \cite{cancer} approach to modelling, on a larger scale, a process which in reality may possess, on a small scale, an intrinsic stochasticity. Within the context of Turing pattern formation, introducing a fixed time delay into the reaction-diffusion mechanism is an oversimplification of the underlying biological process on a microscopic level, and leads us to consider a distribution of time delays at the macroscopic level \cite{bratsun,krausenew}.

In this dissertation, we are interested in conducting a more systematic study of the time-to-pattern properties of models with fixed time delay, and a more careful consideration of sensitivity to initial and boundary conditions. We are also concerned with whether implementing a different form of delay, specifically distributed delay, can alleviate some of the problems caused by the fixed delay case. For the rest of this Chapter, we outline some of the mathematical preliminaries used throughout this dissertation, including an outline of Turing pattern theory, and the numerical methods we use. In Chapter 2, we study the LI variant of the Schnakenberg model, where time-delayed terms are considered only in the activator's dynamics. We use numerical simulations to systematically evaluate the robustness of results in the current literature to variations in initial and boundary conditions. Our results show that, although the type of pattern we see may be affected by these variations, the relationship between time delay and time until onset of patterning is robust. We also extend the current linear analysis presented in \cite{yigaffneyli} and study the effect of fixed time delays on the Turing space, as well as considering in more depth the effects that time delay have on the time lag until onset of patterning. Chapter 3 will focus on the distributed delay model, where we aim to produce novel linear analysis and show that an incorporated distributed delay behaves almost identically to a fixed delay, and thus in a sense, the distributions we use do not matter. Finally, in Chapter 4, we introduce the GM model for pattern formation, and present initial findings showing the effect of a fixed time delay on the Turing space. Furthermore, we highlight the importance of the positioning of the time-delayed terms on the results seen, and thus the importance of understanding the biological processes that lead to gene expression time delays. Since the first three chapters of the dissertation are concerned only with the Schnakenberg model, this is the model we introduce first. The GM model is only considered in Chapter 4.

\section{Model Introduction}
\subsection{Without Time Delay}
The mathematical model we will consider in Chapters 1-3 is the Schnakenberg model \cite{schnakenberg} -- one of the simplest `toy' models that exhibit some of the key behaviours that we are interested in, and a model that has also been studied extensively in the context of fixed time delays. The model describes the evolution and interaction between two reactants, $U$ and $V$. Only considering two reactants is a gross simplification of the underlying biological processes responsible for pattern formation, but it is still a non-trivial case that can admit Turing instabilities. In this dissertation, we restrict our investigation to one spatial domain. The chemical reaction describing the Schnakenberg kinetics \cite{baker} is given by
\begin{equation}\label{chem}
A\xrightleftharpoons[c_{-1}]{c_1} U,\quad B\xrightarrow{c_2} V,\quad 2U+V\xrightarrow{c_3} 3U,
\end{equation}
where the $c_i$ represent reaction rates. The quantities $A$ and $B$ are substances whose evolution is not considered, and we assume a constant supply. We use $u$, $v$, $a$, and $b$ to denote the concentrations of substances $U$, $V$, $A$ and $B$ respectively. Letting the reactants diffuse, and applying the law of mass action with a non-dimensionalisation \cite{murray}, yields the reaction-diffusion system
\begin{equation}\label{system}
    \begin{split}
    \frac{\partial u}{\partial t}&=\frac{\epsilon^2}{L^2}\frac{\partial^2 u}{\partial x^2}+a-u+u^2v,\\
    \frac{\partial v}{\partial t}&=\frac{1}{L^2}\frac{\partial^2 v}{\partial x^2}+b-u^2v,
    \end{split}
\end{equation}
where $x\in\Omega=[0,1]$ is the non-dimensionalised spatial domain and $a,b>0$ are fixed parameters. The parameter $\epsilon^2$ can be thought of as the ratio of diffusion coefficients between the activator $u$ and inhibitor $v$, and $L^2$ a scaling of the domain length on which the problem is being solved. Typical values in the literature \cite{gaffmonk} are $L^2=1/200$ and $\epsilon^2=0.001$. Unless otherwise stated, in this dissertation, we use the same $\epsilon^2=0.001$, and a domain size, $L$, $30$ times of that used in \cite{gaffmonk}, namely $L^2=9/2$. Since we are interested in the pattern formation arising from the self-organisation of cells, we implement no flux (homogeneous Neumann) boundary conditions on the boundary of the spatial domain, namely
\begin{equation}\label{neumannbc}
    \frac{\partial u}{\partial x}=\frac{\partial v}{\partial x}=0, \quad \quad x=0,1.
\end{equation}
As typical when studying Turing patterns, initial conditions $(u_0,v_0)$ are chosen as a small random Gaussian perturbation from the spatially homogeneous steady state $(u_\star,v_\star)$. In this dissertation, unless otherwise stated, the initial conditions we use are
\begin{equation}\label{firstic}
\begin{pmatrix}u_0\\v_0\end{pmatrix}=\begin{pmatrix}u_\star(1+r)\\v_\star(1+r)\end{pmatrix},
\end{equation}
where $r$ is a random variable such that $r\sim\mathcal{N}\left(0,0.01^2\right)$. The notation $r\sim\mathcal{N}\left(\mu_{\text{IC}},\sigma_{\text{IC}}^2\right)$ denotes a Normally distributed random variable $r$ with mean $\mu_{\text{IC}}$ and standard deviation of the initial perturbation $\sigma_{\text{IC}}$.

\subsection{With Fixed Time Delay}\label{section:fixedeq}

The form of the model we consider with fixed time delay is the LI variant of the standard Schnakenberg model. We do not consider the RLB model as it was found in \cite{william}, that under certain conditions, the numerical solutions of activator and inhibitor concentrations became physically infeasible, with negative solutions. The LI model assumes that a reaction at the cell surface is followed by internalisation of a morphogen before the gene expression process can continue and morphogen production can occur \cite{leegaffney,yigaffneyli}, introducing a time delay in the activator's dynamics.  This is based on the assumption that the gene expression process, and thus the source of the time delay, is responsible for autocatalysis of the activator in the reaction-diffusion mechanism \cite{gaffmonk}. As described in \cite{baker}, applying the delay to the final nonlinear term of \eqref{chem} yields the reaction described by
\begin{equation}\label{chem2}
A\xrightleftharpoons[c_{-1}]{c_1} U,\quad B\xrightarrow{c_2} V,\quad 2U+V\xrightarrow{c_3} W,\quad W\xrightarrow{\text{\footnotesize delay } \tau}3U.
\end{equation}
The reaction describes an internalisation of two particles of $U$, and one particle of $V$, which are removed from the reaction, forming substance $W$. However, three particles of $U$ are obtained from a reaction at a time $\tau$ in the past. The reaction-diffusion system describing the LI model is thus written as \cite{leegaffney}
\begin{equation}\label{fixed}
  \begin{split}
  \frac{\partial u}{\partial t}&=\frac{\epsilon^2}{L^2}\frac{\partial^2u}{\partial x^2}+a-u-2u^2v+3\hat{u}^2\hat{v},\\
  \frac{\partial v}{\partial t}&=\frac{1}{L^2}\frac{\partial^2v}{\partial x^2}+b-u^2v,
\end{split}
\end{equation}
where $u=u(x,t)$, $v=v(x,t)$ and $\hat{u}$, $\hat{v}$ are evaluated at some delay $\tau$, so that $\hat{u}=u(x,t-\tau)$ and $\hat{v}=v(x,t-\tau)$.

In order to solve delay differential equations (DDEs), a history function is required to define the solution for $t\in[-\tau,0)$ for the terms with time delay, so that the solutions of $u(x,t-\tau)$ and $v(x,t-\tau)$ are defined for $t\in[0,\tau)$. Throughout this dissertation, unless otherwise stated, a constant history function equal to the initial conditions is used, so that
\begin{equation}\label{hist}
    \begin{split}
u(x,t-\tau)&=u_0,\\
v(x,t-\tau)&=v_0,
\end{split}
\end{equation}
for all $x\in[0, 1]$ and $t\in[0,\tau]$.

\subsection{With Distributed Time Delay}

The stochastic nature of gene expression delays leads us to consider a mean-field approach to modelling the time delay \cite{bratsun,krausenew}. We can thus write the LI model with distributed time delay as

\begin{equation}\label{distmodel}
  \begin{split}
    \frac{\partial u}{\partial t}&=\frac{\epsilon^2}{L^2}\frac{\partial^2u}{\partial x^2}+a-u-2u^2v+3\int_{a}^{b}k(s;\textbf{p})\hat{u}^2\hat{v} \ \text{ds},\\
    \frac{\partial v}{\partial t}&=\frac{1}{L^2}\frac{\partial^2v}{\partial x^2}+b-u^2v,
\end{split}
\end{equation}
where $\hat{u}=u(x,t-s)$ and $\hat{v}=v(x,t-s)$, with $s$ the integration variable ranging over the delays. The function $k(s;\textbf{p})$ denotes some probability distribution function, with $s$ the integration variable, and $\textbf{p}$ the distribution parameters. The integration domain of delays is given by $[a,b]$ with $a>0$ to ensure positive time delays. Choices of different probability density functions will be considered in more detail in Chapter \ref{section:distdel}.

\section{Mathematical Preliminaries}
\subsection{Turing Pattern Formation Without Delay}
Here we give a brief overview of the mathematical theory underpinning Turing pattern formation, closely following the description in \cite{murray}. For further details, the reader should consult \cite{murray,beentjes}. Turing instabilities occur when the spatially homogeneous stable steady state becomes unstable in the presence of diffusion. We therefore first consider the spatially homogeneous model (the system defined in \eqref{system} without diffusive terms), and explore conditions necessary for the steady state to be stable. In the case of the Schnakenberg model, the single steady state occurs at $(u_\star, v_\star)=\left(a+b, \frac{b}{(a+b)^2}\right)$, with $u_\star,v_\star>0$. Following the methodology in \cite{murray}, we perform linear stability analysis. Taking a small perturbation from the steady state, so that $u(x,t)=u_\star+\delta\xi(x,t) $, $v(x,t)=v_\star+\delta\eta(x,t) $ for $|\delta|\ll1$, we consider the evolution of the perturbation. Denoting $\pmb{\xi}=\begin{bmatrix}\xi \\ \eta\end{bmatrix}$ as the vector of perturbations, and Taylor expanding up to $O(\delta)$, the linearised system of \eqref{system} is given as

\begin{equation}\label{linsys}
\frac{d\pmb{\xi}}{dt}=\textbf{J}_{(u_\star,v_\star)}\pmb{\xi},
\end{equation}
where $\textbf{J}_{(u_\star,v_\star)}$ is the Jacobian matrix of the kinetic equations evaluated at the steady state, namely,
$$
\textbf{J}_{(u_\star,v_\star)}=\begin{pmatrix}f_u&f_v\\g_u&g_v\end{pmatrix}\Bigg|_{(u_\star,v_\star)}.
$$
The notation $f_u$ is used to denote the partial derivative of $f$ with respect to $u$. For the Schnakenberg model (without time delay), the kinetic functions are given as
\begin{align*}
f(u,v)&=a-u+u^2v,\\
g(u,v)&=b-u^2v.
\end{align*}
We consider solutions of \eqref{linsys} that are of the form
$$
\pmb{\xi}\propto e^{\lambda t},
$$
for eigenvalues $\lambda$ of $\textbf{J}_{(u_\star,v_\star)}$. The steady state is said to be asymptotically stable if the perturbation decays.
Denoting $\text{spec}(\textbf{M})$ as the set of eigenvalues of some matrix $\textbf{M}$, asymptotic stability occurs when $\Re(\lambda)<0 \ \text{for all }\lambda\in \text{spec}(\textbf{J}_{(u_\star,v_\star)})$. However, if there exists $\lambda\in \text{spec}(\textbf{J}_{(u_\star,v_\star)})\text{ such that } \Re(\lambda)>0$,
then the perturbation will grow with time and the steady state is unstable. The sum and product of the eigenvalues of $\textbf{J}_{(u_\star,v_\star)}$
are given by $\text{Tr}(\textbf{J}_{(u_\star,v_\star)})$ and $\text{det}(\textbf{J}_{(u_\star,v_\star)})$ respectively. The required conditions for stability are therefore
\begin{equation}\label{cond1}
    \begin{split}
\text{Tr}(\textbf{J}_{(u_\star,v_\star)})<0 &\implies (f_u+g_v)\big|_{(u_\star,v_\star)}<0, \\
\text{det}(\textbf{J}_{(u_\star,v_\star)})>0 &\implies (f_ug_v-f_vg_u)\big|_{(u_\star,v_\star)}>0.
\end{split}
\end{equation}

\begin{figure}[H]
    \centering
    \begin{subfigure}[t]{0.45\textwidth}
        \centering
        \includegraphics[width=7cm,height = 6cm]{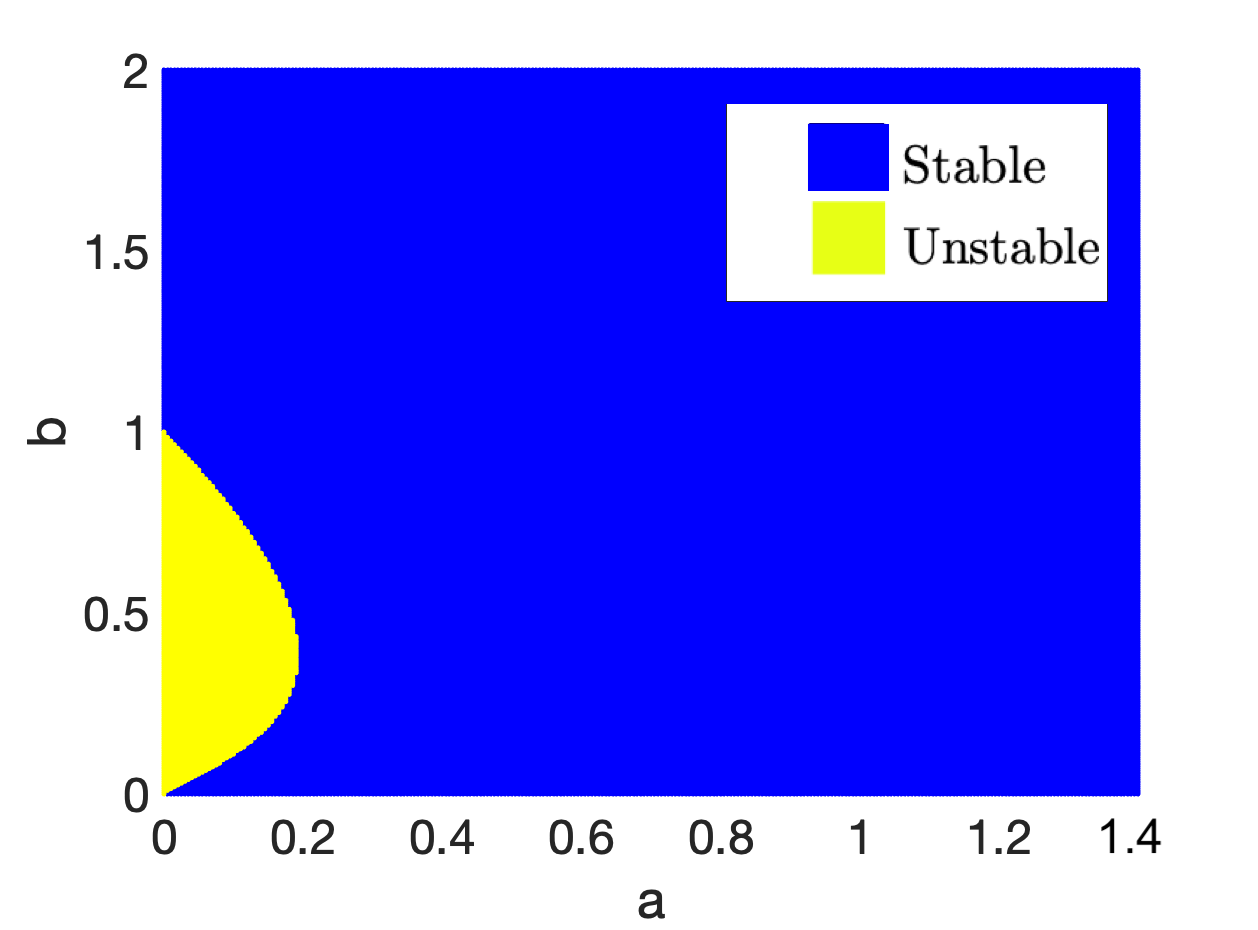}
        \caption{Bifurcation diagram for spatially homogeneous model, no delay.}
        \label{fig:bifsh}
    \end{subfigure}
    \hfill
    \begin{subfigure}[t]{0.45\textwidth}
        \centering
        \includegraphics[width=7cm,height = 6cm]{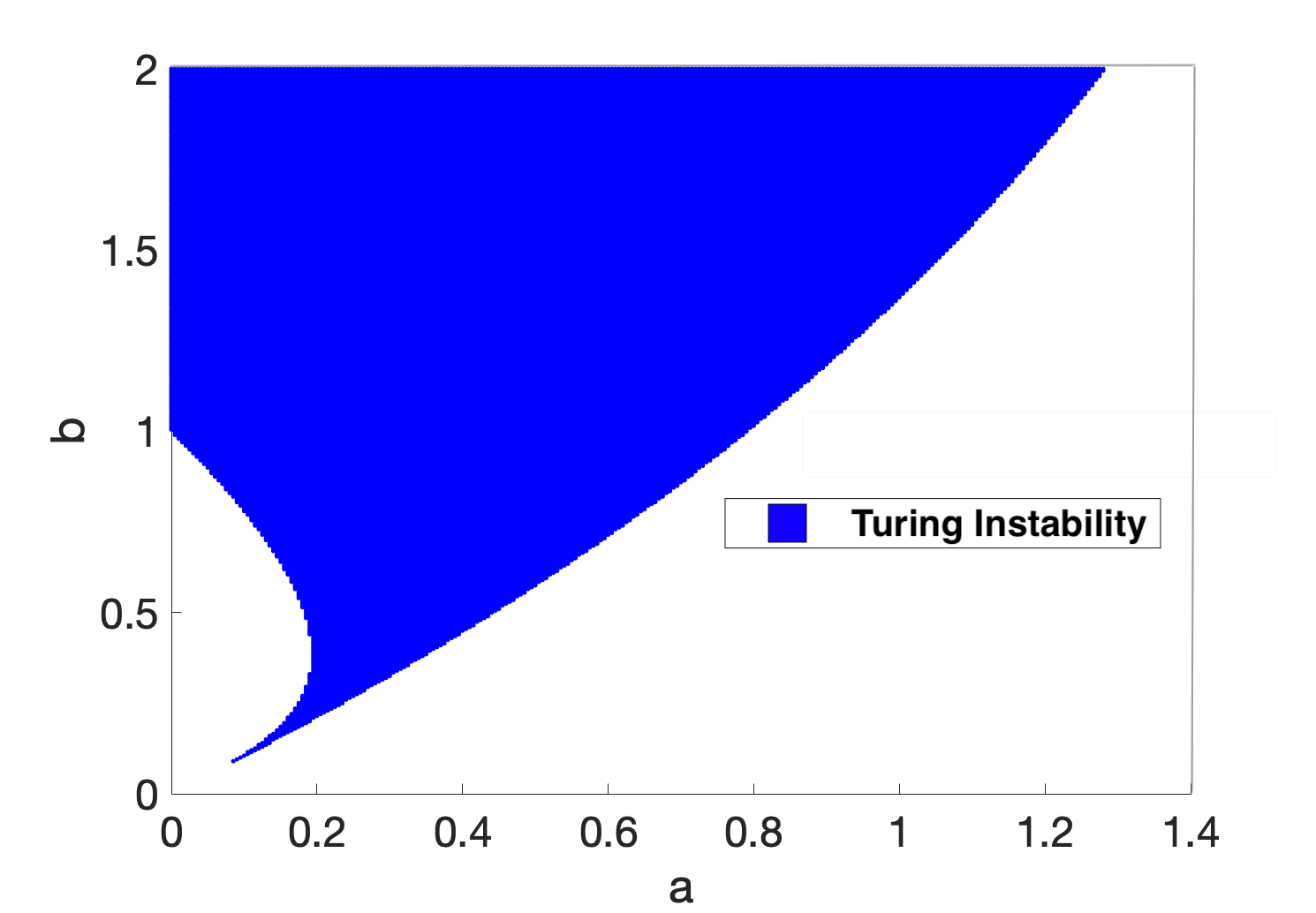}
        \caption{Turing space, no delay. $\epsilon^2=0.001$.}
        \label{fig:turingspace}
    \end{subfigure}
    \caption{Conditions \eqref{cond1} and \eqref{cond2} used to plot bifurcation diagram and Turing space for parameters $(a,b)\in[0,1.4]\times[0,2]$, for the Schnakenberg model.}
    \label{fig:dispfixed}
\end{figure}

We now consider the full diffusive model and look for necessary conditions such that the previously stable steady state is driven to instability. The linearised system is given by
\begin{equation}\label{linsys2}
    \frac{\partial \pmb{\xi}}{\partial t}=\left[\textbf{D}\frac{\partial}{\partial x^2}+\textbf{J}_{(u_\star,v_\star)} \right]\pmb{\xi},
\end{equation}
where $\textbf{D}=\begin{pmatrix}\frac{\epsilon^2}{L^2}&0\\0&\frac{1}{L^2}\end{pmatrix}$ is the matrix containing the diffusion coefficients of reactants. The solution to the spatially dependent eigenvalue problem can be written as a linear combination of the eigenfunctions $w_k$ that satisfy the problem
\begin{equation}\label{eigprob}
\nabla^2w_k=-k^2w_k,\quad \quad \frac{\partial w_k}{\partial x}=0\quad x=0, 1.
\end{equation}
Considering only a regular 1D domain $\Omega=[0,1]$ with no flux boundary conditions, we note that the eigenfunctions will be of the form $w_k=\cos(k\pi x)$, $x\in[0,1]$. We thus look for solutions to \eqref{linsys2} of the form
\begin{equation}\label{perturbgrow}
    \pmb{\xi}=\sum_{k=0}^{\infty} \textbf{c}_ke^{\lambda_k t}w_k(x),
\end{equation}
where the constants $\textbf{c}_k$ are determined by using a Fourier expansion of the initial conditions in terms of the eigenfunctions $w_k$. $\lambda_k$ is the eigenvalue which determines the rate of temporal growth for each mode $k$, and thus determines whether a particular mode of pattern will be unstable and grow. Substituting this form \eqref{perturbgrow} into \eqref{linsys2}, along with using \eqref{eigprob} and simplifying, we obtain, for each $k$ by orthogonality
$$
\lambda_k w_k=\textbf{J}w_k-\textbf{D}k^2w_k \implies (\lambda_k \textbf{I}-\textbf{J}+k^2\textbf{D})w_k=\textbf{0},
$$
with $\textbf{I}$ the identity matrix. Looking for non-trivial solutions for $w_k$, we solve for roots of the characteristic polynomial, namely $\text{det}(\lambda_k \textbf{I}-\textbf{J}+k^2\textbf{D})=0$, which yields a quadratic equation for eigenvalues $\lambda_k(k)$ as a function of $k$. Finding roots of this quadratic such that $\Re(\lambda_k(k))>0$ for some $k\neq0$, we conclude \cite{murray} two necessary conditions for the instability of the steady state in the presence of diffusion, namely
\begin{equation}\label{cond2}
    \begin{split}
    \left(\frac{1}{\epsilon^2}f_u+g_v\right)\bigg|_{(u_\star,v_\star)}>0,&\\
    \left(\left(\frac{1}{\epsilon^2}f_u+g_v\right)^2-\frac{4}{\epsilon^2}(f_ug_v-f_vg_u)\right)\bigg|_{(u_\star,v_\star)}>0.
\end{split}
\end{equation}
We therefore have four necessary conditions in terms of $(a,b,\epsilon^2)$ for Turing patterns to occur. These conditions are only necessary, and not sufficient, because conditions \eqref{cond2} assume $k$ to be a continuous variable, rather than discrete, and this is only strictly valid in the limit $L\to\infty$. Using the first two conditions in \eqref{cond1}, a bifurcation diagram in the $(a,b)$ parameter space can be plotted showing the regions corresponding to a stable or unstable steady state. This can be seen in Figure \ref{fig:bifsh}. Using the additional conditions in \eqref{cond2} and the fixed value $\epsilon^2=0.001$, the parameter region in the $(a,b)$ parameter space in which Turing patterns can occur can also be plotted. This `Turing space' can be seen in Figure \ref{fig:turingspace}. We note that throughout this dissertation, where results are presented for varying parameter values $(a,b)$, the parameter space is discretised at regular intervals of $0.02$, for both $a$ and $b$.

\subsection{Numerical Implementation}\label{section:numimp}
In order to numerically resolve the spatial derivatives $\frac{\partial^2 u}{\partial x^2}$, $\frac{\partial^2 v}{\partial x^2}$ and implement the relevant boundary conditions, a finite-difference scheme is used. Throughout the dissertation, we use $m=500$ equally spaced spatial discretisation points on the domain $x\in\Omega=[0,1]$. This discretisation results in $m=500$ ODEs or DDEs in time, which are solved via built-in time-stepping solvers in Julia. Further details on the derivation and implementation of the finite difference scheme and boundary conditions can be found in Appendix \ref{section:appA}.

Reaction-diffusion systems can be numerically stiff to solve \cite{stiff1, william}, and thus to solve these systems with time delay, we require stiff numerical solvers suitable for DDEs. The inherent stiffness of the problem makes standard DDE solvers in MATLAB such as \emph{dde23} and \emph{ddesd} unsuitable, and past work has been restricted in the progress made through numerical simulations \cite{william} due to the computationally expensive task of solving reaction-diffusion systems with non-stiff solvers. Standard stiff solvers in MATLAB, such as \emph{ode23s} and \emph{ode15s} do not support time delay. For this dissertation, we therefore develop neat and efficient code using the Julia language to numerically solve these systems. Julia has an extensive differential equations solver suite \cite{rodas}, and has the capability to apply the method of steps \cite{methsteps} to a stiff solver, allowing the incorporation of fixed time delays. Throughout the dissertation, we use absolute and relative solver tolerances of $10^{-6}$, with a maximum timestep set as $0.1$. For these tolerances, the default stiff solver implemented by Julia is \emph{Rodas5}, a 5-th order A-stable solver, from the family of Rosenbrock methods \cite{rodas}. An interested reader can find more details on Rosenbrock methods in \cite{rosenbrock}. The Julia code used to generate all numerical solutions throughout this dissertation can be found at \cite{git}.

Finally, since the Schnakenberg model has \textit{cross} reaction kinetics, as discussed in Section \ref{section:background}, we have that when the concentration of the activator $u$ is high, the concentration of the inhibitor $v$ is low, and vice-versa \cite{murray}. The concentration gradients of the two morphogens $u$ and $v$ are thus effectively `out of phase', and so it is sufficient to consider just the numerical solution of the activator $u$. Throughout the dissertation therefore, where relevant, only the numerical solution of the activator $u$ is plotted.

\chapter{Fixed Delay Model}\label{section:fixdel}

In this chapter, we first revisit the analysis of the LI model through a more careful review of the linear theory presented in \cite{yigaffneyli,jiang}. We aim to analytically determine the effects of an increasing time delay on the Turing space and the time lag to onset of patterning in more detail than currently considered in the literature. Results are also confirmed through full numerical simulations.

We find through bifurcation analysis that an increasing time delay increases the size of the Turing space, resulting in a wider parameter range that can exhibit Turing instabilities. We also show that on a small scale, the linear analysis provides a good approximation to the time-to-pattern with an increasing fixed time delay, and in fact the time-to-pattern increases linearly with time delay. The concept of time-to-pattern is one we formalise more rigorously within Section \ref{section:delaypatt}. Using full numerical solutions, we also confirm this linear relationship on a larger time scale. This linear relationship is not one that has been formalised in the current literature.

\section{Linear Analysis}
As defined in \eqref{fixed}, the equations we study for the LI model are
\begin{equation}\label{fixed2}
  \begin{split}
  \frac{\partial u}{\partial t}&=\frac{\epsilon^2}{L^2}\frac{\partial^2u}{\partial x^2}+a-u-2u^2v+3\hat{u}^2\hat{v},\\
  \frac{\partial v}{\partial t}&=\frac{1}{L^2}\frac{\partial^2v}{\partial x^2}+b-u^2v,
\end{split}
\end{equation}
with no flux boundary conditions, and where $u=u(x,t)$, $v=v(x,t)$ and $\hat{u}$, $\hat{v}$ are evaluated at some delay $\tau>0$, so that $\hat{u}=u(x,t-\tau)$ and $\hat{v}=v(x,t-\tau)$. Following the methodology in \cite{yigaffneyli}, we take a small perturbation about the steady state $u(x,t)=u_\star+\delta\xi(x,t)$ and $v(x,t)=v_\star+\delta\eta(x,t)$, where $|\delta|\ll 1$. Taylor expanding up to $O(\delta)$ about the steady state, the linearised dynamics of \eqref{fixed2} are then given by
\begin{equation}\label{linfixed}
  \begin{split}
\frac{\partial\xi}{\partial t}&=\frac{\epsilon^2}{L^2}\frac{\partial^2\xi}{\partial x^2}-\xi-4u_\star v_\star\xi+6u_\star v_\star\hat{\xi}-2u_\star^2\eta+3u_\star^2\hat{\eta},\\
\frac{\partial\eta}{\partial t}&=\frac{1}{L^2}\frac{\partial^2\eta}{\partial x^2}-2u_\star v_\star\xi-u_\star^2\eta,
\end{split}
\end{equation}
with $\hat{\xi}=\xi(x,t-\tau)$ and $\hat{\eta}=\eta(x,t-\tau)$. Substituting into \eqref{linfixed} an ansatz of the form $\begin{pmatrix}\xi\\\eta\end{pmatrix}=\begin{pmatrix}\xi_0e^{\lambda_k t}\cos(k\pi x)\\ \eta_0e^{\lambda_k t}\cos(k\pi x)\end{pmatrix}$, we obtain the characteristic equation, $\mathcal{D}_k=0$, given by
\begin{equation}\label{characfix}
\mathcal{D}_k=\lambda_k^2+\alpha_k\lambda_k+\beta_k+(\gamma_k\lambda_k+\delta_k)e^{-\lambda_k\tau}=0,
\end{equation}
where the coefficients are given as,\footnote{We note the coefficient of $\beta_k$ differs from that of \cite{yigaffneyli} due to a typographical error in the cited paper.}
\begin{equation}\label{fixcoeffs}
    \begin{split}
\alpha_k&=\left(\frac{\epsilon^2}{L^2}+\frac{1}{L^2}\right)k^2\pi^2+u_\star^2+4u_\star v_\star+1,\\
\beta_k&=\left(\frac{1}{L^2}\pi^2k^2+u_\star^2\right)\left(\frac{\epsilon^2}{L^2}\pi^2k^2+4u_\star v_\star+1\right)-4u_\star^3v_\star,\\
\gamma_k&=-6u_\star v_\star,\\
\delta_k&=-\frac{6}{L^2}u_\star v_\star k^2\pi^2.
\end{split}
\end{equation}
This characteristic equation can be used to determine the parameter sets $(a,b,\epsilon^2,L,\tau)$ in which a Turing instability occurs, and hence where we expect pattern formation. From the linear theory, as indicated in \eqref{perturbgrow}, the perturbation is expected to grow like $e^{\lambda_k t}\cos(k\pi x)$, and so if there exists a $k\neq0$ for a given $(a,b,\epsilon^2,L,\tau)$ such that $\max_k(\Re(\lambda_k))>0$, we expect pattern formation, where the final spatial pattern will look like $\cos(k\pi x)$ for the dominating mode $k$. Figure \ref{fig:dispfixed} shows $\max_k(\Re(\lambda_k))$ plotted against $\tau$, for multiple given $(a,b)$ parameter sets.   complex roots for $\lambda_k$ of the characteristic equation were found using the \emph{roots} command of the MATLAB package Chebfun \cite{chebfun}. These plots were produced by varying $k\in\mathbb{Z}$ over $[0,50]$ for a given $\tau$, and for each $k$, the roots of \eqref{characfix} were computed. The maximum over the $k$ of the $\Re(\lambda_k)$ was then taken. This was repeated for time delay varied over $\tau\in[0,1]$ at regular intervals of $0.1$. We do not consider a $k$ larger than $50$ as full numerical solutions for the parameter values used tended towards patterns with four `spikes', so we do not expect large wavenumbers to be excited.

\begin{figure}[H]
    \centering
    \begin{subfigure}[t]{0.45\textwidth}
        \centering
        \includegraphics[width=7cm,height = 5.5cm]{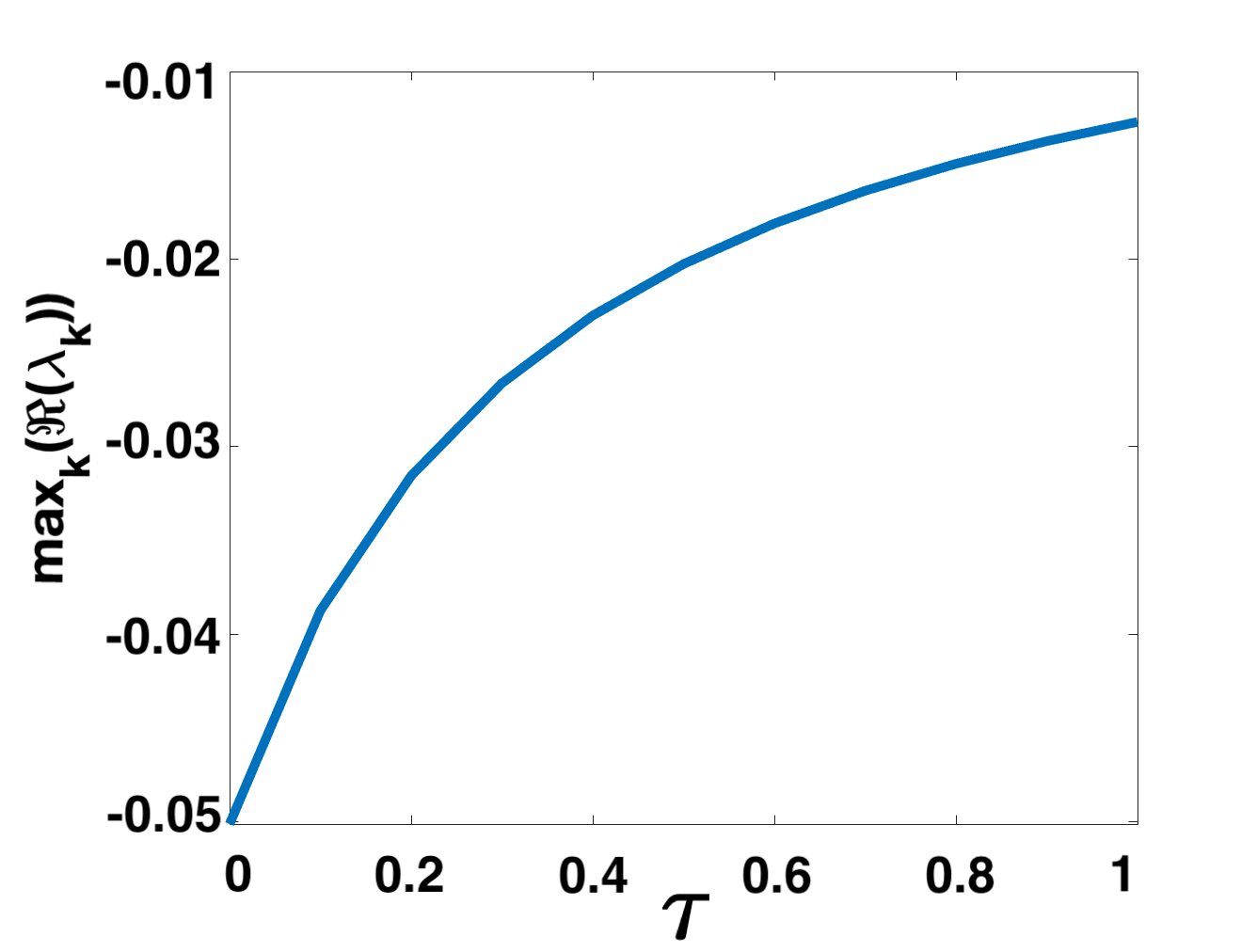}
        \caption{$(a,b)=(0.4,0.4)$. $\max_k(\Re(\lambda_k))<0 \quad \text{for all }\tau\in[0,1]$. Linear theory predicts no pattern formation for all $\tau\in[0,1]$. }
        \label{}
    \end{subfigure}
    \hfill
    \begin{subfigure}[t]{0.45\textwidth}
        \centering
        \includegraphics[width=7cm,height = 5.5cm]{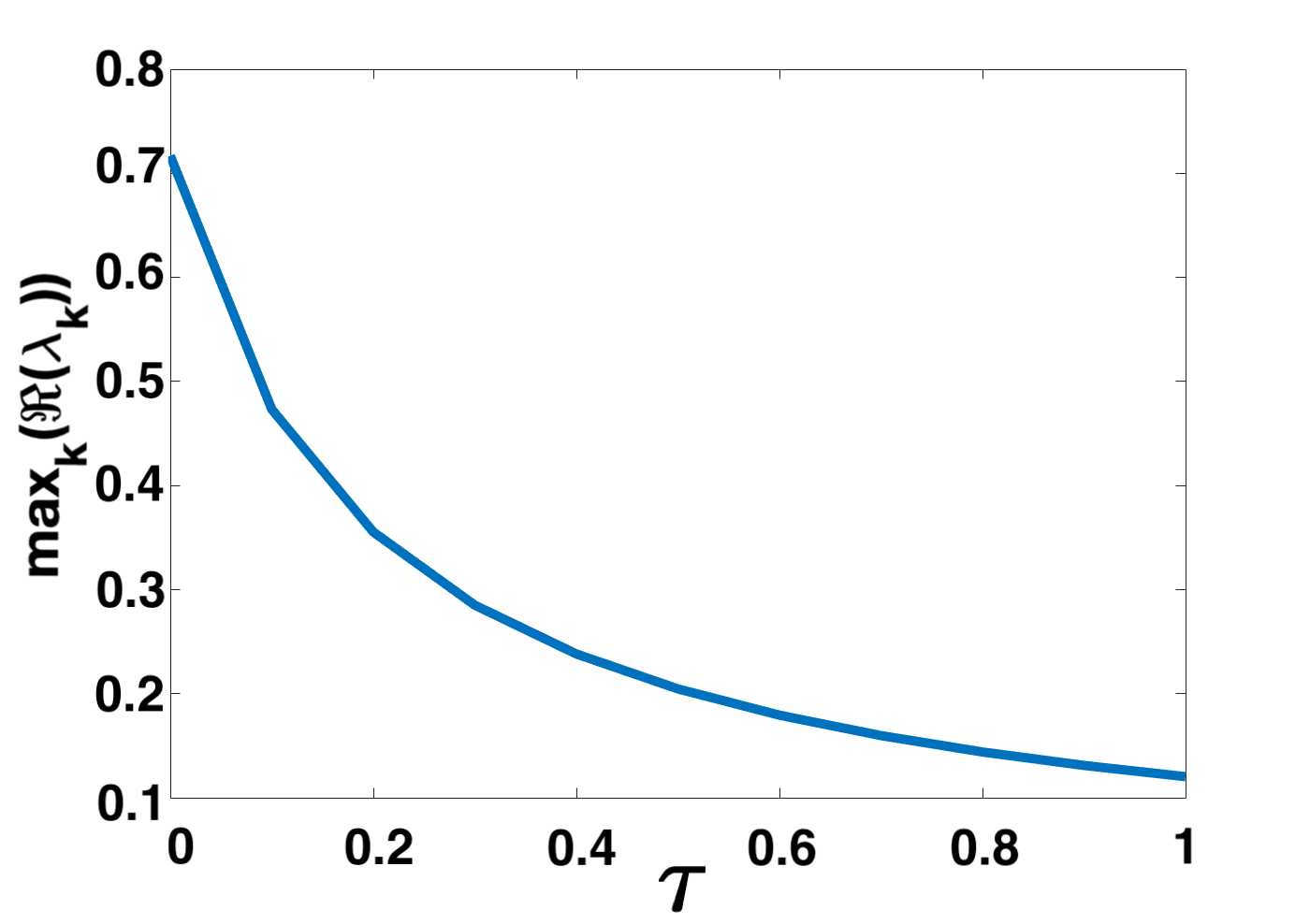}
        \caption{$(a,b)=(0.1,0.9)$. $\max_k(\Re(\lambda_k))>0 \quad \text{for all }\tau\in[0,1]$. Linear theory predicts pattern formation for all $\tau\in[0, 1]$.}
        \label{}
    \end{subfigure}
    \caption{Characteristic equation \eqref{characfix} solved and $\max_k(\Re(\lambda_k))$ plotted against $\tau\in[0,1]$ for two different parameter sets. $\epsilon^2=0.001$ and $L^2=9/2$.}
    \label{fig:dispfixed}
\end{figure}
\begin{figure}[H]
    \centering
    \begin{subfigure}[t]{0.45\textwidth}
        \centering
        \includegraphics[width=6cm,height = 5cm]{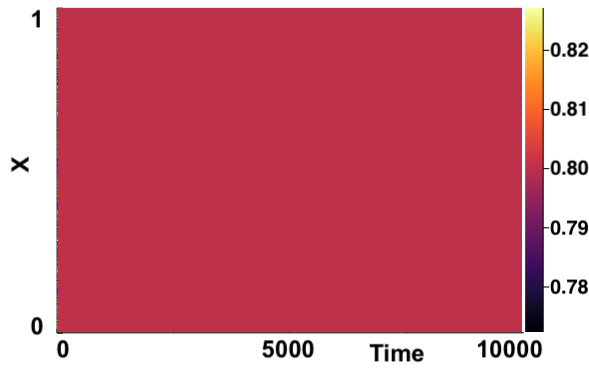}
        \caption{$\tau=0$. No pattern formation after $t=10^4$. }
        \label{}
    \end{subfigure}
    \hfill
    \begin{subfigure}[t]{0.45\textwidth}
        \centering
        \includegraphics[width=6cm,height = 5cm]{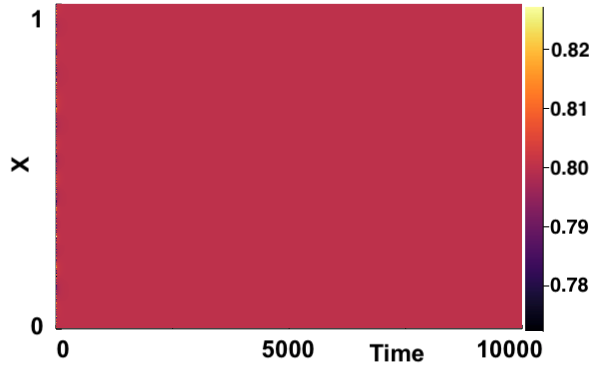}
        \caption{$\tau=1.$ No pattern formation after $t=10^4$.}
        \label{}
    \end{subfigure}
    \caption{Numerical simulations of \eqref{fixed2} showing no pattern formation with $(a,b)=(0.4,0.4)$, $\epsilon^2=0.001$ and $L^2=9/2$. Boundary conditions given by \eqref{neumannbc} and initial conditions by \eqref{firstic}.}
    \label{fig:fixedsim1}
\end{figure}
\begin{figure}[H]
    \centering
    \begin{subfigure}[t]{0.45\textwidth}
        \centering
        \includegraphics[width=6cm,height = 5cm]{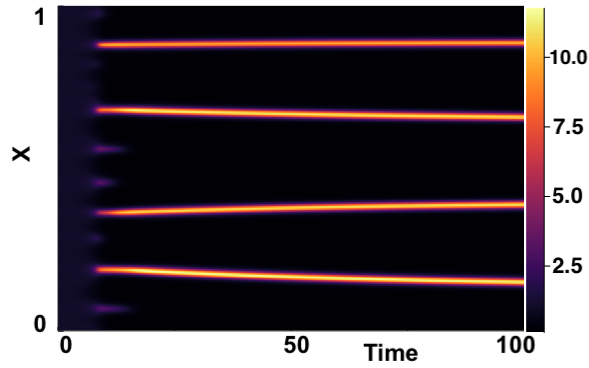}
        \caption{$\tau=0$. Distinct spikes formed at $t\approx7$ }
        \label{}
    \end{subfigure}
    \hfill
    \begin{subfigure}[t]{0.45\textwidth}
        \centering
        \includegraphics[width=6cm,height = 5cm]{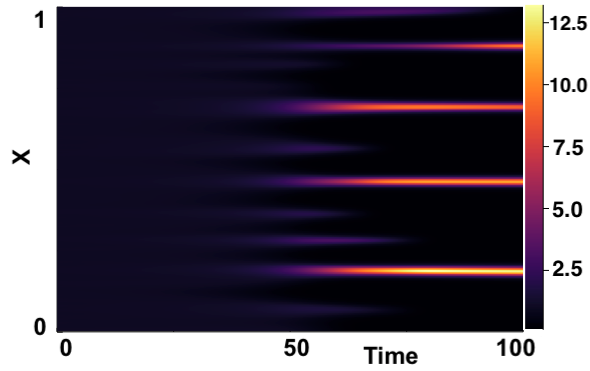}
        \caption{$\tau=1$. Distinct spikes formed at $t\approx50$.}
        \label{}
    \end{subfigure}
    \caption{Numerical simulations of \eqref{fixed2} showing pattern formation for $(a,b)=(0.1,0.9)$, $\epsilon^2=0.001$ and $L^2=9/2$. Boundary conditions given by \eqref{neumannbc} and initial conditions by \eqref{firstic}.}
    \label{fig:fixedsim2}
\end{figure}

Figure \ref{fig:dispfixed} suggests that for all $\tau\in[0,1]$, pattern formation will not occur for $(a,b)=(0.4,0.4)$, but will occur for $(a,b)=(0.1,0.9)$. We also hypothesise that since $\max_k(\Re(\lambda_k))$ at $\tau=0$ is greater than at $\tau=1$, the time taken to pattern formation will be longer at $\tau=1$. This relationship between time-to-pattern and time delay is explored in more detail in section \ref{section:delaypatt}. Numerical results in Figures \ref{fig:fixedsim1} and \ref{fig:fixedsim2} verify the findings fromFigure \ref{fig:dispfixed}, namely that pattern formation does not occur for $(a,b)=(0.4,0.4)$, but does for $(a,b)=(0.1,0.9)$, hence confirming predictions from the linear theory.

We note that by convention, as explained in section \ref{section:numimp}, the numerical solution of only the activator $u$ is plotted.

\subsection{Bifurcation Analysis}\label{section:fixedbif}
\begin{figure}[H]
    \centering
    \begin{subfigure}[t]{0.47\textwidth}
        \centering
        \includegraphics[width=7cm,height = 6cm]{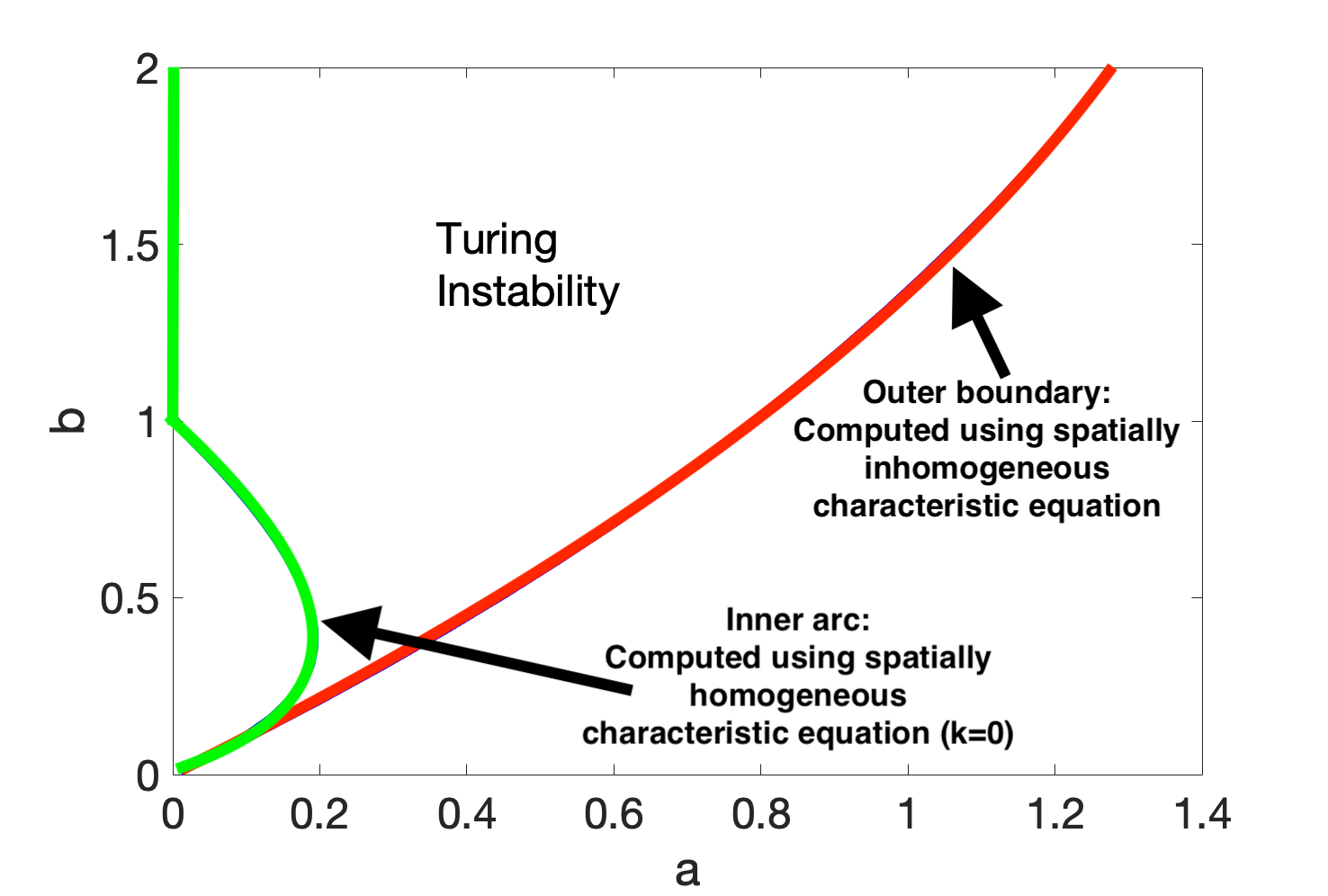}
        \caption{Stability lines for $\tau=0$ computed by solving characteristic equation with $\epsilon^2=0.001$, $L^2=9/2$.}
        \label{fig:bif0}
    \end{subfigure}
    \hfill
    \begin{subfigure}[t]{0.47\textwidth}
        \centering
        \includegraphics[width=7cm,height = 6cm]{turingspace.png}
        \caption{Turing space as in Figure \ref{fig:turingspace} plotted from parameters $(a,b)$ satisfying conditions Turing conditions.}
        \label{}
    \end{subfigure}
    \caption{Comparison of Turing instability region for $\tau=0$ computed via equations (\eqref{realfixbif} and \eqref{complexfixbif}) against Turing instability region computed via Turing conditions (\eqref{cond1} and \eqref{cond2}). Parameter space chosen as $(a,b)\in[0,1.4]\times[0,2]$.}
    \label{fig:tspace1}
\end{figure}

The Turing plot produced in Figure \ref{fig:turingspace}, computed using the conditions in \eqref{cond1} and \eqref{cond2}, is a bifurcation diagram indicating regions of Turing instability. We note two separate curves which separate the parameter space into its distinct regions. These will be referred to as the \textit{stability lines}.
These two curves are indicated in Figure \ref{fig:bif0}. The inner arc corresponds to the $(a,b)$ such that $\Re(\lambda_k)=0$ for the
spatially homogeneous characteristic equation, $\mathcal{D}_k=0$ when $k=0$. For $\tau=0$, this corresponds exactly to equating conditions $\eqref{cond1}$ to 0. The outer boundary is comprised of the points $(a,b)$ such that $\max_k(\Re(\lambda_k))=0$ for the spatially inhomogeneous characteristic equation, $\mathcal{D}_k$ when $k\neq0$. For $\tau=0$, this is identical to equating the conditions $\eqref{cond2}$ to 0. By letting $\lambda_k=x_k+iy_k$ for $x_k,y_k\in\mathbb{R}$, we split the characteristic equation $\mathcal{D}_k=0$ into its real and imaginary parts, $\mathcal{D}_k^{\Re}=0$ and $\mathcal{D}_k^{\Im}=0$, given by
\begin{align}\label{realfix}
\mathcal{D}_k^{\Re}&=x_k^2-y_k^2+\alpha_kx_k+\beta_k+e^{-x_k\tau}[\gamma_kx_k\cos(-y_k\tau)-\gamma_ky_k\sin(-y_k\tau)+\delta_k\cos(-y_k\tau)]=0,\\
\mathcal{D}_k^{\Im}&=2x_ky_k+\alpha_ky_k+e^{-x_k\tau}[\gamma_kx_k\sin(-y_k\tau)+\gamma_ky_k\cos(-y_k\tau)+\delta_k\sin(-y_k\tau)]=0.\label{complexfix}
\end{align}
By setting $\Re(\lambda_k)=x_k=0$ in equations \eqref{realfix} and \eqref{complexfix}, the real and imaginary parts of $\mathcal{D}_k$ can be simplified to
\begin{align}\label{realfixbif}
  \mathcal{D}_k^{\Re}&=-y_k^2\beta_k+[-\gamma_ky_k\sin(-y_k\tau)+\delta_k\cos(-y_k\tau)],\\
  \mathcal{D}_k^{\Im}&=\alpha_ky_k+[\gamma_ky_k\cos(-y_k\tau)+\delta_k\sin(-y_k\tau)].\label{complexfixbif}
\end{align}
For a fixed $\tau$ and $b$, the roots of \eqref{realfixbif} and \eqref{complexfixbif} (at $k=0$) can be found for $a$ and $\Im(\lambda_k)$.
Taking the $\max_k(a)$, a curve can be plotted in the $(a,b)$ parameter space for the outer boundary (and the inner arc) resulting in a bifurcation diagram of distinct regions where Turing instabilities can occur. We use a relatively large $L^2=9/2$, so the bifurcation diagram computed in this manner for $\tau=0$ should be a good approximation to the Turing space plot produced in Figure \ref{fig:turingspace}. Figure \ref{fig:bif0} shows the bifurcation plot produced in this manner for $\tau=0$ alongside the Turing space plot in Figure \ref{fig:turingspace} for comparison.

\begin{figure}[H]
        \centering
        \includegraphics[width=9cm,height = 6cm]{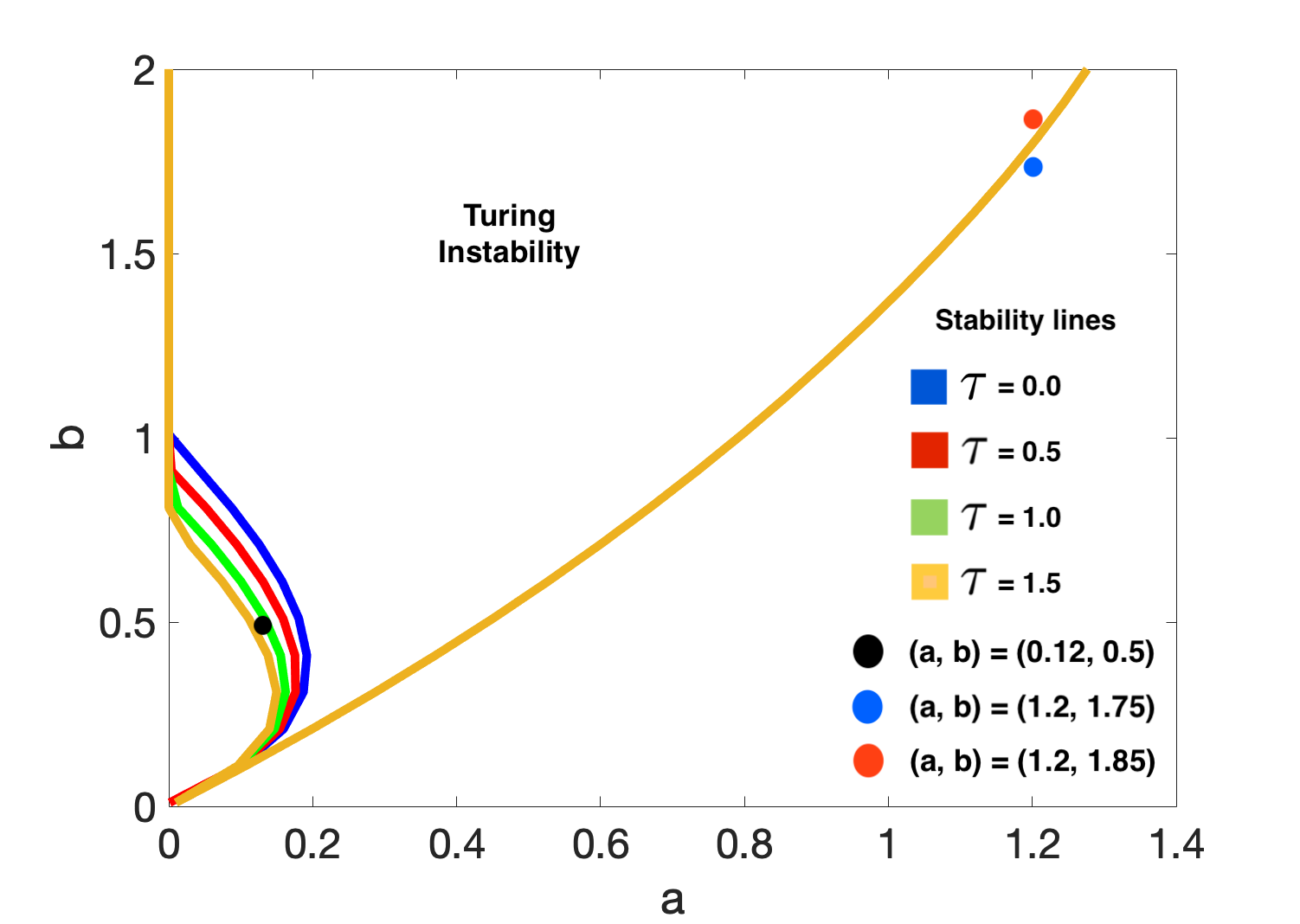}
        \caption{Stability lines for $\tau\in\{0,0.5,1,1.5\}$ computed by solving \eqref{realfixbif} and \eqref{complexfixbif}. $\epsilon^2=0.001$, $L^2=9/2$.}
        \label{fig:tspacetau}
\end{figure}

\begin{figure}[h]
    \centering
    \begin{subfigure}[t]{0.45\textwidth}
        \centering
        \includegraphics[width=7cm,height=4cm]{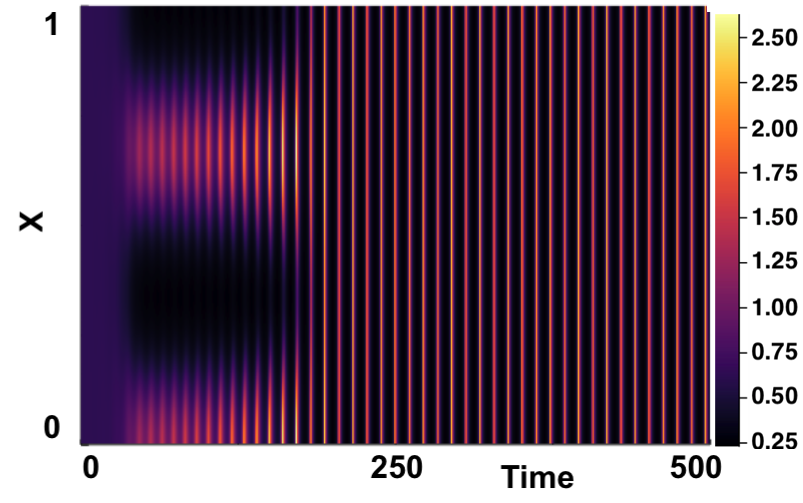}
        \caption{$\tau=0$. Oscillations seen.}
        \label{}
    \end{subfigure}
    \hfill
    \begin{subfigure}[t]{0.45\textwidth}
        \centering
        \includegraphics[width=7cm,height=4cm]{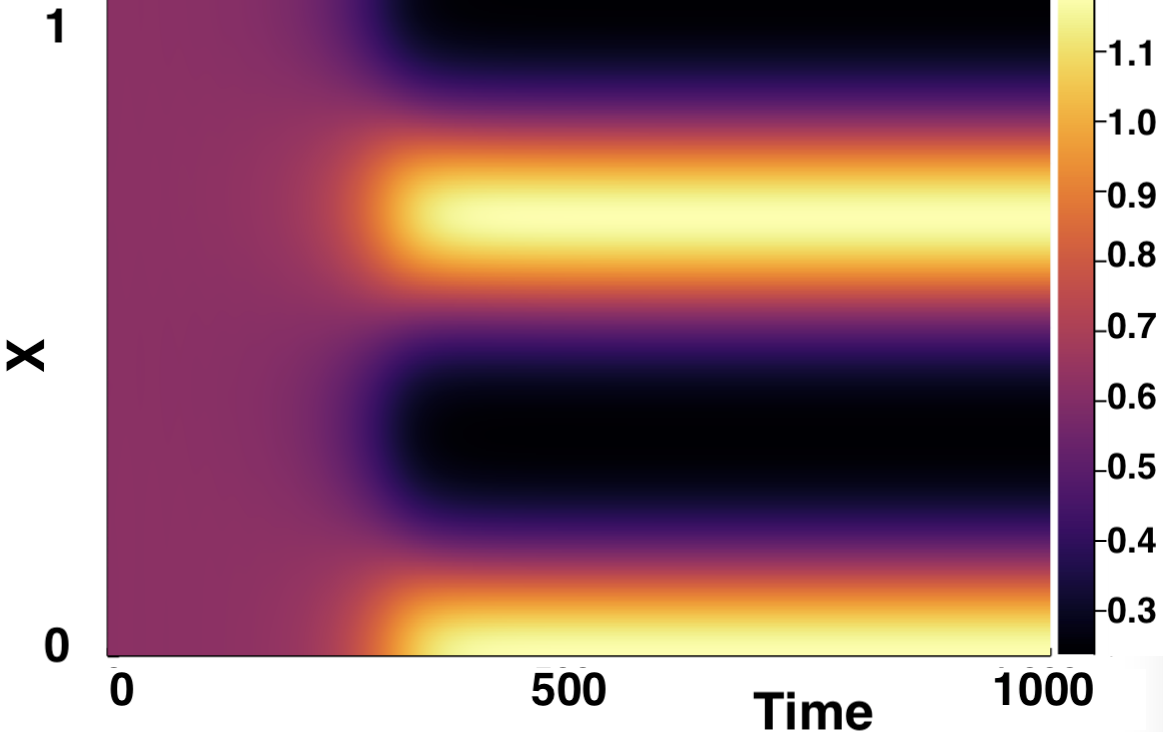}
        \caption{$\tau=1.5$. Pattern formation seen.}
        \label{}
    \end{subfigure}
    \caption{Numerical simulations of \eqref{fixed2} produced with parameters $(a,b)=(0.12,0.5)$, for $\tau=0,1.5$. $\epsilon^2=0.1$ and $L^2=9/2$. Boundary conditions given by \eqref{neumannbc} and initial conditions by \eqref{firstic}. Linear theory in Figure \ref{fig:tspacetau} suggests we see Turing pattern formation at $\tau=1.5$ but not at $\tau=0$.}
    \label{fig:testturing}
\end{figure}

\begin{figure}[H]
    \centering
    \begin{subfigure}[t]{0.45\textwidth}
        \centering
        \includegraphics[width=7cm,height=4cm]{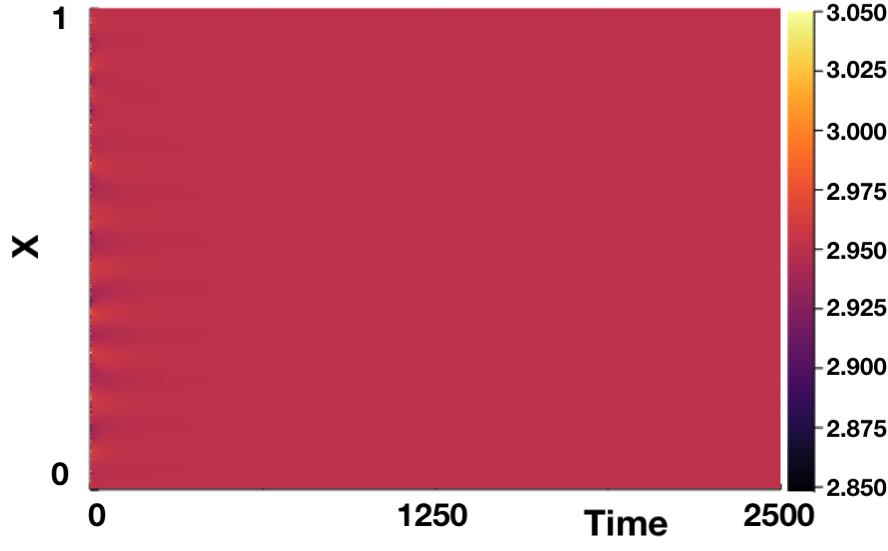}
        \caption{$\tau=0$.}
        \label{}
    \end{subfigure}
    \hfill
    %\begin{subfigure}[t]{0.45\textwidth}
    %    \centering
    %    \includegraphics[width=7cm,height=4cm]{p3t05.png}
    %    \caption{$\tau=0.5$}
    %    \label{}
    %\end{subfigure}
    %\hfill
    %\begin{subfigure}[t]{0.45\textwidth}
%        \centering
%        \includegraphics[width=7cm,height=4cm]{p3t1.png}
%        \caption{$\tau=1$}
%        \label{}
%    \end{subfigure}
%    \hfill
    \begin{subfigure}[t]{0.45\textwidth}
        \centering
        \includegraphics[width=7cm,height=4cm]{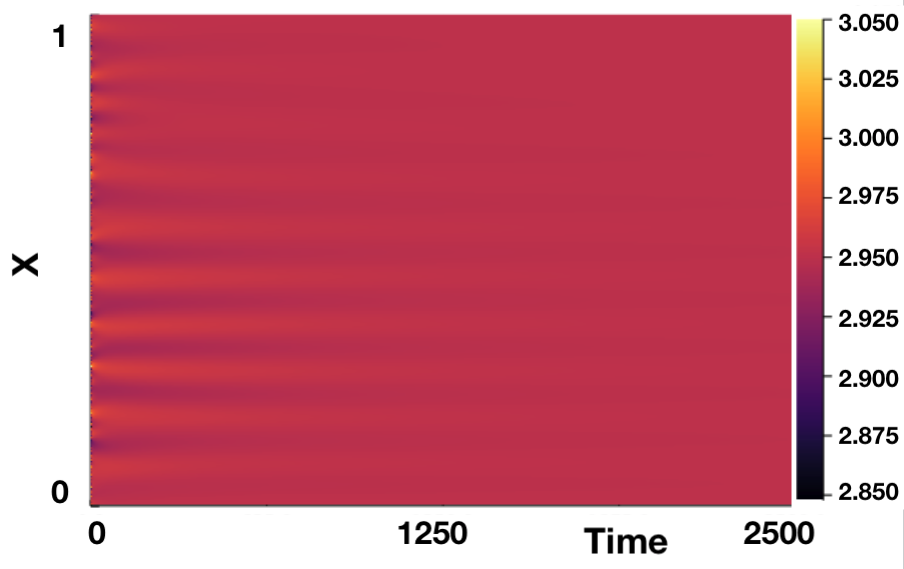}
        \caption{$\tau=1.5$.}
        \label{}
    \end{subfigure}
    \caption{Numerical simulations of \eqref{fixed2} for $(a,b)=(1.2,1.75)$. $\epsilon^2=0.001$ and $L^2=9/2$. Boundary conditions given by \eqref{neumannbc} and initial conditions by \eqref{firstic}. We see no Turing pattern formation for $\tau\in\{0,1.5\}$ as suggested by linear theory, seen in Figure \ref{fig:tspacetau}. Results for $\tau=0.5,1$ can be seen in Appendix \ref{section:Bfix}.}
    \label{fig:testturing2}
\end{figure}

\begin{figure}[H]
    \centering
    \begin{subfigure}[t]{0.45\textwidth}
        \centering
        \includegraphics[width=7cm,height=4cm]{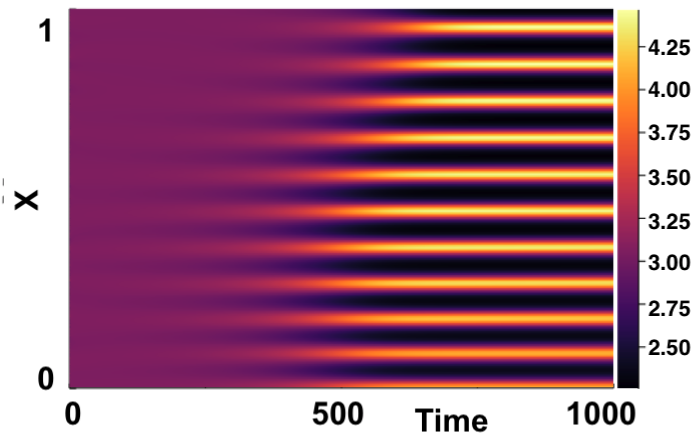}
        \caption{$\tau=0$.}
        \label{}
    \end{subfigure}
    \hfill
%    \begin{subfigure}[t]{0.45\textwidth}
%        \centering
%        \includegraphics[width=7cm,height=4cm]{p2t05.png}
%        \caption{$\tau=0.5$}
%        \label{}
%    \end{subfigure}
%    \hfill
%    \begin{subfigure}[t]{0.45\textwidth}
%        \centering
%        \includegraphics[width=7cm,height=4cm]{p2t1.png}
%        \caption{$\tau=1$}
%        \label{}
%    \end{subfigure}
%    \hfill
    \begin{subfigure}[t]{0.45\textwidth}
        \centering
        \includegraphics[width=7cm,height=4cm]{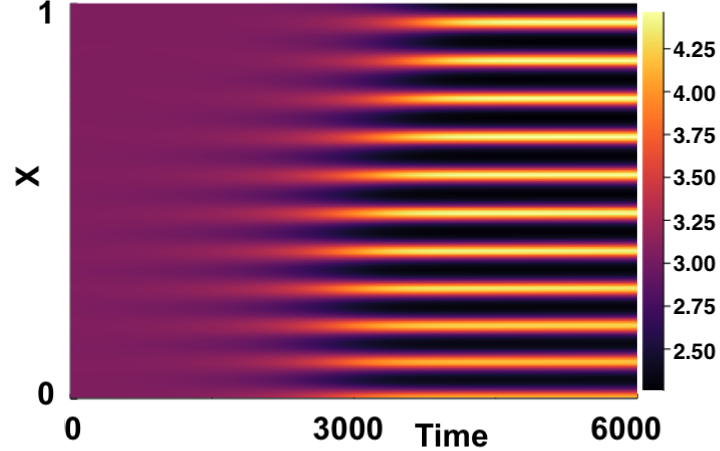}
        \caption{$\tau=1.5$.}
        \label{}
    \end{subfigure}
    \caption{Numerical simulations of \eqref{fixed2} for $(a,b)=(1.2,1.85)$. $\epsilon^2=0.001$ and $L^2=9/2$. Boundary conditions given by \eqref{neumannbc} and initial conditions by \eqref{firstic}. We see Turing pattern formation on an increasing timescale for $\tau\in\{0,1.5\}$ as suggested by linear theory, seen in Figure \ref{fig:tspacetau}. Results for $\tau=0.5,1$ can be seen in Appendix \ref{section:Bfix}.}
    \label{fig:testturing3}
\end{figure}

Figure \ref{fig:tspacetau} shows the stability lines computed for a varying $\tau\in\{0,0.5,1,1.5\}$. It can be seen that the outer boundary computed from the characteristic equation for the spatially inhomogeneous model stays the same, at least at the resolution of plotting. The inner arc computed from using the characteristic equation from the spatially homogeneous model shifts to the left, increasing the region of parameter space for which Turing instabilities can occur. This observation supports the findings in \cite{william}, where bifurcation analysis for the spatially homogeneous model showed an increase in size of the stable parameter region (corresponding to a shifting of the inner arc). We see that for the LI model, where delay-terms are placed solely in the activator dynamics, time delay acts as a promoting agent for pattern formation, expanding the Turing space, and thus increasing the parameter space where Turing instabilities can occur.
We verify the results of Figure \ref{fig:tspacetau} through numerical simulations. Three parameter points, $(a,b)=\{(0.12,0.5),(1.2,1.75),(1.2,1.85)\}$ are indicated in Figure \ref{fig:tspacetau}. At $(a,b,\tau)=(0.12,0.5,0)$, linear theory suggests that there will be no pattern formation, but at $(a,b,\tau)=(0.12,0.5,1.5)$ there will be a Turing instability and thus patterns will form.
The parameter region in the bottom left of the parameter space is a delicate region that can exhibit both Turing and Hopf bifurcations, leading to complex spatio-temporal behaviours. This type of dynamics in reaction-diffusion systems has been studied more extensively in \cite{krausefixed,jiang}. Although the linear theory is unable to provide information about the more intricate nonlinear dynamics, it can predict the expected type of behaviour for certain parameter values. To show a change in behaviour as $\tau$ changes from $0$ to $1.5$, from a temporally oscillating solution, to one exhibiting a Turing pattern, we increase the diffusive ratio to $\epsilon^2=0.1$. This result can be seen in Figure \ref{fig:testturing}. The linear theory also suggests that for all $\tau\in\{0,0.5,1,1.5\}$, pattern formation will occur for $(a,b)=(1.2,1.85)$, but not for $(a,b)=(1.2,1.75)$. Figures \ref{fig:testturing2} and \ref{fig:testturing3} show the results for numerical simulations at $(a,b)=\{(1.2,1.75),(1.2,1.85)\}$ for $\tau=0,1.5$. Results for $\tau=0.5,1$ can be seen in Appendix \ref{section:Bfix}.

\begin{figure}[H]
    \centering
    \begin{subfigure}[t]{0.45\textwidth}
        \centering
        \includegraphics[width=7cm,height=5cm]{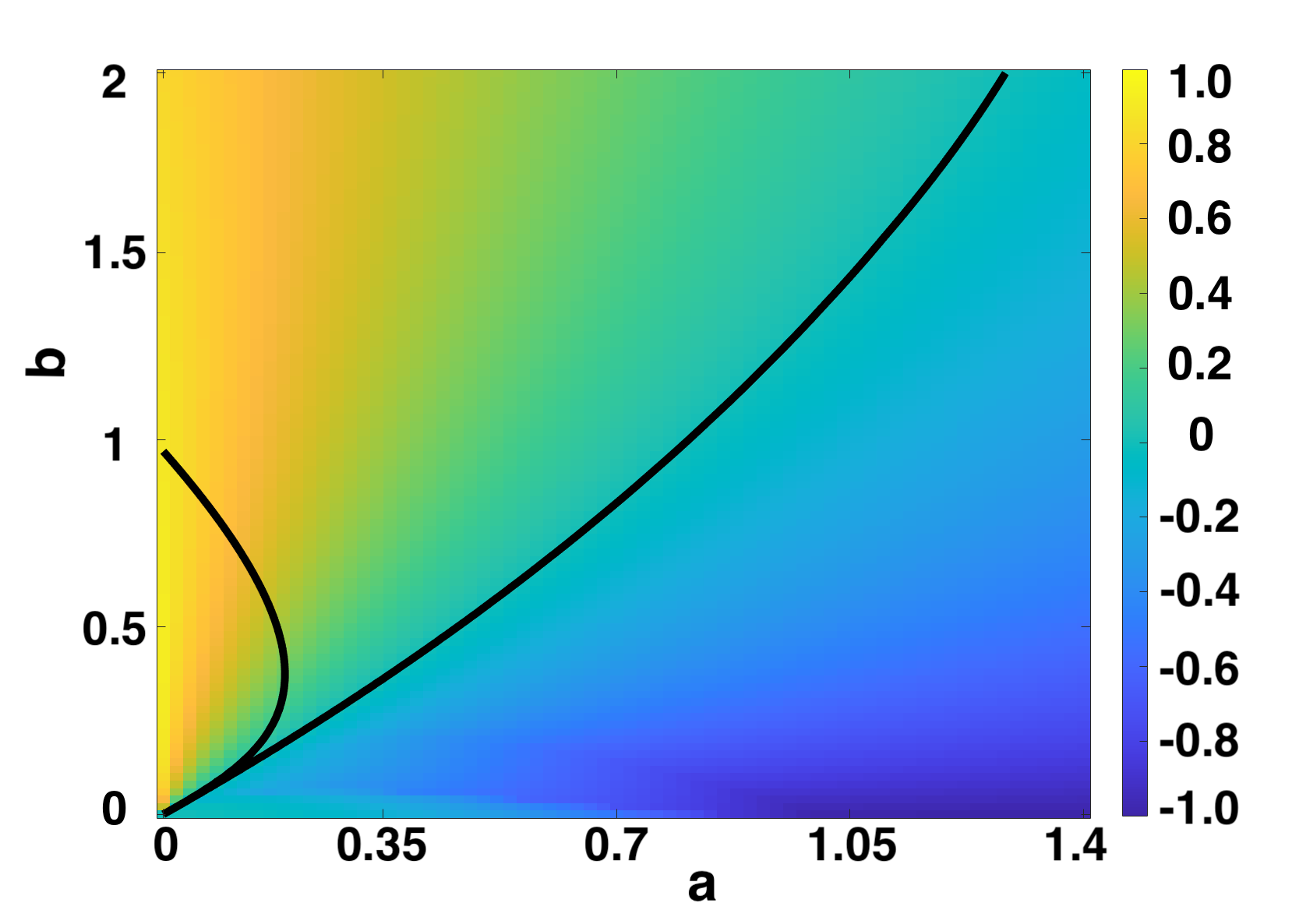}
        \caption{$\tau=0$.}
        \label{}
    \end{subfigure}
    \hfill
%    \begin{subfigure}[t]{0.45\textwidth}
%        \centering
%        \includegraphics[width=7cm,height=5cm]{tau05bif.png}
%        \caption{$\tau=0.5$}
%        \label{}
%    \end{subfigure}
%    \hfill
%    \begin{subfigure}[t]{0.45\textwidth}
%        \centering
%        \includegraphics[width=7cm,height=5cm]{tau1bif.png}
%        \caption{$\tau=1$}
%        \label{}
%    \end{subfigure}
%    \hfill
    \begin{subfigure}[t]{0.45\textwidth}
        \centering
        \includegraphics[width=7cm,height=5cm]{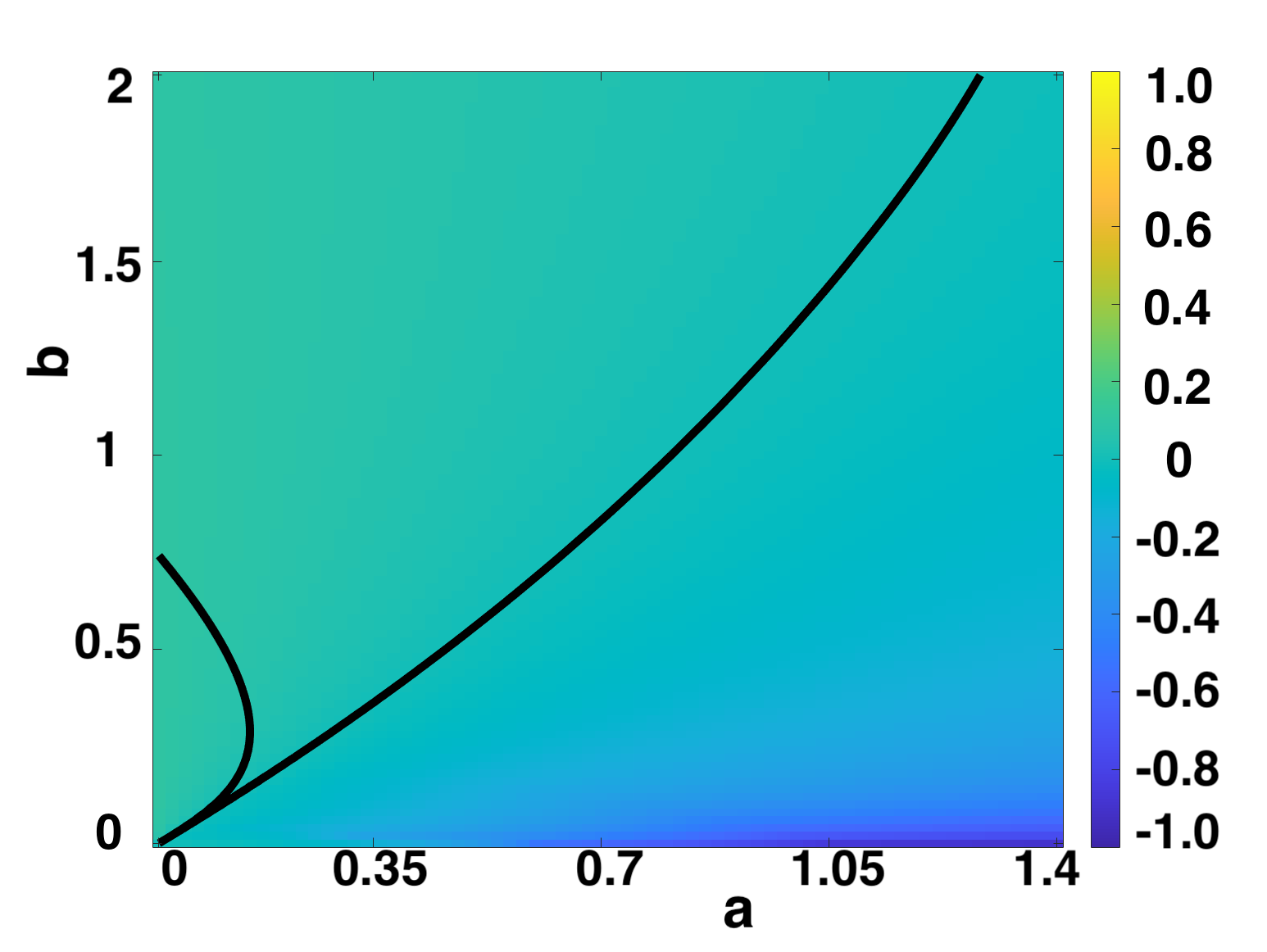}
        \caption{$\tau=1.5$.}
        \label{}
    \end{subfigure}
    \caption{$\max_k(\Re(\lambda_k))$ computed over $(a,b)$ parameter space by solving \eqref{realfixbif} and \eqref{complexfixbif}, with $\epsilon^2=0.001$, $L^2=9/2$. As $\tau$ increases, $|\max_k(\Re(\lambda_k))|$ decreases. Contour lines for $\Re(\lambda_0)=0$ and $\max_k(\Re(\lambda_k))=0$ overlayed, indicated Turing instability region. Results for $\tau=0.5,1$ can be seen in Appendix \ref{section:Bfix}.}
    \label{fig:lambdavary}
\end{figure}

\begin{figure}[H]
    \centering
    \begin{subfigure}[t]{0.45\textwidth}
        \centering
        \includegraphics[width=7cm,height=5cm]{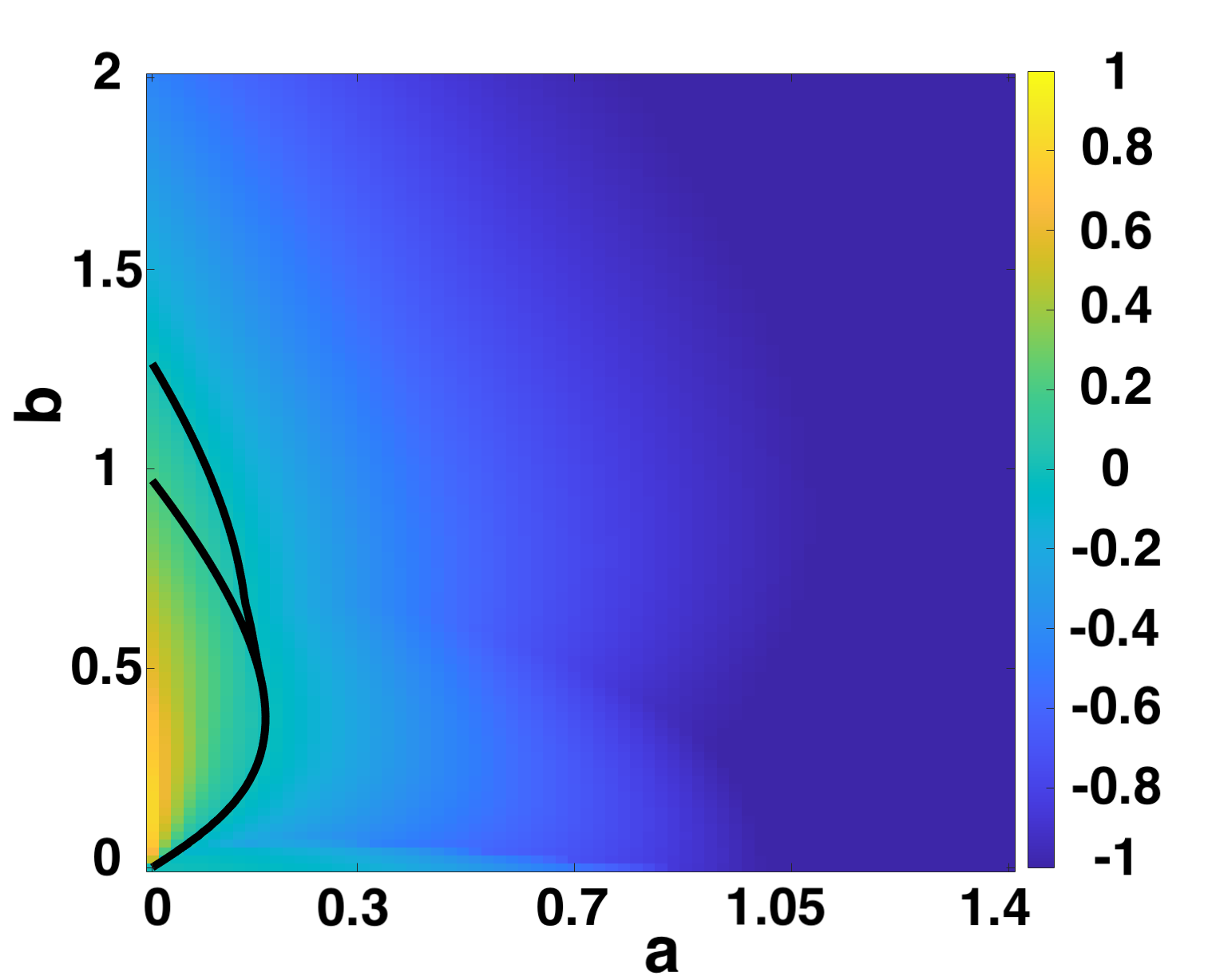}
        \caption{$\tau=0$.}
        \label{}
    \end{subfigure}
    \hfill
%    \begin{subfigure}[t]{0.45\textwidth}
%        \centering
%        \includegraphics[width=7cm,height=5cm]{fixbif22.png}
%        \caption{$\tau=0.5$}
%        \label{}
%    \end{subfigure}
%    \hfill
%    \begin{subfigure}[t]{0.45\textwidth}
%        \centering
%        \includegraphics[width=7cm,height=5cm]{fixbif23.png}
%        \caption{$\tau=1$}
%        \label{}
%    \end{subfigure}
%    \hfill
    \begin{subfigure}[t]{0.45\textwidth}
        \centering
        \includegraphics[width=7cm,height=5cm]{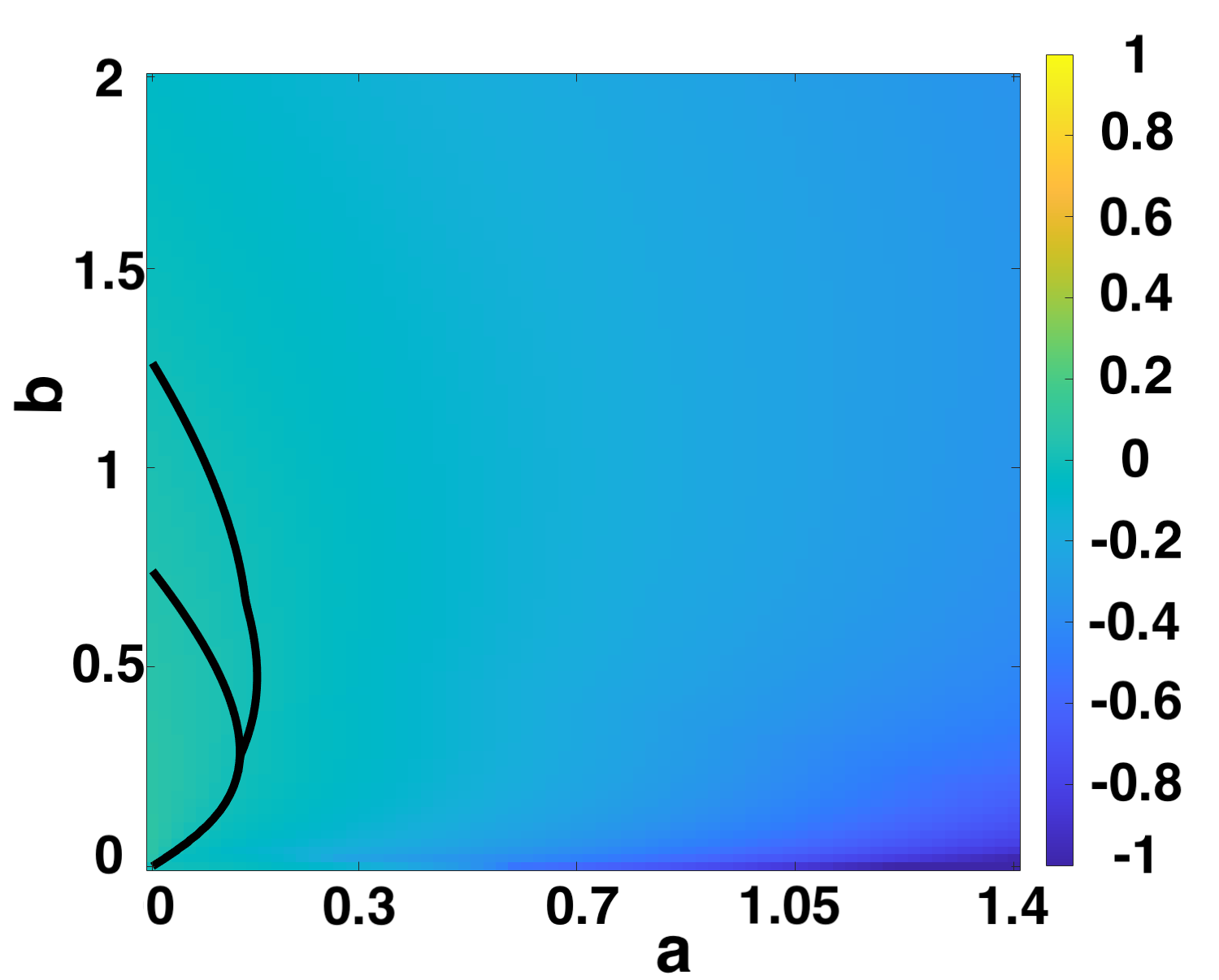}
        \caption{$\tau=1.5$.}
        \label{}
    \end{subfigure}
    \caption{$\max_k(\Re(\lambda_k))$ computed over $(a,b)$ parameter space by solving \eqref{realfixbif} and \eqref{complexfixbif}, with $\epsilon^2=0.1$, $L^2=9/2$. As $\tau$ increases, $|\max_k(\Re(\lambda_k))|$ decreases. Contour lines for $\Re(\lambda_0)=0$ and $\max_k(\Re(\lambda_k))=0$ overlayed, indicated Turing instability region. Results for $\tau=0.5,1$ can be seen in Appendix \ref{section:Bfix}.}
    \label{fig:fixbif2}
\end{figure}

Figure \ref{fig:tspacetau} shows how the time delay affects the region of Turing instability, but it provides no information as to how $\max_k(\Re(\lambda_k))$ varies as $\tau$ increases over the $(a,b)$ parameter space. In Figure \ref{fig:lambdavary} we plot a heatmap of $\max_k(\Re(\lambda_k))$ over the $(a,b)$ parameter space for varying $\tau\in\{0,1.5\}$. Overlayed onto these plots are contour lines corresponding to where $\Re(\lambda_0)=0$ and $\max_k(\Re(\lambda_k))=0$,
highlighting the Turing instability region.
As $\tau$ increases, it can be seen that the absolute value $|\max_k(\Re((\lambda_k)))|$ also decreases. This suggests that for $(a,b)$ values within the Turing instability region, pattern formation will take longer to occur. It also suggests however that for $(a,b)$ such that $\max_k(\Re(\lambda_k))<0$, it will take a longer time for the eigenfunctions with modes $k\neq0$ to decay to a spatially homogeneous steady state. We note this behaviour in Figure \ref{fig:testturing2}, where it can be seen, by carefully considering the timescales, that the time taken for the initial perturbation to fully decay back to a spatially homogeneous steady state increases as $\tau$ increases. Figure \ref{fig:fixbif2} shows analogous bifurcation diagrams as in Figure \ref{fig:lambdavary}, but with $\epsilon^2=0.1$. We note that as the ratio of diffusion constants in the reaction-diffusion system, $\epsilon^2$, moves closer to $1$, the region of parameter space exhibiting Turing instability decreases. It can be observed however, that altering $\epsilon^2$ does not change the effect that an increasing $\tau$ has on $\max_k(\Re(\lambda_k))$, and that increasing the delay $\tau$ continues to act to promote Turing instabilities, with a shifting of the spatially homogeneous inner arc.

\section{Investigation of Variation in Initial and Boundary Conditions}
\begin{figure}[H]
    \centering
    \begin{subfigure}[t]{0.32\textwidth}
        \centering
        \includegraphics[width=5cm,height=4.5cm]{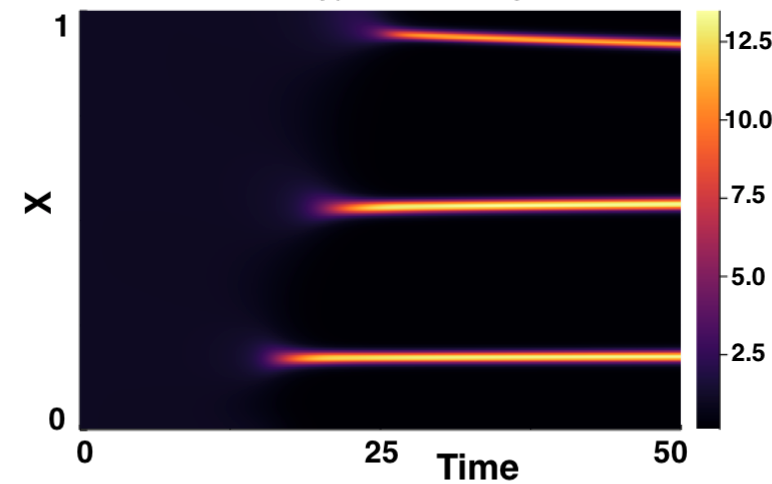}
        \caption{$\text{IC}_1$ given in reference \cite{gaffmonk}.}
        \label{}
    \end{subfigure}
    \hfill
    \begin{subfigure}[t]{0.32\textwidth}
        \centering
        \includegraphics[width=5cm,height=4.5cm]{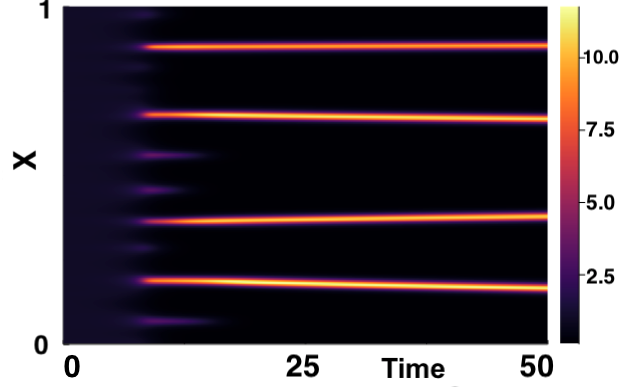}
        \caption{$\text{IC}_2$ given by equation \eqref{firstic}.}
        \label{}
    \end{subfigure}
    \hfill
    \begin{subfigure}[t]{0.32\textwidth}
        \centering
        \includegraphics[width=5cm,height=4.5cm]{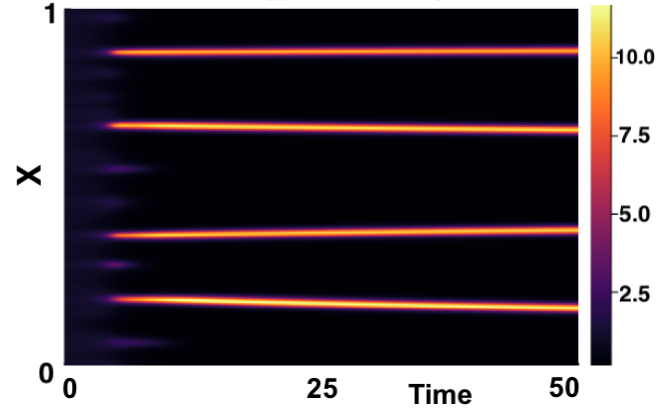}
        \caption{$\text{IC}_3$ given by equation \eqref{ic3}.}
        \label{}
    \end{subfigure}
    \caption{Numerical simulations of \eqref{fixed2} showing comparison of varying ICs for $\tau=0$. Boundary conditions given by \eqref{neumannbc}. $(a,b)=(0.1,0.9)$, $\epsilon^2=0.001$, $L^2=9/2$. }
    \label{fig:figtau0}
\end{figure}

\begin{figure}[H]
    \centering
    \begin{subfigure}[t]{0.32\textwidth}
        \centering
        \includegraphics[width=5cm,height=4.5cm]{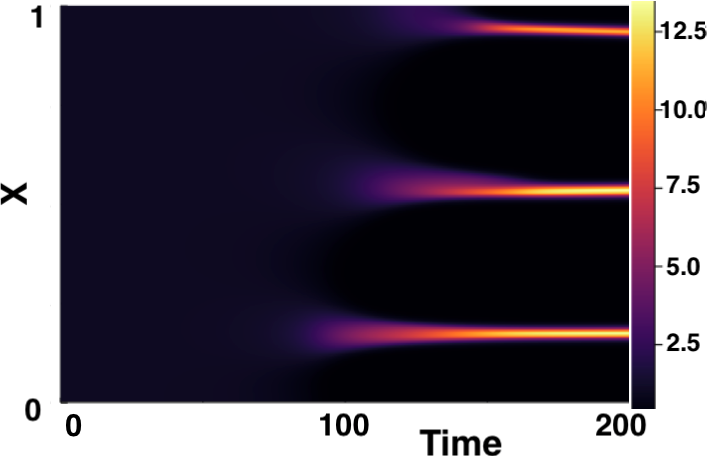}
        \caption{$\text{IC}_1$ given in reference \cite{gaffmonk}.}
        \label{}
    \end{subfigure}
    \hfill
    \begin{subfigure}[t]{0.32\textwidth}
        \centering
        \includegraphics[width=5cm,height=4.5cm]{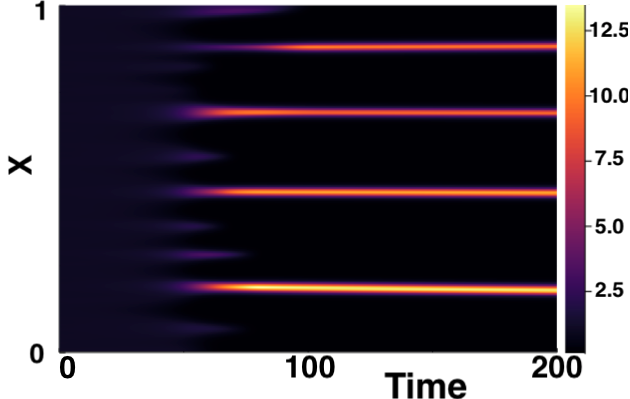}
        \caption{$\text{IC}_2$ given by equation \eqref{firstic}.}
        \label{}
    \end{subfigure}
    \hfill
    \begin{subfigure}[t]{0.32\textwidth}
        \centering
        \includegraphics[width=5cm,height=4.5cm]{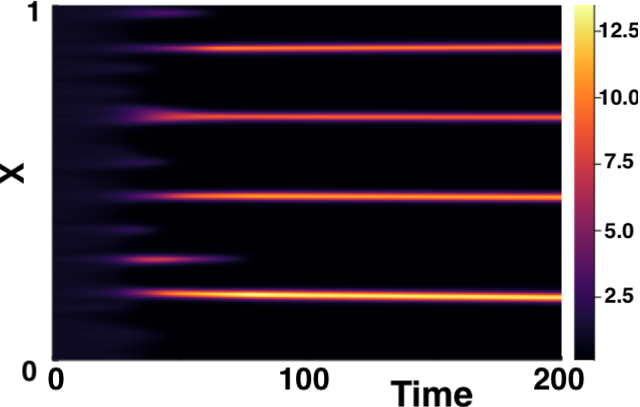}
        \caption{$\text{IC}_3$ given by equation \eqref{ic3}.}
        \label{}
    \end{subfigure}
    \caption{Numerical simulations of \eqref{fixed2} showing comparison of varying ICs for $\tau=1$. Boundary conditions given by \eqref{neumannbc}. $(a,b)=(0.1,0.9)$, $\epsilon^2=0.001$, $L^2=9/2$.}
    \label{fig:figtau1}
\end{figure}

\begin{figure}[H]
    \centering
    \begin{subfigure}[t]{0.32\textwidth}
        \centering
        \includegraphics[width=5cm,height=4.5cm]{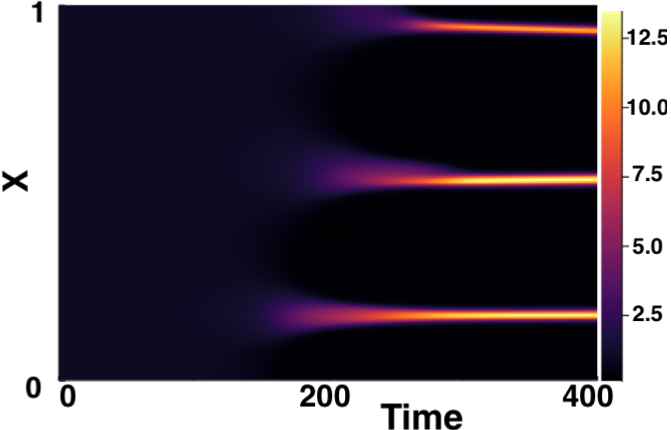}
        \caption{$\text{IC}_1$ given in reference \cite{gaffmonk}.}
        \label{}
    \end{subfigure}
    \hfill
    \begin{subfigure}[t]{0.32\textwidth}
        \centering
        \includegraphics[width=5cm,height=4.5cm]{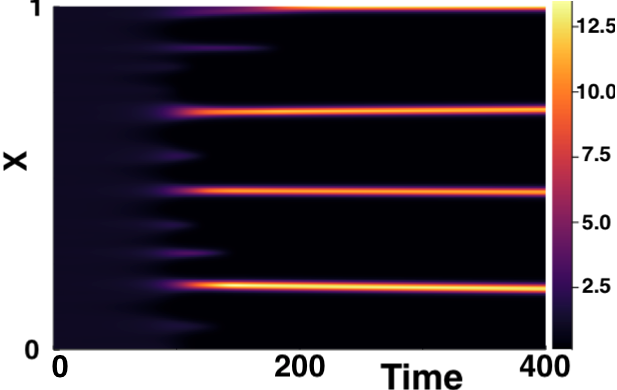}
        \caption{$\text{IC}_2$ given by equation \eqref{firstic}.}
        \label{}
    \end{subfigure}
    \hfill
    \begin{subfigure}[t]{0.32\textwidth}
        \centering
        \includegraphics[width=5cm,height=4.5cm]{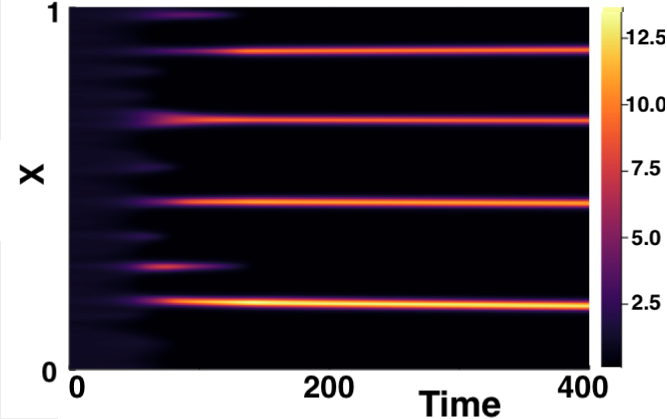}
        \caption{$\text{IC}_3$ given by equation \eqref{ic3}.}
        \label{}
    \end{subfigure}
    \caption{Numerical simulations of \eqref{fixed2} showing comparison of varying ICs for $\tau=2$. Boundary conditions given by \eqref{neumannbc}. $(a,b)=(0.1,0.9)$, $\epsilon^2=0.001$, $L^2=9/2$.}
    \label{fig:figtau2}
\end{figure}

\begin{figure}[H]
    \centering
    \begin{subfigure}[t]{0.32\textwidth}
        \centering
        \includegraphics[width=5cm,height=4.5cm]{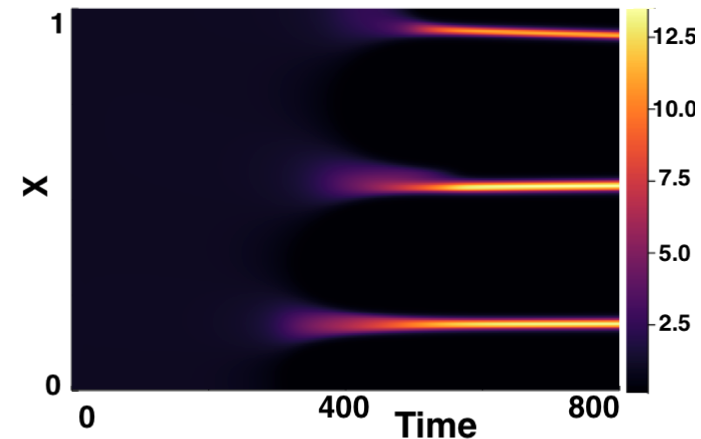}
        \caption{$\text{IC}_1$ given in reference \cite{gaffmonk}.}
        \label{}
    \end{subfigure}
    \hfill
    \begin{subfigure}[t]{0.32\textwidth}
        \centering
        \includegraphics[width=5cm,height=4.5cm]{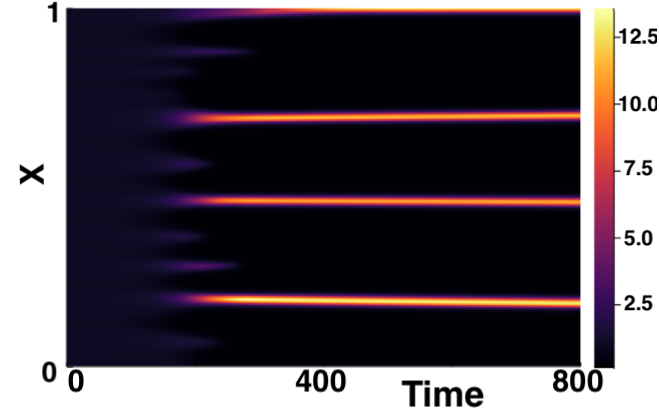}
        \caption{$\text{IC}_2$ given by equation \eqref{firstic}.}
        \label{}
    \end{subfigure}
    \hfill
    \begin{subfigure}[t]{0.32\textwidth}
        \centering
        \includegraphics[width=5cm,height=4.5cm]{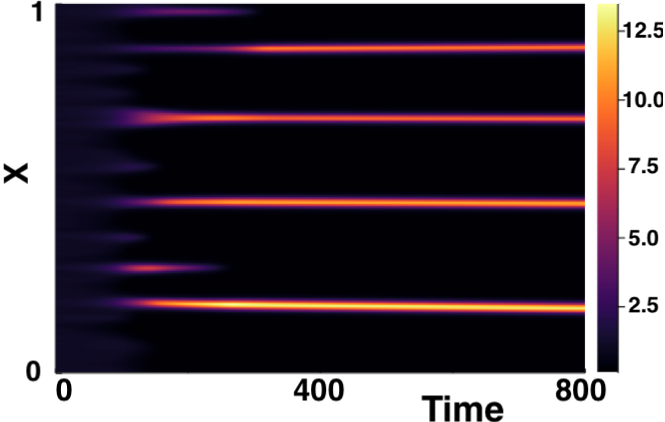}
        \caption{$\text{IC}_3$ given by equation \eqref{ic3}.}
        \label{}
    \end{subfigure}
    \caption{Numerical simulations of \eqref{fixed2} showing comparison of varying ICs for $\tau=4$. Boundary conditions given by \eqref{neumannbc}. $(a,b)=(0.1,0.9)$, $\epsilon^2=0.001$, $L^2=9/2$.}
    \label{fig:figtau4}
\end{figure}

\begin{figure}[H]
    \centering
    \begin{subfigure}[t]{0.32\textwidth}
        \centering
        \includegraphics[width=5cm,height=4.5cm]{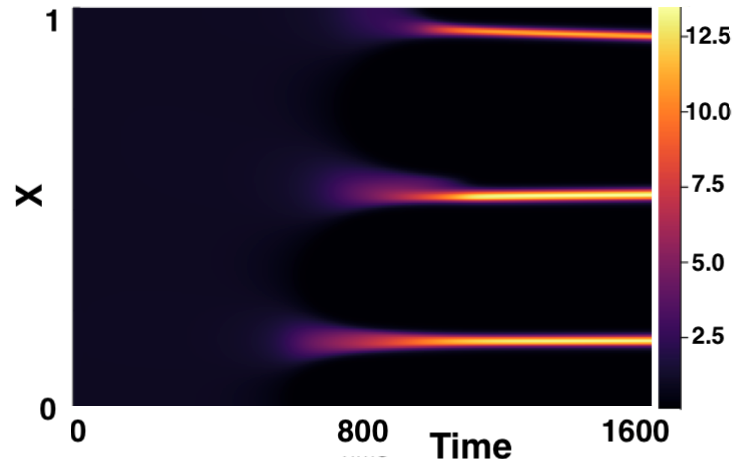}
        \caption{$\text{IC}_1$ given in reference \cite{gaffmonk}.}
        \label{}
    \end{subfigure}
    \hfill
    \begin{subfigure}[t]{0.32\textwidth}
        \centering
        \includegraphics[width=5cm,height=4.5cm]{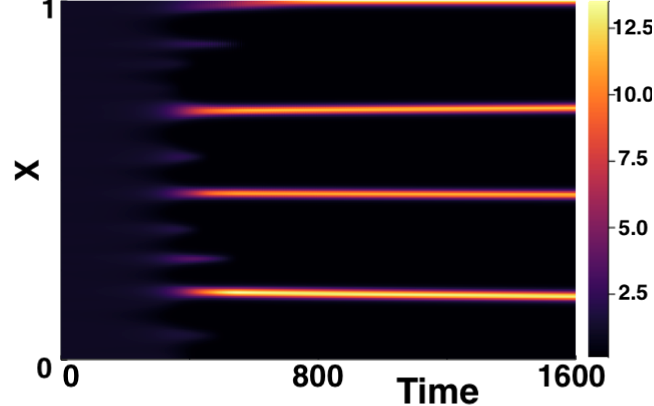}
        \caption{$\text{IC}_2$ given by equation \eqref{firstic}.}
        \label{}
    \end{subfigure}
    \hfill
    \begin{subfigure}[t]{0.32\textwidth}
        \centering
        \includegraphics[width=5cm,height=4.5cm]{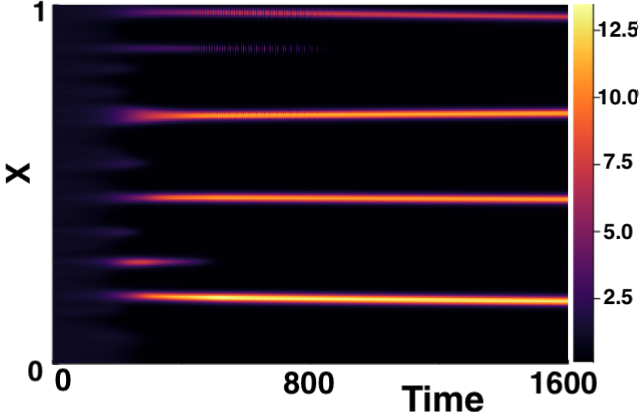}
        \caption{$\text{IC}_3$ given by equation \eqref{ic3}.}
        \label{}
    \end{subfigure}
    \caption{Numerical simulations of \eqref{fixed2} showing comparison of varying ICs for $\tau=8$. Boundary conditions given by \eqref{neumannbc}. $(a,b)=(0.1,0.9)$, $\epsilon^2=0.001$, $L^2=9/2$.}
    \label{fig:figtau8}
\end{figure}

\begin{figure}[H]
    \centering
    \begin{subfigure}[t]{0.32\textwidth}
        \centering
        \includegraphics[width=5cm,height=4.5cm]{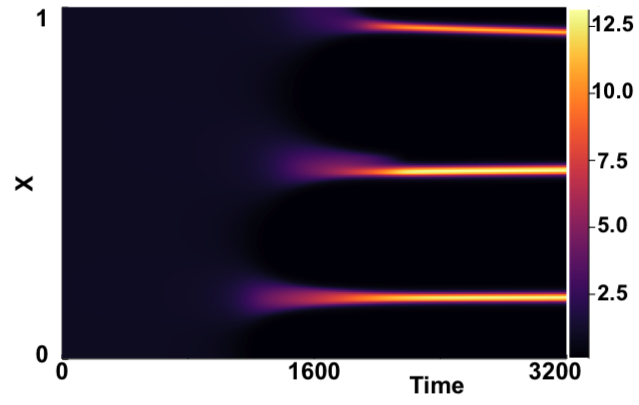}
        \caption{$\text{IC}_1$ given in reference \cite{gaffmonk}.}
        \label{}
    \end{subfigure}
    \hfill
    \begin{subfigure}[t]{0.32\textwidth}
        \centering
        \includegraphics[width=5cm,height=4.5cm]{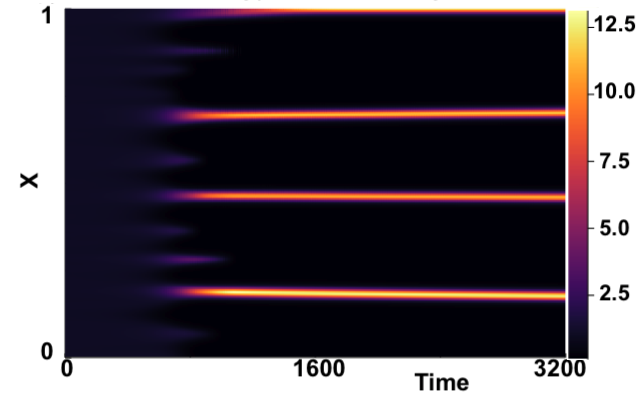}
        \caption{$\text{IC}_2$ given by equation \eqref{firstic}.}
        \label{}
    \end{subfigure}
    \hfill
    \begin{subfigure}[t]{0.32\textwidth}
        \centering
        \includegraphics[width=5cm,height=4.5cm]{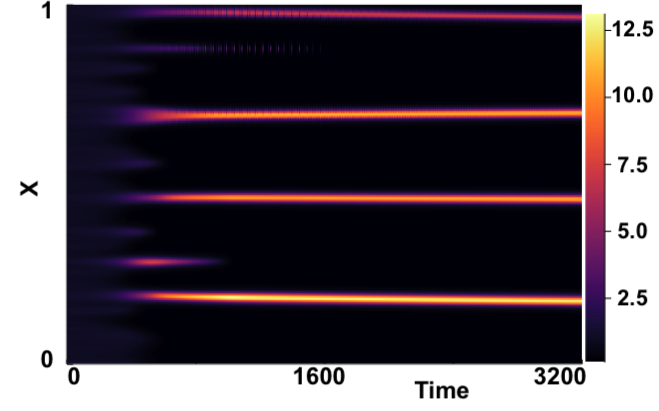}
        \caption{$\text{IC}_3$ given by equation \eqref{ic3}.}
        \label{}
    \end{subfigure}
    \caption{Numerical simulations of \eqref{fixed2} showing comparison of varying ICs for $\tau=16$. Boundary conditions given by \eqref{neumannbc}. $(a,b)=(0.1,0.9)$, $\epsilon^2=0.001$, $L^2=9/2$.}
    \label{fig:figtau16}
\end{figure}

In this section, the robustness of the results obtained in \cite{gaffmonk} are examined numerically under varying of initial conditions and boundary conditions. We first consider the sensitivity of pattern formation in the context of a fixed time delay to varying initial conditions. Three different sets of initial conditions are considered. $\text{IC}_1$ corresponds to the initial conditions used in \cite{gaffmonk}. The functional form of $\text{IC}_1$ can be found in Appendix \ref{section:appA}. $\text{IC}_2$ denotes the same initial conditions defined in \eqref{firstic}, and $\text{IC}_3$ are the initial conditions given by
\begin{equation}\label{ic3}
\text{IC}_3:\quad\quad\quad\begin{pmatrix}u_0\\v_0\end{pmatrix}=\begin{pmatrix}u_\star(1+r)\\v_\star(1+r)\end{pmatrix}\quad r\sim\mathcal{N}\left(0,0.1^2\right).
\end{equation}
We note that computationally a fixed random seed was set, and unless otherwise stated, a constant history function equal to the initial conditions as defined in \eqref{hist} was used. The model parameters used match those used in \cite{gaffmonk}, with $(a,b)=(0.1,0.9)$. The results in Figures \ref{fig:figtau0}, \ref{fig:figtau1}, \ref{fig:figtau2}, \ref{fig:figtau4}, \ref{fig:figtau8}, and \ref{fig:figtau16} show the pattern formation observed for each of the initial conditions for varying fixed time delay $\tau\in\{0,1,2,4,8,16 \}$. This range of time delays was motivated by those used in \cite{gaffmonk}.


It can be seen that the final pattern is sensitive to the choice of initial conditions, and that, intuitively, the larger $\sigma_{\text{IC}}$ used in $\text{IC}_3$, compared to that of $\text{IC}_2$, results in a faster onset of pattern formation. We see from considering the timescales as to which pattern formation occurs however, that although the time taken until onset of patterning varies with different initial conditions, the increase in time-to-pattern with an increasing time delay is consistent independent of the initial conditions chosen. By considering the varying $x$-axis, we also note that in each case, this relationship seems to be linear. We formalise this in section \ref{section:delaypatt}.

Numerical results were also simulated to study the effects of a temporal variation in the history function. A history function was set as $h(t)=u_\star(1+r\sin(\omega t))$ for $t\in[-\tau,0)$, where $r$ is the random variable used in $\text{IC}_2$. Simulations were conducted for varying $\tau$ and $\omega$. Preliminary simulations, which can be found in Appendix \ref{section:Bfix}, show that this type of variation in history does not have a significant effect on the results seen.

\begin{figure}[H]
    \centering
    \begin{subfigure}[t]{0.45\textwidth}
        \centering
        \includegraphics[width=6cm,height=4.5cm]{ic20.png}
        \caption{$\text{BC}_1$ given by equation \eqref{neumannbc}.}
        \label{}
    \end{subfigure}
    \hfill
    \begin{subfigure}[t]{0.45\textwidth}
        \centering
        \includegraphics[width=6cm,height=4.5cm]{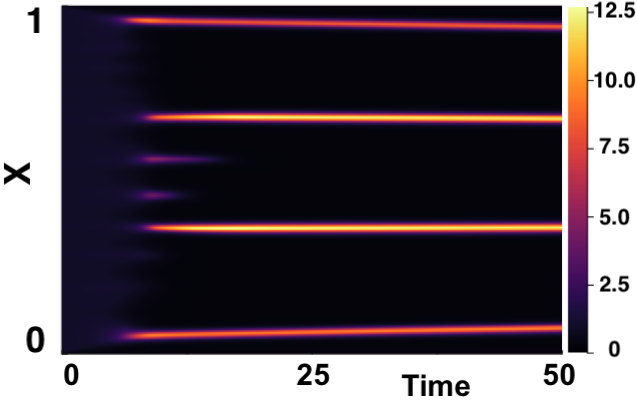}
        \caption{$\text{BC}_2$ given by equation \eqref{homogeneousbc}.}
        \label{}
    \end{subfigure}
    \caption{Comparison of varying BCs for $\tau=0$. $(a,b)=(0.1,0.9)$, $\epsilon^2=0.001$, $L^2=9/2$. Initial conditions given by \eqref{firstic}.}
    \label{fig:bctau1}
\end{figure}

\begin{figure}[H]
    \centering
    \begin{subfigure}[t]{0.45\textwidth}
        \centering
        \includegraphics[width=6cm,height=4.5cm]{ic21.png}
        \caption{$\text{BC}_1$ given by equation \eqref{neumannbc}.}
        \label{}
    \end{subfigure}
    \hfill
    \begin{subfigure}[t]{0.45\textwidth}
        \centering
        \includegraphics[width=6cm,height=4.5cm]{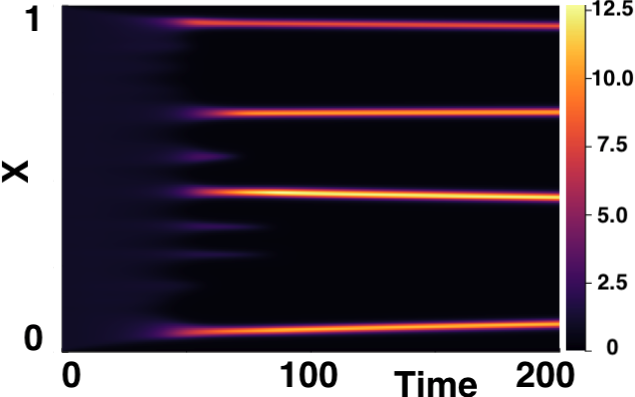}
        \caption{$\text{BC}_2$ given by equation \eqref{homogeneousbc}.}
        \label{}
    \end{subfigure}
    \caption{Comparison of varying BCs for $\tau=1$. $(a,b)=(0.1,0.9)$, $\epsilon^2=0.001$, $L^2=9/2$. Initial conditions given by \eqref{firstic}.}
    \label{fig:bctau2}
\end{figure}

\begin{figure}[H]
    \centering
    \begin{subfigure}[t]{0.45\textwidth}
        \centering
        \includegraphics[width=6cm,height=4.5cm]{ic216.png}
        \caption{$\text{BC}_1$ given by equation \eqref{neumannbc}.}
        \label{}
    \end{subfigure}
    \hfill
    \begin{subfigure}[t]{0.45\textwidth}
        \centering
        \includegraphics[width=6cm,height=4.5cm]{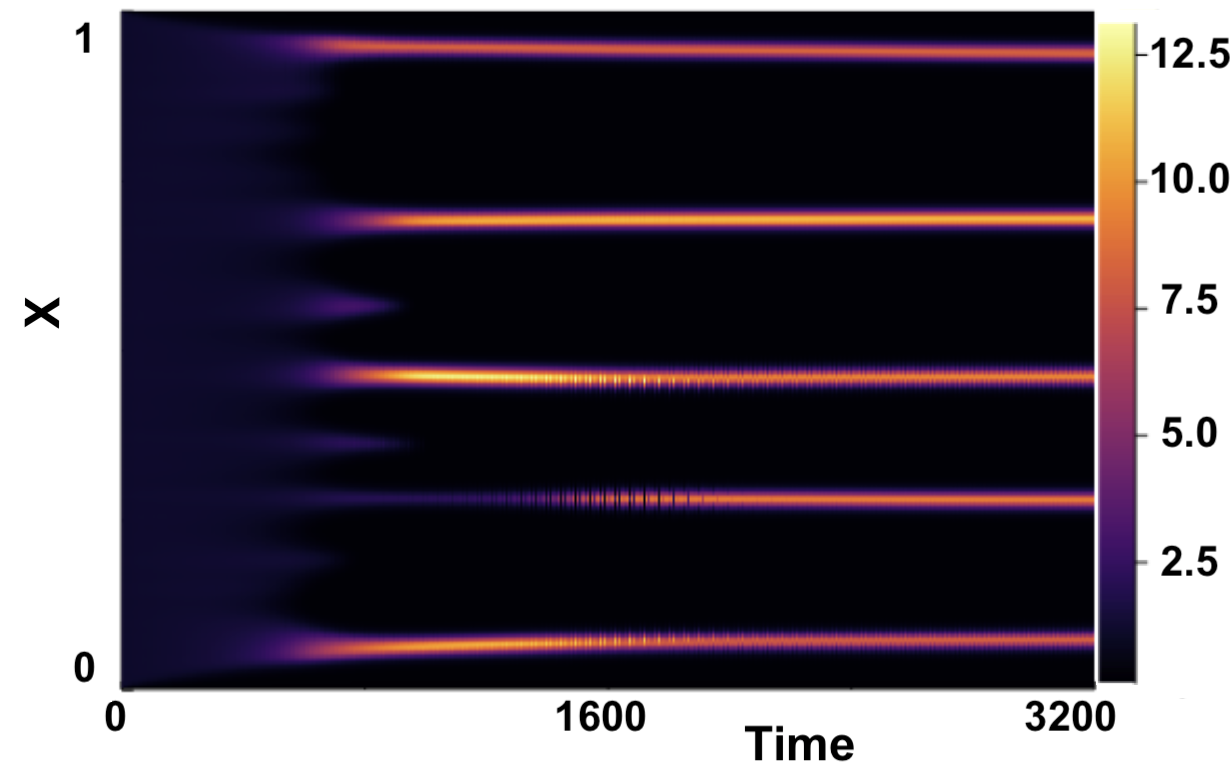}
        \caption{$\text{BC}_2$ given by equation \eqref{homogeneousbc}.}
        \label{}
    \end{subfigure}
    \caption{Comparison of varying BCs for $\tau=16$. $(a,b)=(0.1,0.9)$, $\epsilon^2=0.001$, $L^2=9/2$. Initial conditions given by \eqref{firstic}.}
    \label{fig:bctau3}
\end{figure}

Finally, we consider the effect of varying boundary conditions. motivated by the analysis in \cite{krausemixed}, homogeneous Dirichlet boundary conditions are implemented for the activator term, and homogeneous Neumann boundary conditions implemented for the inhibitor term. Thus, we have that, on the boundaries of the domain $\Omega=[0,1]$,
\begin{equation}\label{homogeneousbc}
u=\frac{\partial v}{\partial x}=0 \quad x=0, 1.
\end{equation}
These conditions are implemented numerically following the methodology outlined in section \ref{section:numimp}. The results in Figures \ref{fig:bctau1}, \ref{fig:bctau2}, and \ref{fig:bctau3} were generated using $\text{IC}_2$, with a varying $\tau\in\{0,1,16\}$, and show the comparison between numerical simulations generated with homogeneous Neumann conditions for both $u$ and $v$, as in \eqref{neumannbc} indicated as $\text{BC}_1$, and those generated with homogeneous Dirichlet conditions for $u$, indicated as $\text{BC}_2$, as in \eqref{homogeneousbc}. Further numerical solutions comparing the boundary conditions for $\tau\in\{2,4,8\}$ can be found in Appendix \ref{section:Bfix}.

We note that, although changing the boundary conditions for the activator term $u$ to homogeneous Dirichlet conditions affects the type of patterns we may see (number and amplitude of spikes), this change does not affect the increased timescales, caused by an increase in time delay, on which onset of patterning occurs.

\section{Relationship Between Time-To-Pattern and Time Delay}\label{section:delaypatt}

We aim to show that, for small $\tau$ and small $L$, the linear theory provides a good approximation to the time taken until pattern formation occurs, and in fact, the relationship between $\tau$ and time-to-pattern under these conditions is linear. We also show that, through full numerical solutions, the relationship between $\tau$ and time-to-pattern on a longer timescale for larger $\tau$ is also linear. We first consider the former.

\begin{figure}[H]
    \centering
    \begin{subfigure}[t]{0.45\textwidth}
        \centering
        \includegraphics[width=7cm,height=5cm]{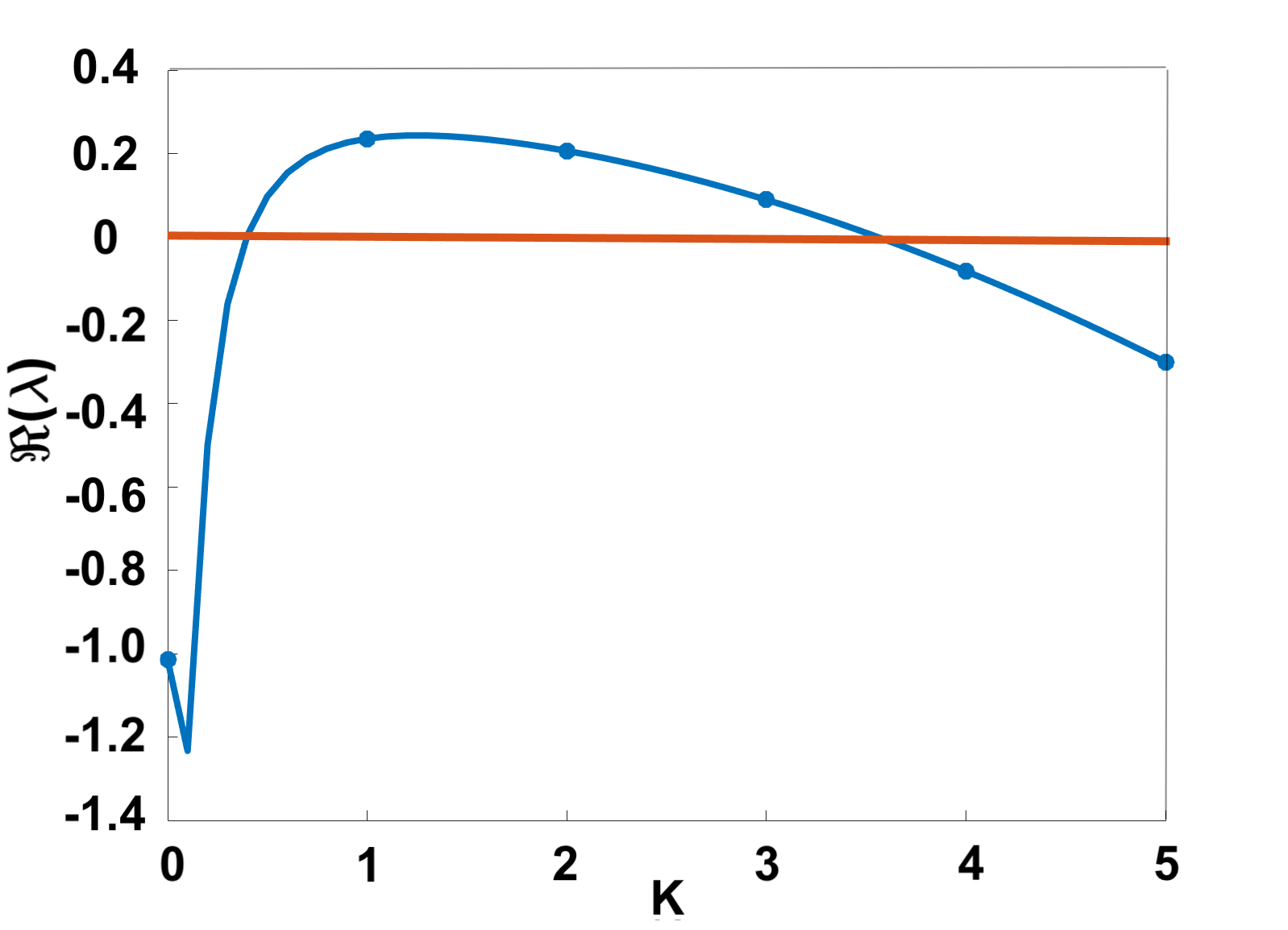}
        \caption{Dispersion curve plotted with domain size $L^2=1/5$. Curve produced by varying $k\in[0,5]$ at regular intervals of $0.1$. Discrete values of $k$ overlayed as scatter points. }
        \label{fig:compdisp1}
    \end{subfigure}
    \hfill
    \begin{subfigure}[t]{0.45\textwidth}
        \centering
        \includegraphics[width=7cm,height=5cm]{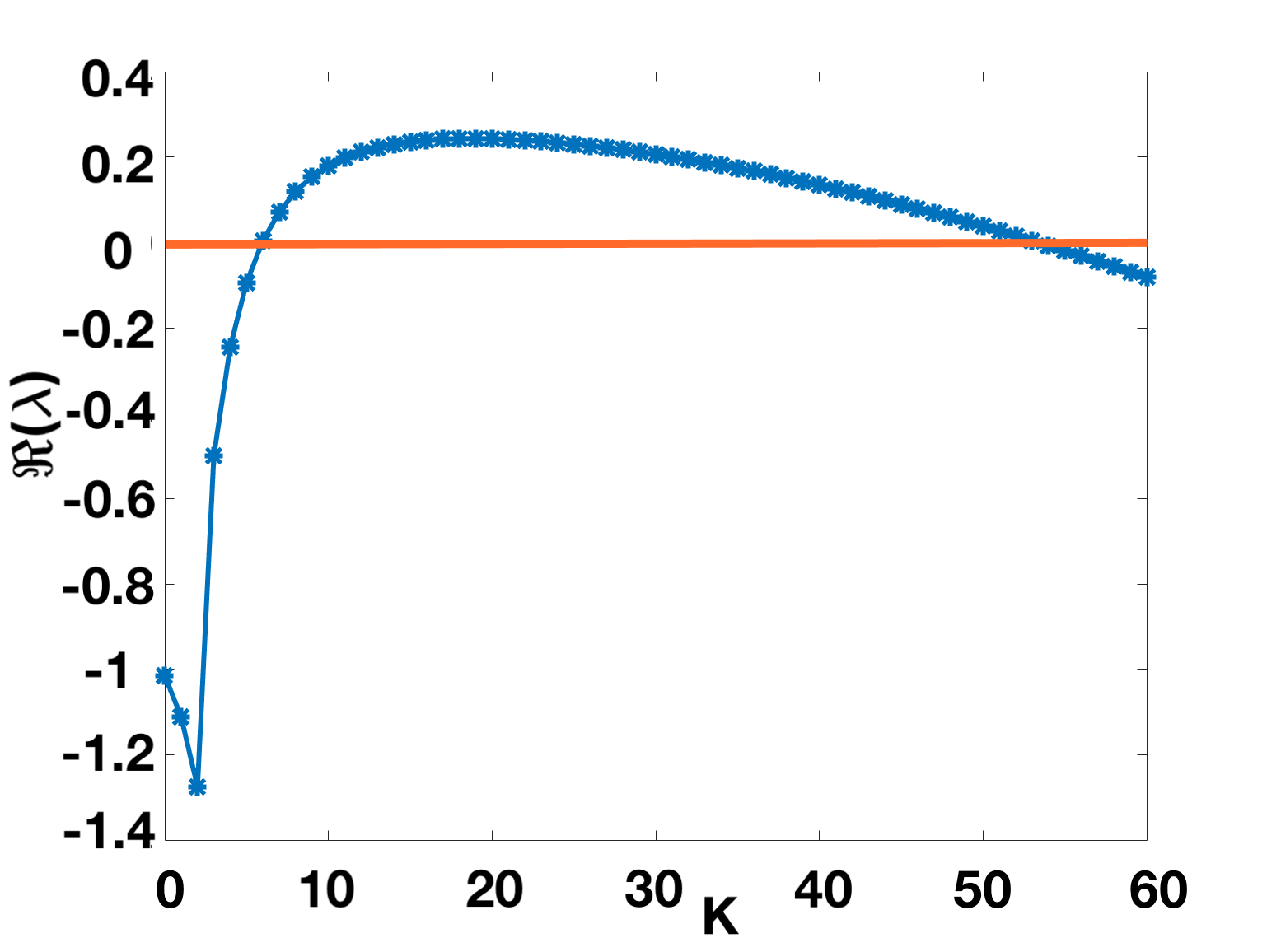}
        \caption{Dispersion curve plotted with domain size $L^2=9/2$. Curve produced by varying $k\in[0,60]$ at regular intervals of $1$.}
        \label{fig:compdisp2}
    \end{subfigure}
    \caption{Dispersion curves of the characterstic equation given in \eqref{characfix} plotted for $(a,b,\tau)=(0.4,1.8,0.2)$ and $\epsilon^2=0.001$. A larger $L$ results in more unstable modes $\lambda_k$ such that $\Re(\lambda_k)>0$. }
    \label{fig:compardisp}
\end{figure}

To minimise the effect of nonlinearity in the dynamics, we restrict the domain size to $L^2=1/5$. Shrinking the domain results in fewer unstable modes and thus less competition for the dominant mode, resulting in a better approximation of the linear theory. This finite size effect can be seen in Figure \ref{fig:compardisp}, where $\Re(\lambda_k)$ is plotted against $k$ for two different domain sizes, for a given $(a,b,\tau)$. Due to numerical restrictions when using Chebfun in finding roots of the characteristic equation \eqref{characfix}, only $\tau\leq1.6$ is considered. Taking a small perturbation in the activator term, $\xi(x,t)$, such that $\xi(x,0)=ru_\star$, where $r$ is a small Gaussian random variable, $r\sim\mathcal{N}\left(0,\sigma_{\text{IC}}^2\right)$,
for some standard deviation of the initial perturbation $\sigma_{\text{IC}}$.

\begin{figure}[H]
    \centering
    \begin{subfigure}[t]{0.45\textwidth}
        \centering
        \includegraphics[width=7cm,height=5cm]{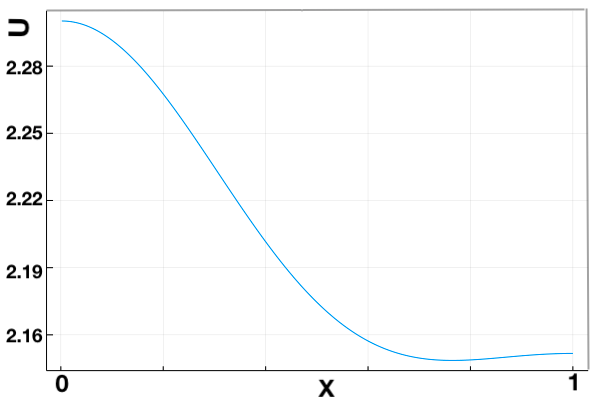}
        \caption{Numerical solution $u(T)$ at $t=T$}
        \label{uT}
    \end{subfigure}
    \hfill
    \begin{subfigure}[t]{0.45\textwidth}
        \centering
        \includegraphics[width=7cm,height=5cm]{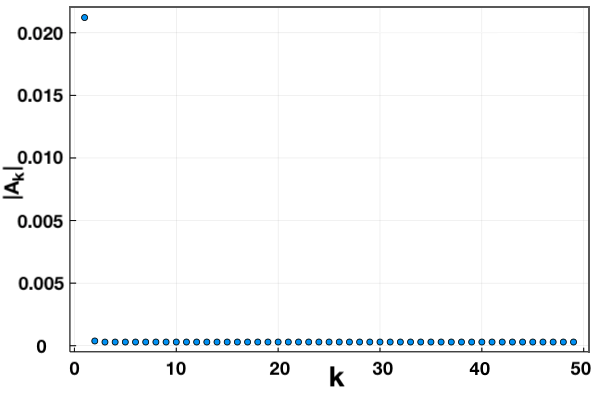}
        \caption{Absolute Fourier coefficients of $u(T)$, for $k\in[1,50]$.}
        \label{fig:uTfc}
    \end{subfigure}
    \caption{Numerical solution of \eqref{fixed2} at $t=T$ with boundary conditions given by \eqref{neumannbc}. Initial perturbation from steady state with $\sigma_{\text{IC}}=10^{-5}$. First $50$ Fourier coefficients for $u(T)$ plotted, with $(a,b)=(0.4,1.8)$, time delay $\tau=0.2$ and $\epsilon^2=0.001$, $L^2=1/5$.}
    \label{fig:Tfc}
\end{figure}

\begin{figure}[H]
    \centering
    \begin{subfigure}[t]{0.32\textwidth}
        \centering
        \includegraphics[width=5cm,height=5cm]{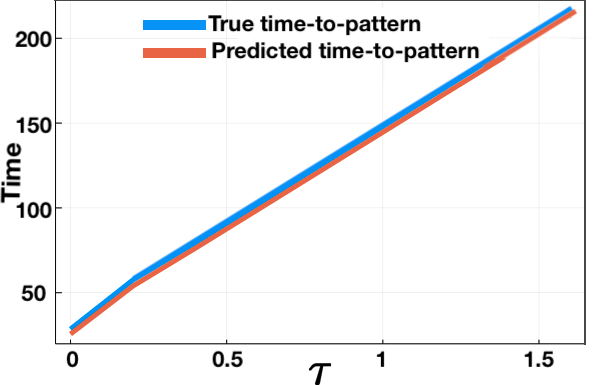}
        \caption{$(a,b)=(0.4,1.8)$.}
        \label{fig:ttp1}
    \end{subfigure}
    \hfill
    \begin{subfigure}[t]{0.32\textwidth}
        \centering
        \includegraphics[width=5cm,height=5cm]{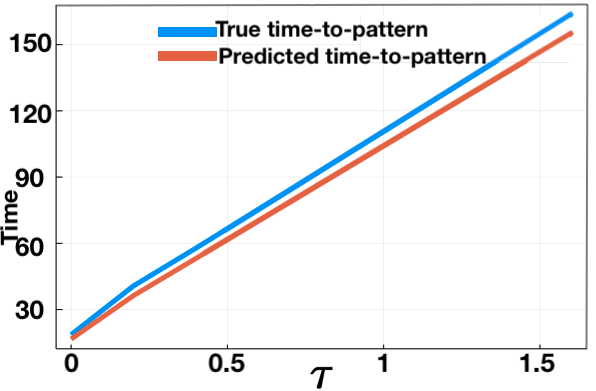}
        \caption{$(a,b)=(0.1,0.9)$.}
        \label{fig:ttp2}
    \end{subfigure}
    \hfill
    \begin{subfigure}[t]{0.32\textwidth}
        \centering
        \includegraphics[width=5cm,height=5cm]{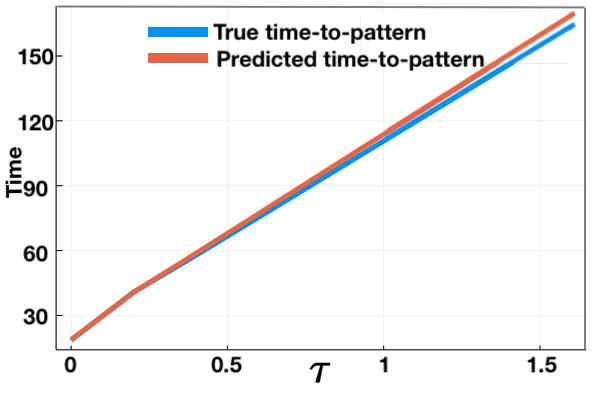}
        \caption{$(a,b)=(0.2,1.3)$.}
        \label{fig:ttp3}
    \end{subfigure}
    \caption{Predicted vs `true' time-to-pattern for numerical solution of \eqref{fixed2} with boundary conditions given by \eqref{neumannbc}. Initial perturbation from steady state with $\sigma_{\text{IC}}= 10^{-5}$ and threshold of $0.1$. Predicted time-to-pattern computed using the relationship \eqref{ttprelation}, for three different parameter sets, with $L^2=1/5$, $\epsilon^2=0.001$, and $\tau\in[0,1.6]$, varied at regular intervals of $0.2$.}
    \label{}
\end{figure}

The linear theory suggests that at some time $t=T$, the perturbation will be of the form $\xi(x,T)\sim A_k(T)\cos(k\pi x)$, where $k$ is the dominant mode and $A_k(T)$ denotes the corresponding Fourier coefficient at time $t=T$. For a given parameter set $(a,b,\epsilon^2,\tau,L)$, we can solve the characteristic equation \eqref{characfix}, and plot $\Re(\lambda_k)$
against $k$, to determine the dominating mode $k$ and the corresponding eigenvalue, or growth rate, $\lambda_k$. We then use this information in the following manner: A Fast Fourier Transform to decompose the initial conditions into a Fourier series is used, and the coefficient $A_k(0)$ for the dominating $k$ is computed. When the perturbation $\xi$ has grown sufficiently, in absolute value, beyond a threshold where pattern formation is considered, we call this time $t=T$, and determine the Fourier coefficient $A_k(T)$ of the fastest-growing mode $k$. More specifically, the time $T$ is the first such that $\max_x|u(T,x)-u_\star|>threshold$, namely the first time such that any solution point across the whole spatial domain is large enough, in absolute difference, from the steady state. Finally, using the relation $A_k(T)\sim A_k(0)e^{\lambda_k T}$, we rearrange for $T$ and thus compute a linear approximation for time-to-pattern as
\begin{equation}\label{ttprelation}
    T=\frac{1}{\lambda_k}\ln\left(\frac{A_k(T)}{A_k(0)}\right).
\end{equation}
We consider an example case for $(a,b,\tau)=(0.4,1.8,0.2)$. The standard deviation for the random variable $r$ is chosen as $\sigma_{\text{IC}}=10^{-5}$, and the threshold value at $0.1$. A very small perturbation was used as a means to improve the accuracy of the linear theory.
Using $\epsilon^2=0.001$, on the domain size $L^2=1/5$, Figure \ref{fig:compdisp1} suggests that, from the linear theory, the dominant mode is $k=1$ with dominant eigenvalue $\lambda_1=0.2356$. Since $k=1$ is the dominant mode, we compute the first Fourier coefficient of the initial conditions, $A_1(0)$, as $A_1(0)=7.95\times10^{-8} (3 s.f.)$. To find $A_1(T)$, a numerical simulation is run until the solution of the activator $u$ has grown, in absolute value, to a threshold value of $0.1$. Figure \ref{fig:Tfc} shows the numerical solution $u(T)$ at the point where this threshold value has been met, as well as a scatter plot of the Fourier coefficients $A_k(T)$, $k\neq0$.
As seen in Figure \ref{fig:uTfc}, the Fourier coefficient corresponding to $k=1$ is given as $0.0262(3 s.f.)$. The approximated time-to-pattern, as predicted by linear theory, for $(a,b,\tau)=(0.4,1.8,0.2)$ and the given initial conditions is thus computed as
\begin{equation}
    T=\frac{1}{\lambda_1}\ln\left(\frac{A_1(T)}{A_1(0)}\right)=\frac{1}{0.2356}\ln\left(\frac{0.0262}{7.95\times10^{-8}}\right)=53.8(3 s.f.).
\end{equation}
It was found through numerical solutions that the `true' time-to-pattern is $\approx57.5(3s.f.)$.
We use `true' time-to-pattern here to mean the time taken for a perturbation to grow above a threshold value found through full numerical solutions, rather than through linear analysis. This process can be repeated for varying $(a,b,\tau)$, and Figures \ref{fig:ttp1}, \ref{fig:ttp2}, \ref{fig:ttp3}, show the predicted time-to-pattern plotted against $\tau$ and compared with the `true' time-to-pattern for three different parameter sets. The time delay is varied here over $\tau\in[0,1.6]$ at intervals of $0.2$.

\begin{figure}[H]
    \centering
    \begin{subfigure}[t]{0.45\textwidth}
        \centering
        \includegraphics[width=6cm,height=5cm]{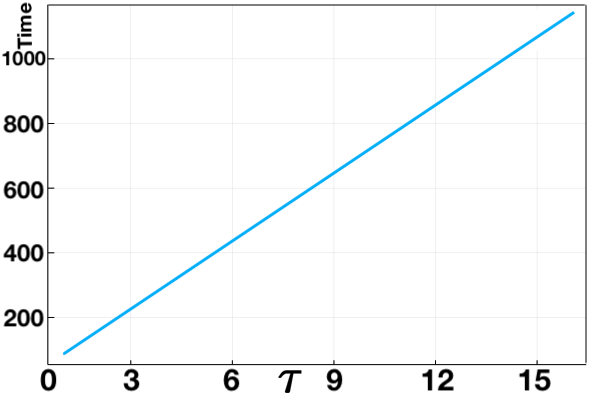}
        \caption{$(a,b)=(0.1,0.9)$}
        \label{fig:linperturb1a}
    \end{subfigure}
    \hfill
    \begin{subfigure}[t]{0.45\textwidth}
        \centering
        \includegraphics[width=6cm,height=5cm]{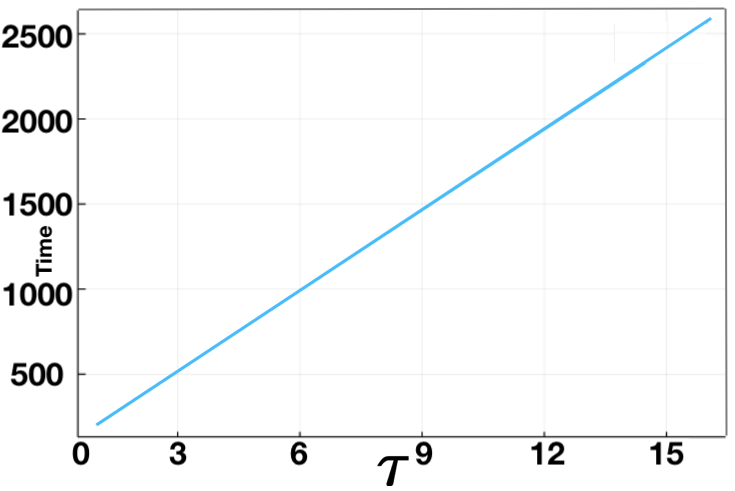}
        \caption{$(a,b)=(0.4,0.8)$}
        \label{fig:linperturb1b}
    \end{subfigure}
    \caption{Time-to-pattern for full numerical solutions of \eqref{fixed2} plotted against $\tau\in[1,16]$ for $\sigma_{\text{IC}}=10^{-5}$ and threshold $0.1$. Parameters used are $\epsilon^2=0.001$ and domain size $L^2=9/2$.}
    \label{fig:linperturb1}
\end{figure}

Finally, through full numerical solutions, we show a linear relationship between $\tau$ and time-to-pattern on a longer timescale. Varying $\tau\in[1,16]$ at regular intervals of $1$, for two different parameter sets $(a,b)=\{(0.1,0.9),(0.4,0.8)\}$, we compute the time taken for a perturbation to grow up to a threshold value $0.1$, from a $\sigma_{\text{IC}}=10^{-5}$, and plot the results. The results can be seen in Figure \ref{fig:linperturb1}, and a linearly increasing relationship can be seen. In order to check this claim further, numerical simulations were also run with a threshold value $2$, from a $\sigma_{\text{IC}}=0.01$. The results for this can be found in Appendix \ref{section:Bfix}, where a linear relationship can also be seen.

%\begin{figure}[H]
%    \centering
%    \begin{subfigure}[t]{0.45\textwidth}
%        \centering
%        \includegraphics[width=6cm,height=5cm]{longlin1.png}
%        \caption{$\sigma_{\text{IC}}=10^{-5}$ and threshold $0.1$}
%        \label{fig:linperturb2a}
%    \end{subfigure}
%    \hfill
%    \begin{subfigure}[t]{0.45\textwidth}
%        \centering
%        \includegraphics[width=7cm,height=5cm]{longlin4.png}
%        \caption{$\sigma_{\text{IC}}=0.01$ and threshold $2$}
%        \label{fig:linperturb2b}
%    \end{subfigure}
%    \caption{Time-to-pattern for full numerical solutions of \eqref{fixed2} plotted against $\tau\in[1,16]$ for two different initial perturbations and threshold values. $(a,b)=(0.4,0.8)$, $\epsilon^2=0.001$ and domain size $L^2=9/2$.}
%    \label{fig:linperturb2}
%\end{figure}

The results show that, on a smaller time scale, smaller spatial domain, and with smaller time delay, the linear theory provides a good approximation of the time-to-pattern compared to full numerical solutions. The full numerical simulations also strongly suggest that there is a linearly increasing relationship between time delay and time-to-pattern.
\section{Summary}

In this chapter, we performed a linear stability analysis of the LI model to determine analytically the effects of an increasing time delay on the Turing space, considered a variation in initial and boundary conditions, and studied the time-to-pattern properties of the model via linear theory and numerical simulations.

The linear theory suggested that for the LI model, time delay can act as a promoting agent for Turing instabilities, expanding the Turing space, and thus increasing the parameter region where Turing instabilities can occur. It was interestingly found that the increase in Turing space comes about solely through the stability of the homogeneous steady state increasing with time delay. Through both linear analysis on a small scale, and full numerical solutions on a larger scale, a linearly increasing relationship between time delay and time-to-pattern was presented.

Numerical results in this chapter were also systematically tested by varying initial and boundary conditions, and by implementing a temporal variation in the history function. Simulations suggested that the increase in time-to-pattern with an increase in delay is robust to these variations. We therefore look for ways to remedy the problems caused by a fixed delay. Considering the complexity and stochastic nature of pattern formation on a cellular level leads us to consider modelling the time delay as a distribution, which we consider in the next chapter.

\chapter{Distributed Delay Model}\label{section:distdel}
A systematic study of distributed delay in the context of Turing instabilities is extremely sparse in the current literature, with little to no systematic analysis of Turing pattern formation in reaction-diffusion systems with distributed delay, in the context of developmental biology. As far as we are aware of, there has been no previous work carried out looking at the Schnakenberg model with incorporated distributed delay. In this chapter, we consider the LI model with time delay modelled as a (skewed) truncated Gaussian distribution. We begin this chapter by outlining the quadrature rule we use to numerically evaluate the integral term in the model \eqref{distmodel}. The linear analysis conducted for the fixed delay case is then extended to the distributed delay model, and we look to show analytically that for small mean time delay $\tau$, using either a symmetric or skewed truncated Gaussian distribution does not have a qualitative difference on the results seen, and thus does not change some of the key problems highlighted in the fixed delay case. These findings are verified through numerical simulations. We also conclude the same results for a larger mean time delay $\tau$ using full numerical solutions.
\section{Composite Simpson's Rule}\label{section:quad}

The LI model with distributed time delay, as defined in \eqref{distmodel}, is given by

\begin{equation}\label{distmodel2}
  \begin{split}
    \frac{\partial u}{\partial t}&=\frac{\epsilon^2}{L^2}\frac{\partial^2u}{\partial x^2}+a-u-2u^2v+3\int_{a}^{b}k(s;\textbf{p})\hat{u}^2\hat{v} \ \text{d}s,\\
    \frac{\partial v}{\partial t}&=\frac{1}{L^2}\frac{\partial^2v}{\partial x^2}+b-u^2v,
\end{split}
\end{equation}
where $\hat{u}=u(x,t-s)$ and $\hat{v}=v(x,t-s)$ and $s$ is the integration variable ranging over the delays. The integration domain $[a, b]$, can be discretised into $N$ sub-intervals of equal length with $N+1$ discretisation points, $s_0,\cdots,s_{N}$, such that $s_0=a$ and $s_N=b$. Using the composite Simpsons's rule \cite{compsimp}, the integral term can be numerically approximated as

\begin{equation}\label{simp}\int_{a}^{b}k(s)\hat{u}^2\hat{v}\  \text{d}s\approx\frac{h}{3}\left[k(s_0)\hat{u}^2_0\hat{v}_0+2\sum_{i=1}^{\frac{N}{2}-1}k(s_{2i})\hat{u}^2_{2i}\hat{v}_{2i}+4\sum_{i=2}^{\frac{N}{2}}k(s_{2i-1})\hat{u}^2_{2i-1}\hat{v}_{2i-1}+k(s_N)\hat{u}^2_N\hat{v}_N\right],
\end{equation}
where $h$ is computed as $h=\frac{b-a}{N}$. We use the notation $\hat{u}_j$ and $\hat{v}_j$ to denote $u(t-s_j)$ and $v(t-s_j$) respectively.

\section{A Symmetric Distribution}\label{section:symmetric}
\subsection{Introduction}

As implemented in \cite{william}, by assuming each individual mechanism within the gene expression process occurs independently and identically, we use the central limit theorem to model the delay as a symmetric Gaussian distribution with parameters $\textbf{p}=(\tau,\sigma)$, for some mean $\tau$ and standard deviation $\sigma$. Throughout Section \ref{section:symmetric}, we use integration limits $a=\tau-n\sigma$ and $b=\tau+n\sigma$ for some $n\in\mathbb{N}$, such that $a=\tau-n\sigma>0$. We can thus write the LI model with distributed time delay as
\begin{equation}\label{symmod}
    \begin{split}
        \frac{\partial u}{\partial t}&=\frac{\epsilon^2}{L^2}\frac{\partial^u}{\partial x^2}+a-u-2u^2v+3\int_{\tau-n\sigma}^{\tau+n\sigma}k(s;\tau,\sigma)\hat{u}^2\hat{v}\ \text{d}s,\\
        \frac{\partial v}{\partial t}&=\frac{1}{L^2}\frac{\partial^2v}{\partial x^2}+b-u^2.
    \end{split}
\end{equation}
The function $k(s;\tau,\sigma)$ is the symmetric truncated Gaussian pdf given by
\begin{equation}
k(s;\tau,\sigma)=\Phi_c\frac{1}{\sigma\sqrt{2\pi}}\exp\left(-\frac{1}{2}\left(\frac{s-\tau}{\sigma}\right)^2\right).
\end{equation}
We use $\Phi_c$ to denote the truncation scaling constant. This constant ensures that $k(s;\tau,\sigma)$ integrates to $1$ over the given integration domain $[a,b]$, and is computed as
\begin{equation}
    \Phi_c=\frac{1}{\phi\left(\frac{b-\tau}{\sigma}\right)-\phi\left(\frac{a-\tau}{\sigma}\right)},
\end{equation}
with $\phi(x)$ the cdf of the (symmetric) standard Gaussian distribution. This is given by
\begin{equation}\label{phi}
    \phi(x)=\frac{1}{2}\left(1+\text{erf}\left(\frac{x}{\sqrt{2}}\right)\right),
\end{equation}
where $\text{erf}(x)$ denotes the error function \footnote{The error function is given by $\text{erf}(x)=\frac{2}{\sqrt{\pi}}\int_0^xe^{-z^2}\ dz$.}. Throughout this section, we use $n=3$ so that the integration limits are $a=\tau-3\sigma$ and $b=\tau+3\sigma$. This was chosen so that a relatively large $\sigma$ value could be used for each $\tau$ while maintaining $a>0$. For each $\tau$ value, a maximum $\sigma$ value can be computed such that $a=\tau-3\sigma\geq0$ as $\sigma_{\max}=\frac{\tau}{3}$. By setting $\sigma<\sigma_{\max}$, we ensure that the integration domain strictly considers positive time delays only.

\begin{figure}[H]
    \centering
    \begin{subfigure}[t]{0.45\textwidth}
        \centering
        \includegraphics[width=7cm,height=4.5cm]{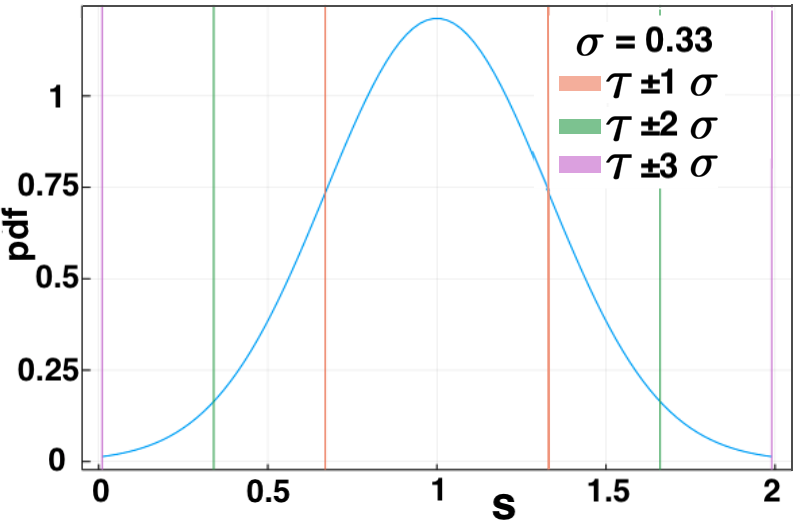}
        \caption{Truncated Gaussian distribution following $\mathcal{N}(1,(\sigma_{\max}\times0.99)^2)$, with $\sigma=\sigma_{\max}\times0.99=0.33$, to 2 decimal places.}
        \label{}
    \end{subfigure}
    \hfill
    \begin{subfigure}[t]{0.45\textwidth}
        \centering
        \includegraphics[width=7cm,height=4.5cm]{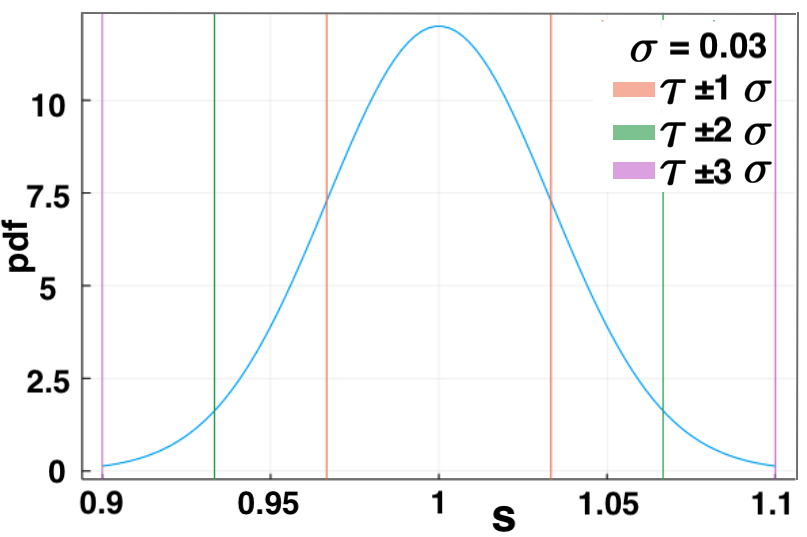}
        \caption{Truncated Gaussian distribution following $\mathcal{N}(1,(\sigma_{\max}\times0.1)^2)$, with $\sigma=\sigma_{\max}\times0.1=0.03$, to 2 decimal places.}
        \label{}
    \end{subfigure}
\caption{PDF of truncated symmetric Gaussian distribution with mean $\tau=1$ and integration domain $[1-3\sigma,1+3\sigma]$. For $\tau=1$, $\sigma_{\max}=1/3$.}
\label{fig:pdf1}
\end{figure}

\begin{figure}[H]
    \centering
    \begin{subfigure}[t]{0.45\textwidth}
        \centering
        \includegraphics[width=7cm,height=4.5cm]{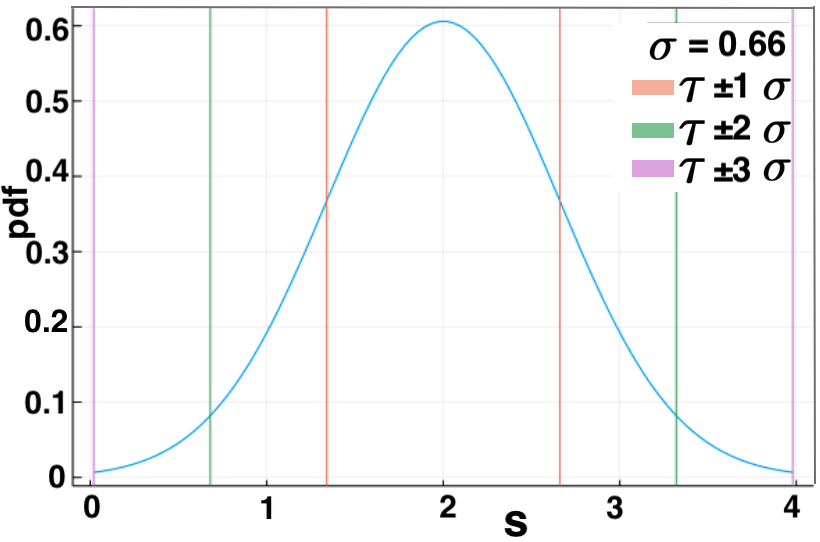}
        \caption{Truncated Gaussian distribution following $\mathcal{N}(2,(\sigma_{\max}\times0.99)^2)$, with $\sigma=\sigma_{\max}\times0.99=0.66$, to 2 decimal places.}
        \label{}
    \end{subfigure}
    \hfill
    \begin{subfigure}[t]{0.45\textwidth}
        \centering
        \includegraphics[width=7cm,height=4.5cm]{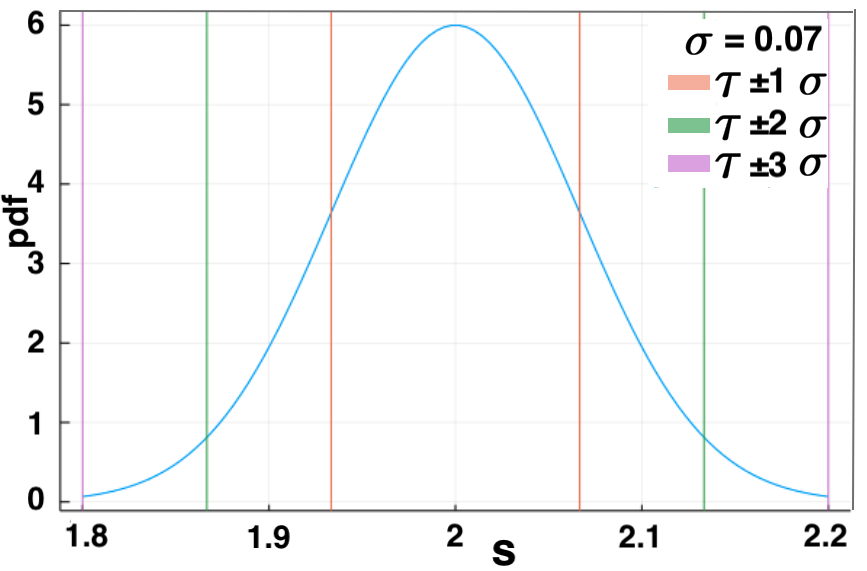}
        \caption{Truncated Gaussian distribution following $\mathcal{N}(2,(\sigma_{\max}\times0.1)^2)$, with $\sigma=\sigma_{\max}\times0.1=0.07 $, to 2 decimal places.}
        \label{}
    \end{subfigure}
    \caption{PDF of symmetric truncated Gaussian distribution with mean $\tau=2$ and integration domain $[2-3\sigma,2+3\sigma]$. For $\tau=2$, $\sigma_{\max}=2/3$.}
    \label{fig:pdf2}
\end{figure}

Figures \ref{fig:pdf1} and \ref{fig:pdf2} show the pdf of a truncated Gaussian distribution centred at a mean $\tau=1,2$ with varying $\sigma$ values as fractions of $\sigma_{\max}$.

We see that $\sigma$ is responsible for scaling on the $x$-axis, while the truncation constant $\Phi_c$, defined in \eqref{distmodel}, scales the $y$-axis, ensuring the pdf integrates to $1$ across the integration domain. Throughout Chapter \ref{section:distdel}, the number of sub-intervals $N$ was chosen to be $N=50$ for the implementation of the quadrature rule applied to the distribution of delay (for both symmetric and skewed distributions). The consideration for such a choice involves both quadrature accuracy and computational efficiency. In this section, we present results obtained in testing the quadrature rule, to show that $N=50$ is a sufficiently large choice. We apply the quadrature to two test integrals: the first is the truncated Gaussian pdf $k(s;\tau,\sigma)$, which should analytically integrate to $1$. For the second test, we apply the quadrature rule to both a spatially and temporally dependent integral
\begin{equation}\label{testint}
\int_a^bk(s;\tau,\sigma)\mathcal{F}(x,t-s)\ \text{d}s,
\end{equation}
where $\mathcal{F}(x,t)=xt$. This can be explicitly evaluated as

\begin{equation}
    \begin{split}
&\int_a^bk(s;\tau,\sigma)\mathcal{F}(x,t-s)\ \text{d}s=xt\int_a^bk(s;\tau,\sigma)\ \text{d}s-x\int_a^bk(s;\tau,\sigma)s\ \text{d}s\\&=xt\left[\frac{\Phi_c}{2}\text{erf}\left(\frac{s-\tau}{\sqrt{2}\sigma}\right)\right]\bigg|_a^b-x\left[\frac{\Phi_c}{2}\text{erf}\left(\frac{s-\tau}{\sqrt{2}\sigma}\right)-\frac{\Phi_c\sigma}{\sqrt{2\pi}}\exp\left(-\frac{1}{2}\left(\frac{s-\tau}{\sigma}\right)^2 \right)\right]\Bigg|_a^b.
    \end{split}
\end{equation}

Figures in Appendix \ref{section:Bdist} show the relative (and absolute error) of the quadrature rule applied to the truncated Gaussian pdf $k(s;\tau,\sigma)$ for different $N$, with varying $\tau$ and $\sigma$. Figures also show the relative error of the quadrature rule applied to \eqref{testint} for $t\in[1.1,500]$ and $x\in[0.1,1]$, for a varying $\tau$ and $\sigma$, with $N=50$. The spatial and temporal domains were chosen to discard the effects of catastrophic cancellation, which is caused by very small solution values for $0\leq t<1.1$ and $0\leq x<0.1$. The results show that the relative error (in absolute value) of the composite Simpson's rule applied to both test integrals, with $N=50$, is of $O(10^{-4})$, namely $<0.1\%$ error, independent of $\tau$ and $\sigma$. We therefore conclude that using $N=50$ quadrature points is sufficiently large.

\subsection{Linear Analysis}\label{section:distlin}
Taking a small perturbation about the steady-state $u=u_\star+\delta\xi$, $v=v_\star+\delta\eta$, where $|\delta|\ll1$, we can write the equation for the activator $u$ in \eqref{distmodel2} as

\begin{equation}\label{perturb}
  \delta\frac{\partial \xi}{\partial t}=\delta \frac{\epsilon^2}{L^2}\frac{\partial^2\xi}{\partial x^2}+f(u_\star+\delta\xi, v_\star+\delta\eta)+g(u_\star+\delta\hat{\xi},v_\star+\delta\hat{\eta}) ,
\end{equation}
where $f(u,v)=a-u+2u^2v$ and $g(\hat{u},\hat{v})=3\int_a^bk(s)\hat{u}^2\hat{v}\ \text{d}s$. The $\hat{\xi}$ notation is used to denote the perturbation evaluated at a delay $\hat{\xi}=\xi(x,t-s)$. Taylor expanding equation \eqref{perturb} for the $f$ term about the steady-state and evaluating the $g$ term, up to $O(\delta)$, yields

\begin{dmath}\label{taylor}
  \delta\frac{\partial \xi}{\partial t}=\delta \frac{\epsilon^2}{L^2}\frac{\partial^2\xi}{\partial x^2}+f(u_\star,v_\star)+3u_\star^2v_\star\int_a^bk(s)\ \text{d}s+\delta\left[\xi f_u(u_\star,v_\star)+\eta f_v(u_\star,v_\star)+6u_\star v_\star\int_a^bk(s)\hat{\xi}\text{d}s+3u_\star^2\int_a^bk(s)\hat{\eta}\ \text{d}s
  \right].
\end{dmath}
We use the notation $f_u$ to denote the derivative of function $f$ with respect $u$. Using the fact that the pdf $k(s;\tau,\sigma)$ integrates to $1$ over $[a,b]$, and evaluating the expressions $f_u(u_\star,v_\star)$ and $f_v(u_\star,v_\star)$, equation \eqref{taylor} can be simplified to

\begin{equation}\label{linu}
  \delta \frac{\partial \xi}{\partial t}=\delta \frac{\epsilon^2}{L^2}\frac{\partial^2\xi}{\partial x^2}+\delta\left[\xi(-1-4u_\star v_\star)-2\eta u_\star^2 +6u_\star v_\star\int_a^bk(s)\hat{\xi}\ \text{d}s+3u_\star^2\int_a^bk(s)\hat{\eta}\ \text{d}s\right].
\end{equation}
The linearised dynamics for $v$ are more simply given by
\begin{equation}\label{linv}
\delta \frac{\partial\eta}{\partial t}=\delta \frac{1}{L^2}\frac{\partial^2\eta}{\partial x^2}-\delta\left[2\xi u_\star v_\star+\eta u_\star^2\right].
\end{equation}
Dividing through by $\delta$ and substituting in an ansatz of the form $\xi=\xi_0e^{\lambda_k t}\cos(k\pi x)$ \cite{yigaffneyli} into \eqref{linu} and $\eta=\eta_0e^{\lambda_k t}\cos(k\pi x)$ into \eqref{linv}, and then dividing through by $e^{\lambda_k t}\cos(k\pi x)$, results in
\begin{equation}\label{sysof}
  \begin{split}
\lambda_k\xi_0&=-\frac{\epsilon^2}{L^2}k^2\pi^2\xi_0+\xi_0(-1-4u_\star v_\star)-2\eta_0u_\star^2+6\xi_0u_\star v_\star E_k+3\xi_0u_\star^2E_k \\
\lambda_k\eta_0&=-\frac{1}{L^2}k^2\pi^2\eta_0-2\xi_0u_\star v_\star-\eta_0u_\star^2,
\end{split}
\end{equation}
where $E_k=\int_a^bk(s;\tau,\sigma)e^{-\lambda_k s}\ \text{d}s$. We can write equation \eqref{sysof} as a homogeneous linear system for $(\xi_0,\eta_0)^T$, given by

\begin{equation}
\underbrace{\begin{pmatrix}-1-4u_\star v_\star-\frac{\epsilon^2}{L^2}k^2\pi^2+6u_\star v_\star E_k-\lambda_k&-2u_\star^2+3u_\star^2E_k\\-2u_\star v_\star&-u_\star^2-\frac{1}{L^2}k^2\pi^2-\lambda_k \end{pmatrix}}_{\textbf{M}}\begin{pmatrix}\xi_0\\\eta_0\end{pmatrix}=\begin{pmatrix}0\\0\end{pmatrix}.
\end{equation}
Looking for non-trivial solutions, we look for roots of the characteristic equation, namely $\mathcal{D}_k=\text{det}(\textbf{M})=0$. The characteristic equation is given as
\begin{equation}\label{characdist}
  \mathcal{D}_k=\lambda_k^2+\alpha_k\lambda_k+\beta_k+(\gamma_k\lambda_k+\delta_k)E_k=0,
\end{equation}
where $\alpha_k,\beta_k,\gamma_k,\delta_k$ are the same as those given in \eqref{fixcoeffs}. Finally, we note that the expression $E_k$ can be evaluated explicitly as
\begin{equation}
E_k=\int_a^bk(s;\tau,\sigma)e^{-\lambda_k s}\ \text{d}s=\frac{\Phi_c}{2}\left[\exp\left(\frac{\lambda_k(\lambda_k\sigma^2-2\tau)}{2}\right) \text{erf} \left(\frac{\lambda_k\sigma^2+s-\tau}{\sqrt{2}\sigma}\right)\right]\Bigg|_a^b.
\end{equation}
The characteristic equation \eqref{characdist} cannot trivially be split into its real and imaginary components, due to the error function term in $E_k$, as was done in the fixed delay case. We therefore cannot explicitly compute the stability lines in $(a,b)$ parameter space. We can however range over $(a,b)$ and compute $\max_k(\Re(\lambda_k))$ for different $\tau$, and produce plots similar to those in Figures \ref{fig:dispfixed} and \ref{fig:lambdavary}. We use these plots to semi-analytically show that using a symmetric Gaussian distribution centred at mean $\tau$ will not change the time-to-pattern seen for a fixed delay of $\tau$, independent of the standard deviation of the distribution $\sigma$. We first plot $\max_k(\Re(\lambda_k))$ against $\tau$, as seen analogously in Figure \ref{fig:dispfixed} for the fixed delay case, for multiple parameters $(a,b,\tau,\sigma)$, and compare these to the fixed delay case.

\begin{figure}[H]
    \centering
    \begin{subfigure}[t]{0.45\textwidth}
        \centering
        \includegraphics[width=7cm,height=5cm]{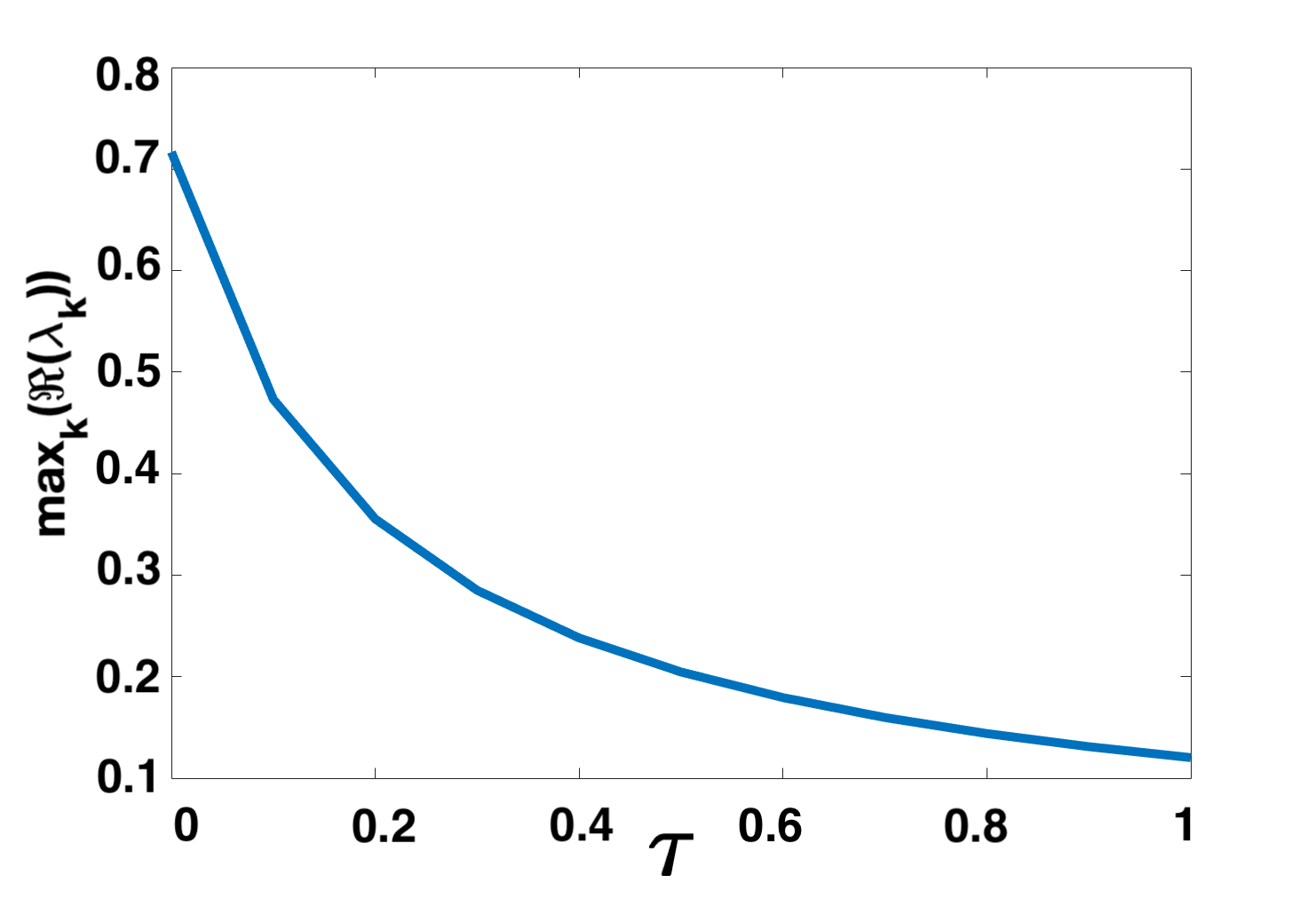}
        \caption{$\max_k(\Re(\lambda_k))$ plotted for fixed delay case.}
        \label{}
    \end{subfigure}
    \hfill
    \begin{subfigure}[t]{0.45\textwidth}
        \centering
        \includegraphics[width=7cm,height=5cm]{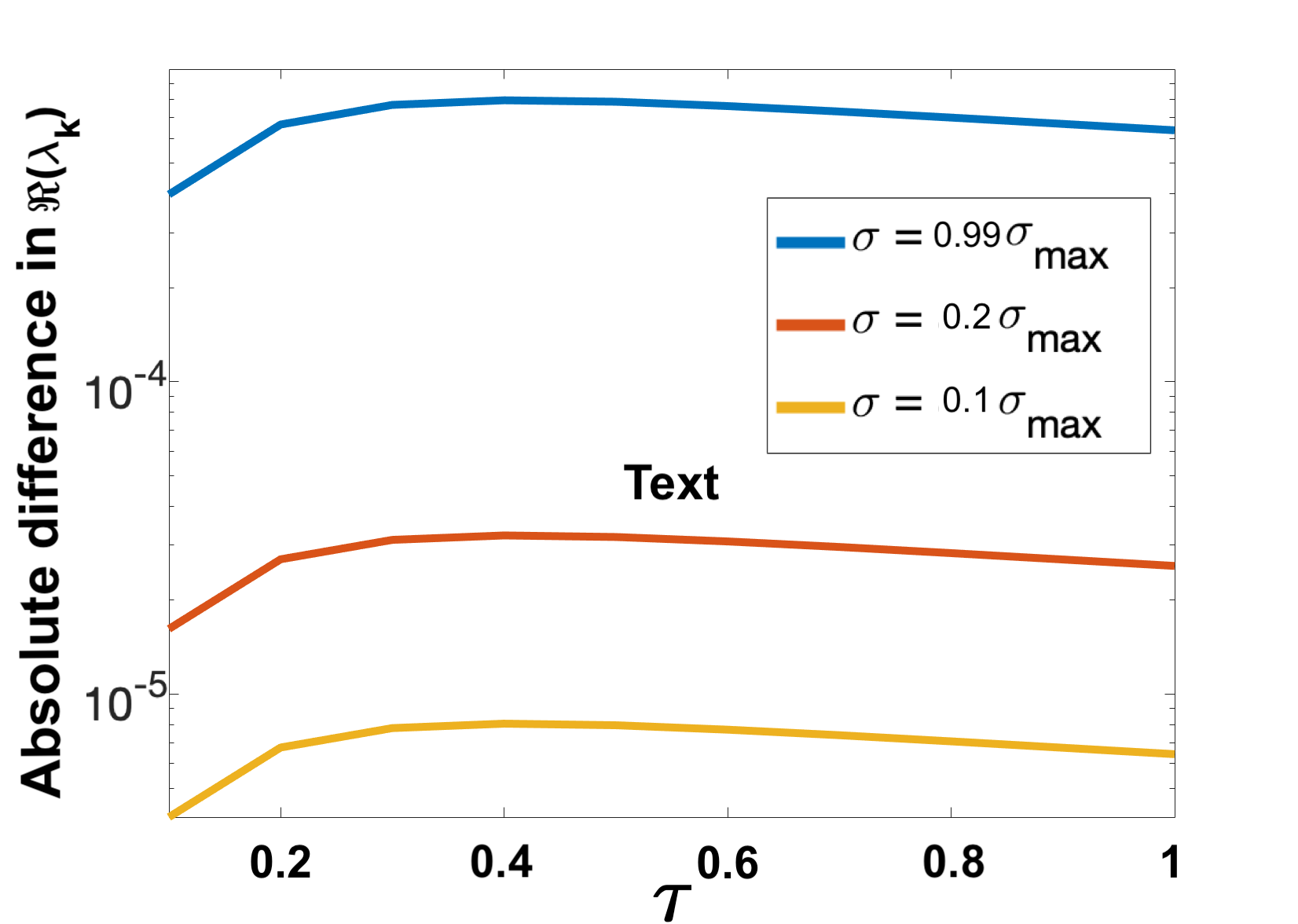}
        \caption{Absolute difference in $\max_k(\Re(\lambda_k))$ as $\sigma$ is varied, between each distributed delay case and the fixed delay case.}
        \label{}
    \end{subfigure}
    \caption{$\max_k(\Re(\lambda_k))$ plotted against $\tau\in[0,1]$ for parameter set $(a,b)=(0.1,0.9)$. $\epsilon^2=0.001$ and $L^2=9/2$. $k\in\mathbb{Z}$ is varied over $[0,50]$. Absolute difference of $\max_k(\Re(\lambda_k))$ between each of the distributed delay cases and fixed delay case plotted.}
    \label{fig:p2}
\end{figure}

\begin{figure}[H]
    \centering
    \begin{subfigure}[t]{0.45\textwidth}
        \centering
        \includegraphics[width=7cm,height=5cm]{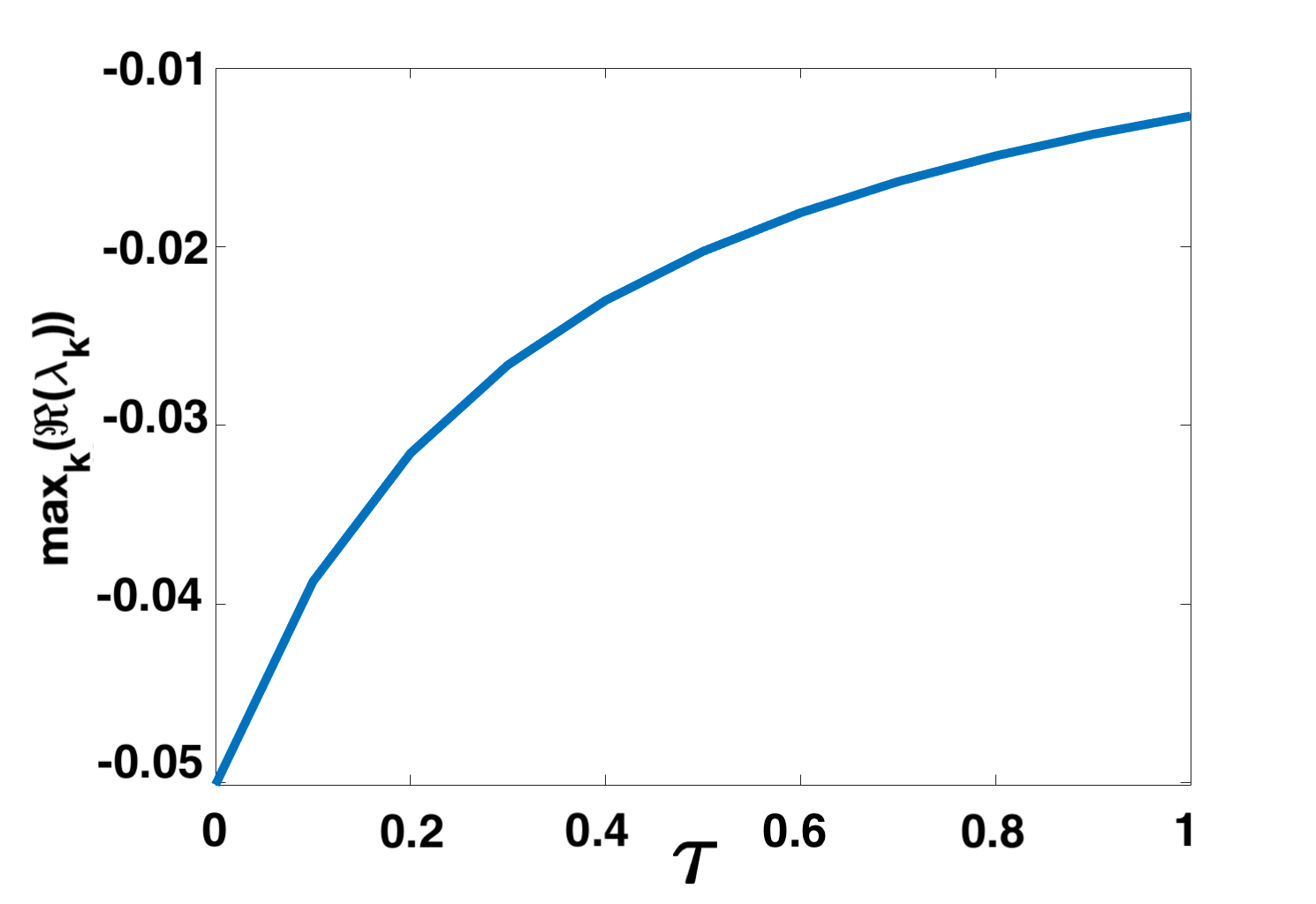}
        \caption{$\max_k(\Re(\lambda_k))$ plotted for fixed delay case.}
        \label{}
    \end{subfigure}
    \hfill
    \begin{subfigure}[t]{0.45\textwidth}
        \centering
        \includegraphics[width=7cm,height=5cm]{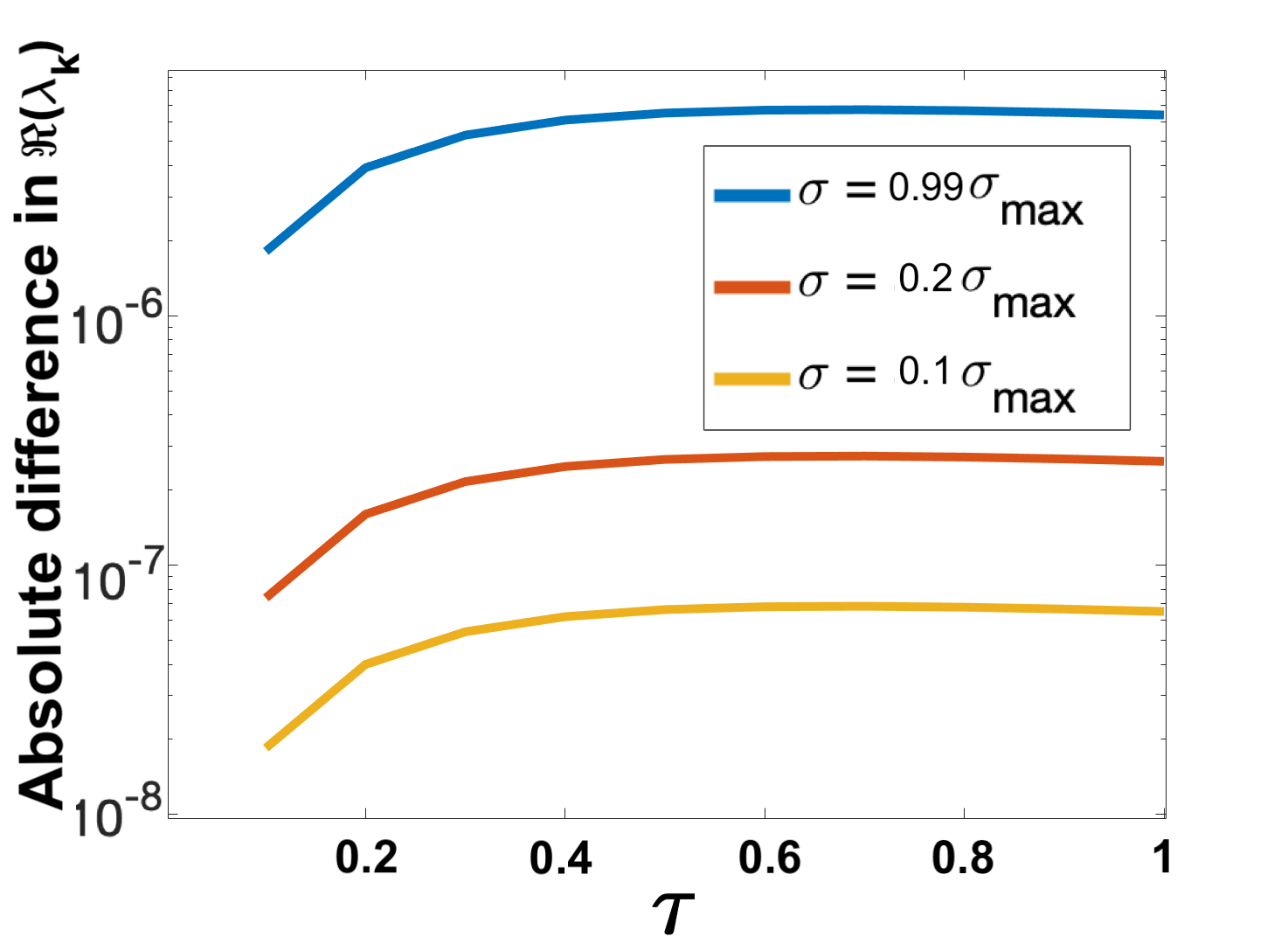}
        \caption{Absolute difference in $\max_k(\Re(\lambda_k))$ as $\sigma$ is varied, between each distributed delay case and the fixed delay case.}
        \label{}
    \end{subfigure}
    \caption{$\max_k(\Re(\lambda_k))$ plotted against $\tau\in[0,1]$ for parameter set $(a,b)=(0.4,0.4)$. $\epsilon^2=0.001$ and $L^2=9/2$. $k\in\mathbb{Z}$ is varied over $[0,50]$. Absolute difference of $\max_k(\Re(\lambda_k))$ between each of the distributed delay cases and fixed delay case plotted.}
    \label{fig:p3}
\end{figure}

Figures \ref{fig:p2} and \ref{fig:p3} show $\max_k(\Re(\lambda_k))$ plotted against $\tau\in[0,1]$ for two different parameter sets $(a,b)=\{(0.1,0.9), (0.4,0.4)\}$, for the fixed delay case. For each parameter set, dispersion curves were computed with different $\sigma$ values as a fraction of $\sigma_{\max}$, and the absolute value of the difference between $\max_k(\Re(\lambda_k))$ for each distributed delay case compared with the fixed delay case is plotted. We note that in the distributed delay case,  $\sigma_{\max}$, and thus the integration limits both change as functions of $\tau$.

% PARAMTER SET 2
% PARAMETER SET 3

Figures \ref{fig:p2} and \ref{fig:p3} show that the largest difference in $\max_k(\Re(\lambda_k))$ occurs when the largest $\sigma$ value is used. The results also suggest that for all $\sigma$ and $\tau\in[0,1]$ considered, that we expect to see pattern formation for $(a,b)=(0.1,0.9)$, but not for $(a,b)=(0.4,0.4)$. We note that the largest absolute difference in $\max_k(\Re(\lambda_k))$ across both parameter sets is of $O(10^{-3})$. This is an extremely small difference and is thus unlikely to make a qualitative difference on the rate of growth of a perturbation, and thus time-to-pattern. We verify these observations numerically in section \ref{section:distsim}.

In order to consider how $\max_k(\Re(\lambda_k))$ varies across a larger parameter plane as $\sigma$ is varied, we consider the absolute difference of $\max_k(\Re(\lambda_k))$ for varying $\sigma$ values as a fraction of $\sigma_{\max}$, for multiple $\tau$, against the fixed delay case, for $\epsilon^2=0.001,0.01$. For each $(\tau,\epsilon^2)$, bifurcation plots were computed for the distributed delay case with varying $\sigma\in\{\sigma_{\max}\times0.99,\sigma_{\max}\times0.2,\sigma_{\max}\times0.1\}$. For each $(\tau,\epsilon^2)$, we consider the absolute difference of $\max_k(\Re(\lambda_k))$ between each distributed delay case and the fixed delay case, across the $(a,b)$ parameter space. These results are summarised in table \ref{tab:tab1}. The bifurcation diagrams of $\max_k(\Re(\lambda_k))$ across the $(a,b)$ space can be found in Appendix \ref{section:Bfix}
% \begin{figure}[H]
%     \centering
%     \begin{subfigure}[t]{0.45\textwidth}
%         \centering
%         \includegraphics[width=7cm,height=4.75cm]{t1f1.png}
%         \caption{$\tau=0.2$.}
%         \label{}
%     \end{subfigure}
%     \hfill
%     \begin{subfigure}[t]{0.45\textwidth}
%         \centering
%         \includegraphics[width=7cm,height=4.75cm]{t2f1.png}
%         \caption{$\tau=1$}
%         \label{}
%     \end{subfigure}
%     \caption{Bifurcation diagrams produced by solving \eqref{characfix} (fixed delay characteristic equation) for $\tau=0.2,1$ and $\epsilon^2=0.001$, on a domain length $L^2=9/2$.}
%     \label{fig:distheat1}
% \end{figure}
% \begin{figure}[H]
%     \centering
%     \begin{subfigure}[t]{0.45\textwidth}
%         \centering
%         \includegraphics[width=7cm,height=4.75cm]{distbif3.png}
%         \caption{$\tau=0.2$.}
%         \label{}
%     \end{subfigure}
%     \hfill
%     \begin{subfigure}[t]{0.45\textwidth}
%         \centering
%         \includegraphics[width=7cm,height=4.75cm]{distbif4.png}
%         \caption{$\tau=0.5$}
%         \label{}
%     \end{subfigure}
%     \caption{Bifurcation diagrams produced by solving \eqref{characfix} (fixed delay characteristic equation) for $\tau=0.2,0.5$ and $\epsilon^2=0.01$, on a domain length $L^2=9/2$.}
%     \label{fig:distheat2}
% \end{figure}

Observing the results in Table \ref{tab:tab1}, the largest absolute difference in $\max_k(\Re(\lambda_k))$ for all $\sigma$, $\tau$ and $\epsilon^2$ considered across the parameter space $(a,b)\in[0,1.4]\times[0,2]$ is $O(10^{-3})$. We therefore expect that for all $(a,b)\in[0,1.4]\times[0,2]$,
using a symmetric Gaussian distribution centred at some mean $\tau$ (for small $\tau$) will not significantly affect the time-taken until pattern formation compared to the fixed delay case, independent of the standard deviation $\sigma$ of the distribution. Absolute differences were considered rather than relative differences to avoid catastrophic cancelling.

\begin{table}[H]
\centering
\begin{tabular}{lrrrr}
\hline
\multicolumn{2}{c}{Parameters Used}    & $\sigma_{\max}\times0.99$ & $\sigma_{\max}\times0.2\ $ & $\sigma_{\max}\times0.1\ $ \\ \hline
$\epsilon^2=0.001$ & \textbf{$\tau=0.2$} & $0.0010$                           & $4.2\times10^{-5}$                & $1.1\times10^{-5}$                \\
$\epsilon^2=0.001$ & $\tau=1.0$          & $0.0078$                           & $3.3\times10^{-4}$                & $8.2\times10^{-5}$                \\
$\epsilon^2=0.01$  & \textbf{$\tau=0.2$} & $0.0025$                           & $9.4\times10^{-5}$                & $2.3\times10^{-5}$                \\
$\epsilon^2=0.01$  & \textbf{$\tau=0.5$} & \textbf{$0.0076$}                  & $2.6\times10^{-4}$                & $6.4\times10^{-5}$               \\ \hline
\end{tabular}
\caption{Table showing $\max_{(a,b)}$ of absolute difference of $\max_k(\Re(\lambda_k))$ between distributed delay cases and fixed delay case, across the $(a,b)\in[0,1.4]\times[0,2]$ parameter space, for multiple $\tau$ and $\epsilon^2$ values. $L^2=9/2$ used. Results displayed to $2 s.f.$}
\label{tab:tab1}
\end{table}

\subsection{Numerical Results}\label{section:distsim}
\begin{figure}[H]
    \centering
    \begin{subfigure}[t]{0.45\textwidth}
        \centering
        \includegraphics[width=7cm,height=5cm]{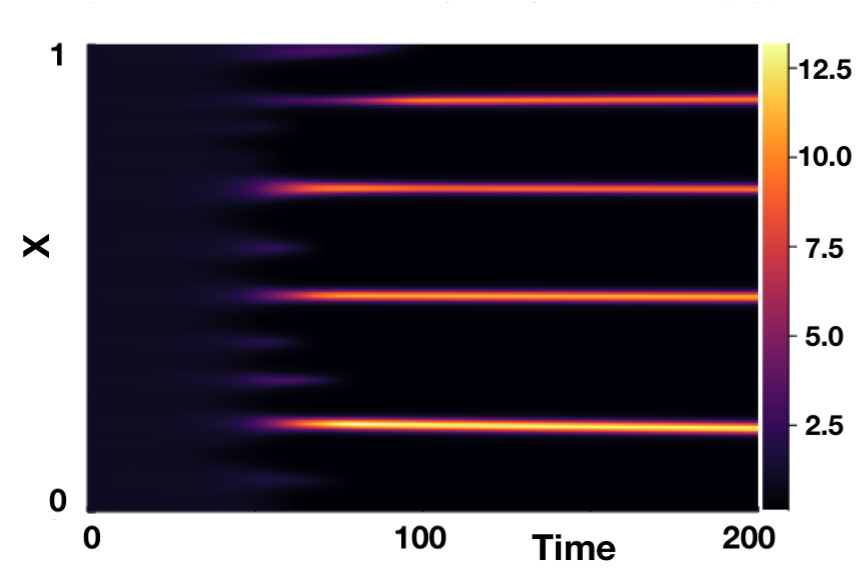}
        \caption{Numerical solution with $\tau=1$ and $\sigma=\sigma_{\max}\times0.99$.}
        \label{}
    \end{subfigure}
    \hfill
    \begin{subfigure}[t]{0.45\textwidth}
        \centering
        \includegraphics[width=7cm,height=5cm]{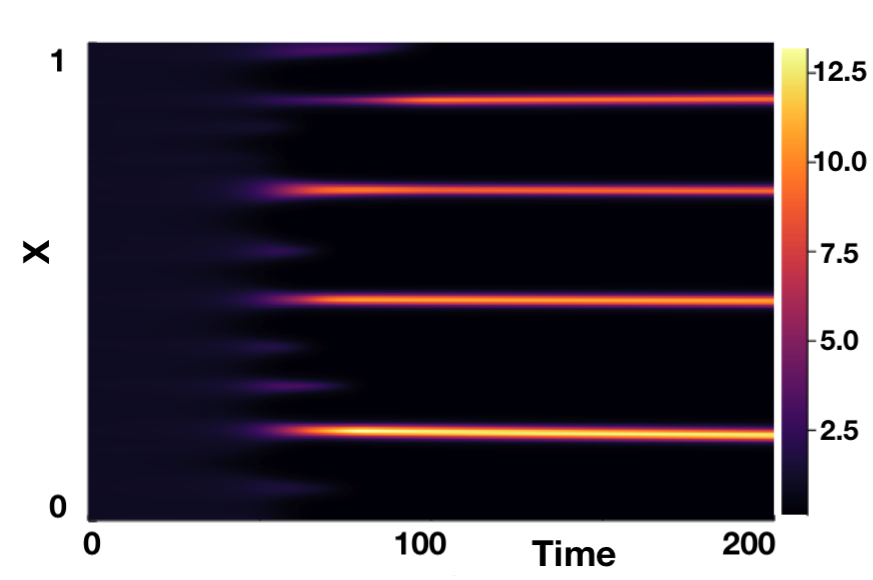}
        \caption{Numerical solution with $\tau=1$ and $\sigma=\sigma_{\max}\times0.1$.}
        \label{}
    \end{subfigure}
    \caption{Numerical solutions produced for $(a,b)=(0.1,0.9)$ with $\tau=1$ and $\sigma=\sigma_{\max}\times0.99, \sigma_{\max}\times0.1$. We use $L^2=9/2$ and $\epsilon^2=0.001$.
    Boundary conditions given by \eqref{neumannbc} and initial conditions by \eqref{firstic}. We see Turing pattern formation, as predicted from linear theory.}
    \label{fig:testdist1}
\end{figure}

\begin{figure}[H]
    \centering
    \begin{subfigure}[t]{0.45\textwidth}
        \centering
        \includegraphics[width=7cm,height=5cm]{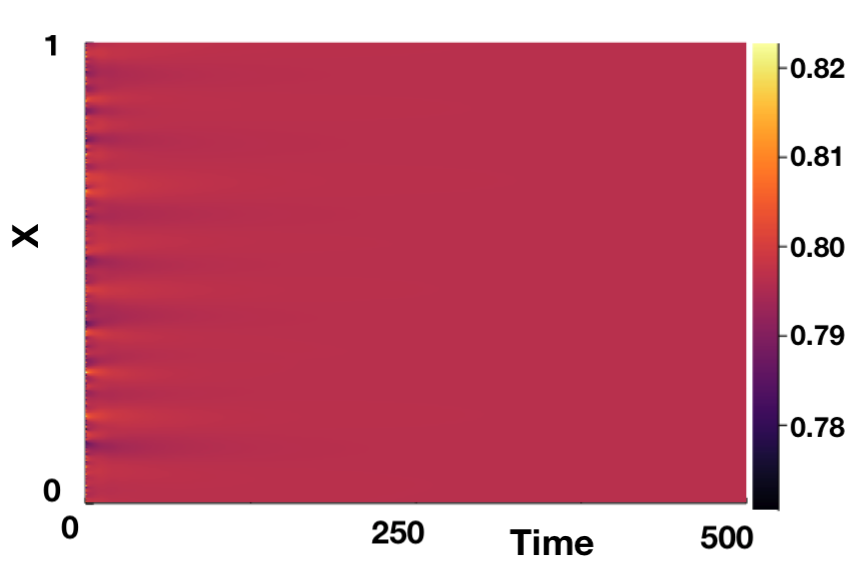}
        \caption{Numerical solution with $\tau=1$ and $\sigma=\sigma_{\max}\times0.99$.}
        \label{}
    \end{subfigure}
    \hfill
    \begin{subfigure}[t]{0.45\textwidth}
        \centering
        \includegraphics[width=7cm,height=5cm]{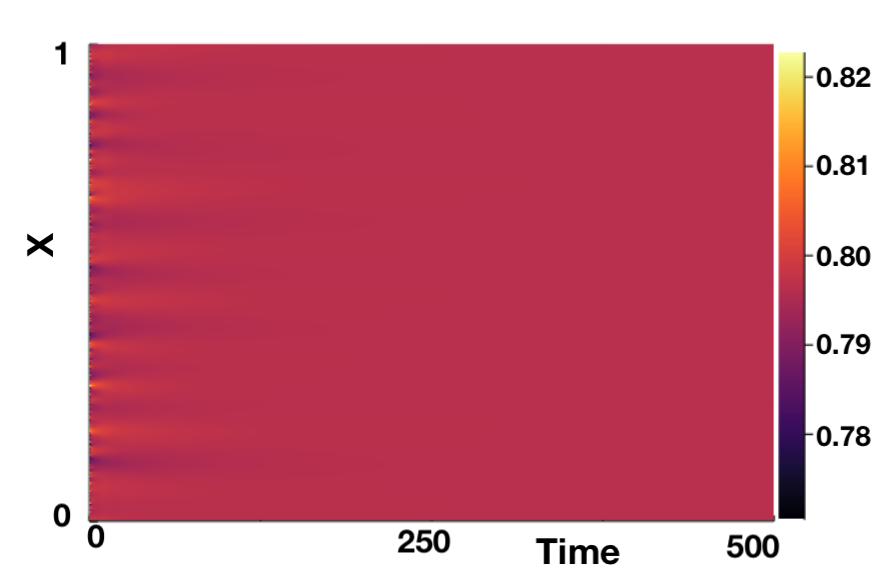}
        \caption{Numerical solution with $\tau=1$ and $\sigma=\sigma_{\max}\times0.1$.}
        \label{}
    \end{subfigure}
    \caption{Numerical solutions produced for $(a,b)=(0.4,0.4)$ with $\tau=1$ and $\sigma=\sigma_{\max}\times0.99, \sigma_{\max}\times0.1$. We use $L^2=9/2$ and $\epsilon^2=0.001$. Boundary conditions given by \eqref{neumannbc} and initial conditions by \eqref{firstic}. We see no Turing pattern formation, as predicted from linear theory.}
    \label{fig:testdist2}
\end{figure}

Numerical simulations are shown here to verify the linear theory presented in section \ref{section:distlin} and to explore the quantitative impacts of distributed delay beyond where the linear theory is valid. We first confirm that the results obtained in Figures \ref{fig:p2} and \ref{fig:p3} are accurate, namely that we find pattern formation for $(a,b)=(0.1,0.9)$ but not for $(a,b)=(0.4,0.4)$, independent of the $\tau\in[0,1]$ and $\sigma$ values considered. We also verify our main result, that modelling time delay as a symmetric Gaussian distribution will not quantitatively change the results seen from that of a fixed delay, independent of the $\sigma$ used. Figures \ref{fig:testdist1} and \ref{fig:testdist2} show the numerical solutions for $(a,b)=\{(0.1,0.9),(0.4,0.4)\}$ for $\tau=1$ and varying $\sigma$. Further numerical results with different $\tau$ and $\sigma$ values can be found in Appendix \ref{section:appB}.

\begin{figure}[H]
    \centering
    \begin{subfigure}[t]{0.32\textwidth}
        \centering
        \includegraphics[width=5cm,height=4.5cm]{ic21.png}
        \caption{Fixed delay model given by \eqref{fixed2}.}
        \label{}
    \end{subfigure}
    \hfill
    \begin{subfigure}[t]{0.32\textwidth}
        \centering
        \includegraphics[width=5cm,height=4.5cm]{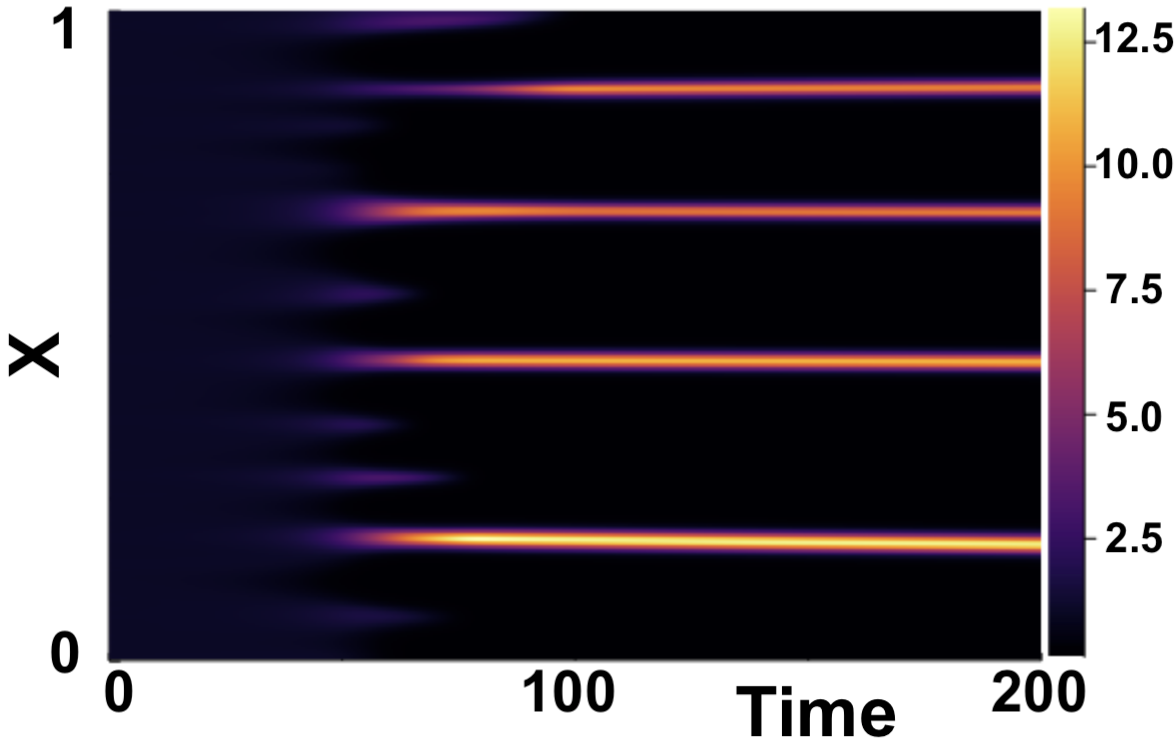}
        \caption{Distributed delay model, \eqref{symmod}, with $\sigma=\sigma_{\max}\times0.99$.}
        \label{}
    \end{subfigure}
    \hfill
    \begin{subfigure}[t]{0.32\textwidth}
        \centering
        \includegraphics[width=5cm,height=4.5cm]{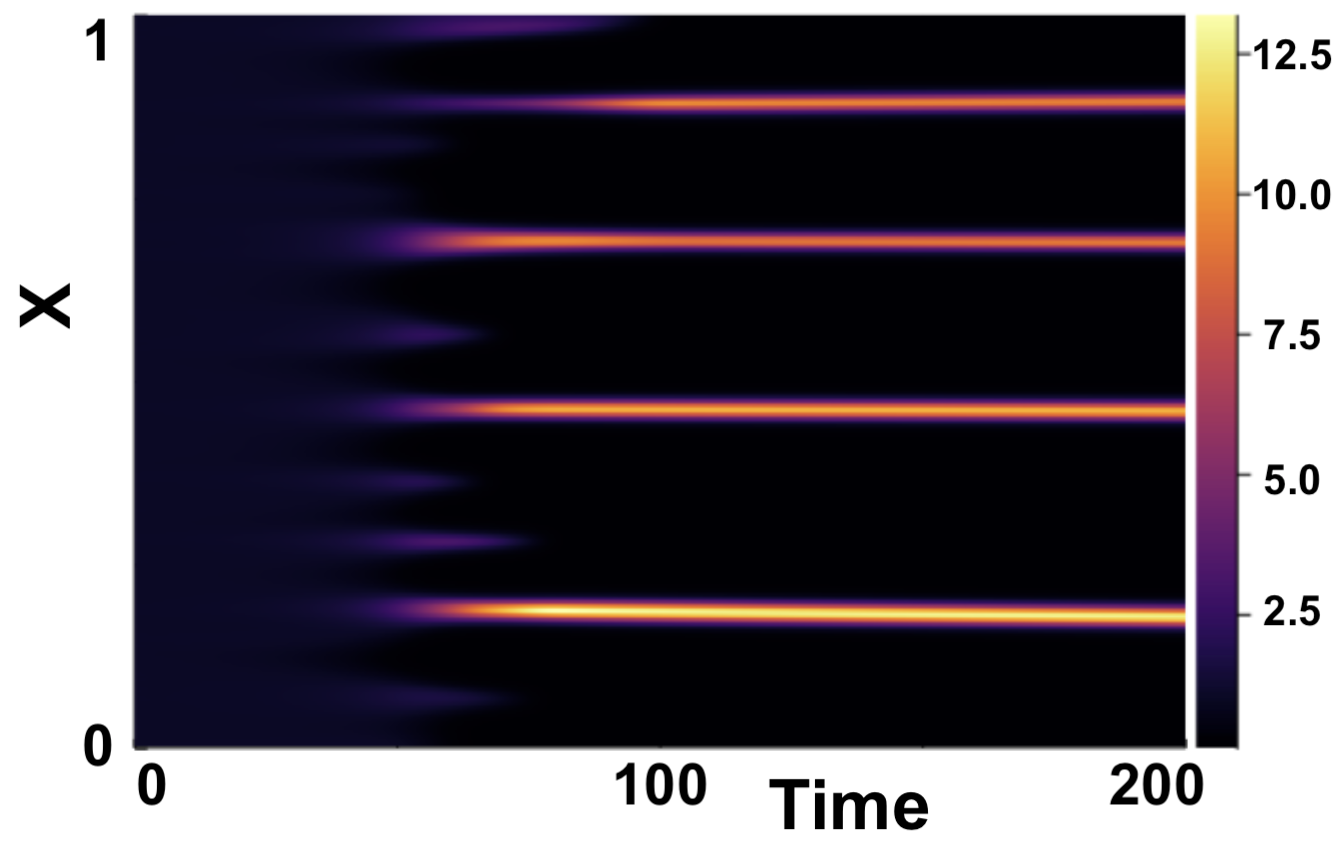}
        \caption{Distributed delay model, \eqref{symmod}, with $\sigma=\sigma_{\max}\times0.1$.}
        \label{}
    \end{subfigure}
    \caption{Numerical simulations showing comparison of fixed delay case vs distributed delay case for $\tau=1$. Boundary conditions given by \eqref{neumannbc} and initial conditions by \eqref{firstic}. $(a,b)=(0.1,0.9)$, $\epsilon^2=0.001$, $L^2=9/2$. }
    \label{fig:distres1}
\end{figure}

\begin{figure}[H]
    \centering
    \begin{subfigure}[t]{0.32\textwidth}
        \centering
        \includegraphics[width=5cm,height=4.5cm]{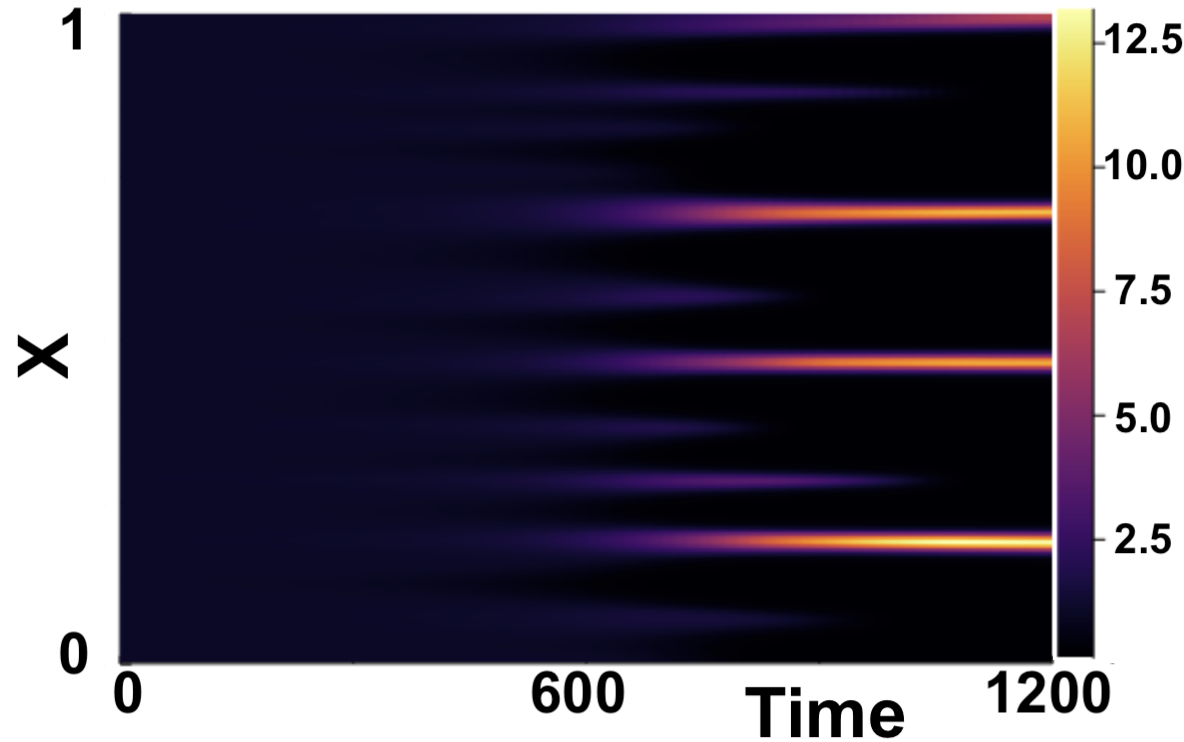}
        \caption{Fixed delay model given by \eqref{fixed2}.}
        \label{}
    \end{subfigure}
    \hfill
    \begin{subfigure}[t]{0.32\textwidth}
        \centering
        \includegraphics[width=5cm,height=4.5cm]{distt16sig10.png}
        \caption{Distributed delay model, \eqref{symmod}, with $\sigma=\sigma_{\max}\times0.99$.}
        \label{}
    \end{subfigure}
    \hfill
    \begin{subfigure}[t]{0.32\textwidth}
        \centering
        \includegraphics[width=5cm,height=4.5cm]{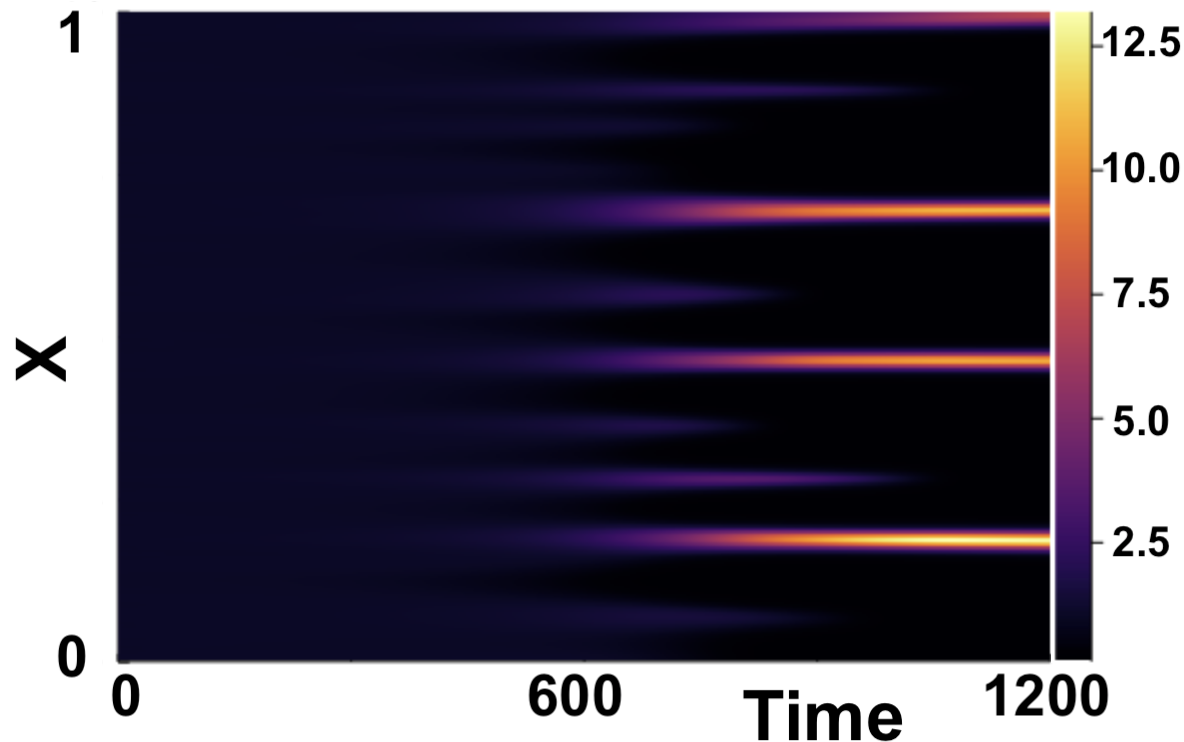}
        \caption{Distributed delay model, \eqref{symmod}, with $\sigma=\sigma_{\max}\times0.1$.}
        \label{}
    \end{subfigure}
    \caption{Numerical simulations showing comparison of fixed delay case vs distributed delay case for $\tau=16$. Boundary conditions given by \eqref{neumannbc} and initial conditions by \eqref{firstic}. $(a,b)=(0.1,0.9)$, $\epsilon^2=0.001$, $L^2=9/2$.}
    \label{fig:distres2}
\end{figure}

Figures \ref{fig:distres1} and \ref{fig:distres2} show numerical solutions using $(a,b)=(0.1,0.9)$ for $\tau=\{1,16\}$ and varying $\sigma$, each compared with the appropriate fixed delay case. The results indicate that the onset of patterning, and the type of pattern we see, is independent of $\sigma$ used. Further numerical solutions for different $(a,b)$ verifying this claim can be found in Appendix \ref{section:Bdist}.


\section{An Asymmetric Distribution}
\subsection{Introduction}
The results in section \ref{section:symmetric} suggest that using a symmetric Gaussian distribution does not have a quantitative effect on the results seen compared to that of the fixed delay case, and thus does not impact the increased time-to-pattern caused by introducing a fixed time delay. We therefore consider how an asymmetric distribution, specifically a skewed truncated Gaussian distribution, affects the results compared to that of a fixed delay. Using the results in \cite{skewed}, and letting $\textbf{p}=(\mu,\omega,\rho)$, the probability density function of the skewed truncated Gaussian distribution, $k(s;\mu,\omega,\rho)$, for some location $\mu$ and scaling $\omega$, is given by
\begin{equation}
    k(s;\mu,\omega,\rho)=\frac{\Psi_c}{\omega}\sqrt{\frac{2}{\pi}}\exp\left(-\frac{1}{2}\left(\frac{s-\mu}{\omega}\right)^2\right)\phi\left(\rho\frac{s-\mu}{\omega}\right),
\end{equation}
where $\phi(x)$ is the same as defined \eqref{phi}. The new parameter $\rho$ is used to denote the skew factor. The distribution is negatively skewed for $\rho<0$ and positively skewed for $\rho>0$. Finally, we have that $\Psi_c$ is the truncation scaling constant. This is given as
\begin{equation}
    \Psi_c=\frac{1}{F\left(\frac{b-\mu}{\omega},\rho\right)-F\left(\frac{a-\mu}{\omega},\rho\right)},
\end{equation}
with $F(x,\rho)$ the cdf of a skewed Gaussian distribution, described by
\begin{equation}
    F(x,\rho)=\phi(x)-2T(x,\rho).
\end{equation}
The function $T(x,\rho)$ denotes the Owen's T function \cite{owenst} and is written as an integral in the form
\begin{equation}
    T(x,\rho)=\frac{1}{2\pi}\int_0^\rho\frac{e^{-\frac{1}{2}x^2(1+s^2)}}{1+s^2}\ \text{d}s\quad -\infty<x,\rho<\infty.
\end{equation}
In the computational implementation of the skewed truncated Gaussian pdf, the integral $T(x,\rho)$ is resolved numerically using the composite Simpson's rule with $100,000$ discretisation points.

Since the distribution is skewed, the parameters $\mu$ and $\omega$ no longer denote the mean and standard deviation of the distribution, but solely the location and scale of the distribution. To compare how the skewed distribution affects the onset of patterning compared to that of the fixed delay case, we consider the mean of the skewed distribution, $\tau$, which is given by
\begin{equation}\label{anmean}
    \tau=\int_a^bs\ k(s;\mu,\omega,\rho)\ \ \text{d}s.
\end{equation}
Following \cite{skewed}, the mean of the skewed truncated Gaussian distribution is computed as
\begin{equation}\label{computetau}
\tau=\mu+\omega\Psi_c\left[k(a;\mu,\omega,\rho)-k(b;\mu,\omega,\rho)+\frac{2\rho}{\hat{\rho}\sqrt{2\pi}}\left(\phi\left(\hat{\rho}\frac{b-\mu}{\omega}\right)-\phi\left(\hat{\rho}\frac{a-\mu}{\omega}\right)\right)\right],
\end{equation}
with $\hat{\rho}=\left(1+\rho^2\right)^{1/2}$. See Appendix \ref{section:appA} for a detailed derivation.

Throughout this section, the integration limits were set to $a=\mu-3\omega$, $b=\mu+3\omega$, where $\omega$ was chosen such that $\omega<\omega_{\max}$, with $\omega_{\max}=\frac{\mu}{3}$ to ensure only positive time delays were considered.
\subsection{Linear Analysis}\label{section:linanalskew}
Conducting an analogous linear analysis to that of the symmetric distributed delay case, we find that the characteristic equation when a skewed distribution is being used is given as
\begin{equation}\label{characskew}
  \mathcal{D}_k=\lambda_k^2+\alpha_k\lambda_k+\beta_k+(\gamma_k\lambda_k+\delta_k)\hat{E}_k=0,
\end{equation}
where $\alpha_k$, $\beta_k$, $\gamma_k$ and $\delta_k$ are the same coefficients as for the symmetric distribution case, also defined in \eqref{fixcoeffs}. The difference is in expression $\hat{E}_k$, which is given by
\begin{equation}\label{Ehat}
    \begin{split}
\hat{E}_k&=\int_a^bk(s;\mu,\omega,\rho)e^{-\lambda_k s}\ \text{d}s\\
&=\frac{\Psi_c}{\omega\sqrt{2\pi}}\int_a^b\left(1+\text{erf}\left(\rho\frac{s-\mu}{\omega\sqrt{2}}\right)\right)\exp\left(-\frac{1}{2}\left(\frac{s-\mu}{\omega}\right)^2-\lambda_ks\right)\ \text{d}s.
    \end{split}
\end{equation}
We numerically compute this integral using the composite Simpson's rule with $100,000$ discretisation points. This allows roots of the characteristic equation \eqref{characskew} to be solved for.

\begin{figure}[H]
    \centering
    \begin{subfigure}[t]{0.45\textwidth}
        \centering
        \includegraphics[width=7cm,height=5cm]{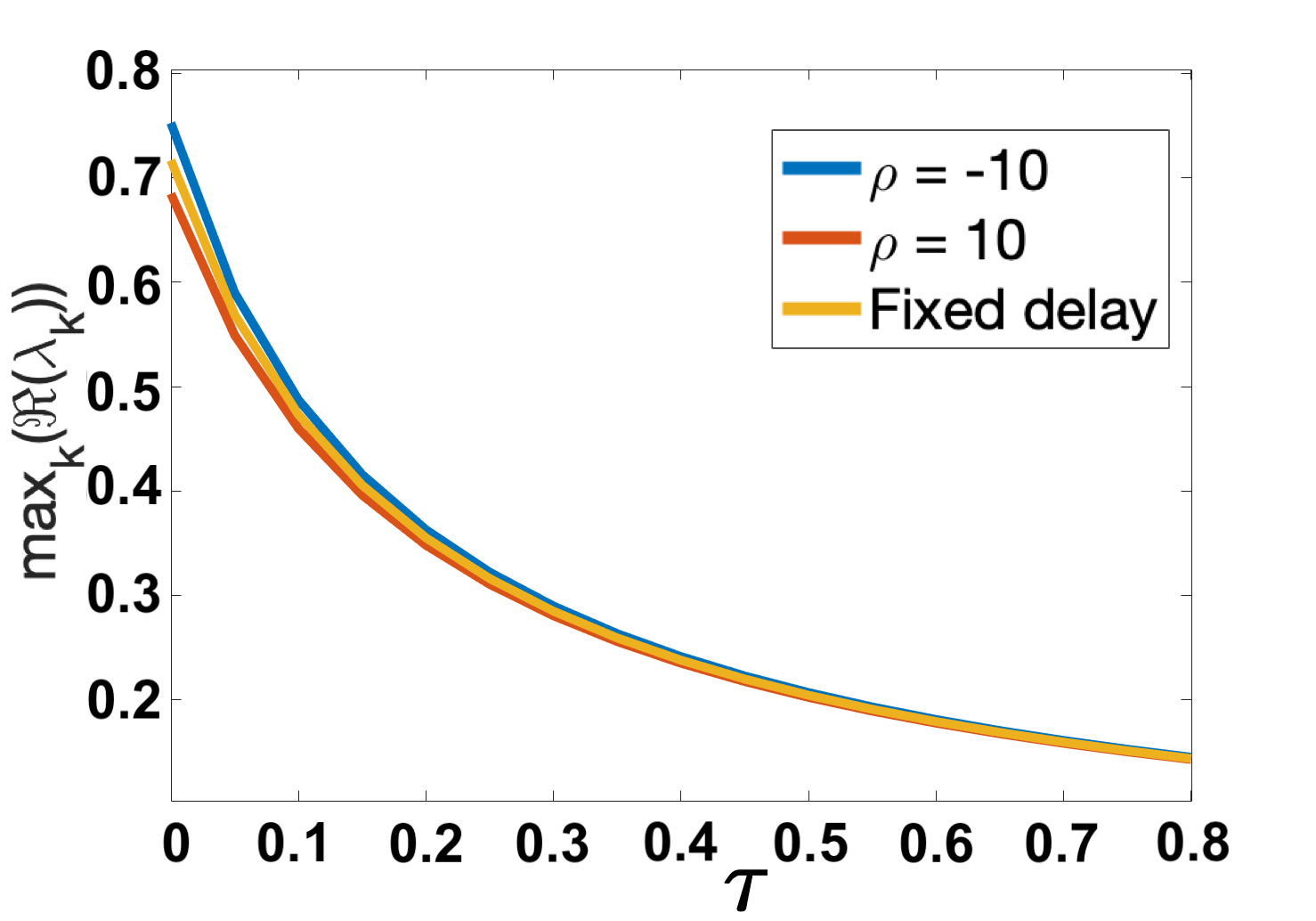}
        \caption{$(a,b)=(0.1,0.9)$.}
        \label{}
    \end{subfigure}
    \hfill
    \begin{subfigure}[t]{0.45\textwidth}
        \centering
        \includegraphics[width=7cm,height=5cm]{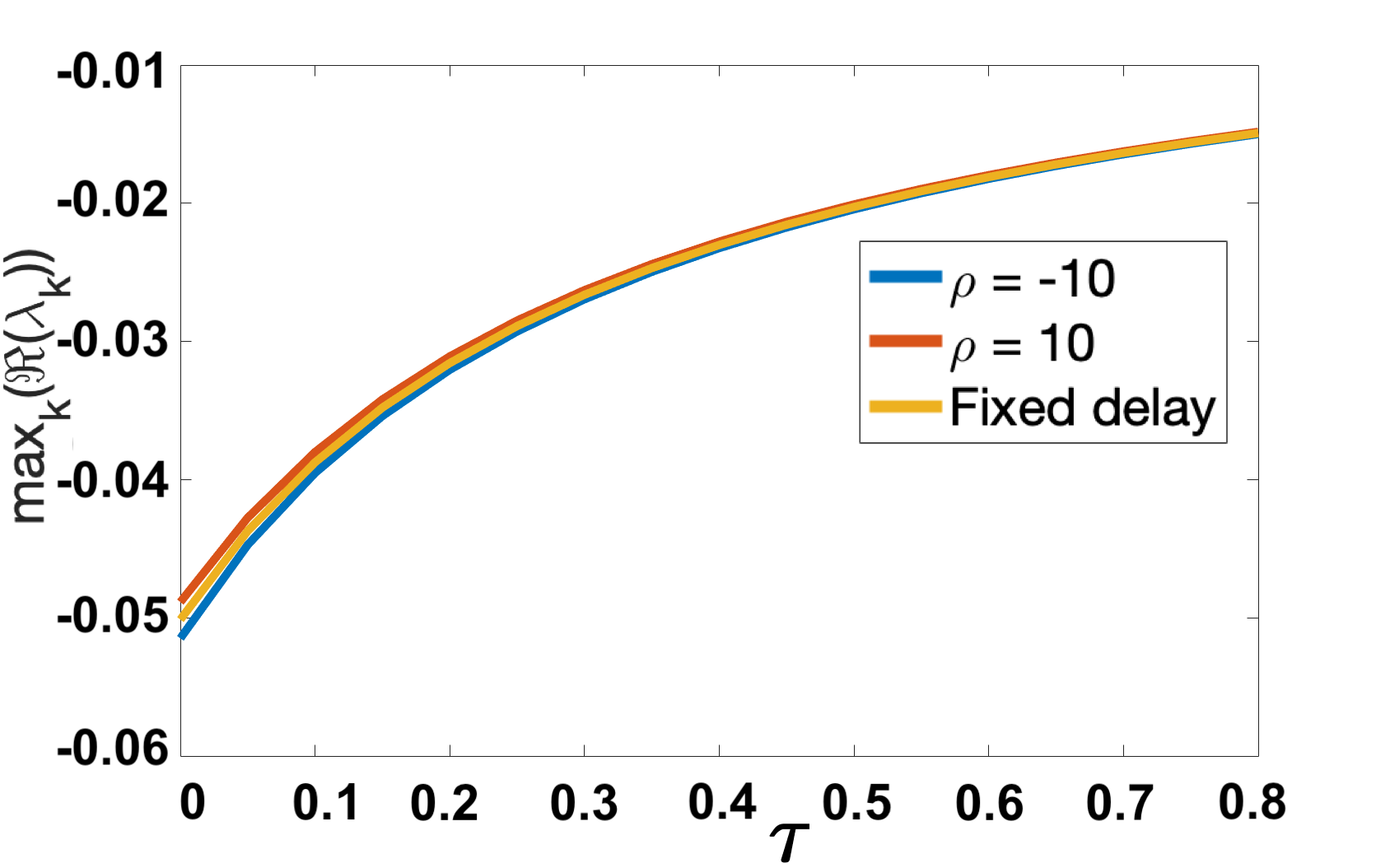}
        \caption{$(a,b)=(0.4,0.4)$.}
        \label{}
    \end{subfigure}
    \caption{Comparison of $\max_k(\Re(\lambda_k))$ plotted against $\tau\in[0,0.8]$ for $\rho=-10,10$ against fixed delay case. Parameter values $\epsilon^2=0.001$ and $L^2=9/2$ used. $\tau$ varied at regular intervals of $0.05$. $k\in\mathbb{Z}$ ranging over $k\in[0,50]$.}
    \label{fig:dispskew}
\end{figure}

Here we present results of the $\max_k(\Re(\lambda_k))$ for varying $\tau$ with a skewed distribution. Namely, we show that for a small mean $\tau$, the skew, positive or negative, does not significantly effect the $\max_k(\Re(\lambda_k))$. We note that, for a given $\rho$, all of the terms on the right-hand side of \eqref{computetau}, namely $\omega$, $\Psi_c$,  and $k(s;\mu,\omega,\rho)$, can be written explicitly in terms of $\mu$. Equation \eqref{computetau} can therefore be solved implicitly for $\mu(\tau)$, for a given $\tau$, using the \textit{fzero} command in MATLAB. For a given $\rho$, and each found $\mu$, we compute $\max_k(\Re(\lambda_k))$ by solving for roots of the characteristic equation \eqref{characskew}. In Figure \ref{fig:dispskew}, we plot $\max_k(\Re(\lambda_k))$ against $\tau\in[0,0.8]$ for skew parameter values of $\rho=-10,10$, and $\omega=\omega_{\max}\times0.99$, with two different $(a,b)$ parameter sets. A plot of $\max_k(\Re(\lambda_k))$ for the fixed delay case is also added for comparison in each case.

From Figure \ref{fig:dispskew} we see that the curves differ slightly for small $\tau$, with $\rho=-10$ having a slightly higher value of the maximum growth rate, and $\rho=10$ a slightly lower value. The overall effect is very small despite the large skew implemented in the distribution. By comparing the results in Figure \ref{fig:dispskew} to those in \ref{fig:p2} and \ref{fig:p3} (dispersion curves for the symmetric distribution vs fixed delay case), we see that the skewed distributions have a larger, but still small, effect on the dispersion curves. We suspect that these effects are still small enough not to have a significant impact on the timescale on which onset of patterning occurs. Numerical simulations confirming these findings from the linear theory for a small are considered in Section \ref{section:numresskew}.

\subsection{Numerical Results}\label{section:numresskew}

\begin{figure}[H]
    \centering
    \begin{subfigure}[t]{0.45\textwidth}
        \centering
        \includegraphics[width=7cm,height=4.75cm]{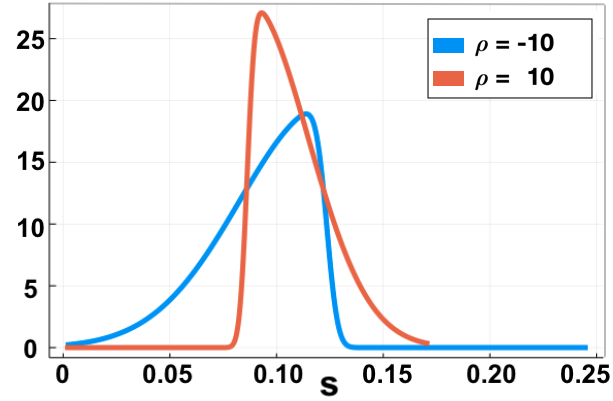}
        \caption{pdfs of skewed truncated Gaussian distributions, with $\rho=-10,10$. Both pdfs have mean $\tau=0.1$.}
        \label{}
    \end{subfigure}
    \hfill
    \begin{subfigure}[t]{0.45\textwidth}
        \centering
        \includegraphics[width=7cm,height=4.75cm]{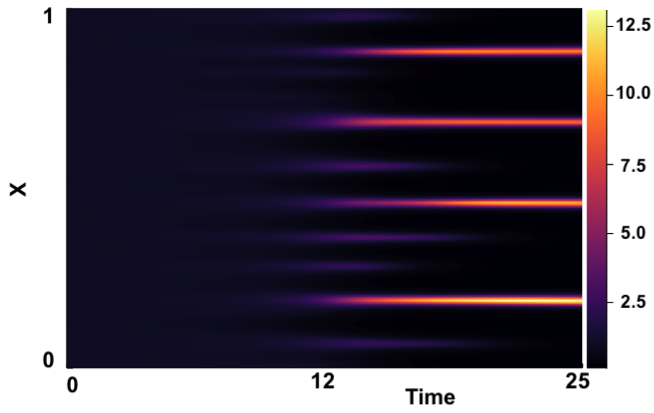}
        \caption{Numerical simulation of fixed delay case with $\tau=0.1$.}
        \label{}
    \end{subfigure}
    \hfill
    \begin{subfigure}[t]{0.45\textwidth}
        \centering
        \includegraphics[width=7cm,height=4.75cm]{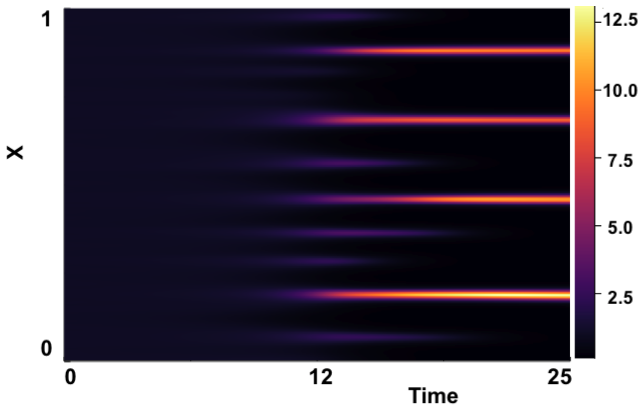}
        \caption{Numerical simulation with skewed distribution of $\rho=-10$. Distribution parameters are $\mu=0.124(3 s.f.)$ and $\omega=0.0408(3 s.f.)$.}
        \label{fig:rhom10}
    \end{subfigure}
    \hfill
    \begin{subfigure}[t]{0.45\textwidth}
        \centering
        \includegraphics[width=7cm,height=4.75cm]{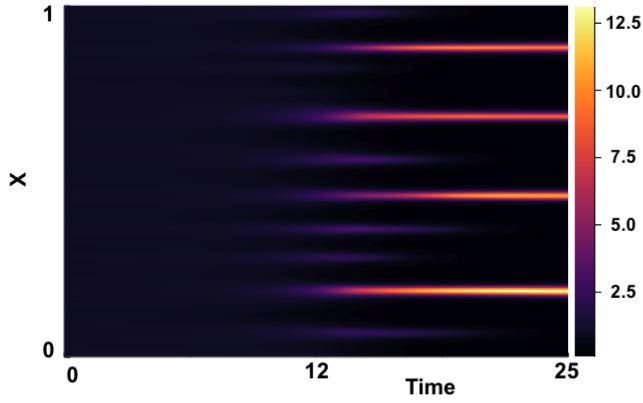}
        \caption{Numerical simulation with skewed distribution of $\rho=10$. Distribution parameters are $\mu=0.0863(3 s.f.)$ and $\omega=0.0285(3 s.f.)$.}
        \label{fig:rho10}
    \end{subfigure}
    \caption{Numerical results for $(a,b)=(0.1,0.9)$ with $\rho=-10,10$ and $\tau=0.1$. Parameters $\epsilon^2=0.01$ and $L^2=9/2$. Initial conditions given by \eqref{firstic} and boundary conditions by \eqref{neumannbc}.}
    \label{fig:linskew1}
\end{figure}

Numerical simulations confirming the linear theory for a small $\tau=0.1$ and $(a,b)=(0.1,0.9)$ can be seen in Figure \ref{fig:linskew1}, where we see a very minor variation in the onset of patterning between the $\rho=-10$ and $\rho=10$ cases (\ref{fig:rhom10} and \ref{fig:rho10}).

We look to verify the results suggested by the linear theory, and show through full numerical solutions that for a larger $\tau$ the results we see are consistent with the finding from linear theory for a smaller $\tau$. Namely, we show that using a skewed truncated Gaussian distribution does not have a significant effect on the onset of patterning. Using an analogous methodology as outlined in the previous section \ref{section:linanalskew}, we present results for $\tau=16$, and $\rho=-10,10$ for a fixed parameter set $(a,b)=(0.1,0.9)$, with a comparison to the fixed delay case. The result for $\tau=16$ is shown in Figure \ref{fig:linskew3}. Further numerical solutions for varying $\tau\in\{1,2,4,8\}$ can be found in Appendix \ref{section:Bdist}.

% \begin{figure}[H]
%     \centering
%     \begin{subfigure}[t]{0.45\textwidth}
%         \centering
%         \includegraphics[width=7cm,height=5cm]{distskew8.png}
%         \caption{pdfs of skewed truncated Gaussian distributions, with $\rho=-10,10$. Both pdfs have mean $\tau=8$.}
%         \label{}
%     \end{subfigure}
%     \hfill
%     \begin{subfigure}[t]{0.45\textwidth}
%         \centering
%         \includegraphics[width=7cm,height=5cm]{fixt8.png}
%         \caption{Numerical simulation of fixed delay case with $\tau=8$.}
%         \label{}
%     \end{subfigure}
%     \hfill
%     \begin{subfigure}[t]{0.45\textwidth}
%         \centering
%         \includegraphics[width=7cm,height=5cm]{skewt8m10.png}
%         \caption{Numerical simulation with skewed distribution of $\rho=-10$. Distribution parameters are $\mu=10.8(3 s.f.)$ and $\omega=3.58(3 s.f.)$.}
%         \label{}
%     \end{subfigure}
%     \hfill
%     \begin{subfigure}[t]{0.45\textwidth}
%         \centering
%         \includegraphics[width=7cm,height=5cm]{skewt810.png}
%         \caption{Numerical simulation with skewed distribution of $\rho=10$. Distribution parameters are $\mu=6.34(3 s.f.)$ and $\omega=2.09(3 s.f.)$.}
%         \label{}
%     \end{subfigure}
%     \caption{Numerical rsesults for $(a,b)=(0.1,0.9)$ with $\rho=-10,10$ and $\tau=8$. Parameters $\epsilon^2=0.01$ and $L^2=9/2$. Initial conditions given by \eqref{firstic} and boundary conditions by \eqref{neumannbc}.}
%     \label{fig:linskew2}
% \end{figure}

\begin{figure}[H]
    \centering
    \begin{subfigure}[t]{0.45\textwidth}
        \centering
        \includegraphics[width=7cm,height=5cm]{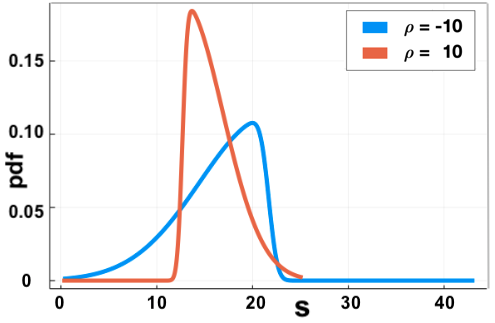}
        \caption{pdfs of skewed truncated Gaussian distributions, with $\rho=-10,10$. Both pdfs have mean $\tau=16$.}
        \label{}
    \end{subfigure}
    \hfill
    \begin{subfigure}[t]{0.45\textwidth}
        \centering
        \includegraphics[width=7cm,height=5cm]{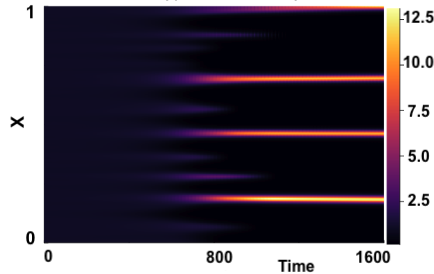}
        \caption{Numerical simulation of fixed delay case with $\tau=16$.}
        \label{}
    \end{subfigure}
    \hfill
    \begin{subfigure}[t]{0.45\textwidth}
        \centering
        \includegraphics[width=7cm,height=5cm]{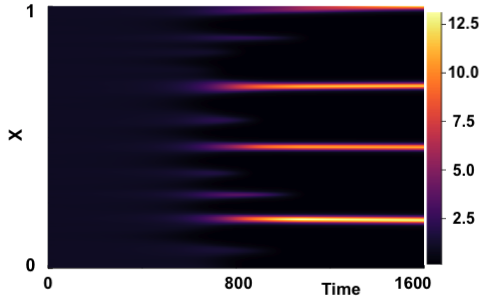}
        \caption{Numerical simulation with distribution of $\rho=-10$. Distribution parameters are $\mu=21.7(3 s.f.)$ and $\omega=7.16(3 s.f.)$.}
        \label{}
    \end{subfigure}
    \hfill
    \begin{subfigure}[t]{0.45\textwidth}
        \centering
        \includegraphics[width=7cm,height=5cm]{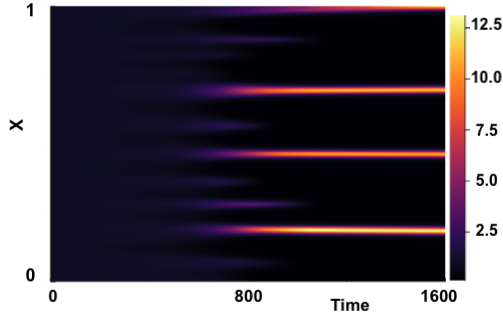}
        \caption{Numerical simulation with distribution of $\rho=10$. Distribution parameters are $\mu=12.7(3 s.f.)$ and $\omega=4.18(3 s.f.)$.}
        \label{}
    \end{subfigure}
    \caption{Numerical results for $(a,b)=(0.1,0.9)$ with $\rho=-10,10$ and $\tau=16$. Parameters $\epsilon^2=0.01$ and $L^2=9/2$. Initial conditions given by \eqref{firstic} and boundary conditions by \eqref{neumannbc}.}
    \label{fig:linskew3}
\end{figure}

\section{Summary}

In this chapter, we performed a linear stability analysis of the LI model with distributed delay to determine analytically the effects of an increasing time delay on the Turing space, for both a symmetric and skewed truncated Gaussian distribution. Through numerical simulations, we also presented the findings that modelling the time delay as a continuous distribution does not significantly change the results seen compared to that of the fixed delay case, where the mean delay of the distribution is used as the fixed time delay. We therefore conclude that fixed delay and distributed delay models have almost identical dynamics. This allows one to use simpler fixed delay models rather than the more complicated distributed delay variants.

\chapter{Gierer-Meinhardt Model}

\begin{figure}[H]
    \centering
    \begin{subfigure}[t]{0.45\textwidth}
        \centering
        \includegraphics[width=7cm,height = 5cm]{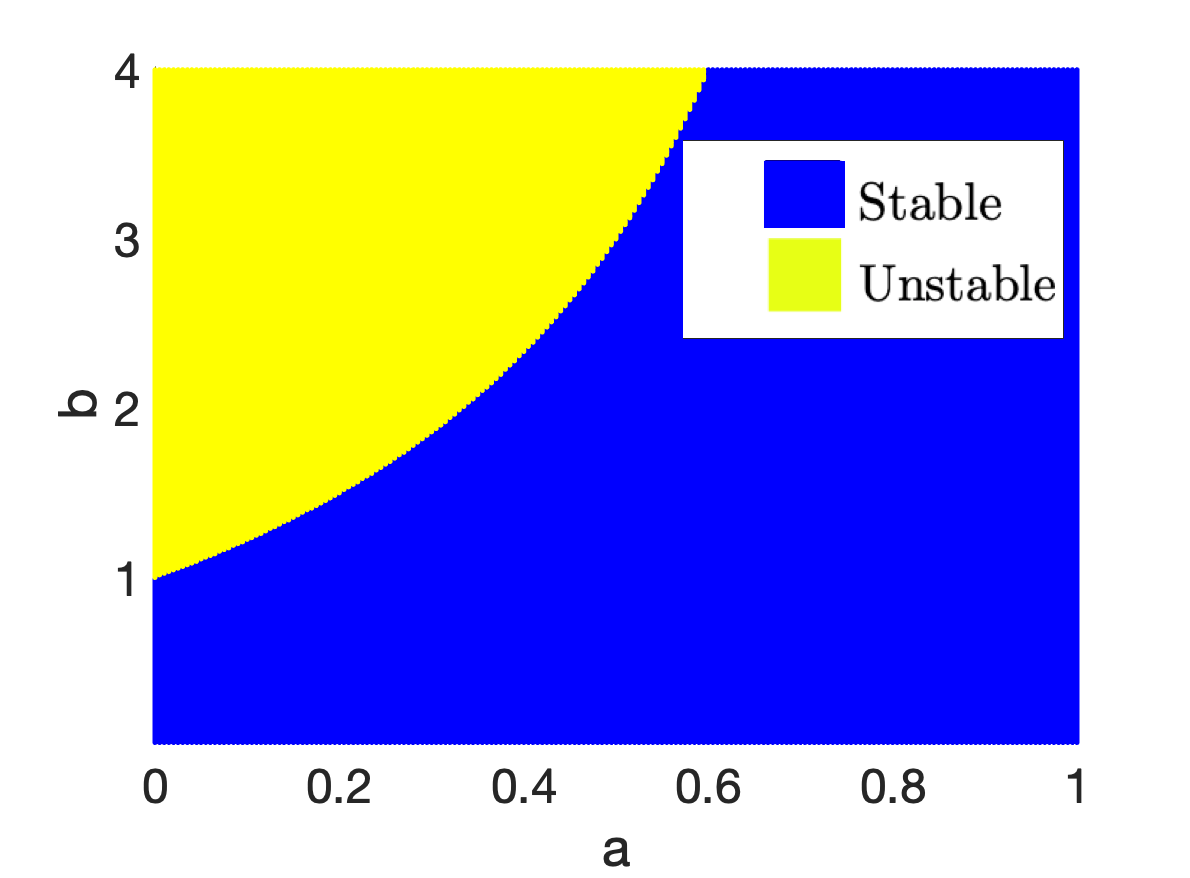}
        \caption{Bifurcation diagram for spatially homogeneous model, no delay.}
        \label{fig:bifgm}
    \end{subfigure}
    \hfill
    \begin{subfigure}[t]{0.45\textwidth}
        \centering
        \includegraphics[width=7cm,height = 5cm]{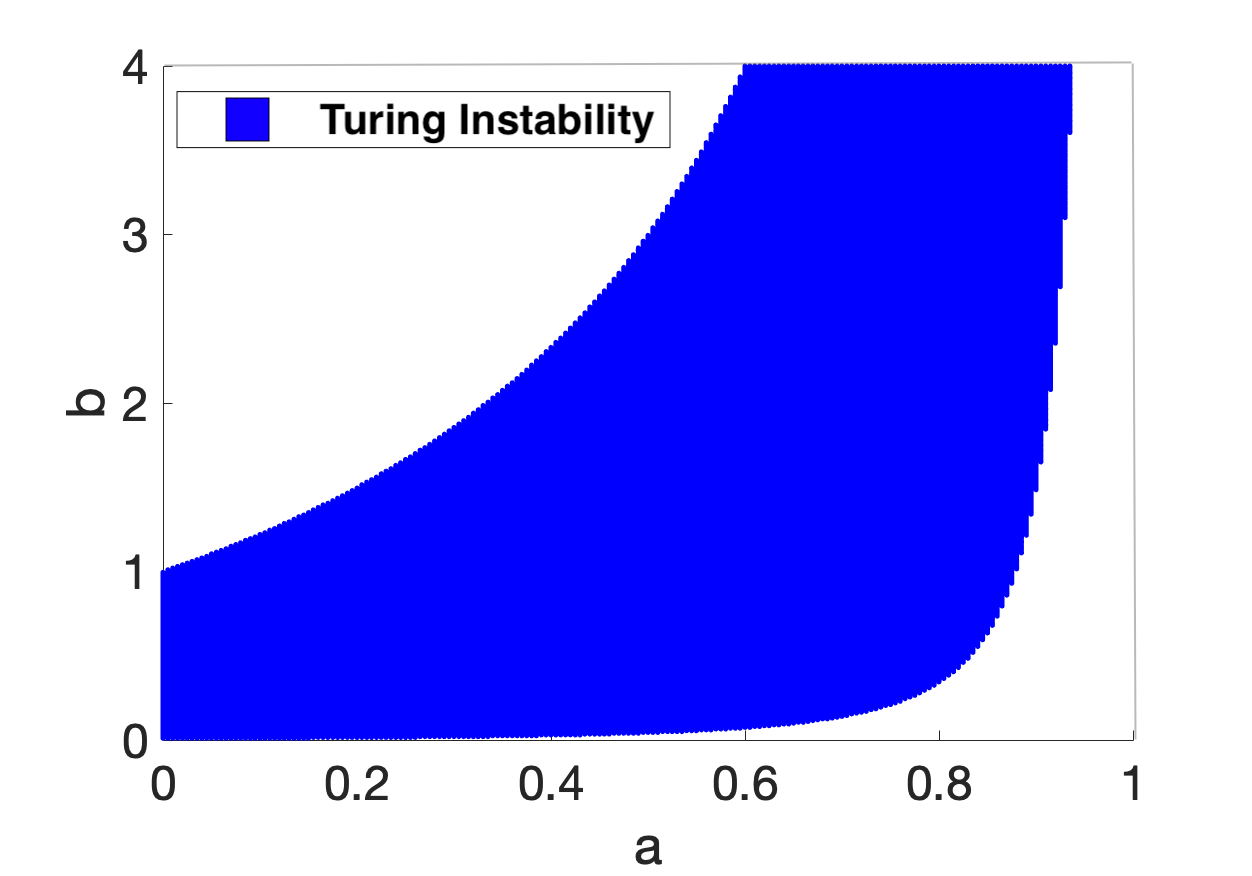}
        \caption{Turing space, no delay. $\epsilon^2=0.001$.}
        \label{fig:turingspacegm}
    \end{subfigure}
    \caption{Conditions \eqref{cond1} and \eqref{cond2} used to plot bifurcation diagram and Turing space for parameters $(a,b)\in[0,1]\times[0,4]$ for GM model.}
    \label{fig:gmspace}
\end{figure}

In this chapter, we introduce two fixed time delay variants of the Gierer-Meinhardt (GM) model. Through a similar linear analysis to that in Chapter \ref{section:fixdel} considered for the LI variant of the Schnakenberg model, we look to examine the effect of an increasing time delay on the Turing space for the GM model. The model descriptions we use here are motivated by the analysis conducted in \cite{fadai1,fadai2}. The results in these papers indicate that the placement of time-delayed terms in the model is of extreme importance in the context of pattern formation. A chemical interpretation of the kinetic reactions for the GM Model can be found in \cite{leegaffmonk}. The two non-dimensionalised model descriptions we consider, with kinetic reactions taken from \cite{murray}, and time-delayed terms motivated by \cite{fadai1} and \cite{fadai2}, are given by \eqref{fadai1} and \eqref{fadai2} respectively, and to be consistent with the notation in Chapter 2, we use $u$ and $v$ to denote the activator and inhibitor concentrations.\footnote{We note that the papers \cite{fadai1,fadai2} consider $v$ as the activator and $u$ as the inhibitor.}
\begin{multicols}{2}
\begin{equation}\label{fadai1}
  \left.\begin{split}
\frac{\partial u}{\partial t}&=\frac{\epsilon^2}{L^2}\frac{\partial^2 u}{\partial x^2}+a-bu+\frac{\hat{u}^2}{\hat{v}},\\
\frac{\partial v}{\partial t}&=\frac{1}{L^2}\frac{\partial^2 v}{\partial x^2}+\hat{u}^2-v,
\end{split}\right\}
\end{equation}
\break
\begin{equation}\label{fadai2}
  \left.\begin{split}
\frac{\partial u}{\partial t}&=\frac{\epsilon^2}{L^2}\frac{\partial^2 u}{\partial x^2}+a-bu+\frac{\hat{u}^2}{v},\\
\frac{\partial v}{\partial t}&=\frac{1}{L^2}\frac{\partial^2 v}{\partial x^2}+\hat{u}^2-v,
\end{split}\right\}
\end{equation}
\end{multicols}
We note the difference in the two models being the $v$ term in the activator's kinetics. The notation $\hat{u}$ and $\hat{v}$ denote the activator and inhibitor evaluated at some fixed time delay $\tau$, namely $\hat{u}=u(x,t-\tau)$, $\hat{v}=v(x,t-\tau)$. The results in \cite{fadai1,fadai2} showed that an increasing time delay in \eqref{fadai1} acted as an antagonistic effect, and shrunk the parameter space exhibiting stable spike solutions. In contrast, an increasing time delay incorporated as in \eqref{fadai2} caused an expansion of the stable spike solution parameter regime. In this chapter, we use linear analysis of the spatially homogeneous steady states to examine how an increasing time delay will affect the Turing space seen for each of these variants. Figure \ref{fig:gmspace} shows the stable and unstable parameter regimes as well as the Turing space for a $\tau=0$. The parameter space considered is $(a,b)\in[0,1]\times[0,4]$, and the unique steady state for the GM model is given as $(u_\star,v_\star)=\left(\frac{a+1}{b},\left(\frac{a+1}{b}\right)^2\right)$.

\section{Linear Analysis}

Using an analogous methodology to that in Chapter \ref{section:fixdel}, we take perturbations $u=u_\star+\delta\xi(x,t)$ and $v=v_\star+\delta\eta$ for $|\delta|\ll1$. The linearised dynamics of \eqref{fadai1} and \eqref{fadai2} are then respectively given by equations \eqref{linf1} and \eqref{linf2}.

\begin{multicols}{2}
\begin{equation}\label{linf1}
    \left.\begin{split}
\frac{\partial \xi}{\partial t}&=\frac{\epsilon^2}{L^2}\frac{\partial^2 \xi}{\partial x^2}-b\xi+2\frac{u_\star}{v_\star}\hat{\xi}-\frac{u_\star^2}{v_\star^2}\hat{\eta},\\
\frac{\partial \eta}{\partial t}&=\frac{1}{L^2}\frac{\partial^2 \eta}{\partial x^2}+2u_\star\hat{\xi}-\eta;
\end{split}\right\}
\end{equation}\break
\begin{equation}
    \left.\begin{split}
\frac{\partial \xi}{\partial t}&=\frac{\epsilon^2}{L^2}\frac{\partial^2 \xi}{\partial x^2}-b\xi+2\frac{u_\star}{v_\star}\hat{\xi}-\frac{u_\star^2}{v_\star^2}\eta,\\
\frac{\partial \eta}{\partial t}&=\frac{1}{L^2}\frac{\partial^2 \eta}{\partial x^2}+2u_\star\hat{\xi}-\eta.\label{linf2}
\end{split}\right\}
\end{equation}
\end{multicols}
with $\hat{\xi}=\xi(x,t-\tau)$ and $\hat{\eta}=\eta(x,t-\tau)$. Substituting into \eqref{linf1} and \eqref{linf2} an ansatz of the form $\begin{pmatrix}\xi\\ \eta\end{pmatrix}=\begin{pmatrix}\xi_0e^{\lambda_kt}\cos(k\pi x)\\\eta_0e^{\lambda_kt}\cos(k\pi x)\end{pmatrix}$ and simplifying, yields a homogeneous system for $(\xi_0,\eta_0)^T$ for each set of linearised dynamics. Finding non-trivial solutions for these systems results in characteristic equations. The characteristic equations, $\mathcal{D}_k=0$ and $\hat{\mathcal{D}}_k=0$, for the linearised dynamics in \eqref{linf1} and \eqref{linf2} respectively are given by
\begin{align}\label{characf1}
\mathcal{D}_k&=\lambda_k^2+\alpha_k\lambda+\beta_k+(\gamma_k\lambda_k+\delta_k)e^{-\lambda_k\tau}+\chi_ke^{-2\lambda_k\tau}=0,\\
\hat{\mathcal{D}}_k&=\lambda_k^2+\alpha_k\lambda+\beta_k+(\gamma_k\lambda_k+(\chi_k+\delta_k))e^{-\lambda_k\tau}=0.\label{characf2}
\end{align}
The coefficients for these characteristic equations are given by
\begin{equation}
    \begin{split}
\alpha_k&=\left(\frac{\epsilon^2}{L^2}+\frac{1}{L^2}\right)k^2\pi^2+b+1,\\
\beta_k&=\left(\frac{\epsilon^2}{L^2}\pi^2k^2+b\right)\left(\frac{1}{L^2}\pi^2k^2+1\right),\\
\gamma_k&=-2\frac{u_\star}{v_\star},\\
\delta_k&=-2\frac{u_\star}{v_\star}\left(\frac{1}{L^2}k^2\pi^2+1\right),\\
\chi_k&=2\frac{u_\star^3}{v_\star^2}.
    \end{split}
\end{equation}

\begin{figure}[H]
    \centering
    \begin{subfigure}[t]{0.32\textwidth}
        \centering
        \includegraphics[width=5.5cm,height = 5cm]{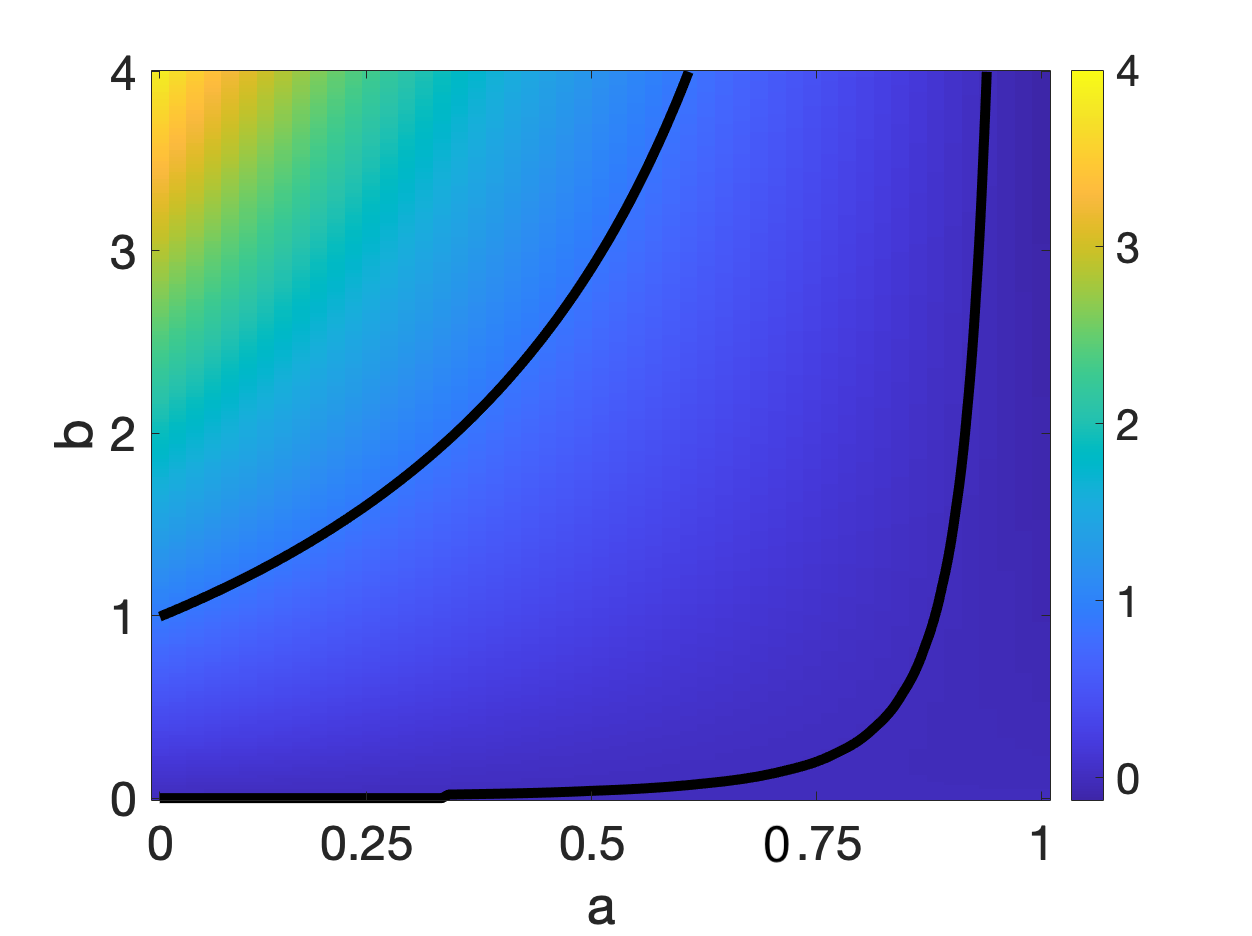}
        \caption{$\tau=0$.}
        \label{}
    \end{subfigure}
    \hfill
    \begin{subfigure}[t]{0.32\textwidth}
        \centering
        \includegraphics[width=5.5cm,height = 5cm]{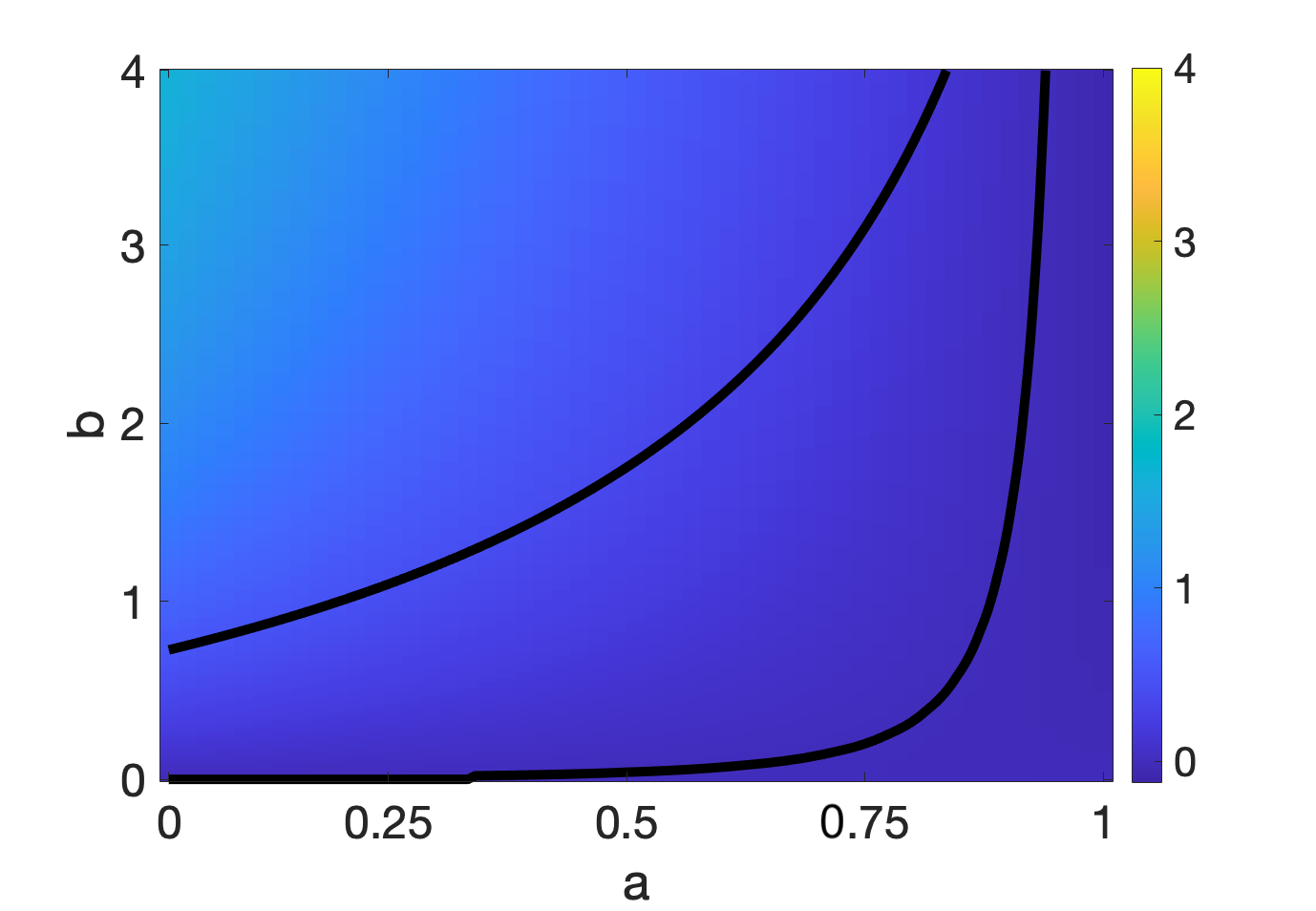}
        \caption{$\tau=0.2$.}
        \label{}
    \end{subfigure}
    \hfill
    \begin{subfigure}[t]{0.32\textwidth}
        \centering
        \includegraphics[width=5.5cm,height = 5cm]{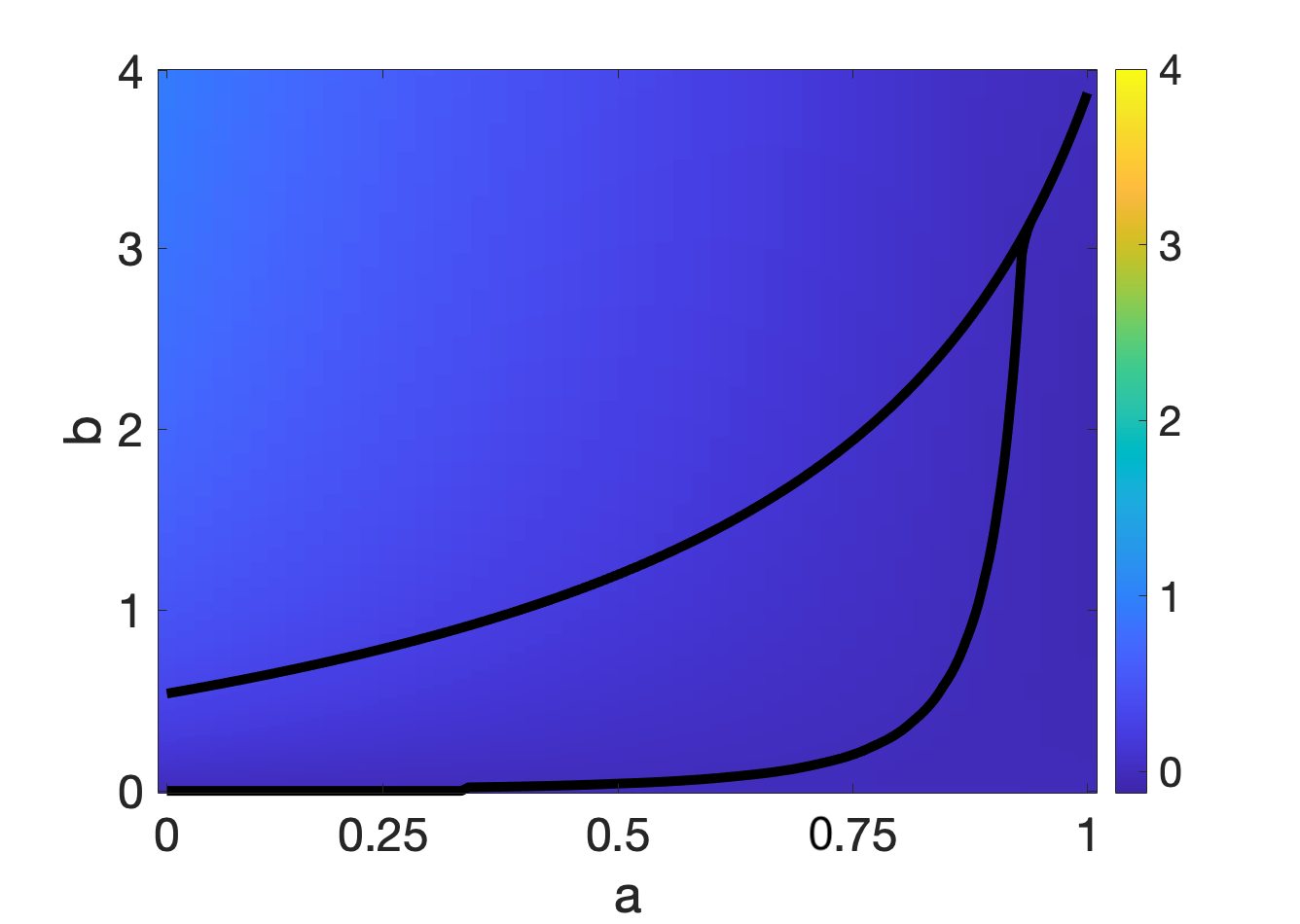}
        \caption{$\tau=0.5$.}
        \label{f}
    \end{subfigure}
    \caption{The maximum growth rate, $\max_k(\Re(\lambda_k))$, corresponding to the linearisation of \eqref{fadai1} plotted for $(a,b)\in[0,1]\times[0,4]$, with varying $\tau$. $\max_k$ taken over $k\in[0,50]$ for $k\in\mathbb{Z}$. Parameters $\epsilon^2=0.001$ and $L^2=9/2$ used.}
    \label{fig:fad1}
\end{figure}

\begin{figure}[H]
    \centering
    \begin{subfigure}[t]{0.32\textwidth}
        \centering
        \includegraphics[width=5.5cm,height = 5cm]{f1t0.png}
        \caption{$\tau=0$.}
        \label{}
    \end{subfigure}
    \hfill
    \begin{subfigure}[t]{0.32\textwidth}
        \centering
        \includegraphics[width=5.5cm,height = 5cm]{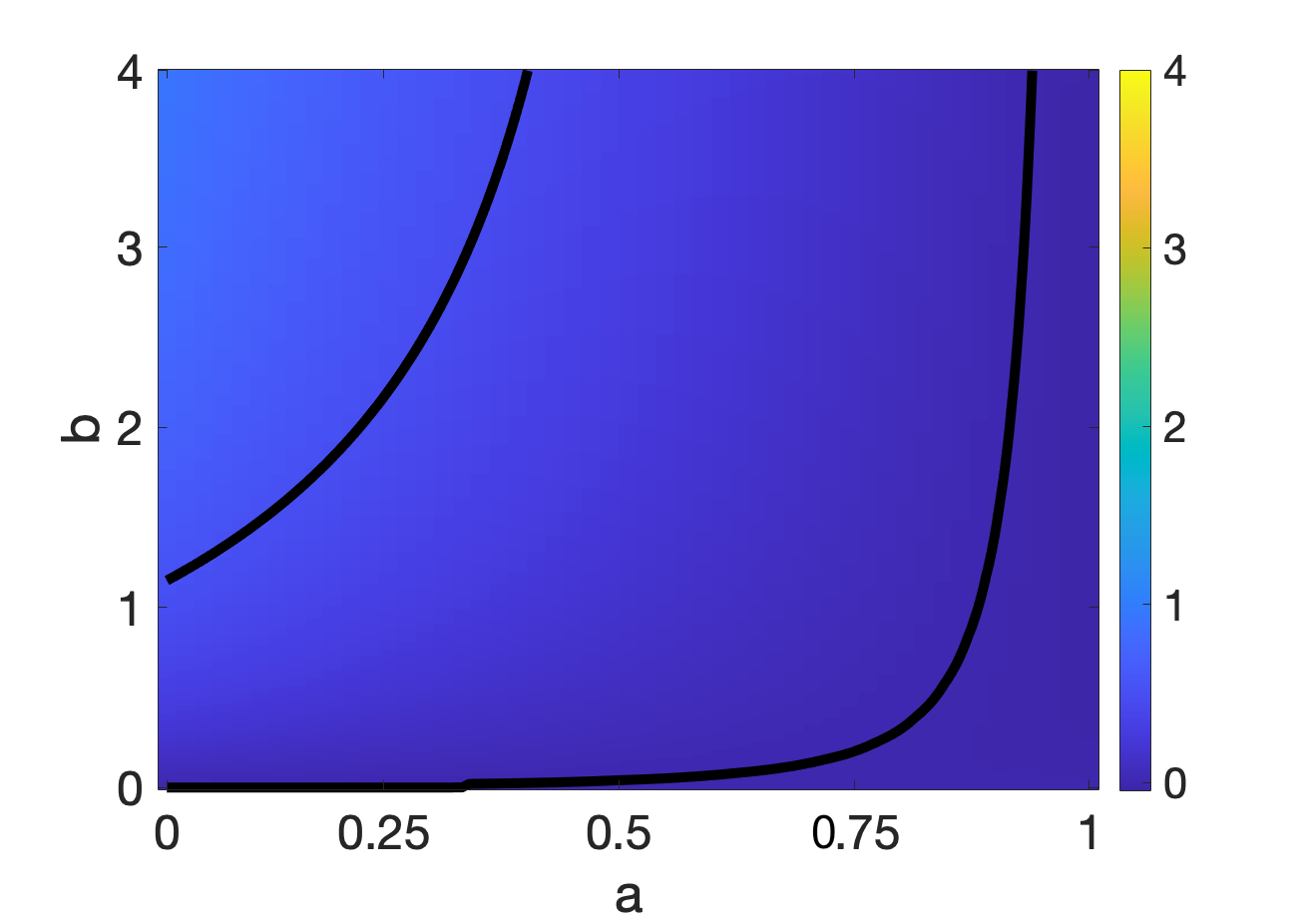}
        \caption{$\tau=0.5$.}
        \label{}
    \end{subfigure}
    \hfill
    \begin{subfigure}[t]{0.32\textwidth}
        \centering
        \includegraphics[width=5.5cm,height = 5cm]{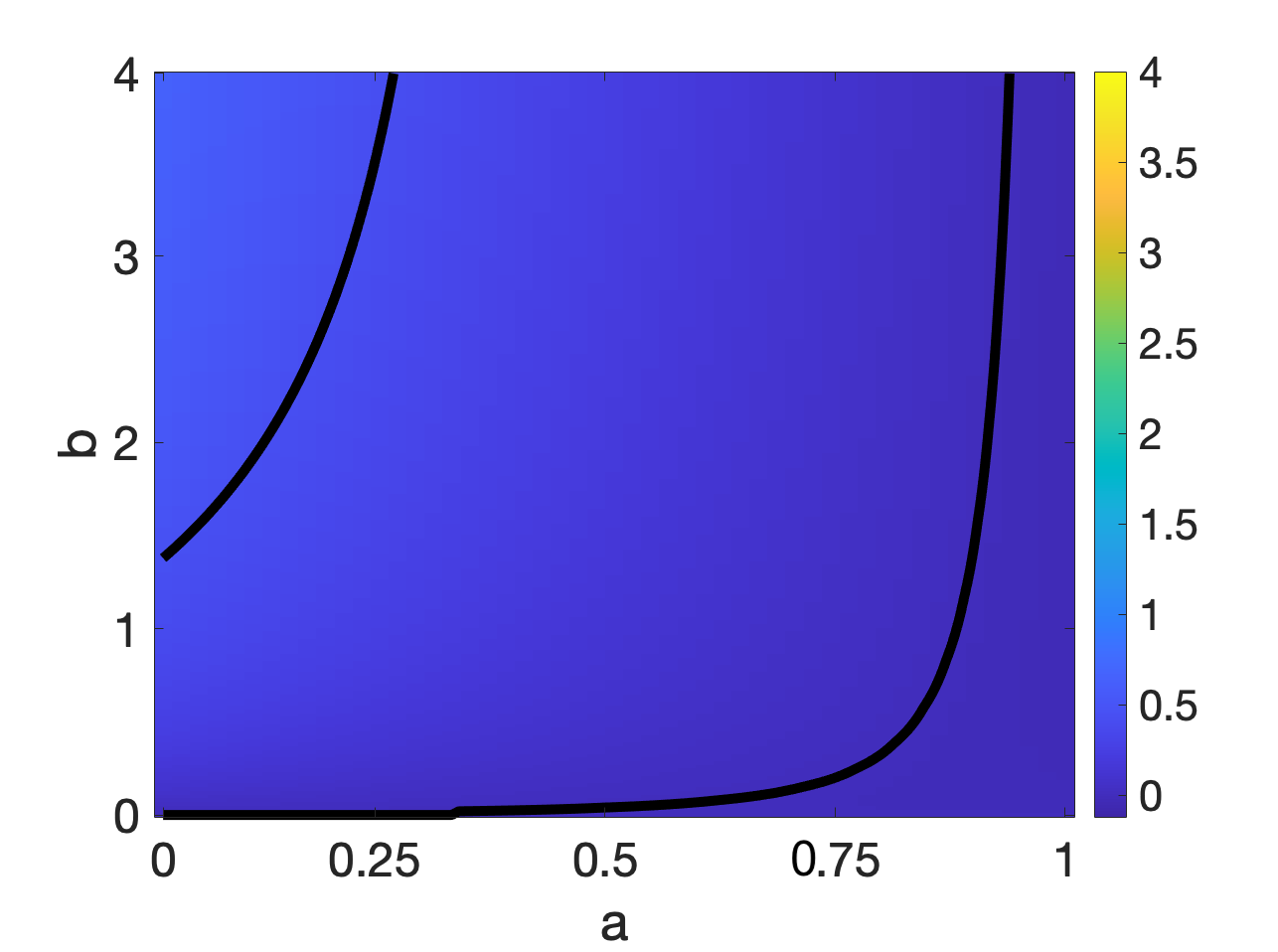}
        \caption{$\tau=0.8$.}
        \label{}
    \end{subfigure}
    \caption{The maximum growth rate, $\max_k(\Re(\lambda_k))$, corresponding to the linearisation of \eqref{fadai2} plotted for $(a,b)\in[0,1]\times[0,4]$, with varying $\tau$. $\max_k$ taken over over $k\in[0,50]$ for $k\in\mathbb{Z}$. Parameters $\epsilon^2=0.001$ and $L^2=9/2$ used.}
    \label{fig:fad2}
\end{figure}

The roots of \eqref{characf1} and \eqref{characf2} can thus be solved, and plots of $\max_k(\Re(\lambda_k))$ across the parameter plane $(a,b)\in[0,1]\times[0,4]$ for each of the models \eqref{fadai1} and \eqref{fadai2}, can be generated. Figures \ref{fig:fad1} and \ref{fig:fad2} show these results for varying $\tau$, where we have also added the contour lines of $\Re(\lambda_0)=0$ and $\max_k(\Re(\lambda_k))=0$ to indicate the Turing space.
For the model given in \eqref{fadai1}, where analysis in \cite{fadai1} showed a de-stabilisation of the stable spike solution parameter space with an increasing $\tau$, the results in Figure \ref{fig:fad1} show a similar result for Turing instabilities. Namely, the Turing space shrinks for increasing $\tau$. Similarly, \cite{fadai2} showed a stabilising effect of increasing $\tau$ on the stable spike solution parameter space for the model in \eqref{fadai2}, and we find an analogous result for the Turing space, as seen in Figure \ref{fig:fad2}. We note here that in all models considered, namely the LI model, and the two GM model variants, the changing size of the Turing space with an increasing $\tau$ is solely dependent on the movement of the curve produced by considering the homogeneous characteristic equation (when $k=0$). This similarity could suggest a particular mechanism by which stabilising or de-stabilising effects of time delay on the Turing space arise.

\begin{figure}[H]
    \centering
    \begin{subfigure}[t]{0.45\textwidth}
        \centering
        \includegraphics[width=6cm,height = 5cm]{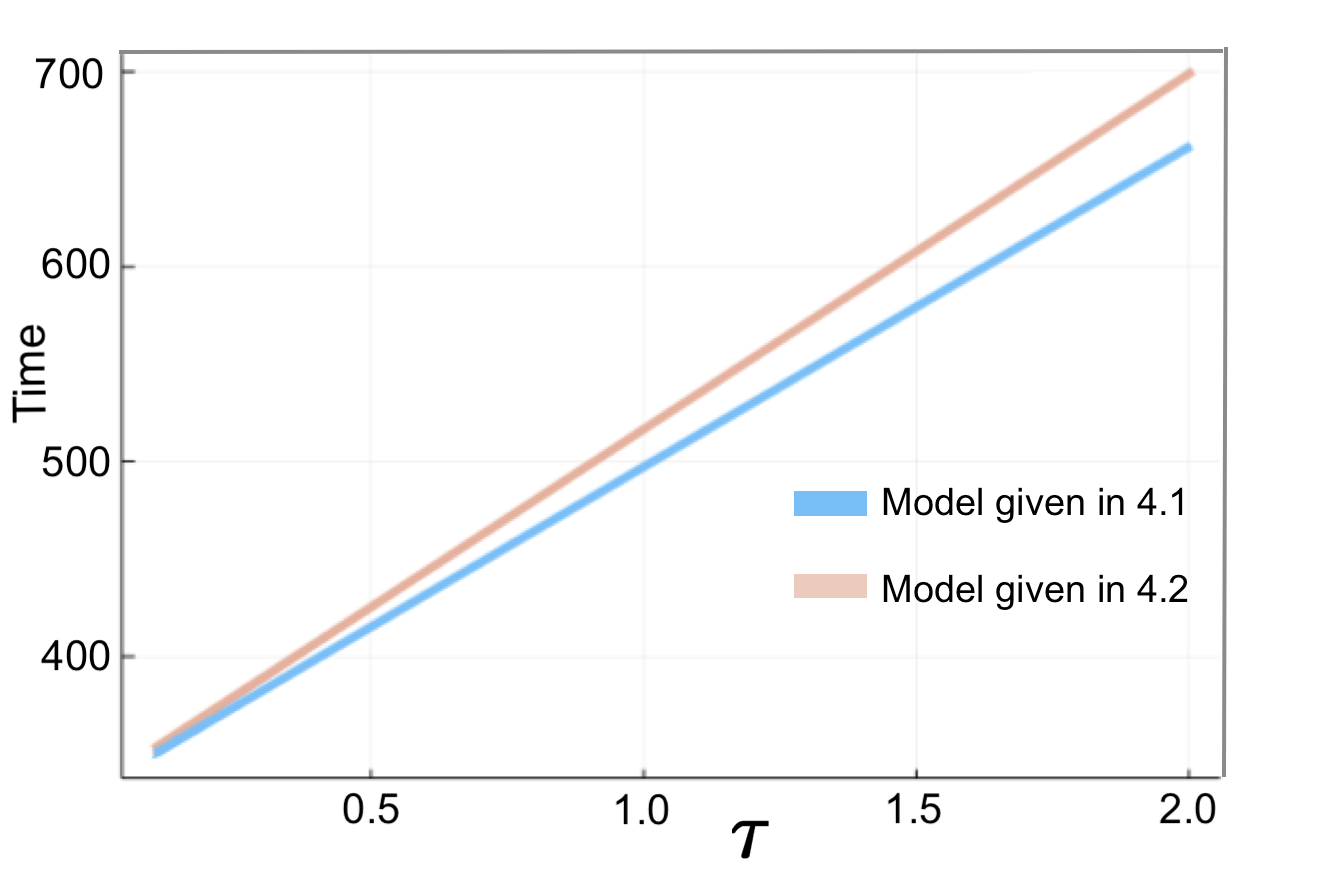}
        \caption{Time-to-pattern vs $\tau$ for $\tau\in[0.1,2]$ varied at intervals of $0.1$. The blue line shows results for the model in \eqref{fadai1}, and the red line for \eqref{fadai2}.}
        \label{fig:gmlin1}
    \end{subfigure}
    \hfill
    \begin{subfigure}[t]{0.45\textwidth}
        \centering
        \includegraphics[width=6cm,height = 5cm]{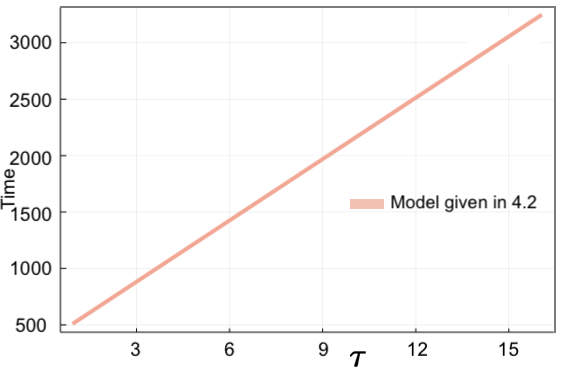}
        \caption{Time-to-pattern vs $\tau$ $\tau\in[1,16]$ varied at intervals of $1$. Model given in \eqref{fadai2}.}
        \label{fig:gmlin2}
    \end{subfigure}
    \caption{Time-to-pattern results for the two GM variants given in \eqref{fadai1} and \eqref{fadai2}. Initial random perturbation given with $\sigma_{\text{IC}}=0.001$, and threshold value given as $10$. Parameters $(a,b)=(0.75,0.5)$, $\epsilon^2=0.001$ and $L^2=9/2$ used. Boundary conditions set as in \eqref{neumannbc}.}
    \label{fig:fadlin}
\end{figure}

Finally, through numerical simulations, we present a linearly increasing relationship between time delay and time-to-pattern for both variants of the GM model. For the model in \eqref{fadai1}, as a result of the shrinking Turing space, we only consider small $\tau\in[0.1,1]$, varied at regular intervals of $0.1$. For the model in \eqref{fadai2}, we consider both small $\tau\in[0.1,2]$ and larger $\tau\in[1,16]$, varied at regular intervals of $0.1$ and $1$ respectively. Using a similar methodology to compute time-to-pattern as in Chapter 2, we set an initial perturbation from the steady state as $\sigma_{\text{IC}}=0.001$, and a threshold value of $10$. These larger values were chosen to improve computational time, and should not impact the relationship we see. The figures for the time-to-pattern results for models \eqref{fadai1} and \eqref{fadai2} can be seen in Figure \ref{fig:fadlin}, where a linear relationship can be deduced for both models.

The results in Figure \ref{fig:gmlin1} suggest a linear relationship between time-to-pattern and $\tau$ for both models for a small $\tau$. We also note a difference in line slope, suggesting that the positioning of time-delayed terms within the kinetics can also affect the sensitivity of the time delay until onset of patterning of a model, to an increasing time delay. Figure \ref{fig:gmlin2} shows that the results for model \eqref{fadai2} hold for larger $\tau$.

In this chapter, we have shown that for the GM model, the linear stability of homogeneous steady states is similar to the linear stability of spike solutions in the literature \cite{fadai1,fadai2}. A linear dependence between time-to-pattern and time delay was also presented for both small and large $\tau$. Finally, the importance of the positioning of time-delayed terms within the kinetic reactions has been highlighted. A difference in positioning of time-delayed terms has been shown to both expand and contract the Turing space, as well as vary the sensitivity of the time lag until onset of patterning to an increasing time delay. Interestingly, it appears that the changing size of the Turing space occurs due to the stability against homogeneous perturbations, as in the LI model.

\chapter{Conclusion}
\section{Summary of Findings}
By using linear analysis, bifurcation theory, and numerical simulations, we have performed a thorough study of reaction-diffusion mechanisms with gene expression time delays. In this dissertation, kinetic equations associated with both Schnakenberg and Gierer-Meinhardt reaction kinetics were considered, and gene expression delays modelled as both a fixed parameter, and as a continuous distribution. The results in this dissertation help facilitate our understanding of the effect that time delay has on Turing pattern formation, and highlight the importance of their consideration in biological patterning events.

By combining Fourier analysis and numerical simulations, we presented a strong linearly increasing relationship, between fixed time delay, and the time-lag until onset of patterning for the Schnakenberg model. Through numerical simulations, a linearly increasing relationship was also shown between time delay and time-to-pattern for both GM variants considered. These results suggest that the impact of time delay on slowing pattern formation processes in reaction-diffusion systems is a general phenomenon, and scales linearly with time delay. A systematic review of the robustness of pattern formation to varying initial conditions was conducted and motivated by the biology considered in \cite{krausemixed}, Dirichlet boundary conditions were implemented for the activator dynamics. It was found that, for the Schnakenberg model, although the \textit{type} of pattern seen changes with these variations, the increase in lag until onset of patterning as a result of time delay is robust and consistent.

For the Schnakenberg kinetics, where fixed gene expression delays were motivated by ligand internalisation models, we have noted that increased time delays act to expand the Turing space. This is in contrast to the results concluded in \cite{yigaffneyli}. This effect was displayed through the use of bifurcation diagrams, and was confirmed through numerical solutions.
Motivated by the stability analysis of spike solutions of the GM model in \cite{fadai1,fadai2}, we demonstrated the importance of the positioning of time-delayed terms within a reaction-diffusion mechanism. This in turn highlights the importance of understanding the biological processes we are considering. Linear and bifurcation analysis showed that for the GM model, increasing of time delay can either act to expand or contract the Turing space. We also note that the expansion and contraction of all the Turing spaces considered were solely dependent on the spatially homogeneous models. This yields an interesting question as to whether this is a more general mechanism by which time delay can affect the parameter space exhibiting Turing instabilities, or whether it is a specific attribute of the Schnakenberg and GM kinetics.

Finally, driven by the inherent stochasticity of the molecular processes underpinning gene expression \cite{raj,elowitz,mcadams,paulsson}, gene expression time delays were modelled as both a symmetric and skewed Gaussian distribution. Through linear analysis, and verified by numerical simulations, it was presented that the distribution used does not matter. Namely, the pattern formation process of the Schnakenberg model seems to be dependent on the mean delay of the distribution used, irrespective of standard deviation or skew, and thus can effectively be modelled as purely a fixed delay.

\section{Future Work}

Our findings, that a distributed representation of time delay does not alleviate the increased timescales of patterning events caused with a fixed delay, calls into question how relevant and applicable Turing mechanisms are for describing biological patterning events. Despite these results, empirical evidence suggests that such Turing instabilities do exist to explain biological phenomena \cite{yigaffneyli,molecular,miura,miura2,sick}. Our research does not close the door to applying Turing's models, but in fact yields an abundance of new and unanswered questions. We first note the extreme simplicity of the reaction-diffusion models we consider in this dissertation, in contrast to the complexity of the biological processes, whose behaviour we attempt to capture. In \cite{mainigeneral}, work was implemented to develop Turing conditions for a system describing the interaction of $n$ morphogens, for any $n\geq2$. However, as far as we are aware, there is no systematic study of how time delay may affect pattern formation events for a reaction-diffusion mechanism with greater than two morphogens. Therefore, one potentially important avenue for further research would be to investigate the effect of time delay on Turing mechanisms encapsulating a larger number of reactants. This would aid in improving the possible applicability and similarity of Turing's models to the more intricate biological dynamics.

Throughout this dissertation, we considered two different types of kinetics, namely those derived from Schnakenberg and Gierer-Meinhardt kinetic reactions. Our results indicate some general attributes that are common to both sets of kinetics. The first being a linearly increasing relationship between incorporated time delay and time until onset of patterning. The second being that the effect of time delay on the exhibited Turing space is only dependent on the spatially homogeneous model, irrespective of whether the Turing space is growing or shrinking. A clear extension to these observations would be to explore these effects for different reaction-diffusion systems that can exhibit Turing patterns. Typical models that could be examined include the Gray-Scott \cite{grayscott} or Thomas \cite{murray} models.

We note that many simplifying assumptions were made in the models considered. Representing time delay as a continuous distribution is a novel field of interest, and thus has not previously been explored in depth. The use of other forms of distribution, such as the gamma or exponential distributions could be considered, in order to verify our findings that, when onset of patterning is being considered, the only relevant modelling parameter required is the mean delay. We also only considered the problem on a one-dimensional stationary spatial domain. Previous research has been conducted on higher-dimensional spatial domains, and growing domains, with fixed delay \cite{gaffmonk,krausefixed}. Although we hypothesise that our results with a distributed delay will be consistent across variations in the spatial domain considered, there is room to explore these possibilities. Finally, we note that due to numerical limitations when using \textit{chebfun} to find roots of the transcendental characteristic equations, the linear theory could only be tested for small-time delays. We found that, for these small-time delays, the linear theory generally provided a good approximation to the time-to-pattern, and all conclusions from the linear theory were able to be verified through full numerical solutions. This however yields an interesting question of, what the limitations of the linear theory are, and for how large of a timescale can the linear theory still be applied. Further work could therefore be considered to solve the characteristic equations derived in this dissertation for larger time delay values, and examine whether the linear theory still provides good approximations to the model behaviour.

%now enable appendix numbering format and include any appendices
\appendix
\chapter{Further Mathematical Details}\label{section:appA}
\section{Finite Difference Scheme}
With the $m=500$ spatially discretised points given as $\textbf{x}=[x_1,\cdots,x_m]^T$ where $x_i=(i-1)\Delta x$, with $\Delta x=\frac{1}{m-1}$. This ensures $x_1=0$ and $x_m=1$. Letting $U_i^t$ denote the numerical approximation to $u(x_i,t)$, we use a second-order central difference approximation \cite{finitediff} to evaluate the second-order derivatve $\frac{\partial^2 u}{\partial x^2}(x_i,t)$. Namely,
\begin{equation}
\frac{\partial^2 u}{\partial x^2}(x_i,t)\approx\frac{1}{\Delta x^2}\left(U_{i+1}^t-2U_i^t+U_{i-1}^t\right).
\end{equation}
By denoting $\textbf{U}^t$ as the vector of numerical approximations to $u(x,t)$ at some time $t$ across the whole spatial domain $\textbf{x}$, so that $\textbf{U}^t=\left[U_1^t,\cdots,U_m^t\right]^T$, the numerical approximation of the second-order derivative across the whole spatial domain at some time $t$ can be computed in matrix form as
\begin{equation}
    \frac{\partial^2 u}{\partial \textbf{x}^2}\approx A\textbf{U}^t.
\end{equation}
$A$ is the discrete second-order differential operator, and is given by
\begin{equation}\label{A}
A=\frac{1}{\Delta x^2}\begin{bmatrix}
   -2&  1&  &  & 0\\
   1&  -2&  1&  & \\
   &  \ddots&  \ddots&  \ddots& \\
   &  &  1&  -2& 1\\
   0&  &  &  1& -2
  \end{bmatrix}.
\end{equation}
The sparse nature of $A$ allows for computational advantages when implementing the finite-difference scheme. In order to implement homogeneous Neumann boundary conditions, a first-order central difference approximation is used with `ghost' nodes appended at $x_{-1}$ and $x_{m+1}$. This results in altering entries $A_{1,2}$ and $A_{m,m-1}$ from a $1$ to a $2$, as seen in \eqref{Aneumann}. To implement homogeneous Dirichlet conditions, the first and last rows of $A$ are set to $0$, as seen in \eqref{Adirichlet}. Since homogeneous Dirichlet conditions require $u(x)=0$ at $x=0,1$ for all $t>0$, we also set the initial conditions and kinetic functions to equal $0$ at the end nodes.
\begin{multicols}{2}
\begin{equation}\label{Aneumann}
    \begin{split}
A&=\frac{1}{\Delta x^2}\begin{bmatrix}
   -2&  2&  &  & 0\\
   1&  -2&  1&  & \\
   &  \ddots&  \ddots&  \ddots& \\
   &  &  1&  -2& 1\\
   0&  &  &  2& -2
  \end{bmatrix}.\\
  A &\textit{ with homogeneous Neumann conditions}
    \end{split}
\end{equation}
\break
\begin{equation}\label{Adirichlet}
    \begin{split}
A&=\frac{1}{\Delta x^2}\begin{bmatrix}
   -0&  0& \cdots &  & 0\\
   1&  -2&  1&  & \\
   &  \ddots&  \ddots&  \ddots& \\
   &  &  1&  -2& 1\\
   0&\cdots  &  &  0& 0
  \end{bmatrix}.\\
  A & \textit{ with homogeneous Dirichlet conditions}
    \end{split}
\end{equation}
\end{multicols}

\section{Functional Form of $\text{IC}_1$.}
The form of $\text{IC}_1$ is taken from those used in \cite{gaffmonk}, and given by
\begin{equation}\label{ic1}
    \begin{split}
    u_0(x)&=u_\star+E_ux^7(1-x^2)[A_ux^3+B_ux^2+C_ux+D_u]\\
    v_0(x)&=v_\star+E_vx^5(1-x^2)[A_vx^3+B_vx^2+C_vx+D_v],
    \end{split}
\end{equation}
for $x\in[0,1]$. The parameter values $(a,b)=(0.1,0.9)$ are used, yielding $(u_\star,v_\star)=(1,0.9)$. The coefficients in \eqref{ic1} are given as
\begin{align*}
A_u&=-130.8444445,\ \ B_u=337.0666669,\ \ C_u=-281.6000002,\ \ D_u=75.3777778,\\
A_v&=-170.6666682,\ \ B_v=412.4444479,\ \ C_v=-312.8888910,\ \ D_v=71.1111113,\\
\end{align*}
and $E_u=0.00600$, $E_v=0.00125$.

\section{Derivation of Mean of Skewed Truncated Gaussian Distribution}

Using a result from \cite{skewed}, we have that the $m$-th moment of a random variable $X$ following a truncated skewed Gaussian distribution, on the domain $[a,b]$ is given by
\begin{equation}\label{mth}
\mathbb{E}[X^m]=\sum_{r=0}^mC_m^r\mu^{m-r}\omega^rs_{\rho,r}(u,v).
\end{equation}
Here, $u=\frac{a-\mu}{\omega}$, $v=\frac{b-\mu}{\omega}$, and $C_m^r=\begin{pmatrix}m\\r\end{pmatrix}$ is a binomial coefficient. The function $s_{\rho,r}$ is given by
\begin{equation}\label{s}
    s_{\rho,r}(u,v)=(r-1)s_{\rho,r-2}(u,v)+q_{\rho,r}(u,v),
\end{equation}
with $s_{\rho,0}(u,v)=1$, and the function $q_{\rho,r}$ defined by
\begin{equation}\label{q}
    q_{\rho,r}(u,v)=-\frac{\left[x^{r-1}f_\rho(x)\right]|_u^v}{\left[F_\rho(x)\right]|_u^v}+\frac{2}{\sqrt{2\pi}}\frac{\rho}{\hat{\rho}^r}\frac{\left[\phi(\hat{\rho x})\right]|_u^v}{\left[F_\rho(x)\right]|_u^v}m_{r-1}(\hat{\rho}u, \hat{\rho}v).
\end{equation}
The functions $f_\rho$ and $F_\rho$ denote the pdf and cdf of the standard skew Gaussian distribution,

 To compute the expection, $\mathbb{E}[X]$, we take the first moment, namely $m=1$. Equation \eqref{mth} can therefore be considerably simplified to
\begin{equation}
    \begin{split}
\mathbb{E}[X]&=C_1^0\mu s_{\rho,0}(u,v)+C_1^1\omega s_{\rho,1}(u,v)\\
&=\mu+\omega q_{\rho,1}(u,v),
\end{split}
\end{equation}
since $s_{\rho,0}(u,v)=1$ and $s_{\rho,1}(u,v)=q_{\rho,1}(u,v)$. We also use the fact that $m_1(u,v)=1$, from \cite{skewed}. Evaluating $q_{\rho,1}(u,v)$ from \eqref{q} leads to the desired result presented in \eqref{computetau}.

\chapter{Further Numerical Results}\label{section:appB}

\section{For Chapter 2}\label{section:Bfix}

\begin{figure}[H]
  \centering
\begin{subfigure}[t]{0.45\textwidth}
    \centering
    \includegraphics[width=7cm,height=5cm]{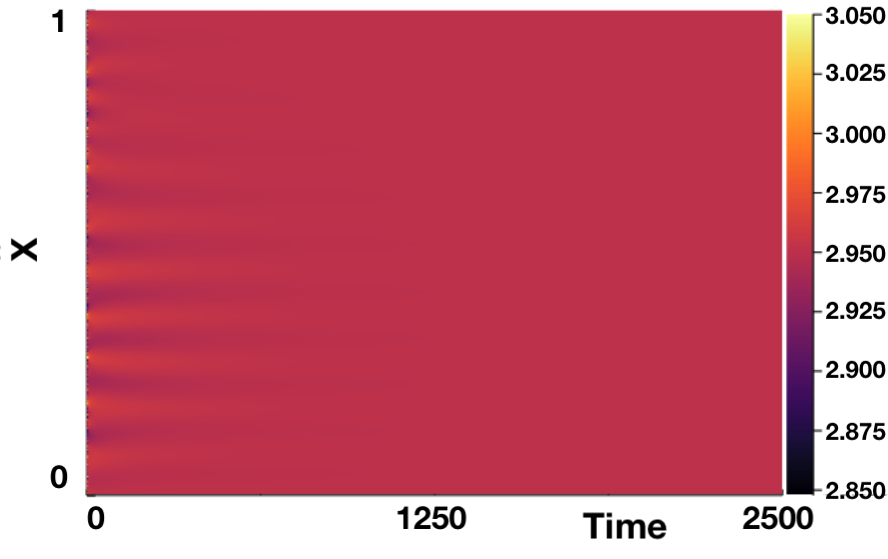}
    \caption{$\tau=0.5$}
    \label{}
\end{subfigure}
\hfill
\begin{subfigure}[t]{0.45\textwidth}
    \centering
    \includegraphics[width=7cm,height=5cm]{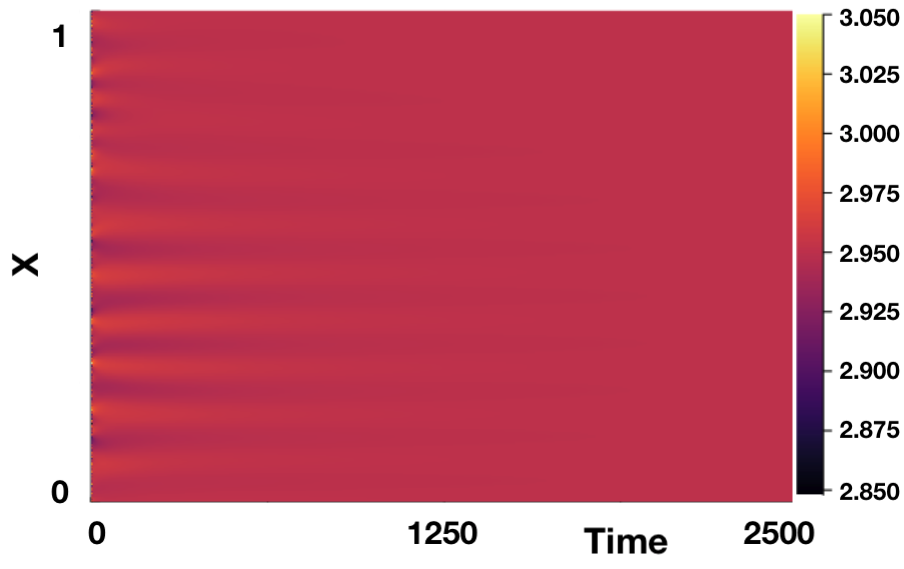}
    \caption{$\tau=1$}
    \label{}
\end{subfigure}
\caption{Numerical simulations of \eqref{fixed2} for $(a,b)=(1.2,1.75)$. $\epsilon^2=0.001$ and $L^2=9/2$. Boundary conditions given by \eqref{neumannbc} and initial conditions by \eqref{firstic}. We see no Turing pattern formation for $\tau\in\{0.5,1\}$ as suggested by linear theory, seen in Figure \ref{fig:tspacetau}. }
\label{fig:appfig1}
\end{figure}

\begin{figure}[H]
  \centering
\begin{subfigure}[t]{0.45\textwidth}
    \centering
    \includegraphics[width=7cm,height=5cm]{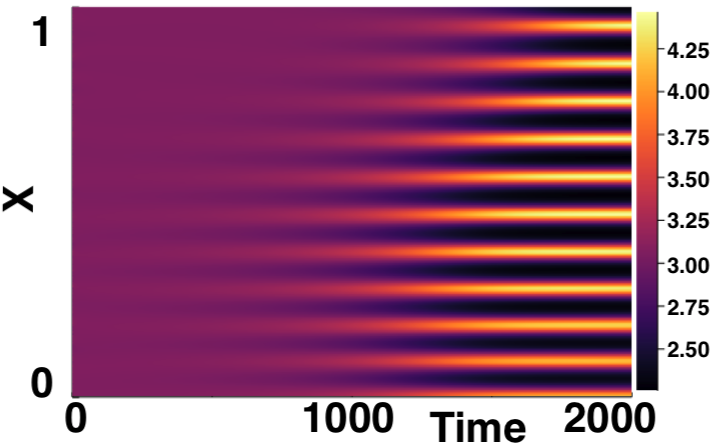}
    \caption{$\tau=0.5$}
    \label{}
\end{subfigure}
\hfill
\begin{subfigure}[t]{0.45\textwidth}
    \centering
    \includegraphics[width=7cm,height=5cm]{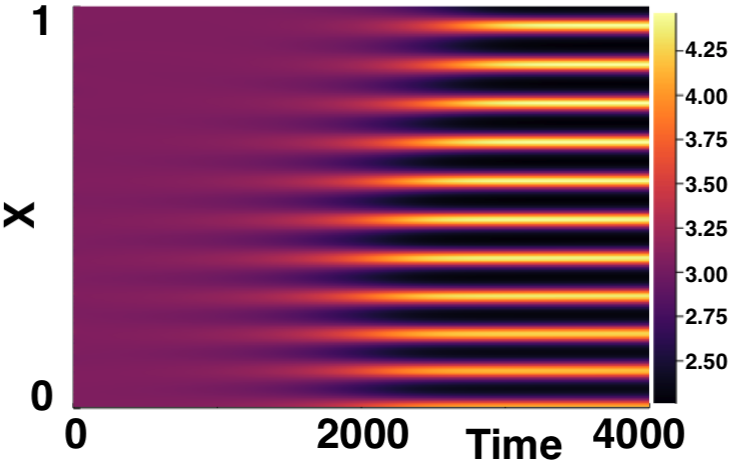}
    \caption{$\tau=1$}
    \label{}
\end{subfigure}
\caption{Numerical simulations of \eqref{fixed2} for $(a,b)=(1.2,1.85)$. $\epsilon^2=0.001$ and $L^2=9/2$. Boundary conditions given by \eqref{neumannbc} and initial conditions by \eqref{firstic}. We see Turing pattern formation on an increasing timescale for $\tau\in\{0.5,1\}$ as suggested by linear theory, seen in Figure \ref{fig:tspacetau}. }
\label{fig:appfig2}
\end{figure}

Figure \ref{fig:appfig1} and \ref{fig:appfig2} show numerical solutions for $\tau=\{0.5,1\}$ further verifying the linear theory presented in Figure \ref{fig:tspacetau}.


\begin{figure}[H]
  \centering
\begin{subfigure}[t]{0.45\textwidth}
    \centering
    \includegraphics[width=7cm,height=5cm]{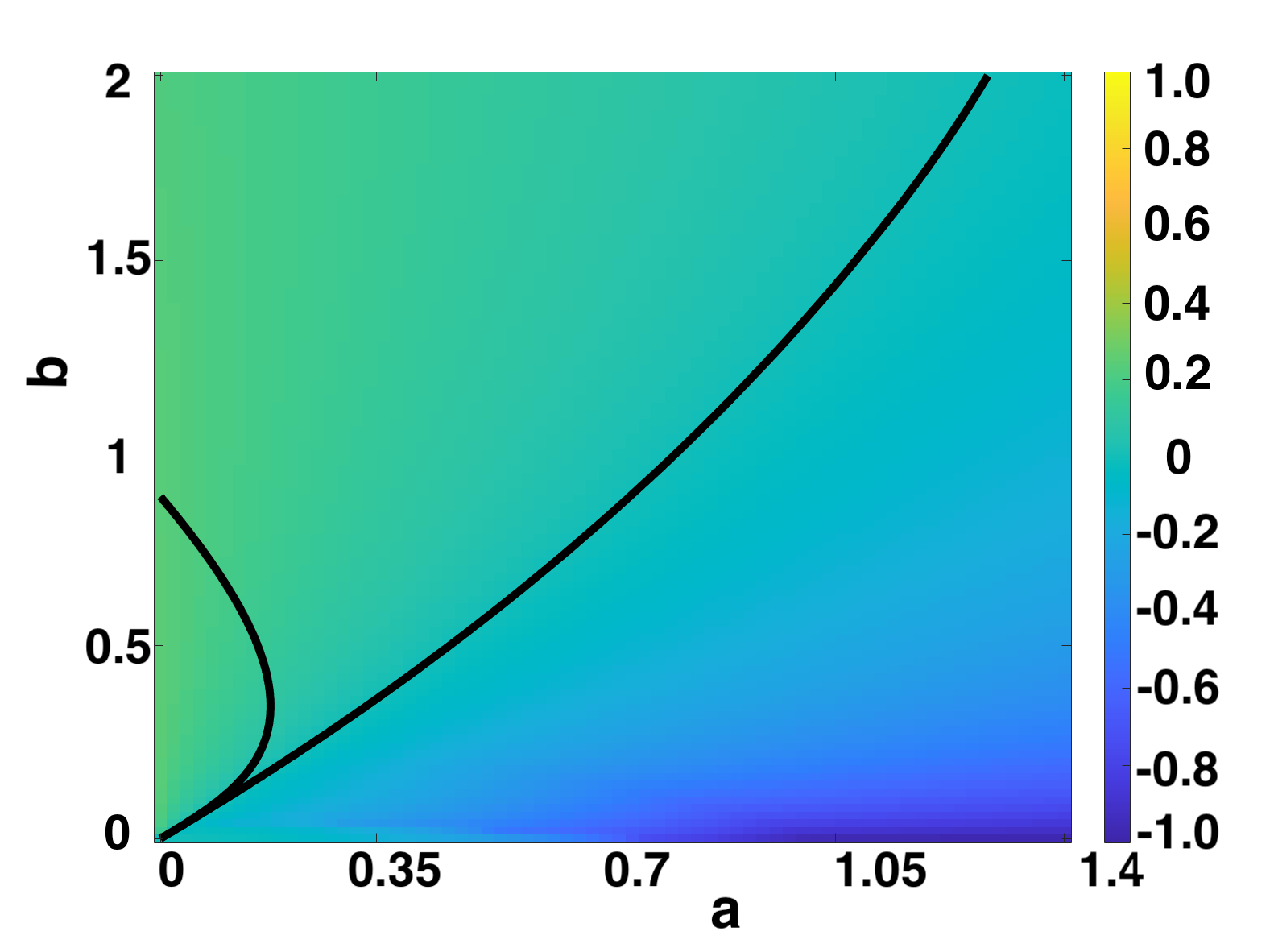}
    \caption{$\tau=0.5$}
    \label{}
\end{subfigure}
\hfill
\begin{subfigure}[t]{0.45\textwidth}
    \centering
    \includegraphics[width=7cm,height=5cm]{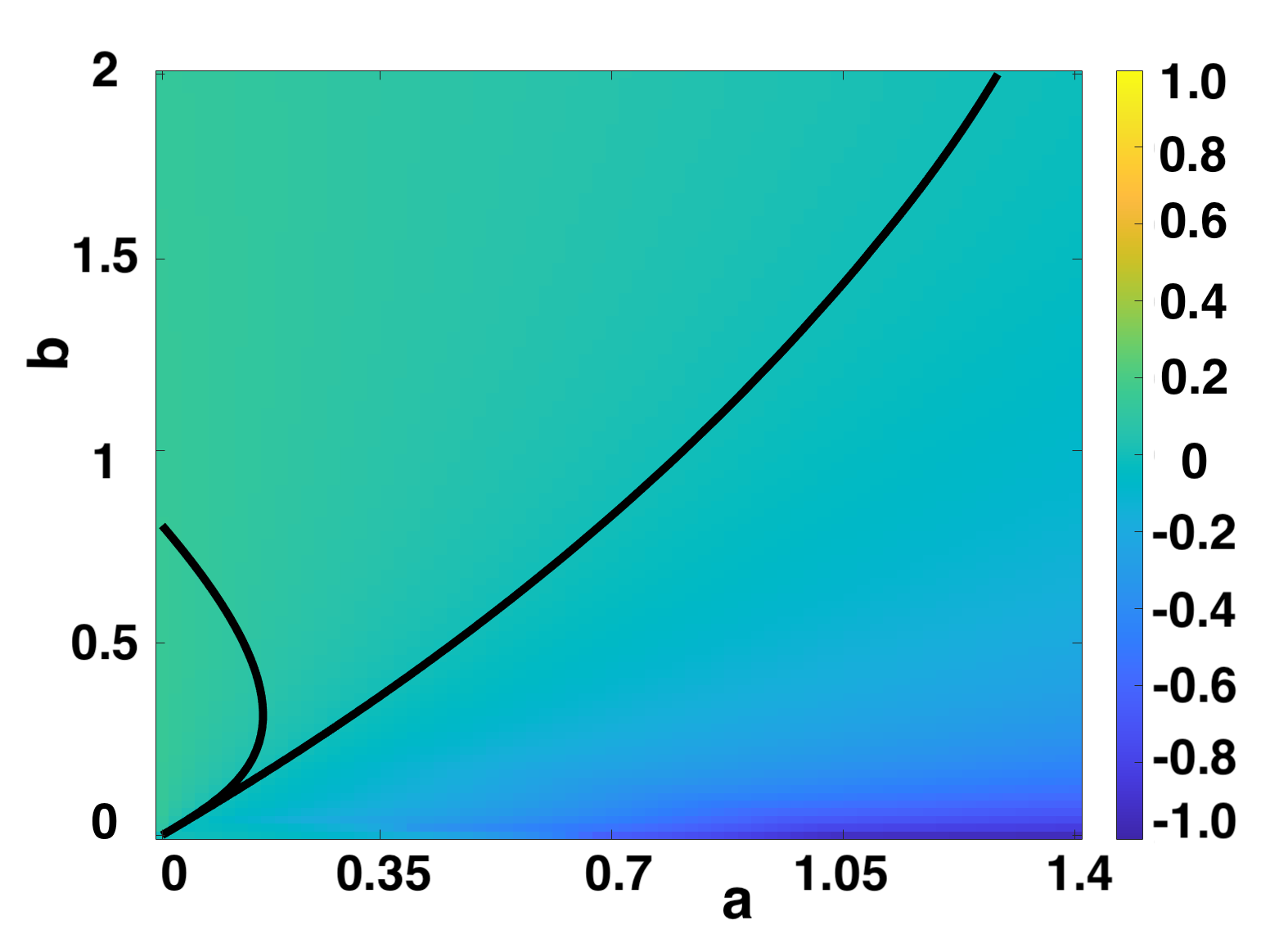}
    \caption{$\tau=1$}
    \label{}
\end{subfigure}
\caption{$\max_k(\Re(\lambda_k))$ computed over $(a,b)$ parameter space by solving \eqref{realfixbif} and \eqref{complexfixbif}, with $\epsilon^2=0.001$, $L^2=9/2$. As $\tau$ increases, $|\max_k(\Re(\lambda_k))|$ decreases. Contour lines for $\Re(\lambda_0)=0$ and $\max_k(\Re(\lambda_k))=0$ overlayed, indicated Turing instability region. }
\label{fig:Bbif1}
\end{figure}

\begin{figure}[H]
  \centering
\begin{subfigure}[t]{0.45\textwidth}
    \centering
    \includegraphics[width=7cm,height=5cm]{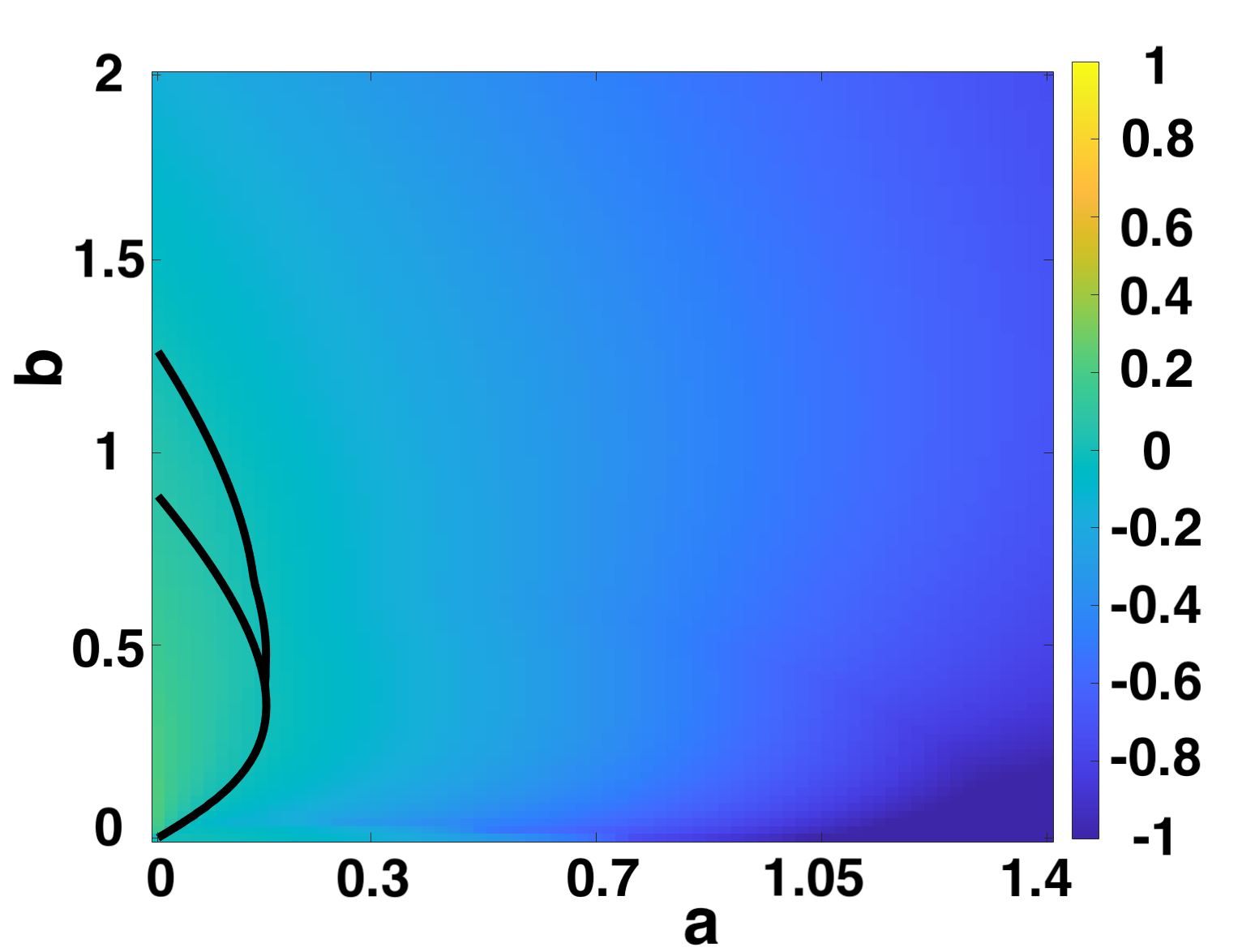}
    \caption{$\tau=0.5$}
    \label{}
\end{subfigure}
\hfill
\begin{subfigure}[t]{0.45\textwidth}
    \centering
    \includegraphics[width=7cm,height=5cm]{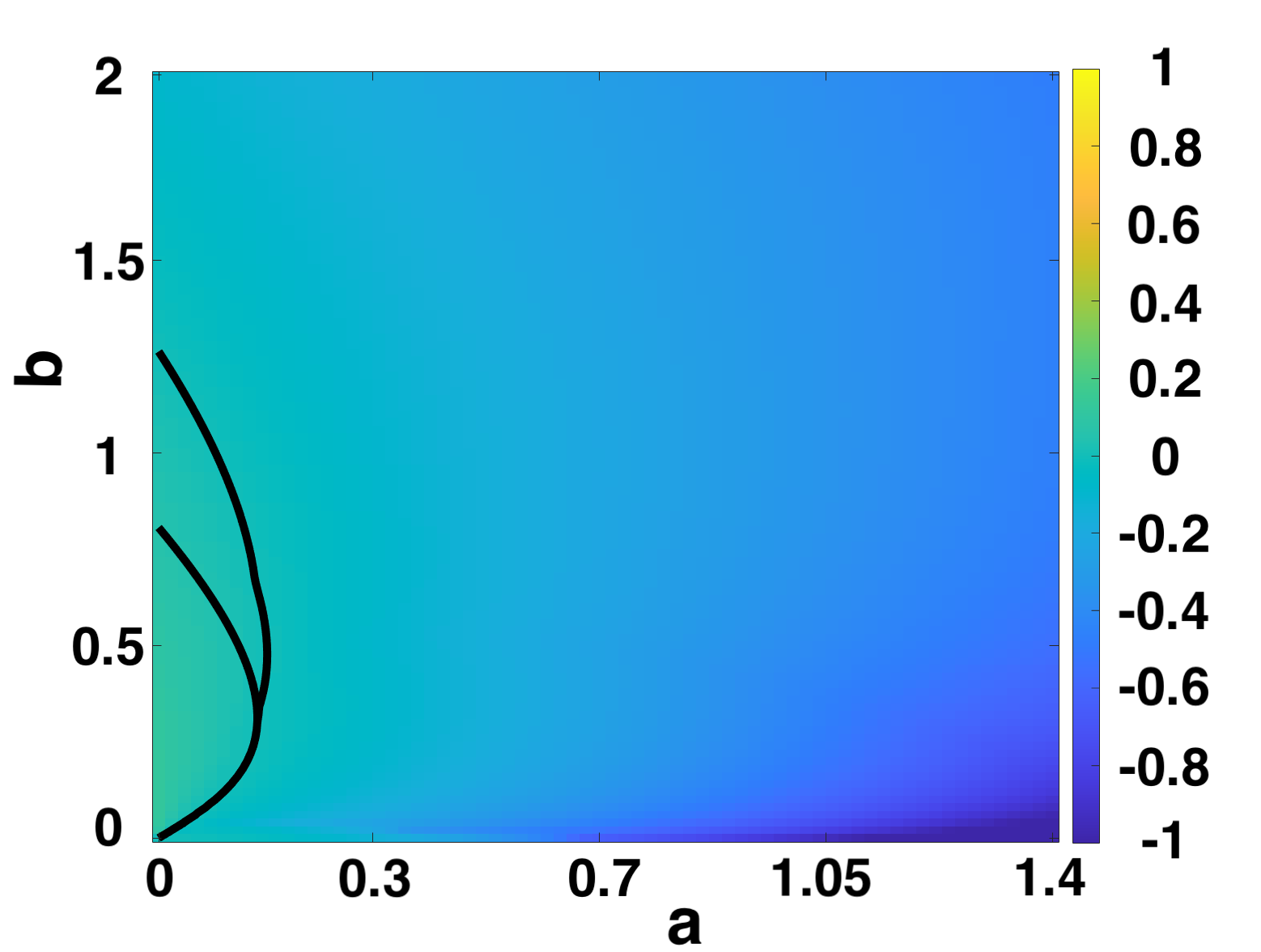}
    \caption{$\tau=1$}
    \label{}
\end{subfigure}
\caption{$\max_k(\Re(\lambda_k))$ computed over $(a,b)$ parameter space by solving \eqref{realfixbif} and \eqref{complexfixbif}, with $\epsilon^2=0.1$, $L^2=9/2$. As $\tau$ increases, $|\max_k(\Re(\lambda_k))|$ decreases. Contour lines for $\Re(\lambda_0)=0$ and $\max_k(\Re(\lambda_k))=0$ overlayed, indicated Turing instability region. }
\label{fig:Bbif2}
\end{figure}

Figures \ref{fig:Bbif1} and \ref{fig:Bbif2} show analogous bifurcation diagrams to those produced in Figures \ref{fig:lambdavary} and \ref{fig:fixbif2}, with $\tau=0.5,1$.


\begin{figure}[H]
    \centering
    \begin{subfigure}[t]{0.45\textwidth}
        \centering
        \includegraphics[width=6cm,height=4.5cm]{ic22.png}
        \caption{$\text{BC}_1$ given by equation \eqref{neumannbc}.}
        \label{}
    \end{subfigure}
    \hfill
    \begin{subfigure}[t]{0.45\textwidth}
        \centering
        \includegraphics[width=6cm,height=4.5cm]{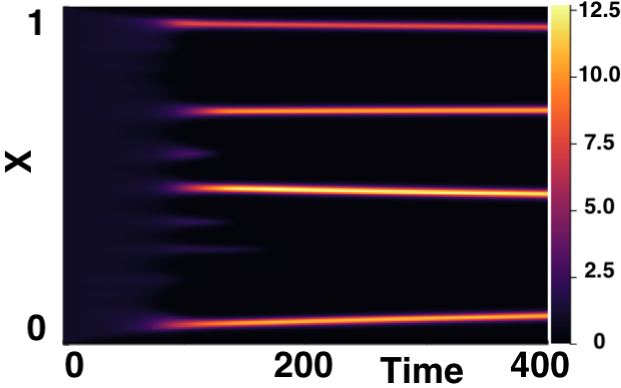}
        \caption{$\text{BC}_2$ given by equation \eqref{homogeneousbc}.}
        \label{}
    \end{subfigure}
    \caption{Comparison of varying BCs for $\tau=2$. $(a,b)=(0.1,0.9)$, $\epsilon^2=0.001$, $L^2=9/2$. Initial conditions given by \eqref{firstic}.}
    \label{fig:Bbc2}
\end{figure}

\begin{figure}[H]
    \centering
    \begin{subfigure}[t]{0.45\textwidth}
        \centering
        \includegraphics[width=6cm,height=4.5cm]{ic24.png}
        \caption{$\text{BC}_1$ given by equation \eqref{neumannbc}.}
        \label{}
    \end{subfigure}
    \hfill
    \begin{subfigure}[t]{0.45\textwidth}
        \centering
        \includegraphics[width=6cm,height=4.5cm]{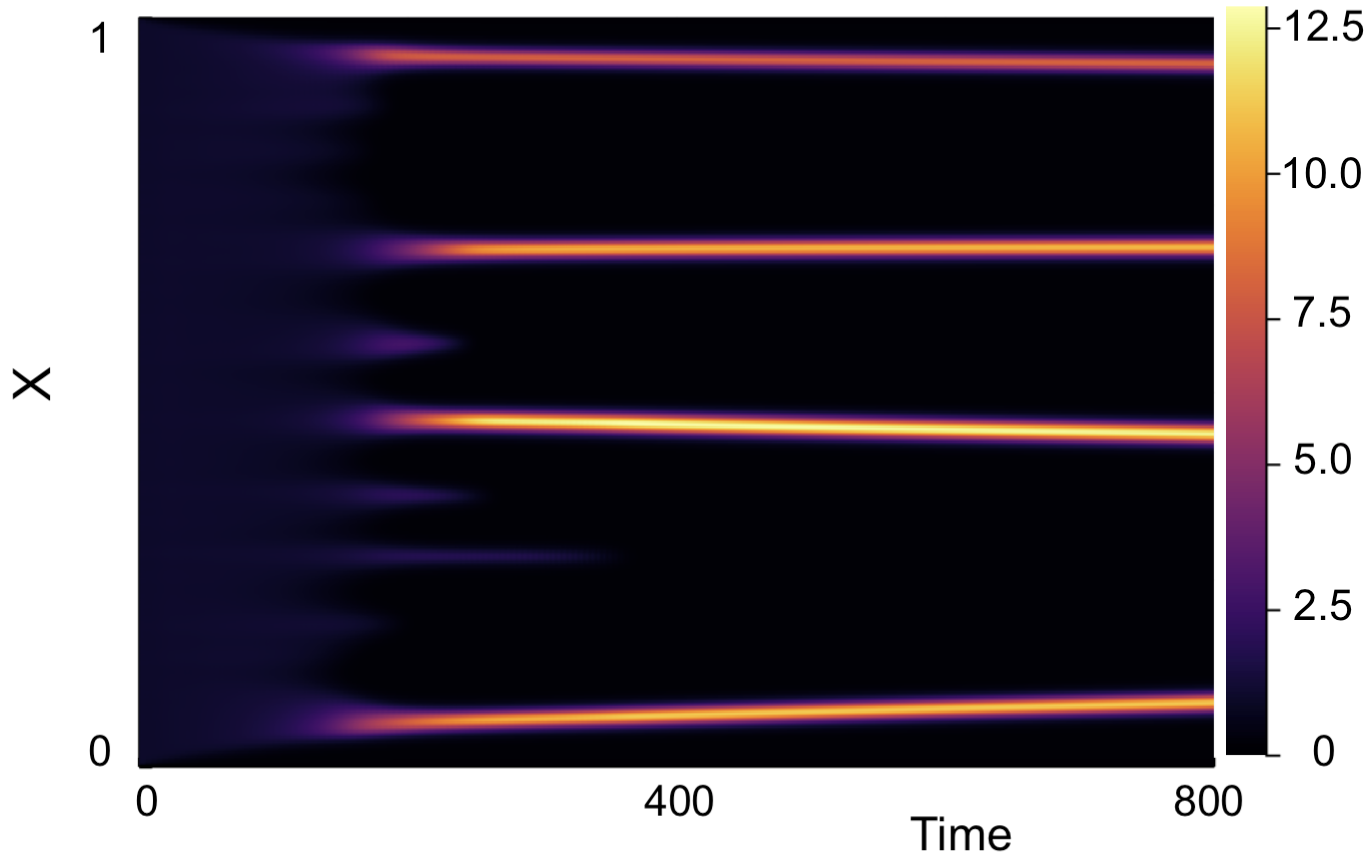}
        \caption{$\text{BC}_2$ given by equation \eqref{homogeneousbc}.}
        \label{}
    \end{subfigure}
    \caption{Comparison of varying BCs for $\tau=4$. $(a,b)=(0.1,0.9)$, $\epsilon^2=0.001$, $L^2=9/2$. Initial conditions given by \eqref{firstic}.}
    \label{fig:Bbc4}
\end{figure}

\begin{figure}[H]
    \centering
    \begin{subfigure}[t]{0.45\textwidth}
        \centering
        \includegraphics[width=6cm,height=4.5cm]{ic28.png}
        \caption{$\text{BC}_1$ given by equation \eqref{neumannbc}.}
        \label{}
    \end{subfigure}
    \hfill
    \begin{subfigure}[t]{0.45\textwidth}
        \centering
        \includegraphics[width=6cm,height=4.5cm]{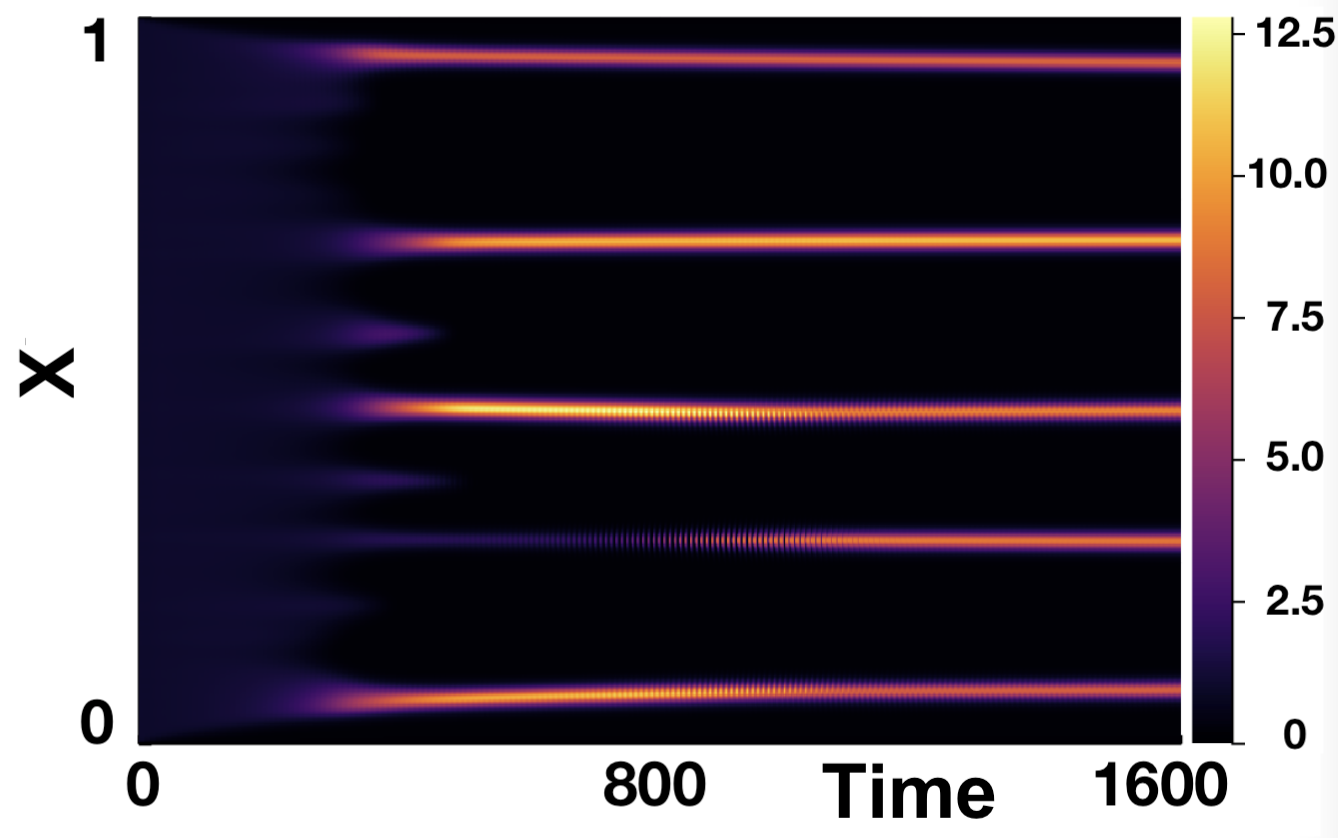}
        \caption{$\text{BC}_2$ given by equation \eqref{homogeneousbc}.}
        \label{}
    \end{subfigure}
    \caption{Comparison of varying BCs for $\tau=8$. $(a,b)=(0.1,0.9)$, $\epsilon^2=0.001$, $L^2=9/2$. Initial conditions given by \eqref{firstic}.}
    \label{fig:Bbc8}
\end{figure}

In Figures \ref{fig:Bbc2}, \ref{fig:Bbc4} and \ref{fig:Bbc8}, we present the comparison of numerical solutions between boundary conditions $BC_1$ and $BC_2$, for $\tau\in\{2,4,8\}$.

\begin{figure}[H]
    \centering
    \begin{subfigure}[t]{0.32\textwidth}
        \centering
        \includegraphics[width=5cm,height=4.5cm]{ic21.png}
        \caption{History function given as in \eqref{hist}.}
        \label{}
    \end{subfigure}
    \hfill
    \begin{subfigure}[t]{0.32\textwidth}
        \centering
        \includegraphics[width=5cm,height=4.5cm]{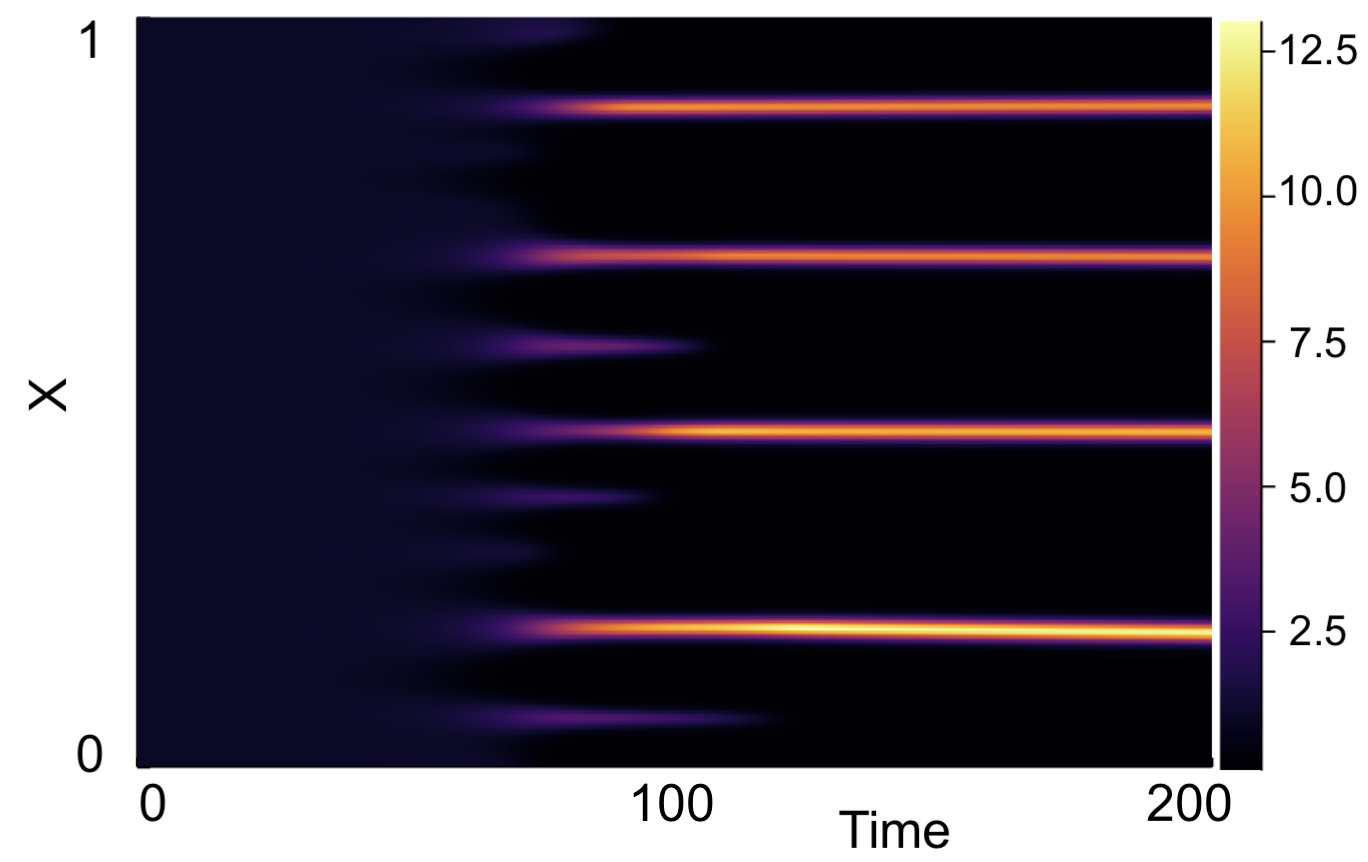}
        \caption{History function $h_1(t)$.}
        \label{}
    \end{subfigure}
    \hfill
    \begin{subfigure}[t]{0.32\textwidth}
        \centering
        \includegraphics[width=5cm,height=4.5cm]{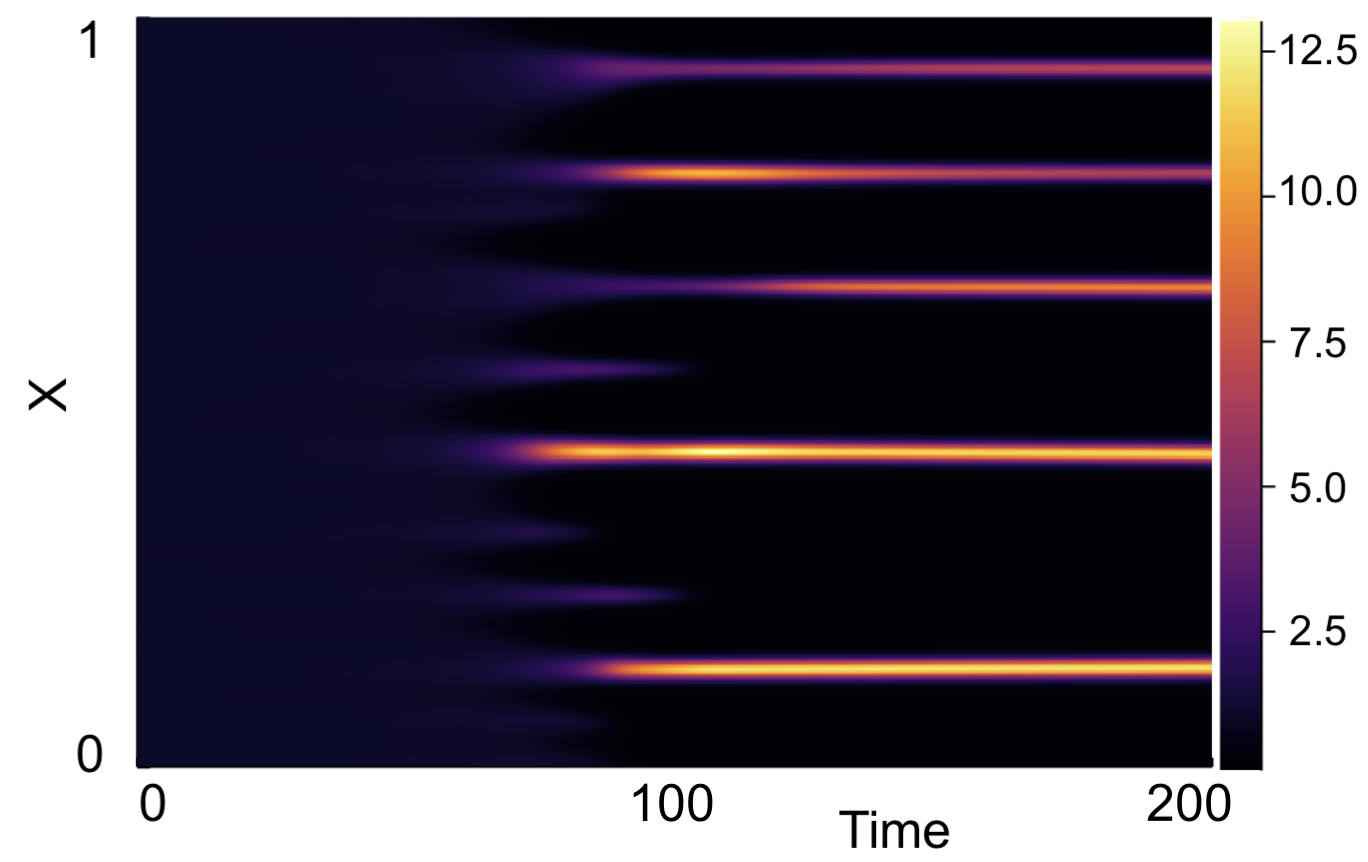}
        \caption{History function $h_2(t)$.}
        \label{}
    \end{subfigure}
    \caption{Numerical simulations of \eqref{fixed2} showing comparison of varying history functions for $\tau=1$. Boundary conditions given by \eqref{neumannbc} and initial conditions by \eqref{firstic}. Parameters $(a,b)=(0.1,0.9)$, $\epsilon^2=0.001$, $L^2=9/2$ used.}
    \label{fig:temp1}
\end{figure}

\begin{figure}[H]
    \centering
    \begin{subfigure}[t]{0.32\textwidth}
        \centering
        \includegraphics[width=5cm,height=4.5cm]{ic22.png}
        \caption{History function given as in \eqref{hist}.}
        \label{}
    \end{subfigure}
    \hfill
    \begin{subfigure}[t]{0.32\textwidth}
        \centering
        \includegraphics[width=5cm,height=4.5cm]{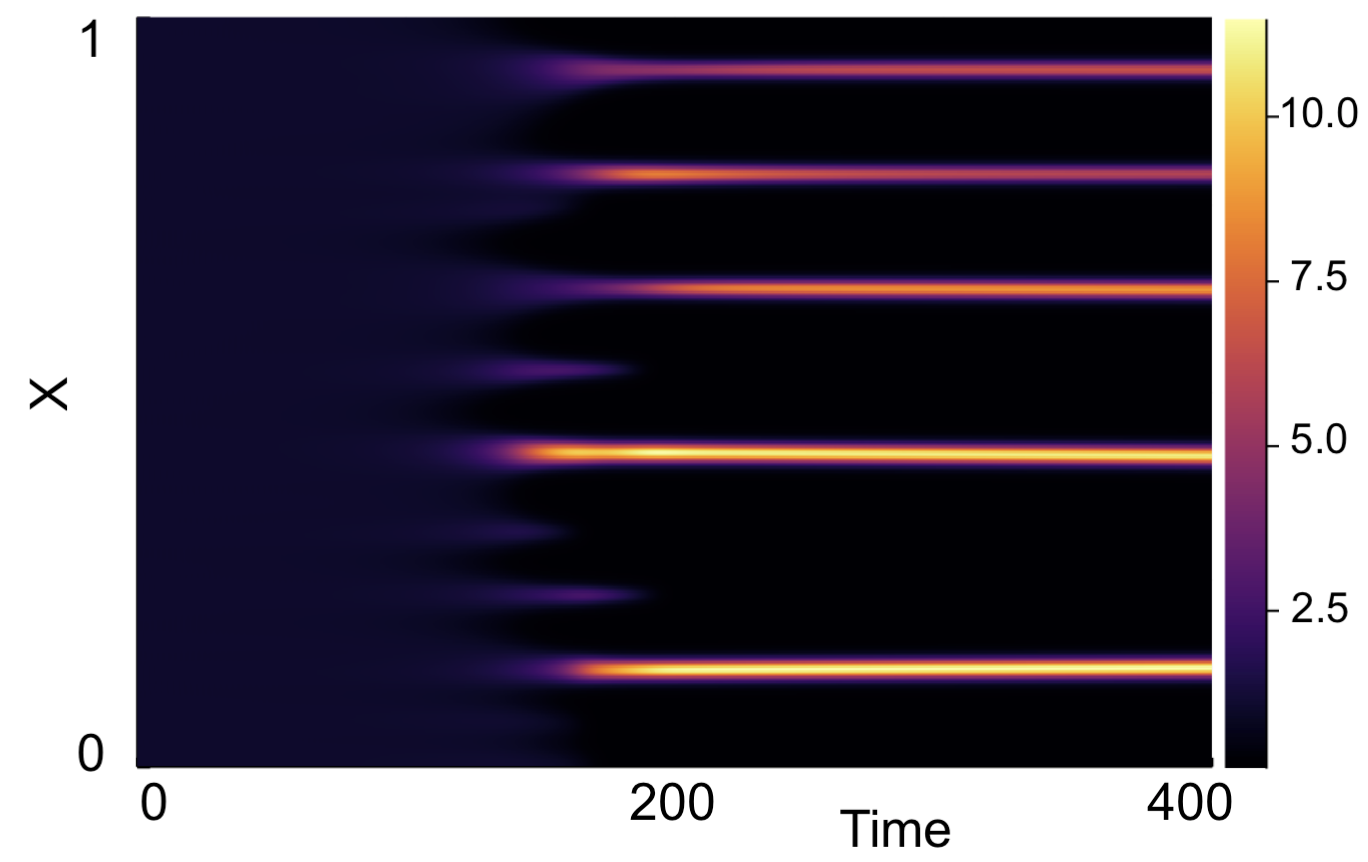}
        \caption{History function $h_1(t)$.}
        \label{}
    \end{subfigure}
    \hfill
    \begin{subfigure}[t]{0.32\textwidth}
        \centering
        \includegraphics[width=5cm,height=4.5cm]{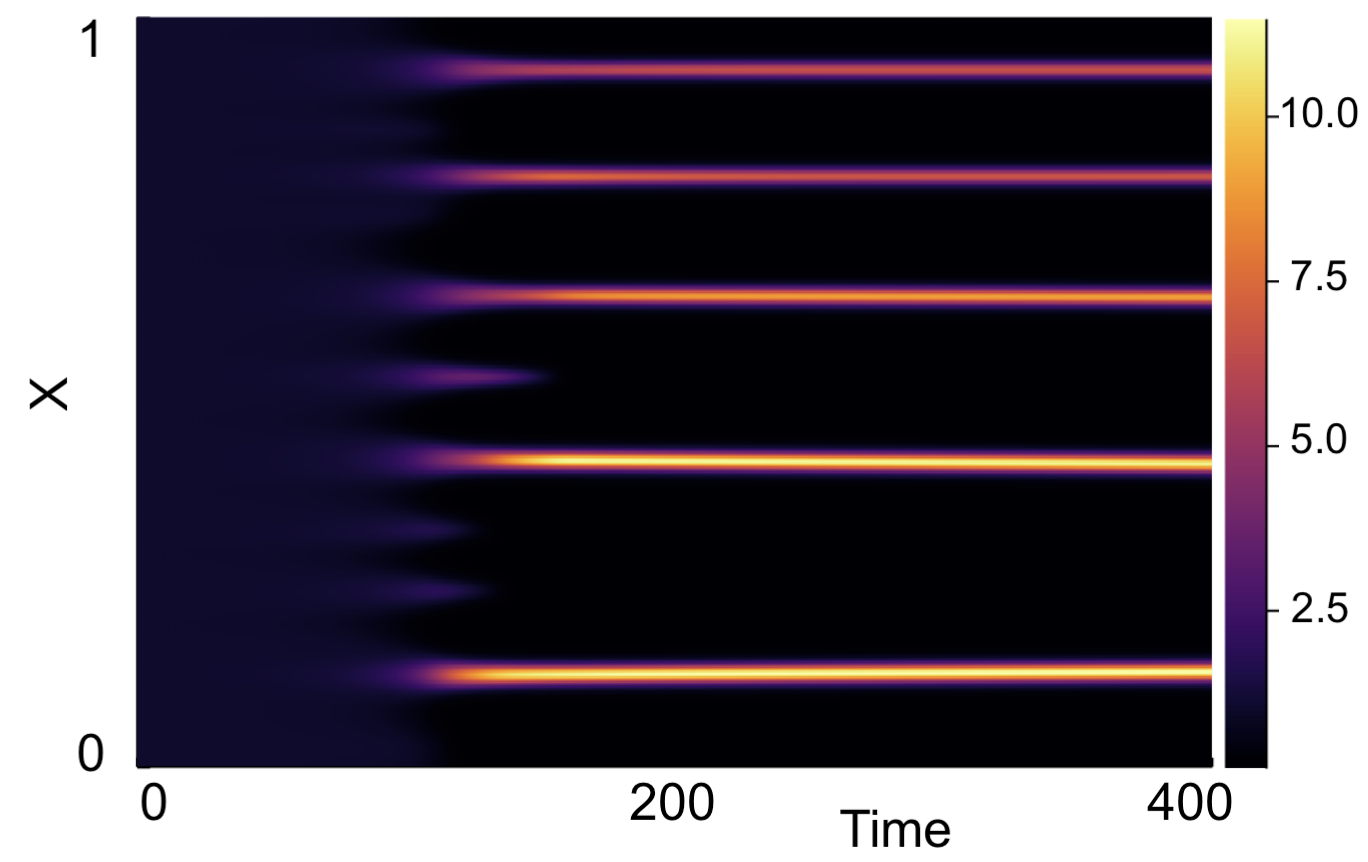}
        \caption{History function $h_2(t)$.}
        \label{}
    \end{subfigure}
    \caption{Numerical simulations of \eqref{fixed2} showing comparison of varying history functions for $\tau=2$. Boundary conditions given by \eqref{neumannbc} and initial conditions by \eqref{firstic}. Parameters $(a,b)=(0.1,0.9)$, $\epsilon^2=0.001$, $L^2=9/2$ used.}
    \label{fig:temp2}
\end{figure}

\begin{figure}[H]
    \centering
    \begin{subfigure}[t]{0.32\textwidth}
        \centering
        \includegraphics[width=5cm,height=4.5cm]{ic24.png}
        \caption{History function given as in \eqref{hist}.}
        \label{}
    \end{subfigure}
    \hfill
    \begin{subfigure}[t]{0.32\textwidth}
        \centering
        \includegraphics[width=5cm,height=4.5cm]{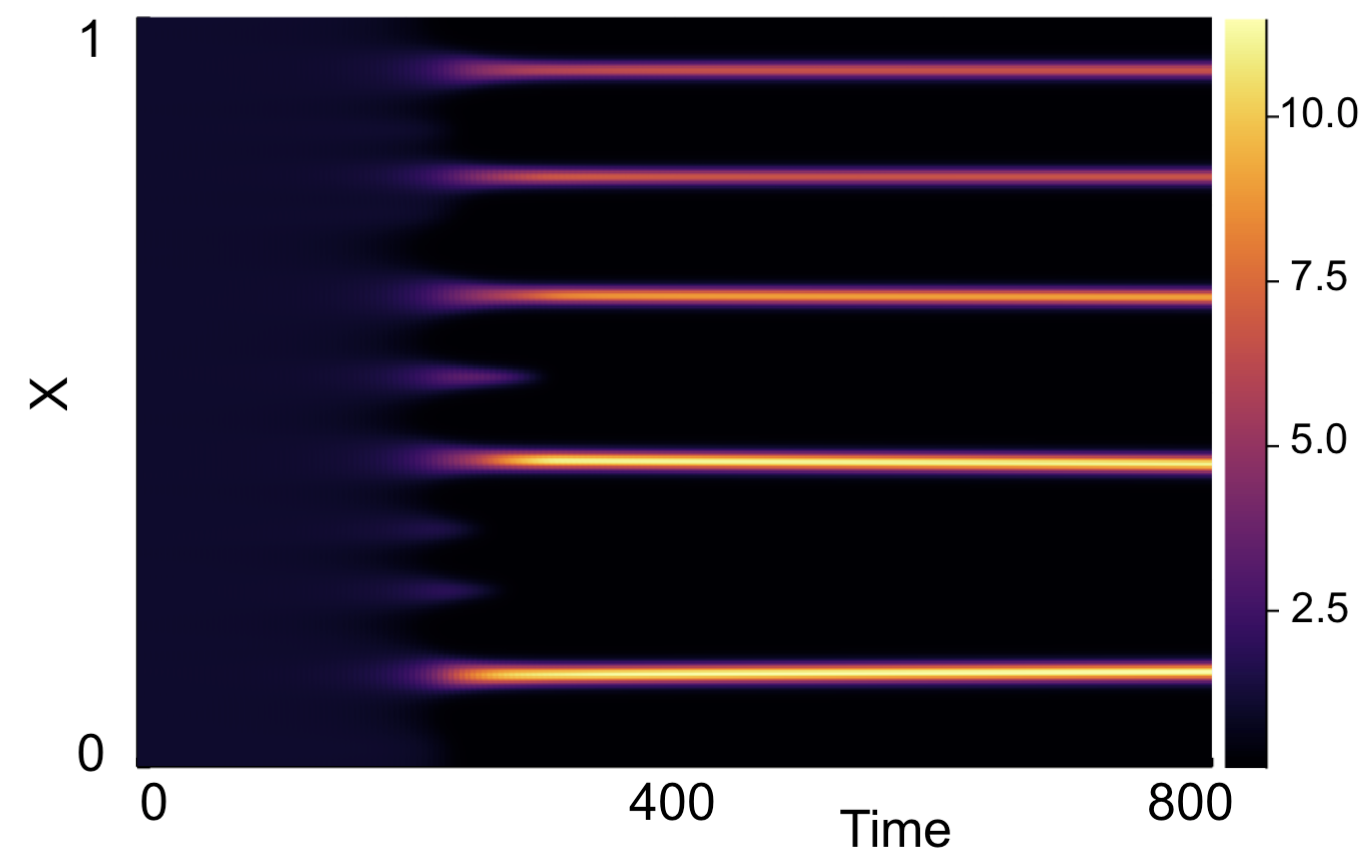}
        \caption{History function $h_1(t)$.}
        \label{}
    \end{subfigure}
    \hfill
    \begin{subfigure}[t]{0.32\textwidth}
        \centering
        \includegraphics[width=5cm,height=4.5cm]{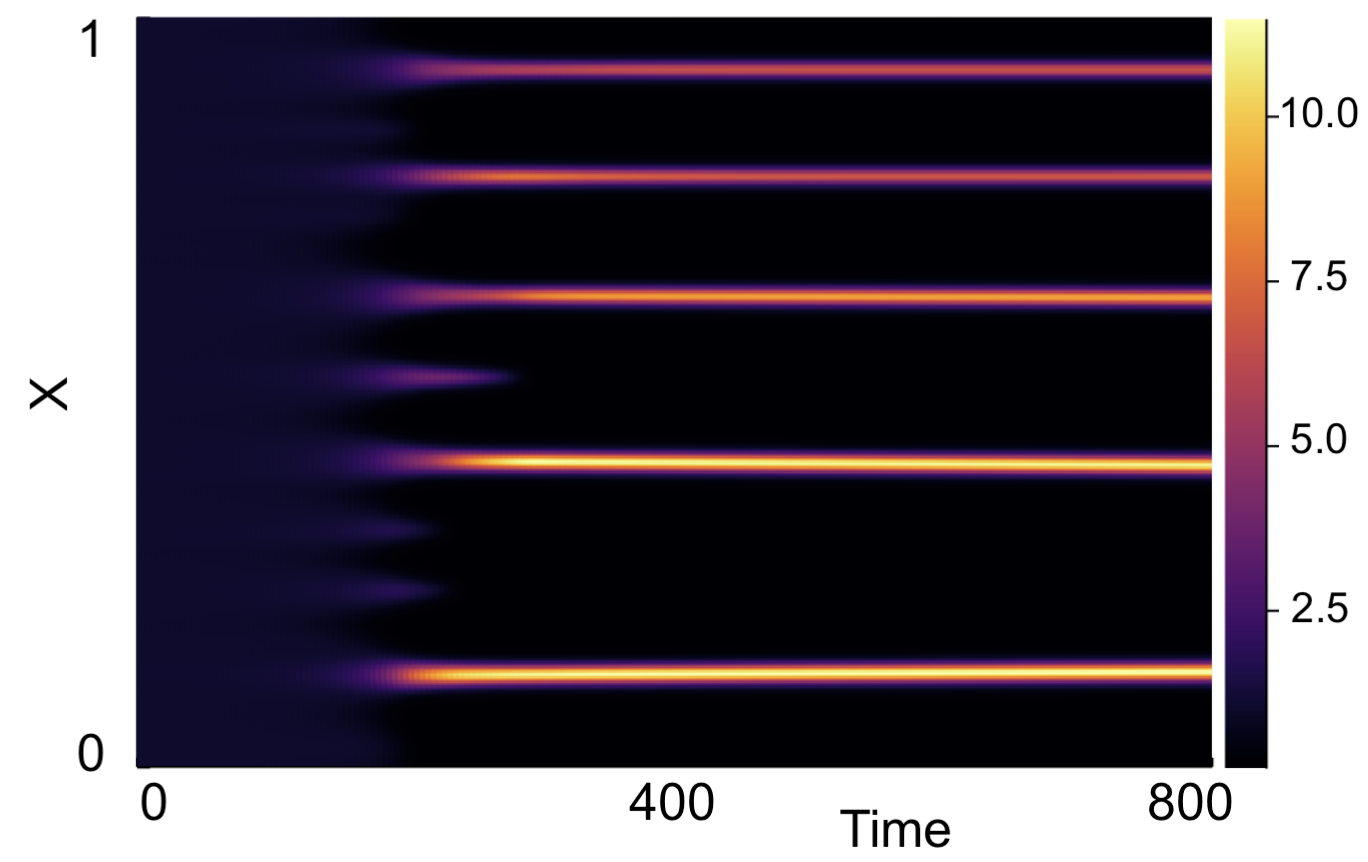}
        \caption{History function $h_2(t)$.}
        \label{}
    \end{subfigure}
    \caption{Numerical simulations of \eqref{fixed2} showing comparison of varying history functions for $\tau=4$. Boundary conditions given by \eqref{neumannbc} and initial conditions by \eqref{firstic}. Parameters $(a,b)=(0.1,0.9)$, $\epsilon^2=0.001$, $L^2=9/2$ used.}
    \label{fig:temp4}
\end{figure}

\begin{figure}[H]
    \centering
    \begin{subfigure}[t]{0.32\textwidth}
        \centering
        \includegraphics[width=5cm,height=4.5cm]{ic28.png}
        \caption{History function given as in \eqref{hist}.}
        \label{}
    \end{subfigure}
    \hfill
    \begin{subfigure}[t]{0.32\textwidth}
        \centering
        \includegraphics[width=5cm,height=4.5cm]{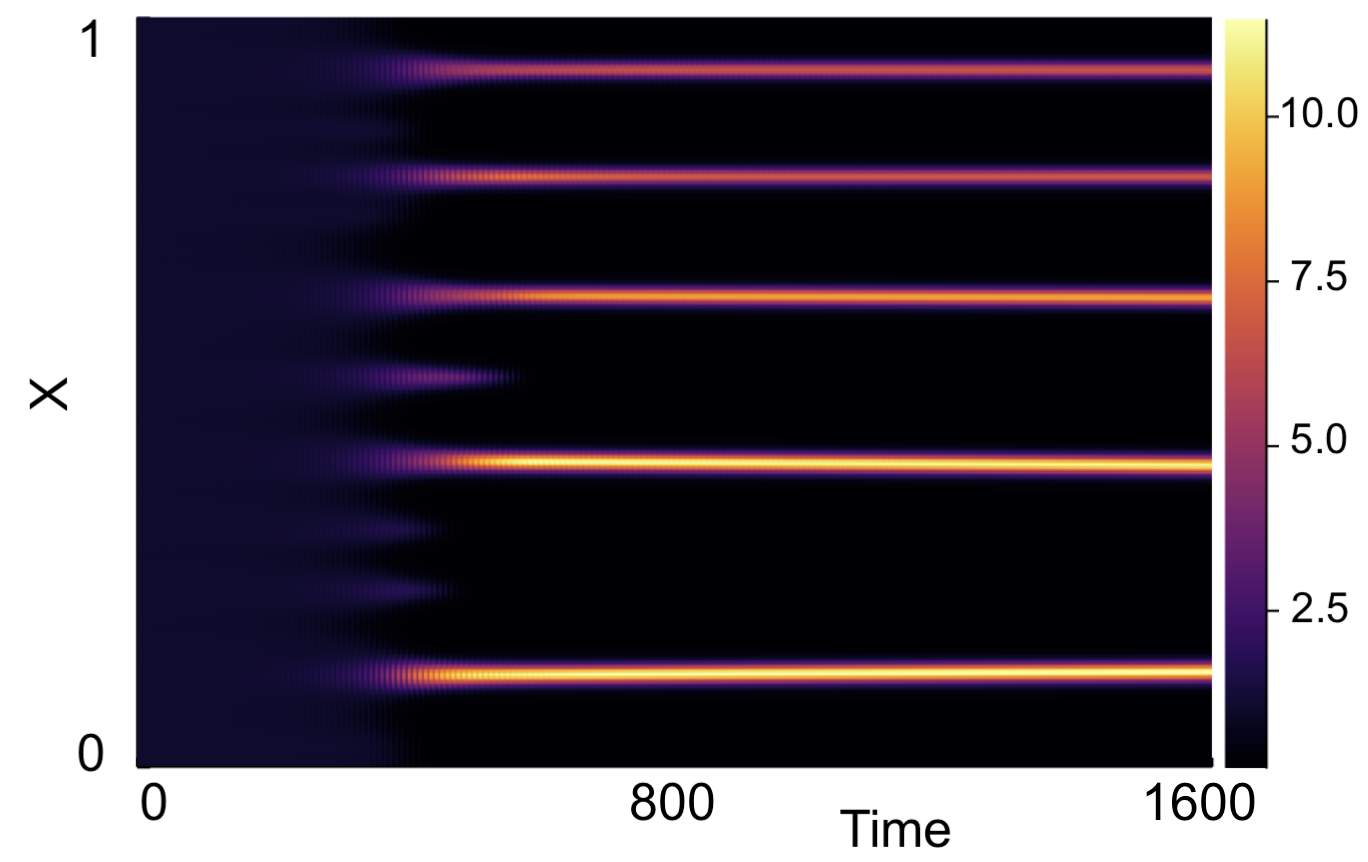}
        \caption{History function $h_1(t)$.}
        \label{}
    \end{subfigure}
    \hfill
    \begin{subfigure}[t]{0.32\textwidth}
        \centering
        \includegraphics[width=5cm,height=4.5cm]{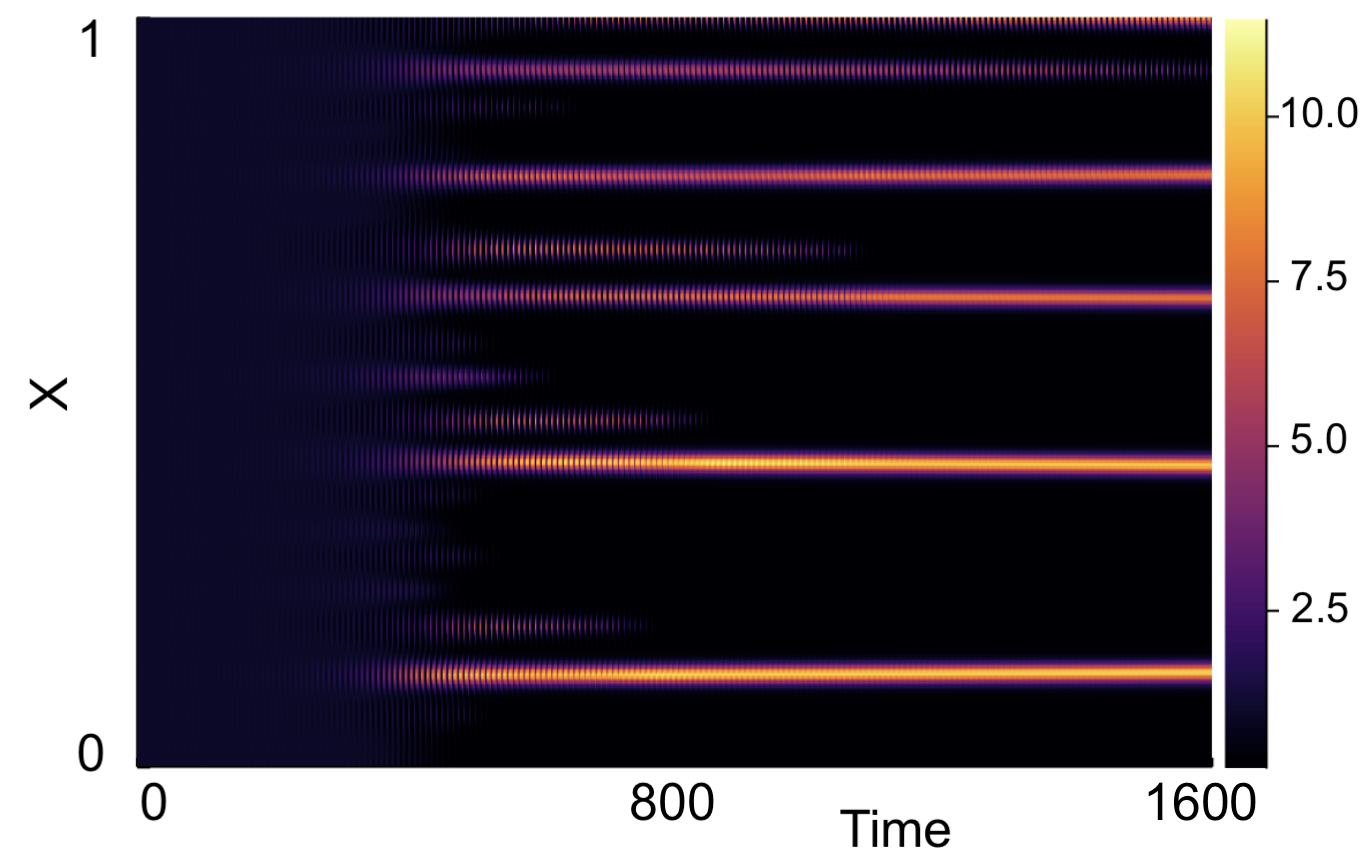}
        \caption{History function $h_2(t)$.}
        \label{}
    \end{subfigure}
    \caption{Numerical simulations of \eqref{fixed2} showing comparison of varying history functions for $\tau=8$. Boundary conditions given by \eqref{neumannbc} and initial conditions by \eqref{firstic}. Parameters $(a,b)=(0.1,0.9)$, $\epsilon^2=0.001$, $L^2=9/2$ used.}
    \label{fig:temp8}
\end{figure}

\begin{figure}[H]
    \centering
    \begin{subfigure}[t]{0.32\textwidth}
        \centering
        \includegraphics[width=5cm,height=4.5cm]{ic216.png}
        \caption{History function given as in \eqref{hist}.}
        \label{}
    \end{subfigure}
    \hfill
    \begin{subfigure}[t]{0.32\textwidth}
        \centering
        \includegraphics[width=5cm,height=4.5cm]{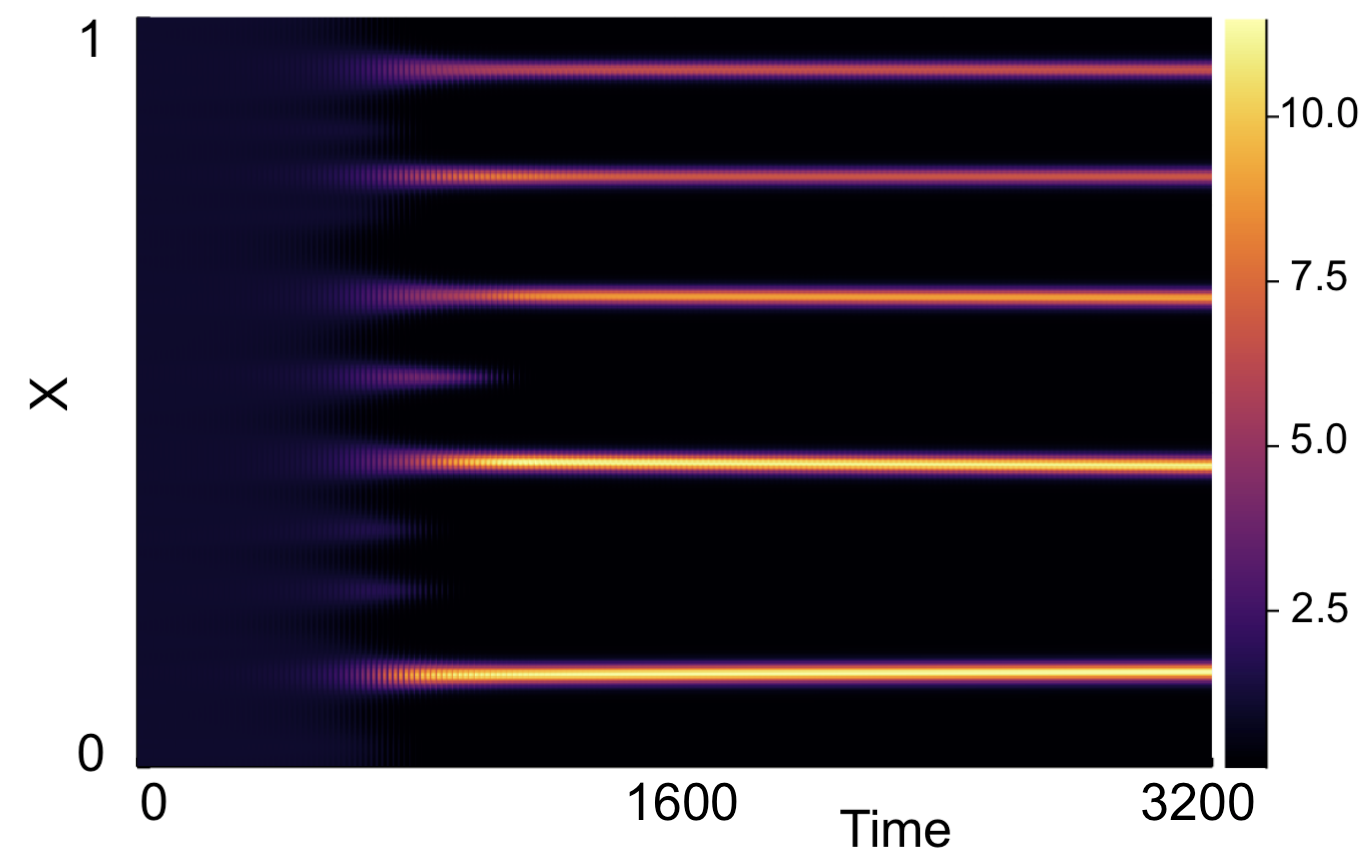}
        \caption{History function $h_1(t)$.}
        \label{}
    \end{subfigure}
    \hfill
    \begin{subfigure}[t]{0.32\textwidth}
        \centering
        \includegraphics[width=5cm,height=4.5cm]{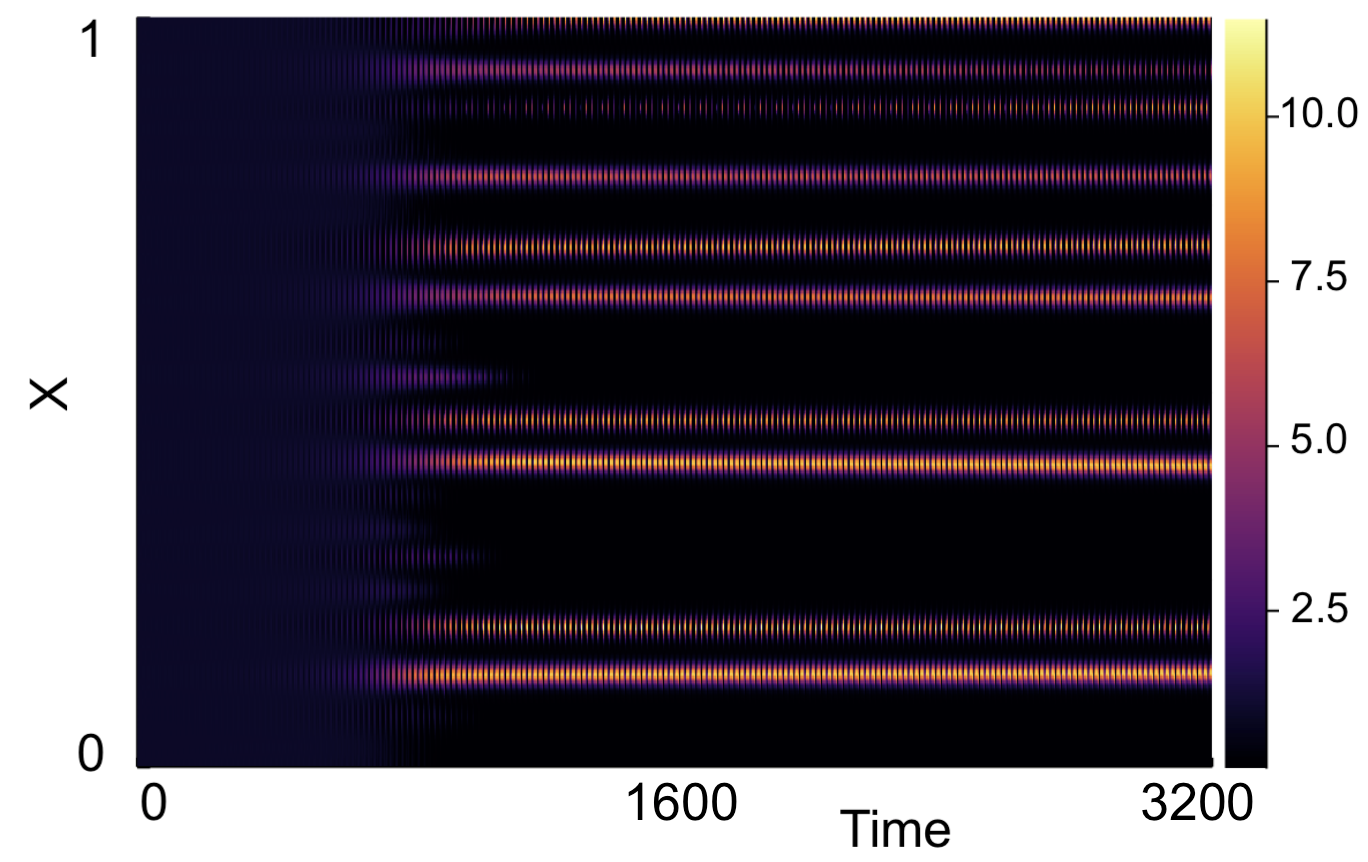}
        \caption{History function $h_2(t)$.}
        \label{}
    \end{subfigure}
    \caption{Numerical simulations of \eqref{fixed2} showing comparison of varying history functions for $\tau=16$. Boundary conditions given by \eqref{neumannbc} and initial conditions by \eqref{firstic}. Parameters $(a,b)=(0.1,0.9)$, $\epsilon^2=0.001$, $L^2=9/2$ used.}
    \label{fig:temp16}
\end{figure}

In Figures \ref{fig:temp1}, \ref{fig:temp2}, \ref{fig:temp4}, \ref{fig:temp8} and \ref{fig:temp16} show the preliminary results for a temporal variation in the history function on the time-to-pattern properties. We consider two history functions, namely $h(t)=u_\star(1+r\sin(\omega t))$, for $\omega=1/7,4/7$ for $t\in[-\tau,0)$, where $r$ is the random variable used in $\text{IC}_2$. The history functions with $\omega=1/7,4/7$ will be denoted $h_1(t)$, and $h_2(t)$ respectively. For each $\tau\in\{1,2,4,8,16\}$, we compare the results for each of these variations in history function with the numerical results simulated with the history function equal to the initial conditions, as given in \eqref{hist}. We see that these variations in the history function do not significantly affect the timescales on which pattern formation occurs.


\begin{figure}[H]
   \centering
   \begin{subfigure}[t]{0.45\textwidth}
       \centering
       \includegraphics[width=7cm,height=5cm]{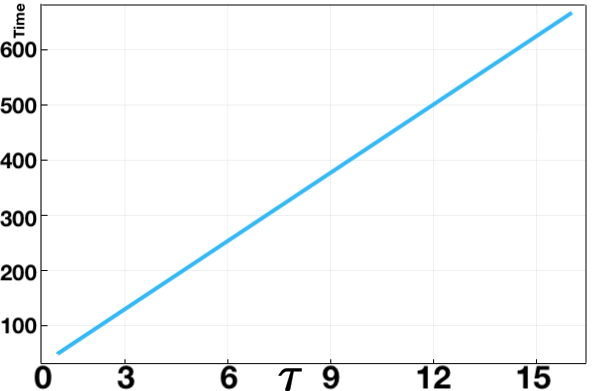}
       \caption{$(a,b)=(0.1,0.9)$}
       \label{}
   \end{subfigure}
   \hfill
   \begin{subfigure}[t]{0.45\textwidth}
       \centering
       \includegraphics[width=7cm,height=5cm]{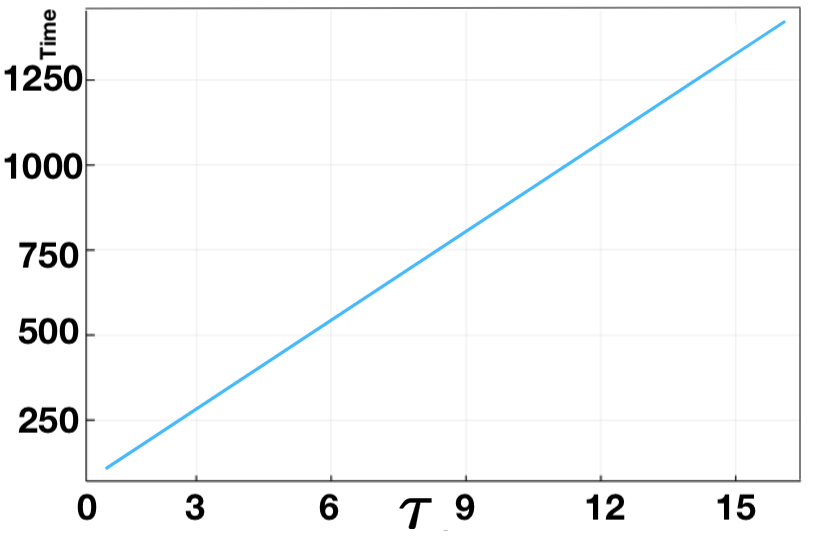}
       \caption{$(a,b)=(0.4,0.8)$}
       \label{}
   \end{subfigure}
   \caption{Time-to-pattern for full numerical solutions of \eqref{fixed2} plotted against $\tau\in[1,16]$ for $\sigma_{\text{IC}}=0.01$ and threshold $2$. Parameters used are $\epsilon^2=0.001$ and domain size $L^2=9/2$.}
   \label{fig:Bbif4}
\end{figure}

In Figure \ref{fig:Bbif4}, we present results showing a linear relationship between time delay and time-to-pattern for the same parameter values used in Figure \ref{fig:linperturb1}, but with a threshold value $2$, from a $\sigma_{\text{IC}}=0.01$.

\section{For Chapter 3}\label{section:Bdist}
\subsection{A Symmetric Distribution}
\begin{figure}[H]
    \centering
    \begin{subfigure}[t]{0.32\textwidth}
        \centering
        \includegraphics[width=5cm,height=4.5cm]{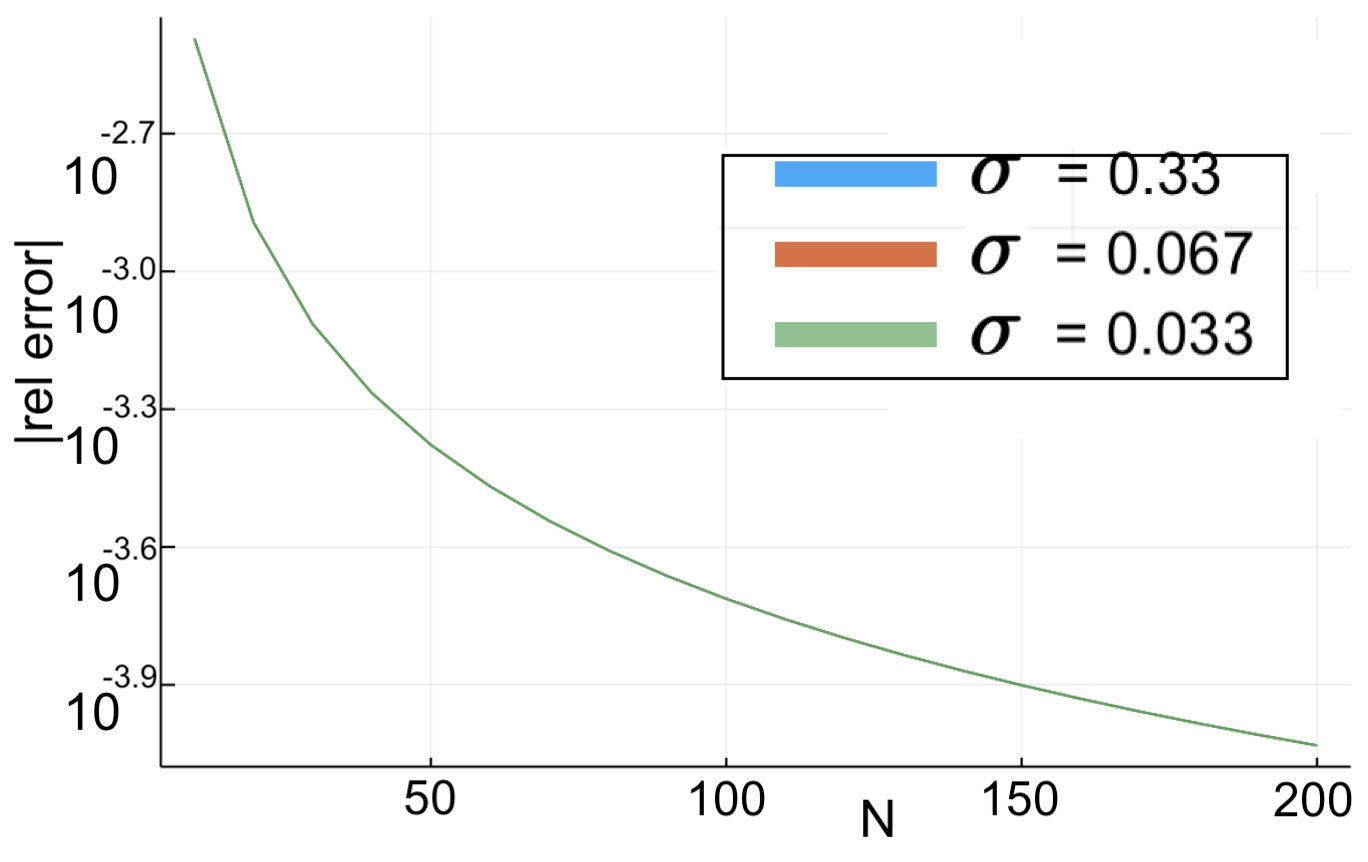}
        \caption{$\tau=1$.}
        \label{}
    \end{subfigure}
    \hfill
    \begin{subfigure}[t]{0.32\textwidth}
        \centering
        \includegraphics[width=5cm,height=4.5cm]{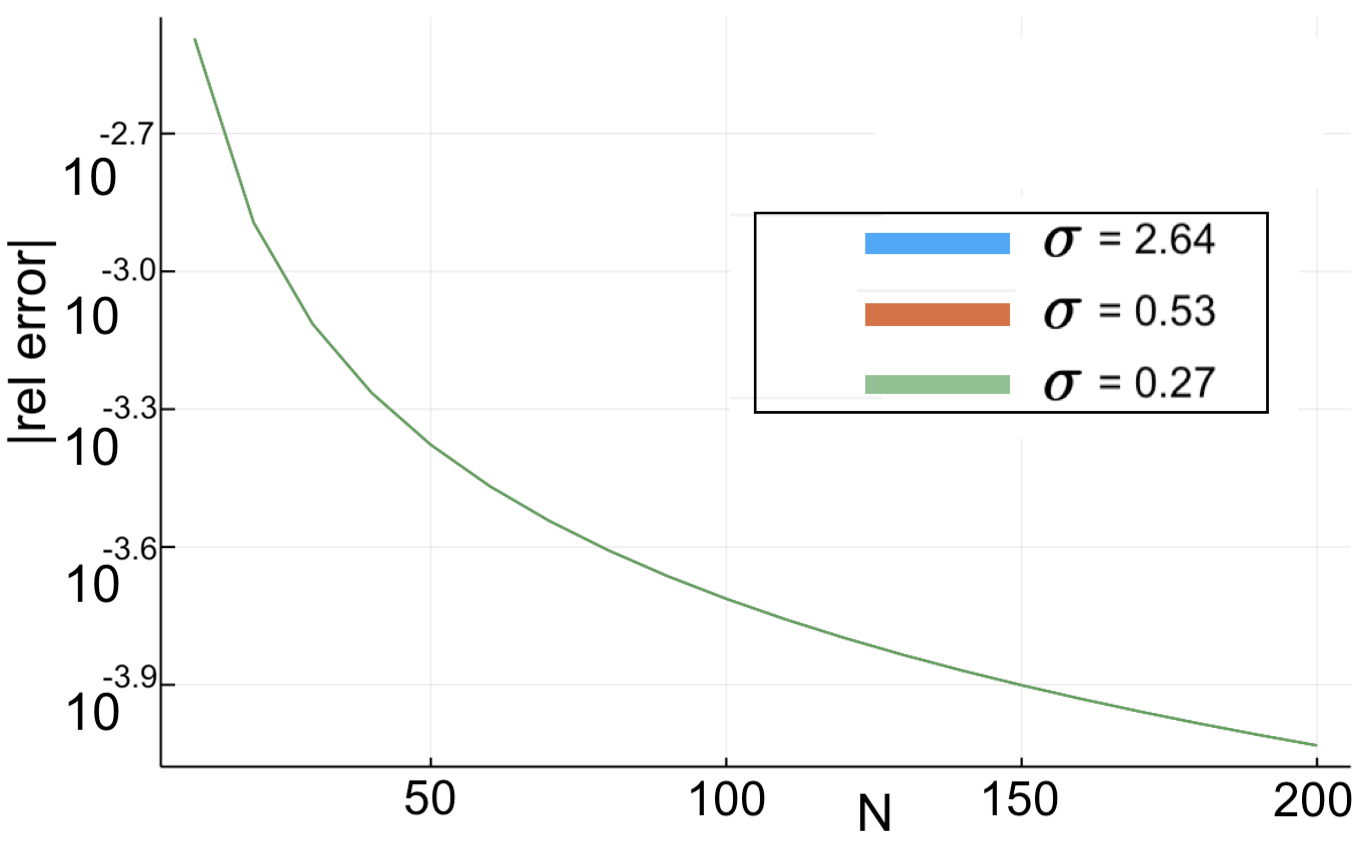}
        \caption{$\tau=8$.}
        \label{}
    \end{subfigure}
    \hfill
    \begin{subfigure}[t]{0.32\textwidth}
        \centering
        \includegraphics[width=5cm,height=4.5cm]{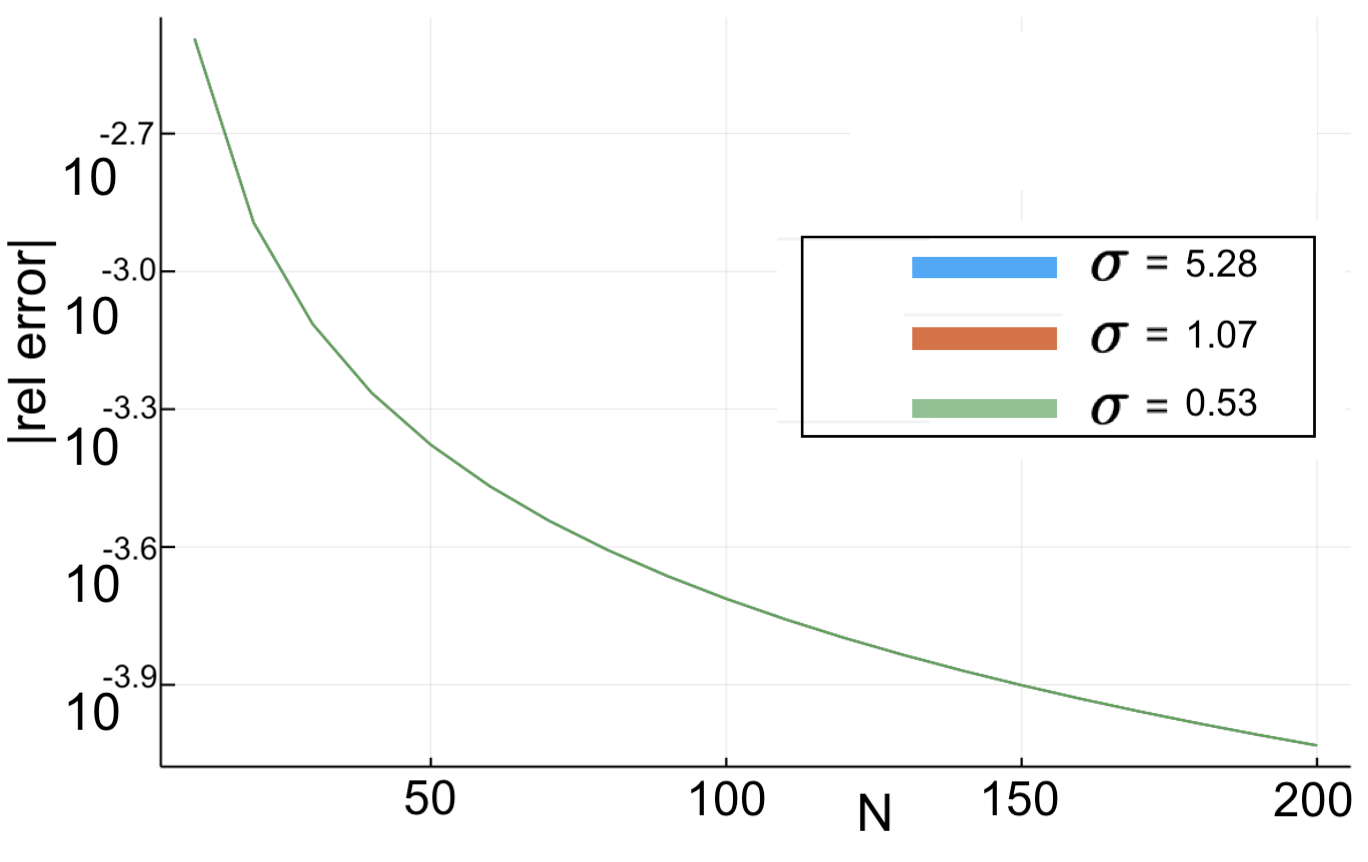}
        \caption{$\tau=16$.}
        \label{}
    \end{subfigure}
    \caption{Relative error of composite Simpson's rule applied to integrating $k(s;\tau,\sigma)$ for varying $\tau\in\{1,8,16\}$ and $\sigma\in\{\sigma_{\max}\times0.99,\sigma_{\max}\times0.2,\sigma_{\max}\times0.1\}$. Number of discretisation points varied $N\in[10,200]$ at regular intervals of $10$.}
    \label{fig:quad}
\end{figure}

\begin{figure}[H]
    \centering
    \begin{subfigure}[t]{0.32\textwidth}
        \centering
        \includegraphics[width=5cm,height=4.5cm]{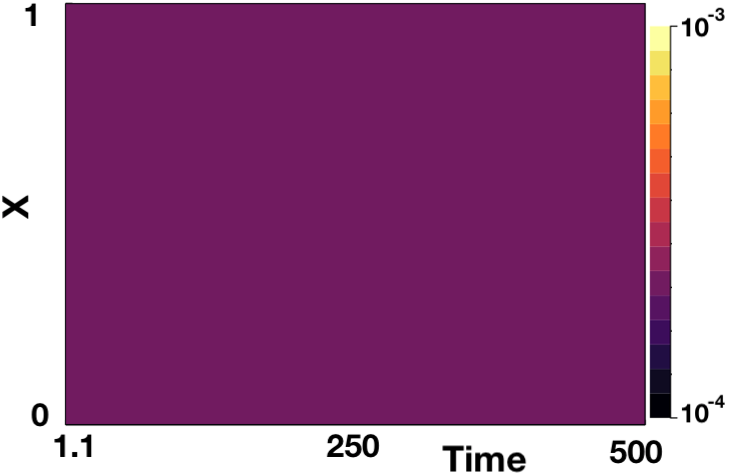}
        \caption{$\tau=1$.}
        \label{}
    \end{subfigure}
    \hfill
    \begin{subfigure}[t]{0.32\textwidth}
        \centering
        \includegraphics[width=5cm,height=4.5cm]{disterr.png}
        \caption{$\tau=8$.}
        \label{}
    \end{subfigure}
    \hfill
    \begin{subfigure}[t]{0.32\textwidth}
        \centering
        \includegraphics[width=5cm,height=4.5cm]{disterr.png}
        \caption{$\tau=16$.}
        \label{}
    \end{subfigure}
    \caption{Relative error of composite Simpson's rule applied to integrating test integral given in \eqref{testint} for varying $\tau\in\{1,8,16\}$, and $\sigma=\sigma_{\max}\times0.99$. $N=50$ discretisation points used.}
    \label{fig:tempquad}
\end{figure}

We present results for the composite Simpson's rule to motivate a choice of $N=50$ temporal discretisation points. Figure \ref{fig:quad} shows the relative error for $N\in[10,200]$ varied at regular intervals of $10$, for the quadrature rule applied to the symmetric truncated Gaussian pdf $k(s;\tau,\sigma)$. We vary $\tau\in\{1,8,16\}$, and for each $\tau$ consider $\sigma\in\{\sigma_{\max}\times0.99,\sigma_{\max}\times0.2,\sigma_{\max}\times0.1\}$. Figure \ref{fig:tempquad} shows the relative error across both the spatial and temporal domains for the quadrature rule applied to the test integral, given in \eqref{testint}, for a the same varying $\tau$, and $\sigma=\sigma_{\max}\times0.99$. Results are shown here for only one $\sigma$ value, as it was found that the relative error was the same, independent of the $\sigma$ used.

\begin{figure}[H]
    \centering
    \begin{subfigure}[t]{0.45\textwidth}
        \centering
        \includegraphics[width=7cm,height=4.75cm]{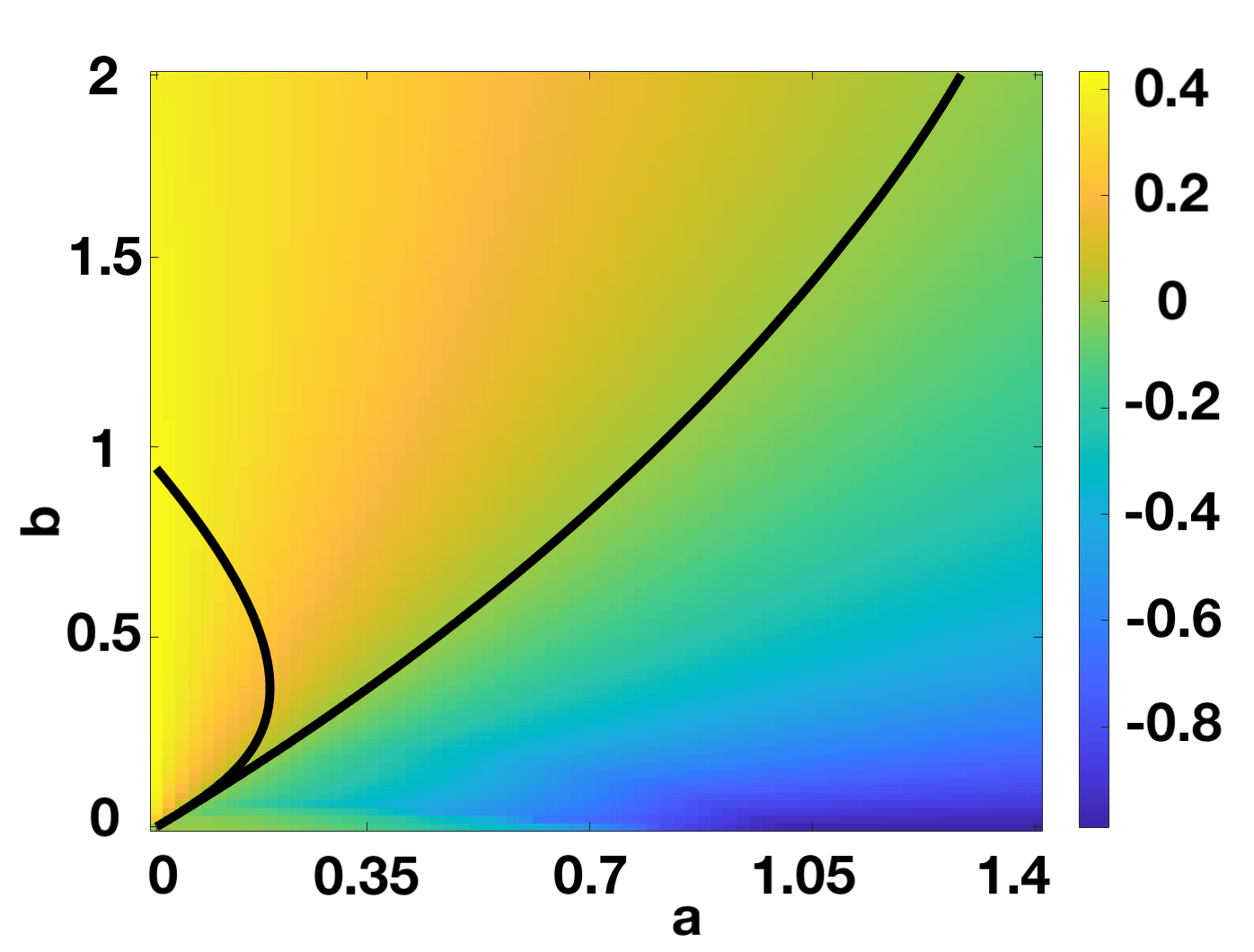}
        \caption{$\tau=0.2$.}
        \label{}
    \end{subfigure}
    \hfill
    \begin{subfigure}[t]{0.45\textwidth}
        \centering
        \includegraphics[width=7cm,height=4.75cm]{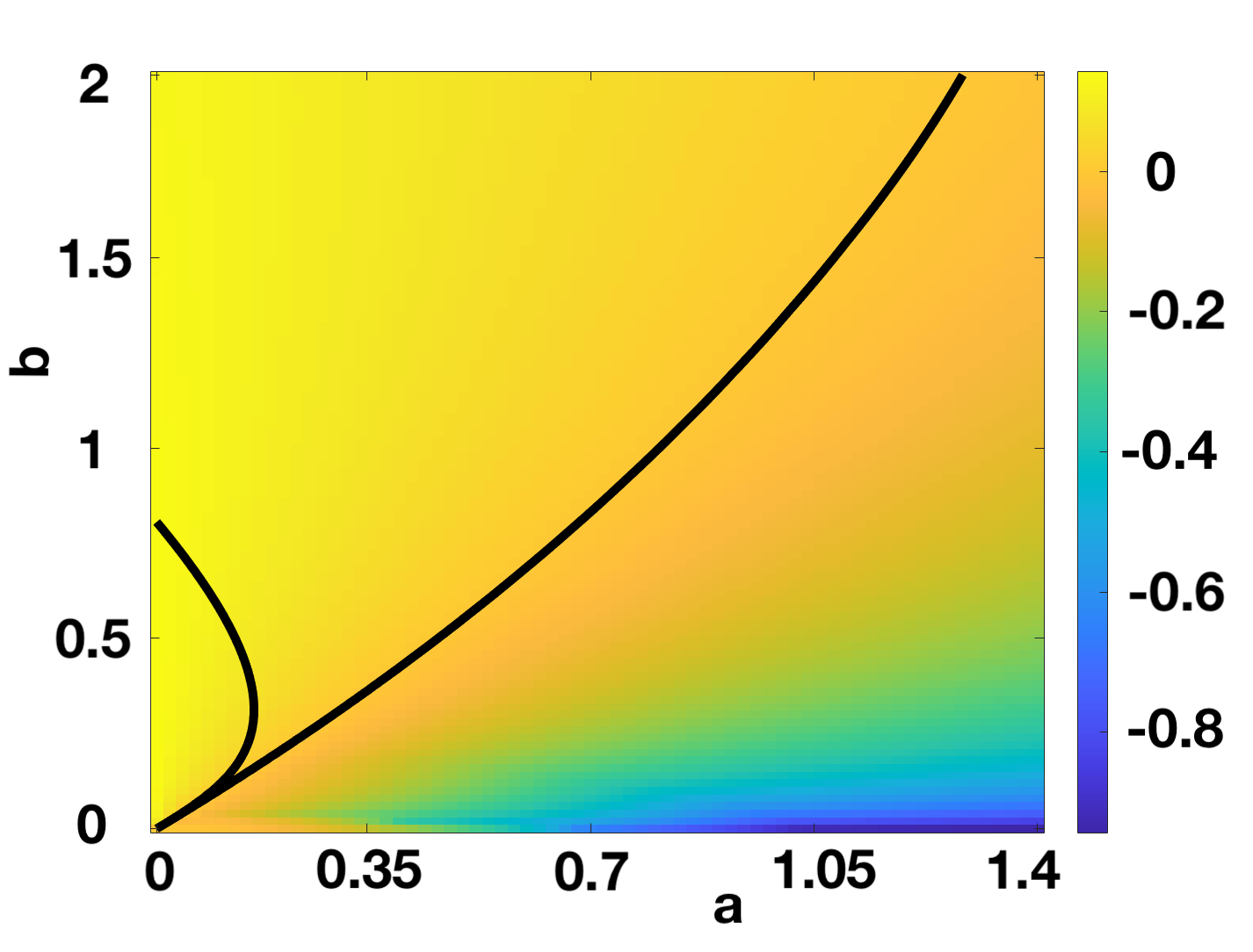}
        \caption{$\tau=1$}
        \label{}
    \end{subfigure}
    \caption{Bifurcation diagrams produced by solving \eqref{characfix} (fixed delay characteristic equation) for $\tau=0.2,1$ and $\epsilon^2=0.001$, on a domain length $L^2=9/2$.}
    \label{fig:distheat1}
\end{figure}

\begin{figure}[H]
    \centering
    \begin{subfigure}[t]{0.45\textwidth}
        \centering
        \includegraphics[width=7cm,height=4.75cm]{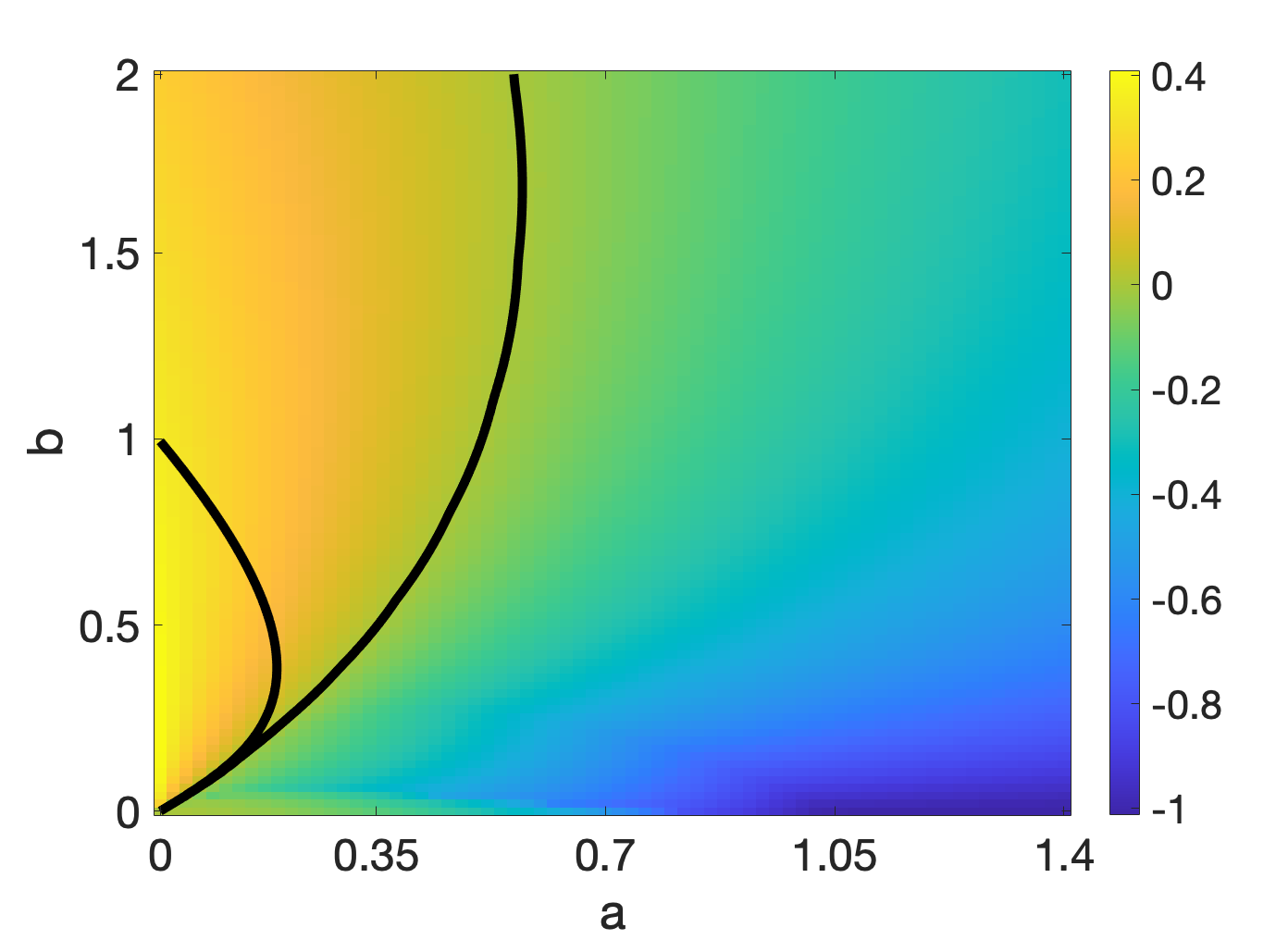}
        \caption{$\tau=0.2$.}
        \label{}
    \end{subfigure}
    \hfill
    \begin{subfigure}[t]{0.45\textwidth}
        \centering
        \includegraphics[width=7cm,height=4.75cm]{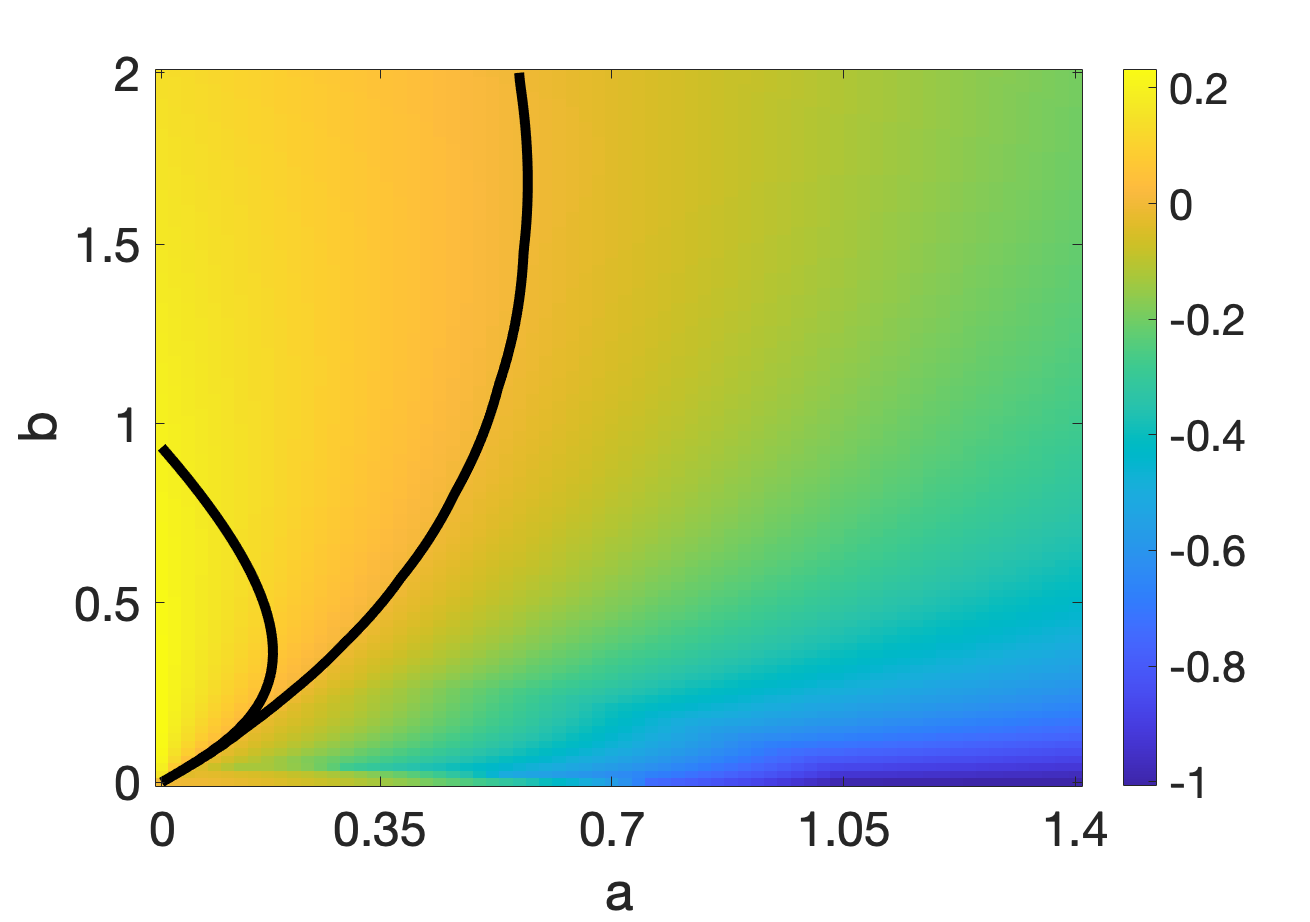}
        \caption{$\tau=0.5$}
        \label{}
    \end{subfigure}
    \caption{Bifurcation diagrams produced by solving \eqref{characfix} (fixed delay characteristic equation) for $\tau=0.2,0.5$ and $\epsilon^2=0.01$, on a domain length $L^2=9/2$.}
    \label{fig:distheat2}
\end{figure}

\begin{figure}[H]
    \centering
    \begin{subfigure}[t]{0.45\textwidth}
        \centering
        \includegraphics[width=7cm,height=4.75cm]{t1f1.png}
        \caption{Fixed delay case}
        \label{}
    \end{subfigure}
    \hfill
    \begin{subfigure}[t]{0.45\textwidth}
        \centering
        \includegraphics[width=7cm,height=4.75cm]{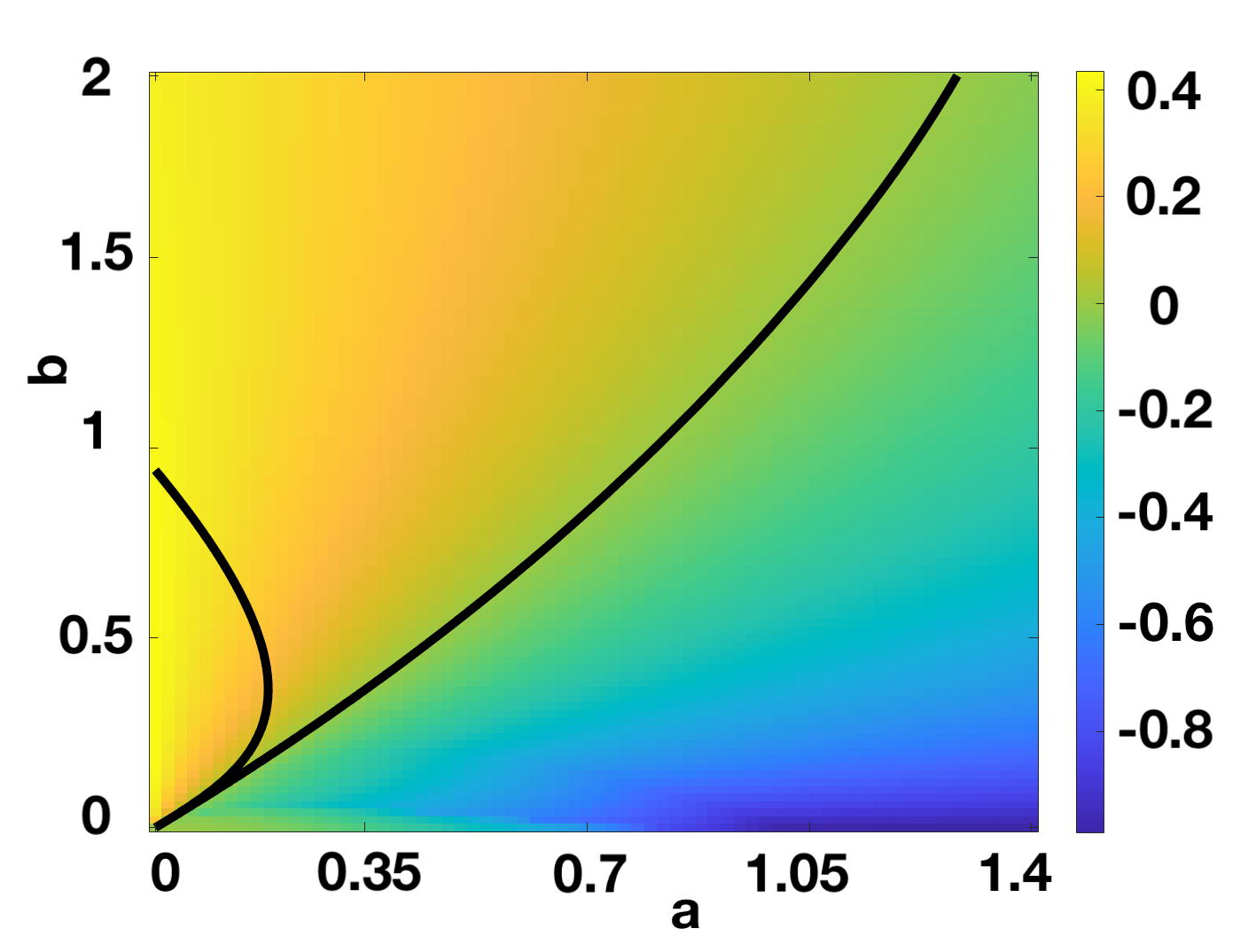}
        \caption{$\sigma=\sigma_{\max}\times0.99$}
        \label{}
    \end{subfigure}
    \hfill
    \begin{subfigure}[t]{0.45\textwidth}
        \centering
        \includegraphics[width=7cm,height=4.75cm]{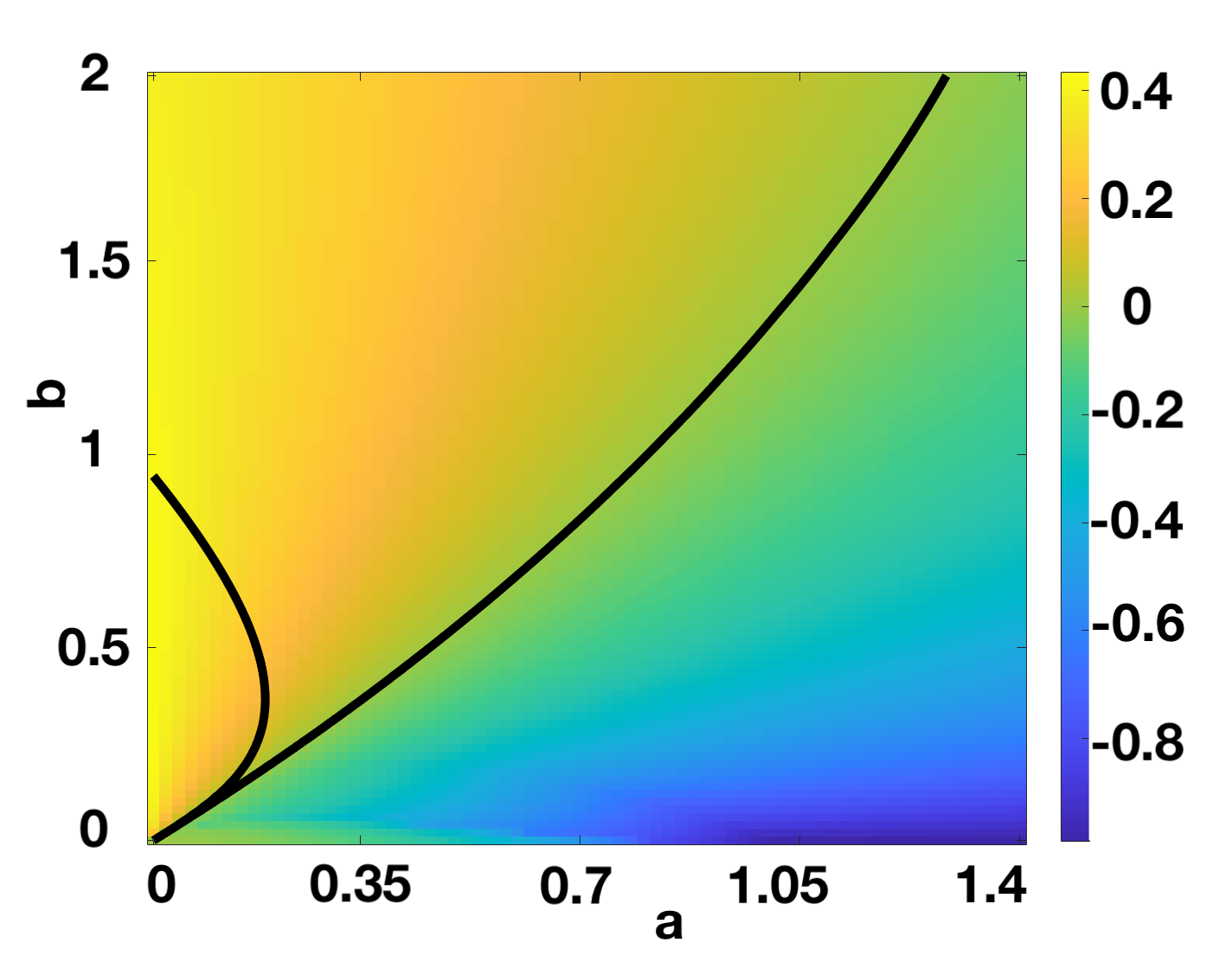}
        \caption{$\sigma=\sigma_{\max}\times0.2$}
        \label{}
    \end{subfigure}
    \hfill
    \begin{subfigure}[t]{0.45\textwidth}
        \centering
        \includegraphics[width=7cm,height=4.75cm]{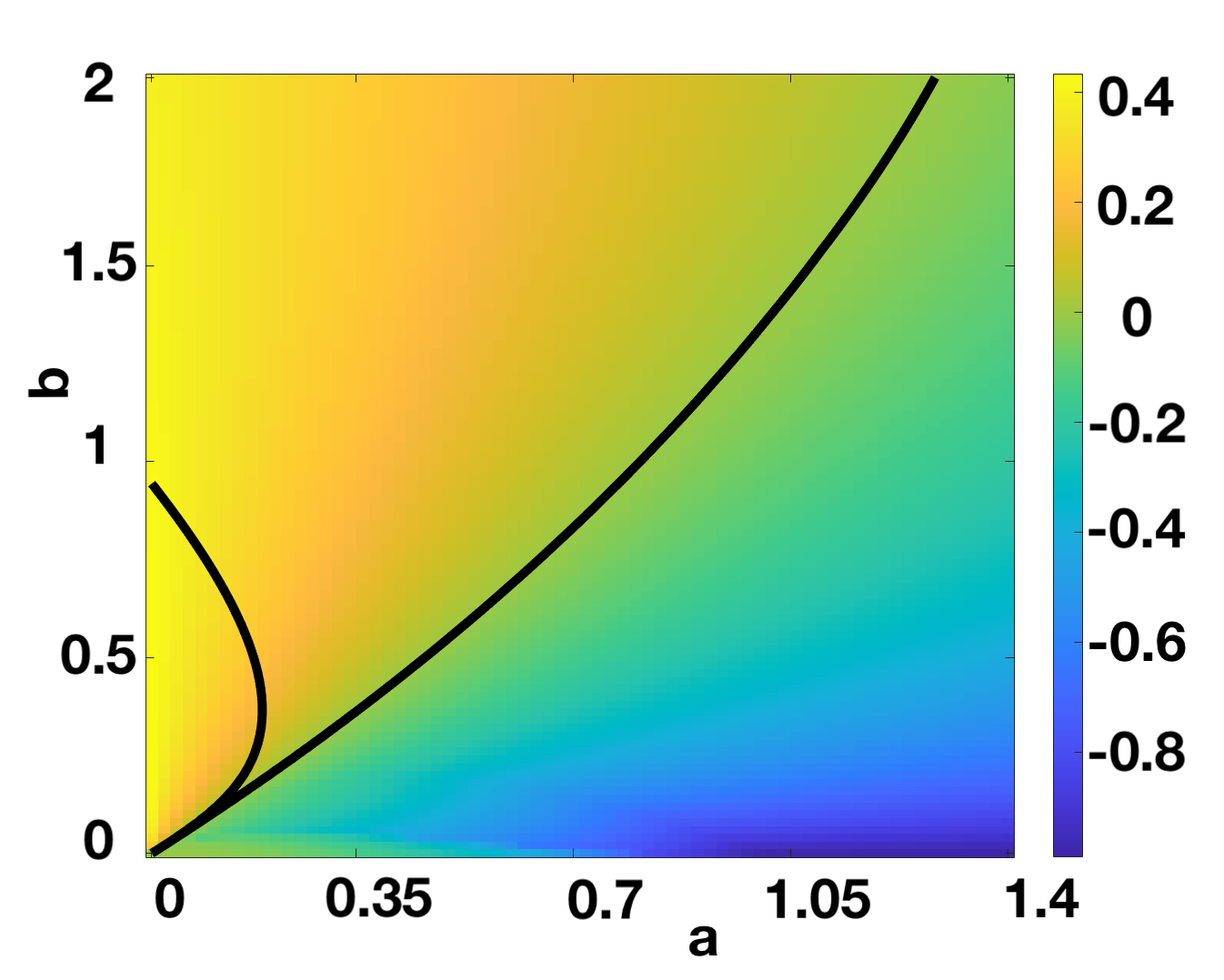}
        \caption{$\sigma=\sigma_{\max}\times0.1$}
        \label{}
    \end{subfigure}
    \caption{Bifurcation diagrams for varying $\sigma$ for $\tau=0.2$ and $\epsilon^2=0.001$, on a domain length $L^2=9/2$.}
    \label{fig:distmap1}
\end{figure}

\begin{figure}[H]
    \centering
    \begin{subfigure}[t]{0.45\textwidth}
        \centering
        \includegraphics[width=7cm,height=4.75cm]{t2f1.png}
        \caption{Fixed delay case}
        \label{}
    \end{subfigure}
    \hfill
    \begin{subfigure}[t]{0.45\textwidth}
        \centering
        \includegraphics[width=7cm,height=4.75cm]{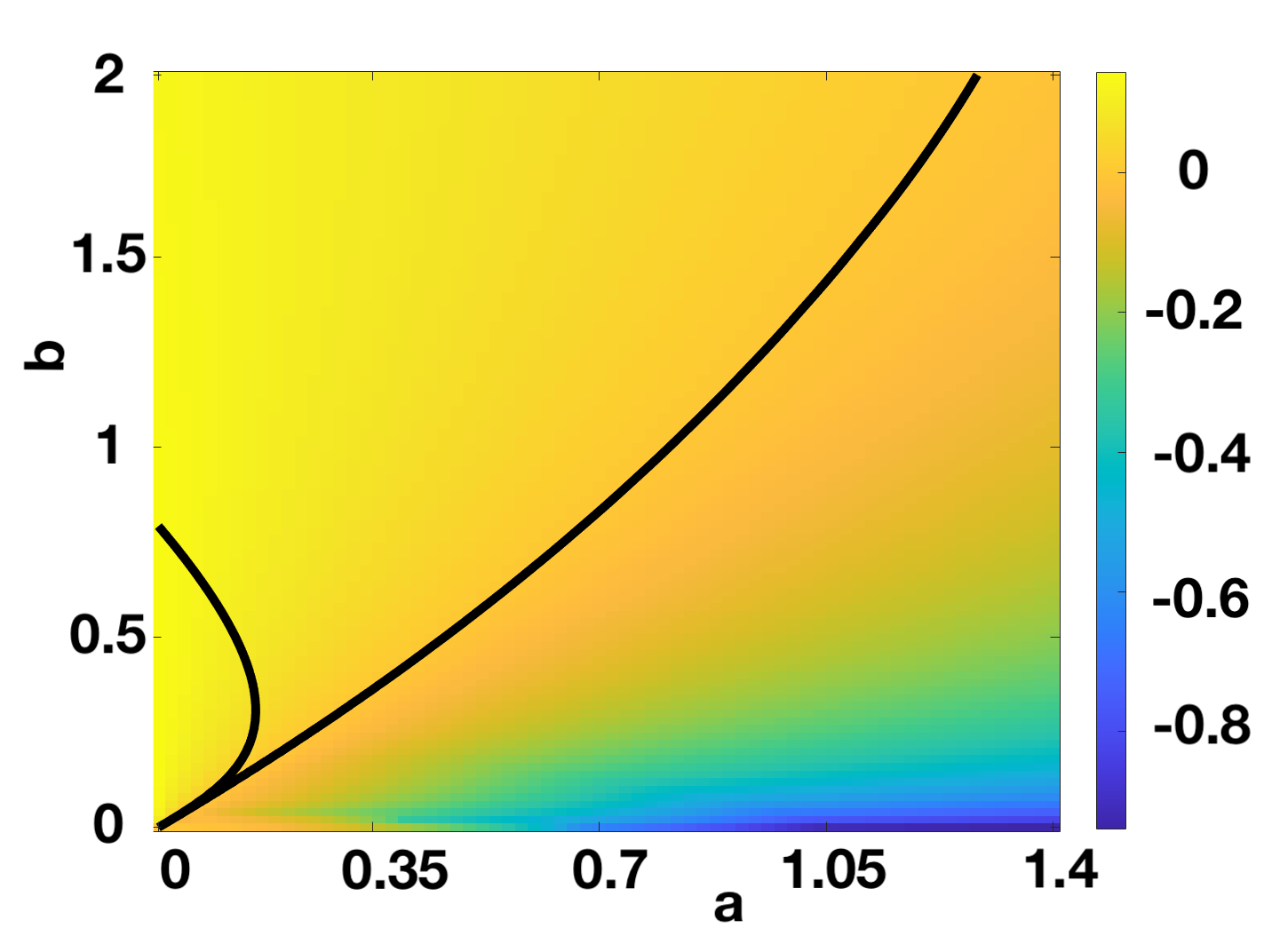}
        \caption{$\sigma=\sigma_{\max}\times0.99$}
        \label{}
    \end{subfigure}
    \hfill
    \begin{subfigure}[t]{0.45\textwidth}
        \centering
        \includegraphics[width=7cm,height=4.75cm]{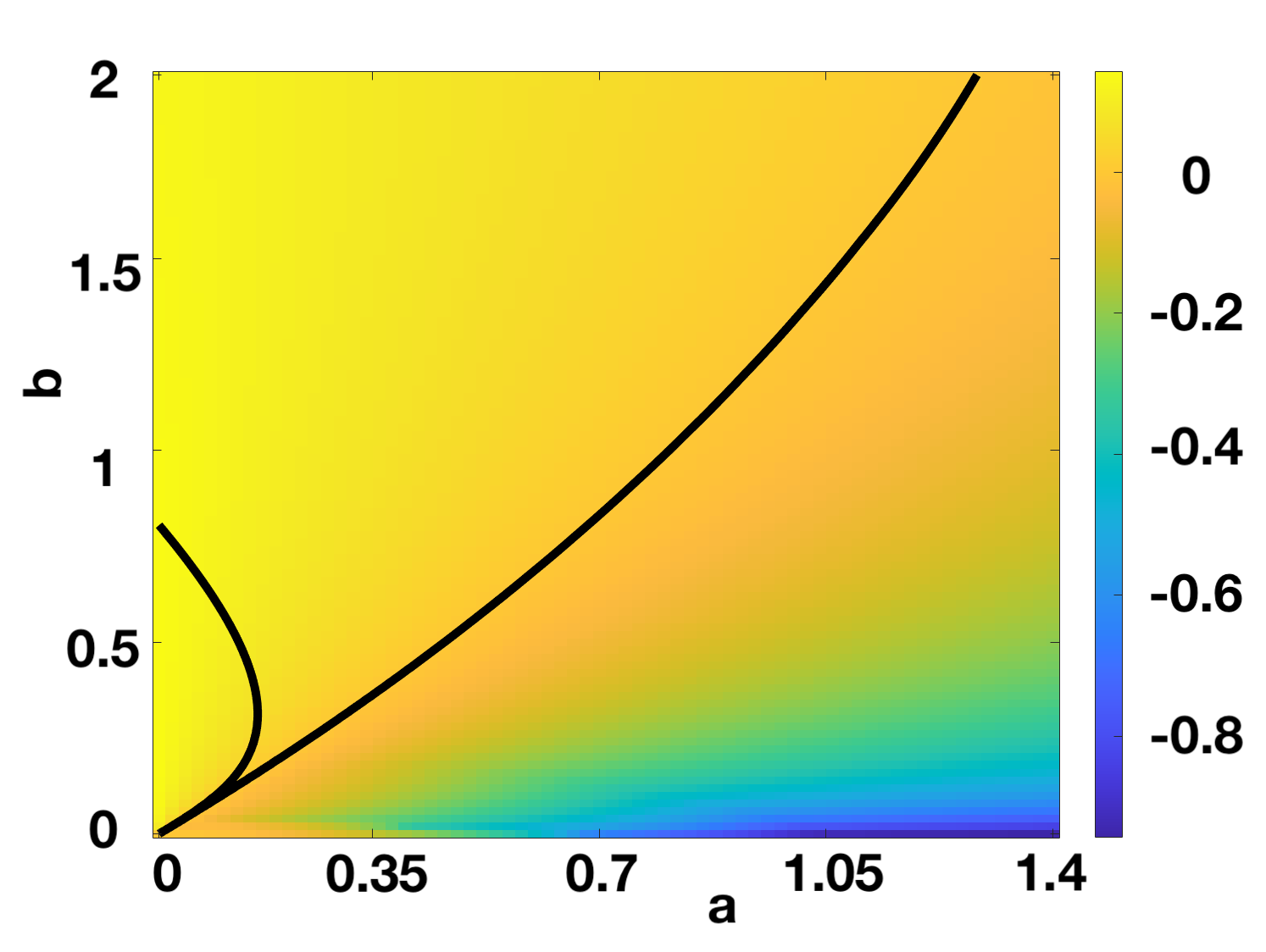}
        \caption{$\sigma=\sigma_{\max}\times0.2$}
        \label{}
    \end{subfigure}
    \hfill
    \begin{subfigure}[t]{0.45\textwidth}
        \centering
        \includegraphics[width=7cm,height=4.75cm]{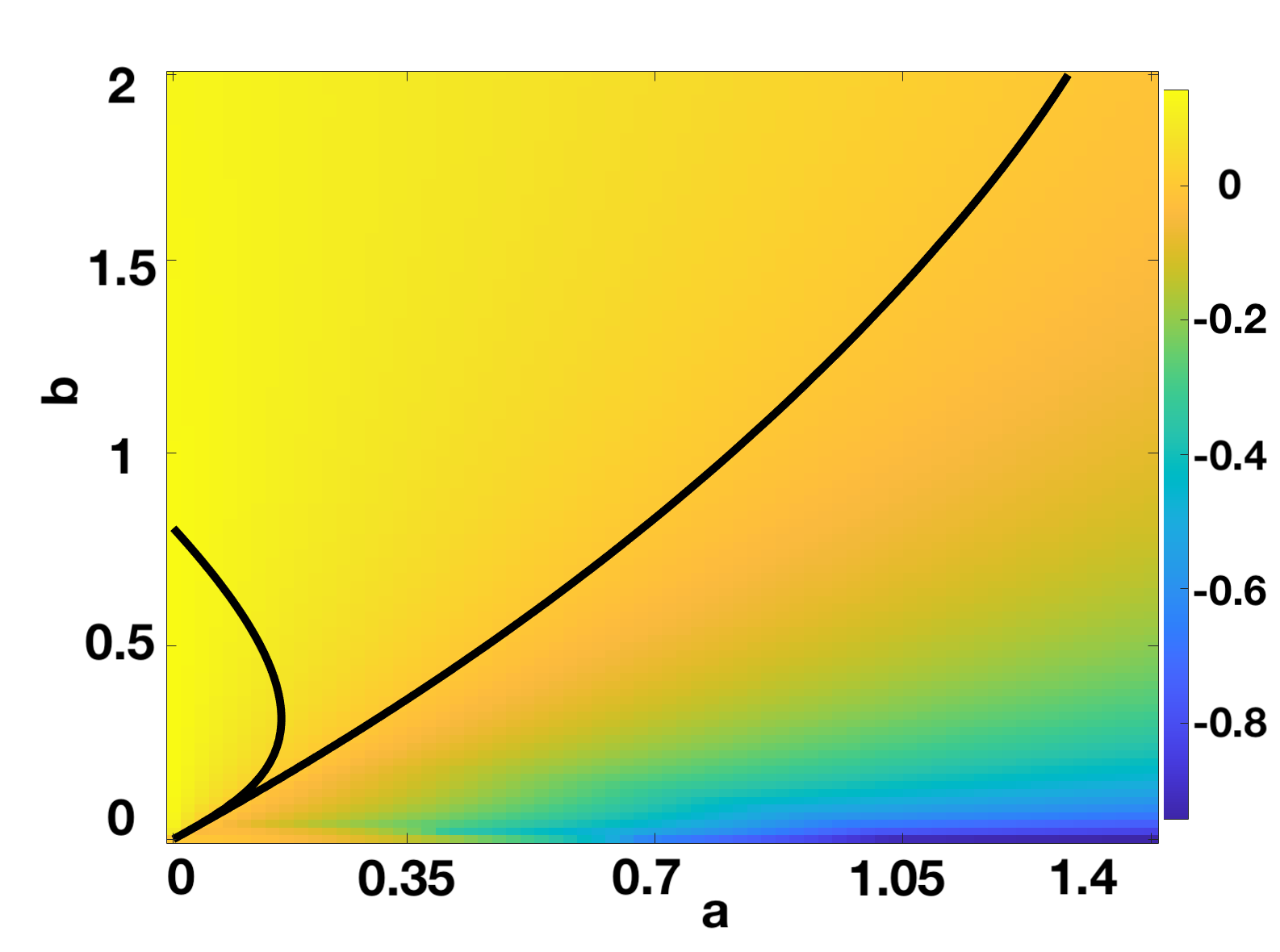}
        \caption{$\sigma=\sigma_{\max}\times0.1$}
        \label{}
    \end{subfigure}
    \caption{Bifurcation diagrams for varying $\sigma$ for $\tau=1$ and $\epsilon^2=0.001$, on a domain length $L^2=9/2$.}
    \label{fig:distmap2}
\end{figure}

In Figures \ref{fig:distheat1} and \ref{fig:distheat2}, we show the fixed delay bifurcation diagrams for varying $\tau$ values for $\epsilon^2=0.001$. Figures \ref{fig:distmap1} and \ref{fig:distmap2} show the distributed delay bifurcation diagrams for varying $\sigma$, for selected $\tau$ and $\epsilon^2$.

\begin{figure}[H]
    \centering
    \begin{subfigure}[t]{0.45\textwidth}
        \centering
        \includegraphics[width=7cm,height=5cm]{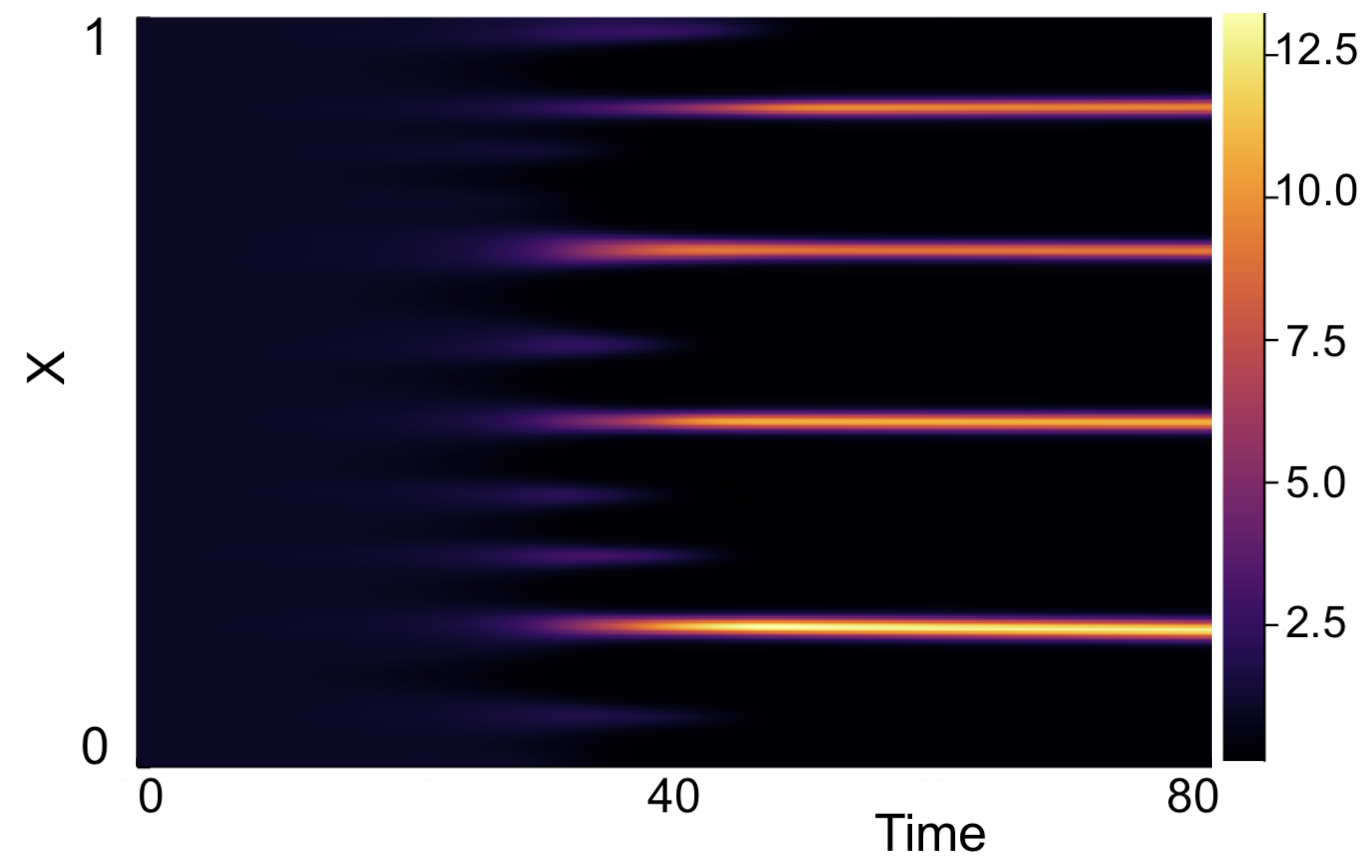}
        \caption{Numerical solution with $\tau=0.5$ and $\sigma=\sigma_{\max}\times0.99$.}
        \label{}
    \end{subfigure}
    \hfill
    \begin{subfigure}[t]{0.45\textwidth}
        \centering
        \includegraphics[width=7cm,height=5cm]{applin1.png}
        \caption{Numerical solution with $\tau=0.5$ and $\sigma=\sigma_{\max}\times0.1$.}
        \label{}
    \end{subfigure}
    \caption{Numerical solutions produced for $(a,b)=(0.1,0.9)$ with $\tau=0.5$ and $\sigma=\sigma_{\max}\times0.99, \sigma_{\max}\times0.1$. We use $L^2=9/2$ and $\epsilon^2=0.001$. Boundary conditions given by \eqref{neumannbc} and initial conditions by \eqref{firstic}. We see Turing pattern formation, as predicted from linear theory.}
    \label{fig:linapp1}
\end{figure}

\begin{figure}[H]
    \centering
    \begin{subfigure}[t]{0.45\textwidth}
        \centering
        \includegraphics[width=7cm,height=5cm]{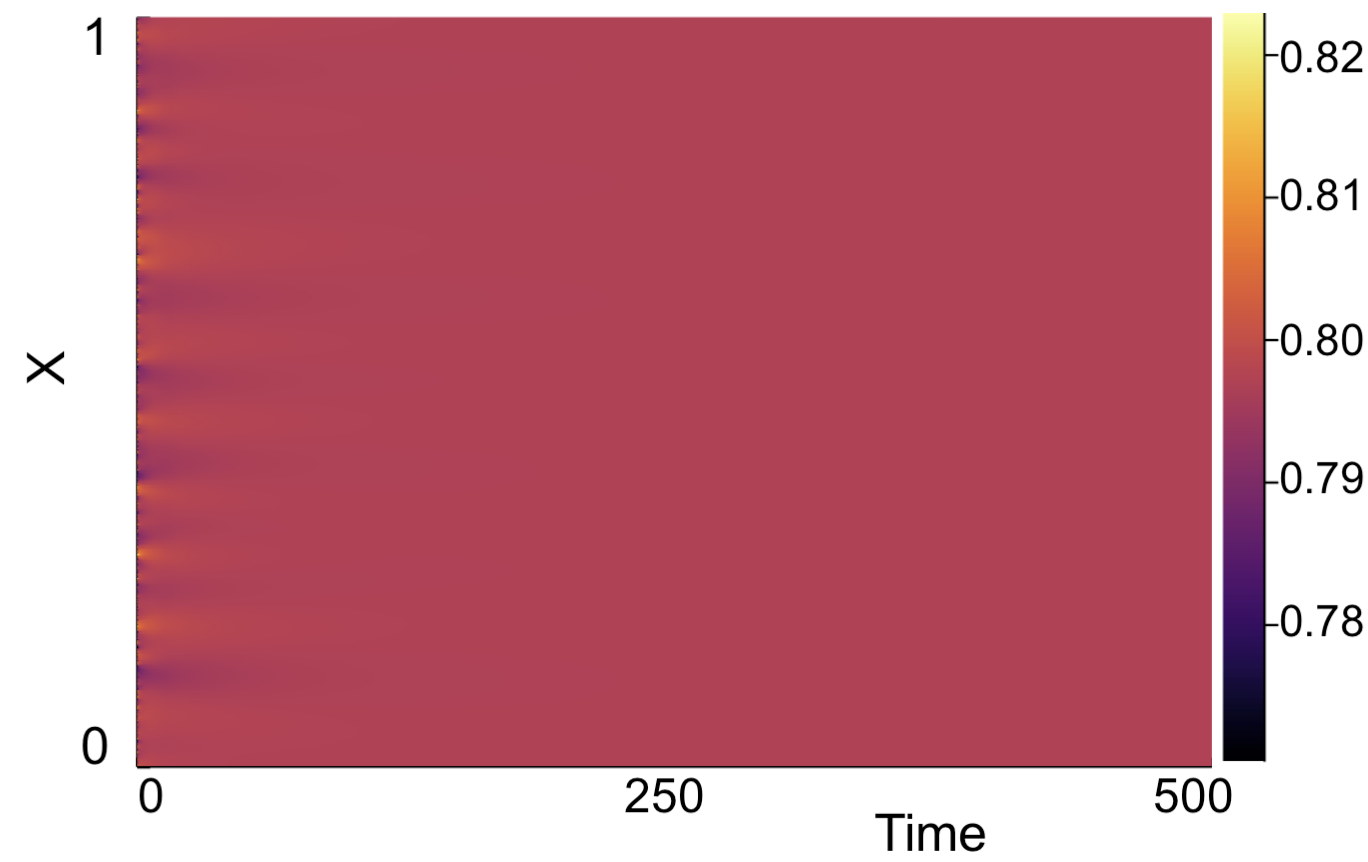}
        \caption{Numerical solution with $\tau=0.5$ and $\sigma=\sigma_{\max}\times0.99$.}
        \label{}
    \end{subfigure}
    \hfill
    \begin{subfigure}[t]{0.45\textwidth}
        \centering
        \includegraphics[width=7cm,height=5cm]{linapp2.png}
        \caption{Numerical solution with $\tau=0.5$ and $\sigma=\sigma_{\max}\times0.1$.}
        \label{}
    \end{subfigure}
    \caption{Numerical solutions produced for $(a,b)=(0.4,0.4)$ with $\tau=0.5$ and $\sigma=\sigma_{\max}\times0.99, \sigma_{\max}\times0.1$. We use $L^2=9/2$ and $\epsilon^2=0.001$. Boundary conditions given by \eqref{neumannbc} and initial conditions by \eqref{firstic}. We see no Turing pattern formation, as predicted from linear theory.}
    \label{fig:linapp2}
\end{figure}

The linear theory from Figures \ref{fig:p2} and \ref{fig:p3} suggests that we have pattern formation for all $\tau\in[0,1]$ independent of the $\sigma$ used for $(a,b)=(0.1,0.9)$, but not for $(a,b)=(0.4,0.4)$. Figures \ref{fig:linapp1} and \ref{fig:linapp2} show similar results to those in \ref{fig:testdist1} and \ref{fig:testdist2}, but with $\tau=0.5$.

Finally, we show further numerical results, for the parameter set $(a,b)=(0.3,1.2)$, to support the conclusion that the onset of patterning, and the type of pattern we see, are independent of $\sigma$ used.

\begin{figure}[H]
   \centering
   \begin{subfigure}[t]{0.32\textwidth}
       \centering
       \includegraphics[width=5cm,height=4.5cm]{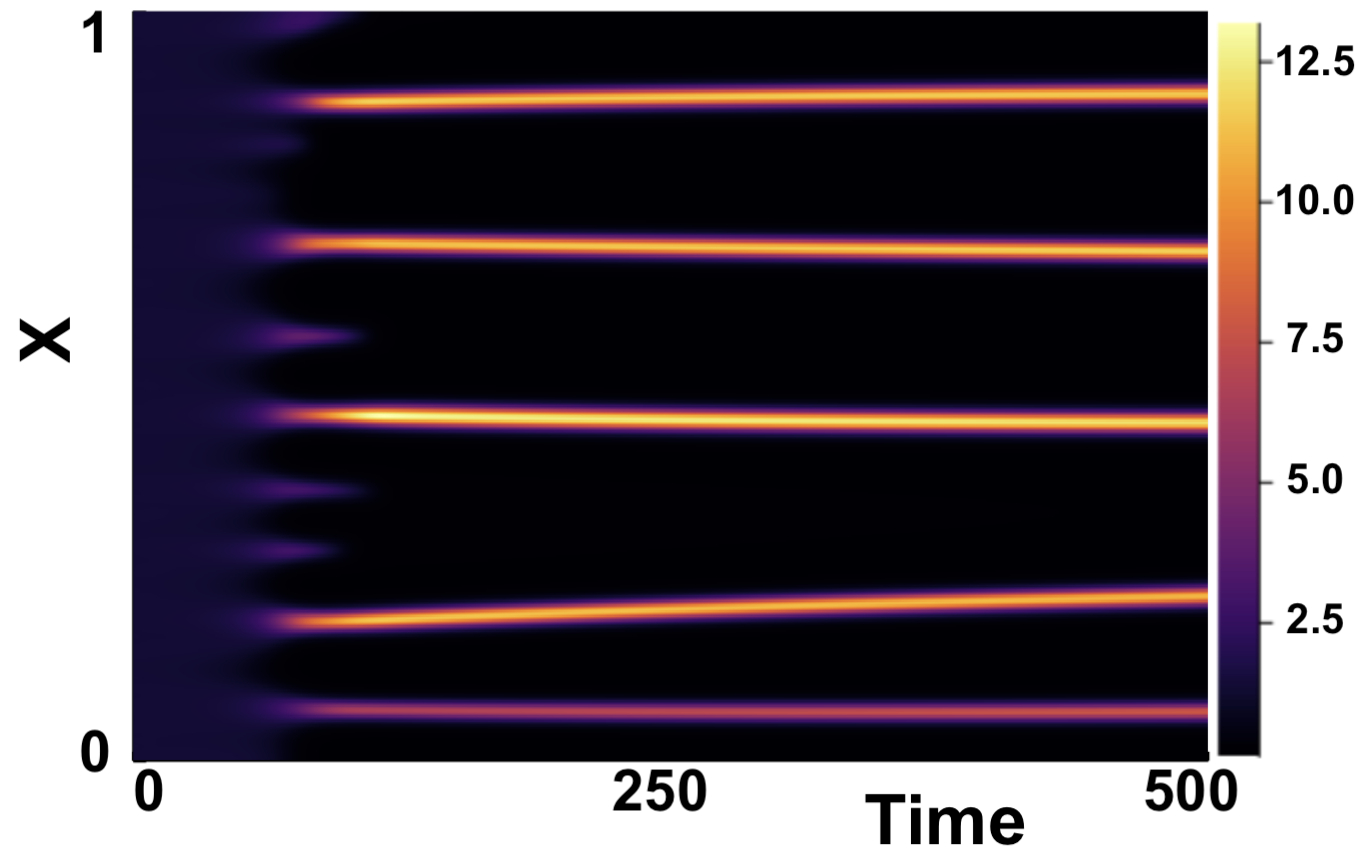}
       \caption{Fixed delay model given by \eqref{fixed2}.}
       \label{}
   \end{subfigure}
   \hfill
   \begin{subfigure}[t]{0.32\textwidth}
       \centering
       \includegraphics[width=5cm,height=4.5cm]{dist2t1sigmax.png}
       \caption{Distributed delay model, \eqref{symmod}, with $\sigma=\sigma_{\max}\times0.99$.}
       \label{}
   \end{subfigure}
   \hfill
   \begin{subfigure}[t]{0.32\textwidth}
       \centering
       \includegraphics[width=5cm,height=4.5cm]{dist2t1sigmax.png}
       \caption{Distributed delay model, \eqref{symmod}, with $\sigma=\sigma_{\max}\times0.1$.}
       \label{}
   \end{subfigure}
   \caption{Numerical simulations showing comparison of fixed delay case vs distributed delay case for $\tau=1$. Boundary conditions given by \eqref{neumannbc} and initial conditions by \eqref{firstic}. $(a,b)=(0.3,1.2)$, $\epsilon^2=0.001$, $L^2=9/2$.}
   \label{fig:distres3}
\end{figure}
\begin{figure}[H]
    \centering
    \begin{subfigure}[t]{0.32\textwidth}
        \centering
        \includegraphics[width=5cm,height=4.5cm]{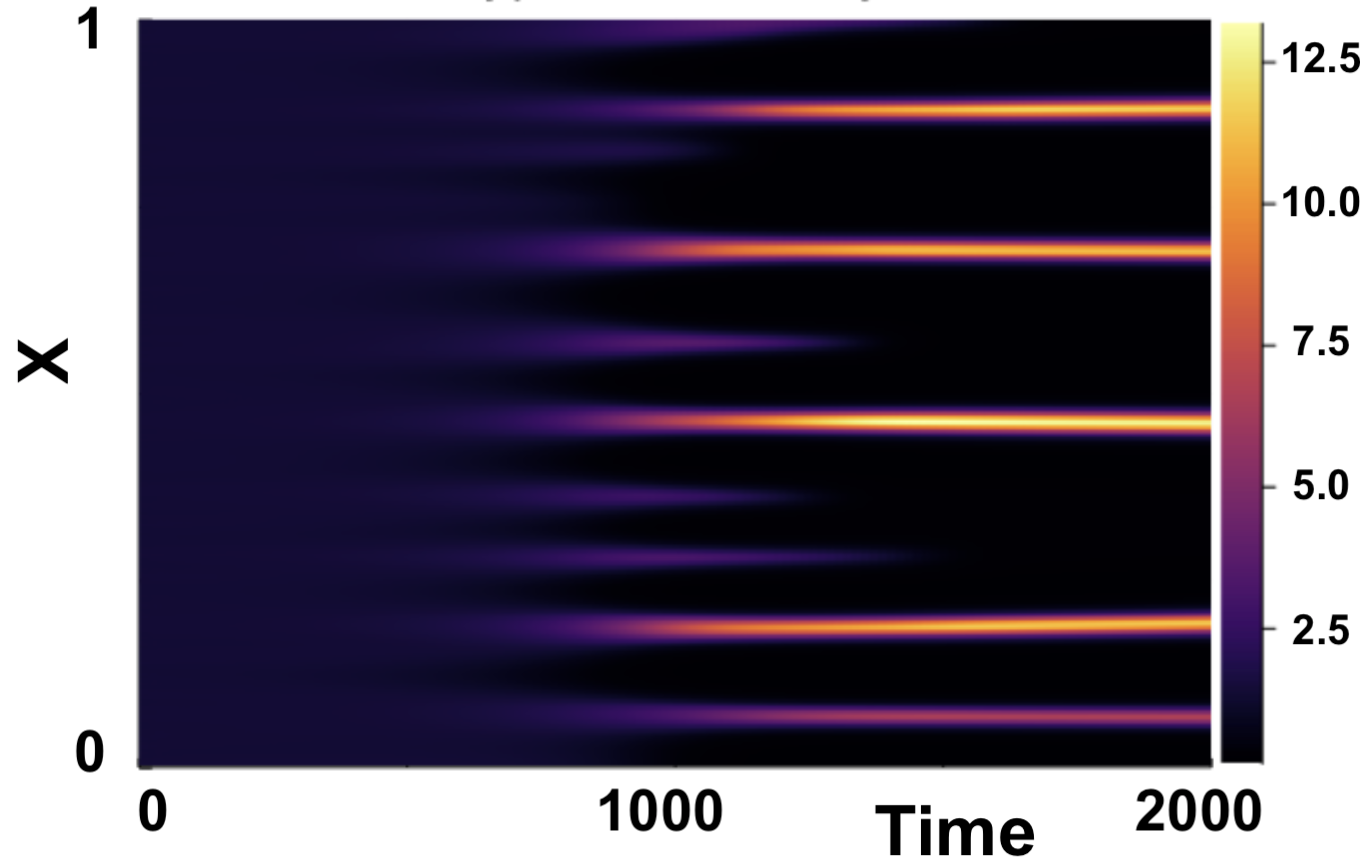}
        \caption{Fixed delay model given by \eqref{fixed2}.}
        \label{}
    \end{subfigure}
    \hfill
    \begin{subfigure}[t]{0.32\textwidth}
        \centering
        \includegraphics[width=5cm,height=4.5cm]{dist2t16sigmax.png}
        \caption{Distributed delay model, \eqref{symmod}, with $\sigma=\sigma_{\max}\times0.99$.}
        \label{}
    \end{subfigure}
    \hfill
    \begin{subfigure}[t]{0.32\textwidth}
        \centering
        \includegraphics[width=5cm,height=4.5cm]{dist2t16sigmax.png}
        \caption{Distributed delay model, \eqref{symmod}, with $\sigma=\sigma_{\max}\times0.1$.}
        \label{}
    \end{subfigure}
    \caption{Numerical simulations showing comparison of fixed delay case vs distributed delay case for $\tau=16$. Boundary conditions given by \eqref{neumannbc} and initial conditions by \eqref{firstic}. $(a,b)=(0.3,1.2)$, $\epsilon^2=0.001$, $L^2=9/2$.}
    \label{fig:distres4}
\end{figure}

\subsection{An Asymmetric Distribution}

We present results here to verify that for larger $\tau$, the skewed distribution does not significantly change the results seen compared to that of the fixed delay case. Figures here show the results for $\tau\in\{1,2,4,8\}$.

% tau = 1
\begin{figure}[H]
    \centering
    \begin{subfigure}[t]{0.45\textwidth}
        \centering
        \includegraphics[width=7cm,height=5cm]{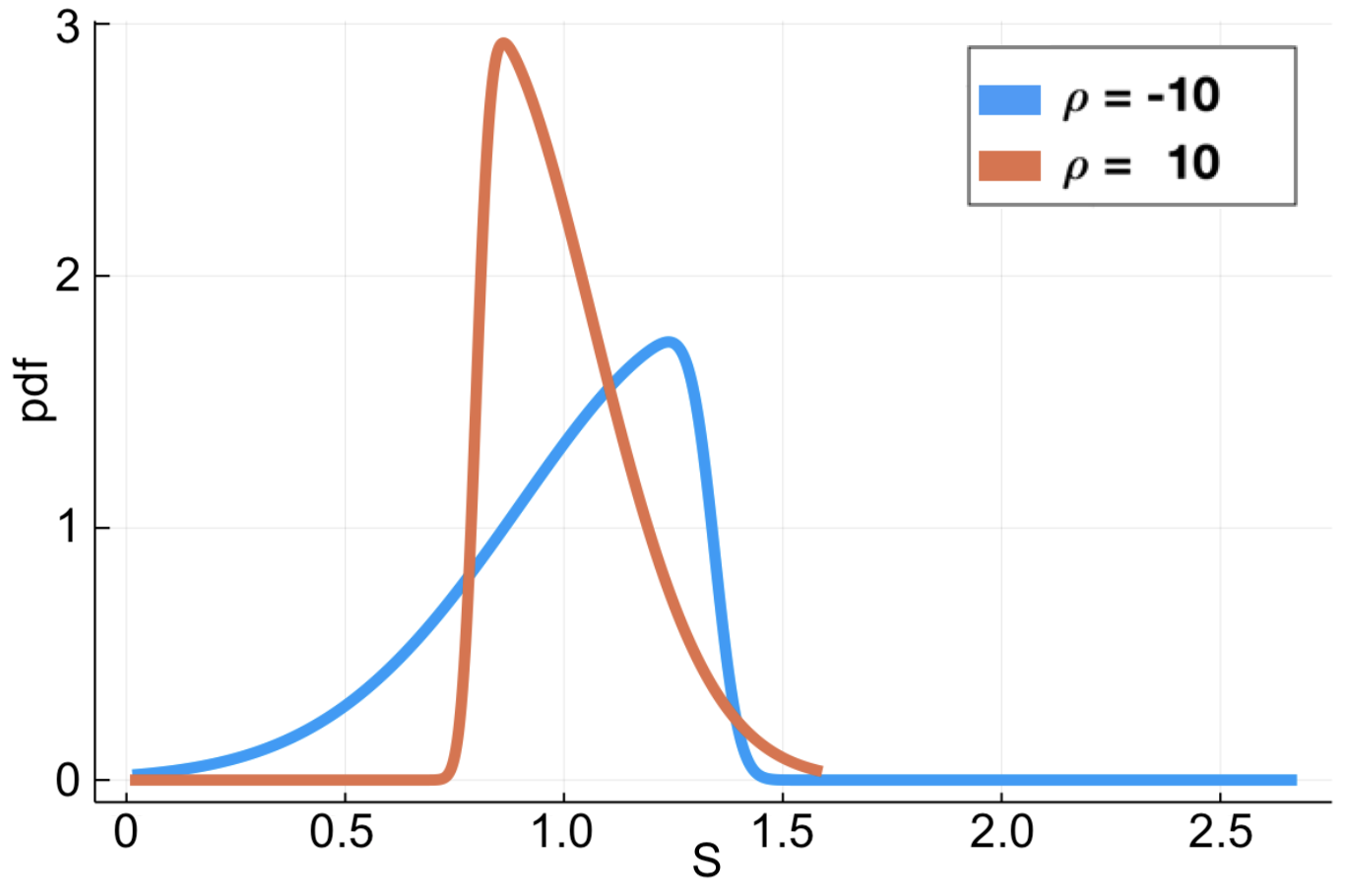}
        \caption{pdfs of skewed truncated Gaussian distributions, with $\rho=-10,10$. Both pdfs have mean $\tau=1$.}
        \label{}
    \end{subfigure}
    \hfill
    \begin{subfigure}[t]{0.45\textwidth}
        \centering
        \includegraphics[width=7cm,height=5cm]{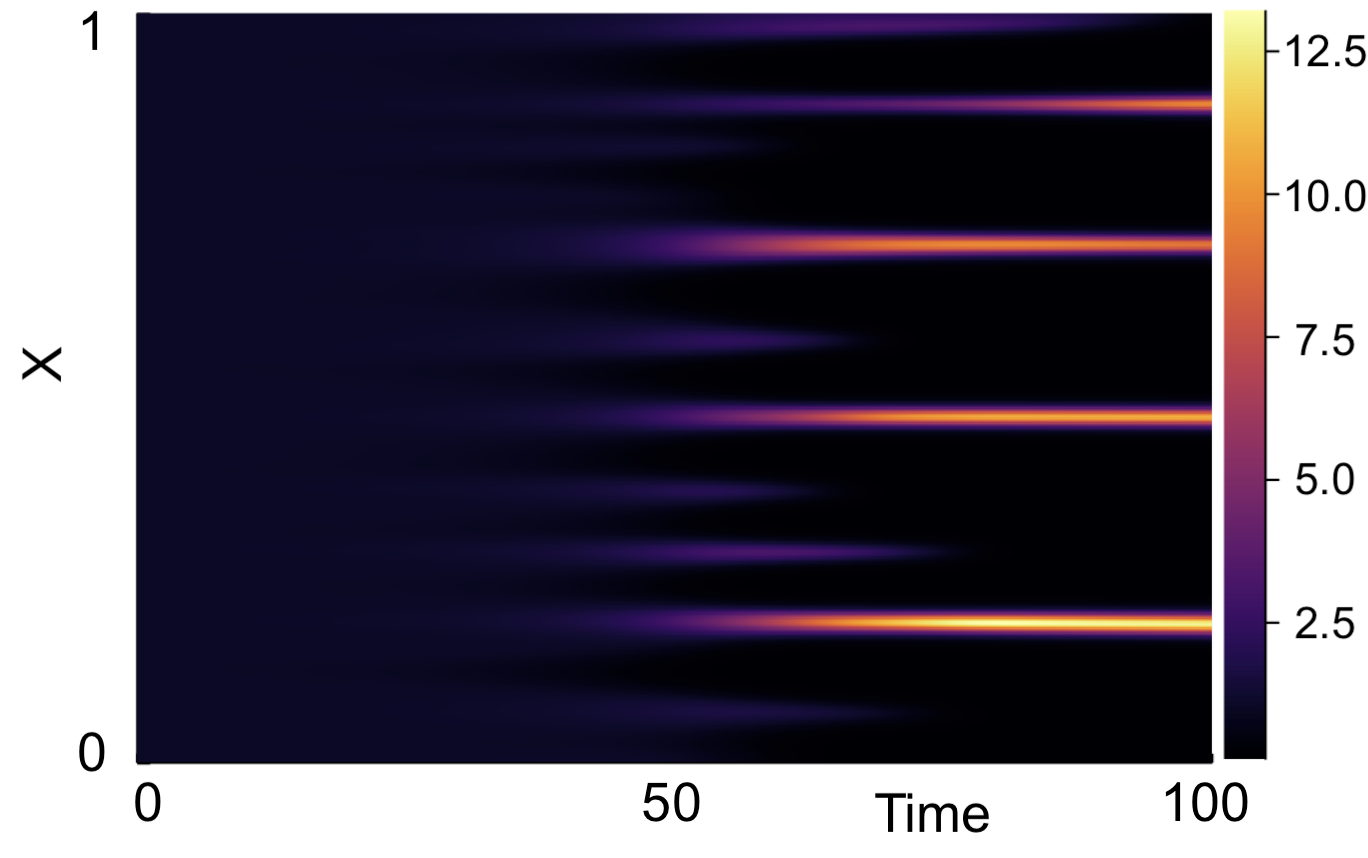}
        \caption{Numerical simulation of fixed delay case with $\tau=1$.}
        \label{}
    \end{subfigure}
    \hfill
    \begin{subfigure}[t]{0.45\textwidth}
        \centering
        \includegraphics[width=7cm,height=5cm]{fixt1.png}
        \caption{Numerical simulation with skewed distribution of $\rho=-10$. Distribution parameters are $\mu=2.70(3 s.f.)$ and $\omega=0.891(3 s.f.)$.}
        \label{}
    \end{subfigure}
    \hfill
    \begin{subfigure}[t]{0.45\textwidth}
        \centering
        \includegraphics[width=7cm,height=5cm]{fixt1.png}
        \caption{Numerical simulation with skewed distribution of $\rho=10$. Distribution parameters are $\mu=1.59(3 s.f.)$ and $\omega=0.525(3 s.f.)$.}
        \label{}
    \end{subfigure}
    \caption{Numerical rsesults for $(a,b)=(0.1,0.9)$ with $\rho=-10,10$ and $\tau=2$. Parameters $\epsilon^2=0.001$ and $L^2=9/2$. Initial conditions given by \eqref{firstic} and boundary conditions by \eqref{neumannbc}.}
    \label{fig:linskew2}
\end{figure}

% tau = 2
\begin{figure}[H]
    \centering
    \begin{subfigure}[t]{0.45\textwidth}
        \centering
        \includegraphics[width=7cm,height=5cm]{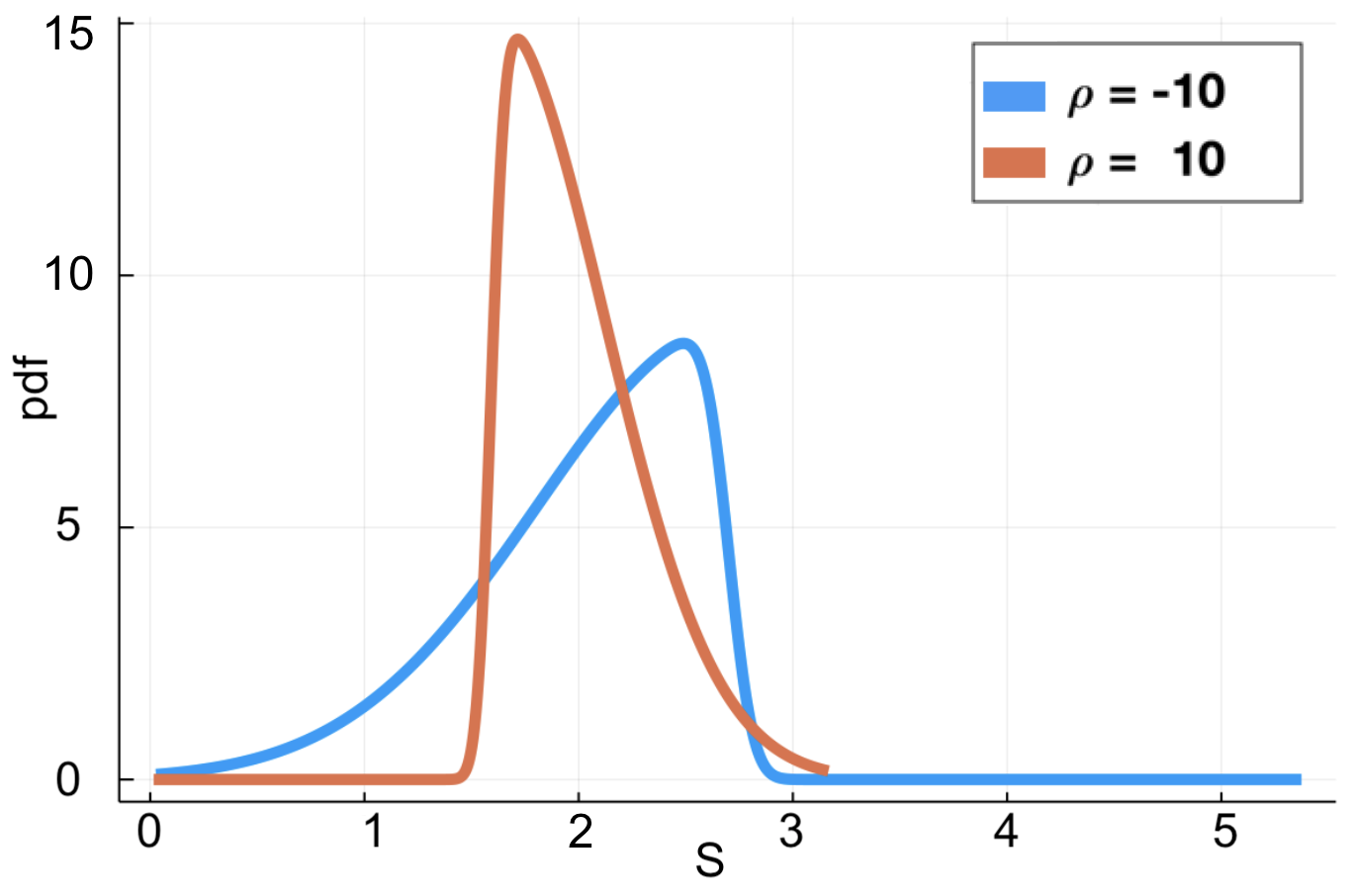}
        \caption{pdfs of skewed truncated Gaussian distributions, with $\rho=-10,10$. Both pdfs have mean $\tau=2$.}
        \label{}
    \end{subfigure}
    \hfill
    \begin{subfigure}[t]{0.45\textwidth}
        \centering
        \includegraphics[width=7cm,height=5cm]{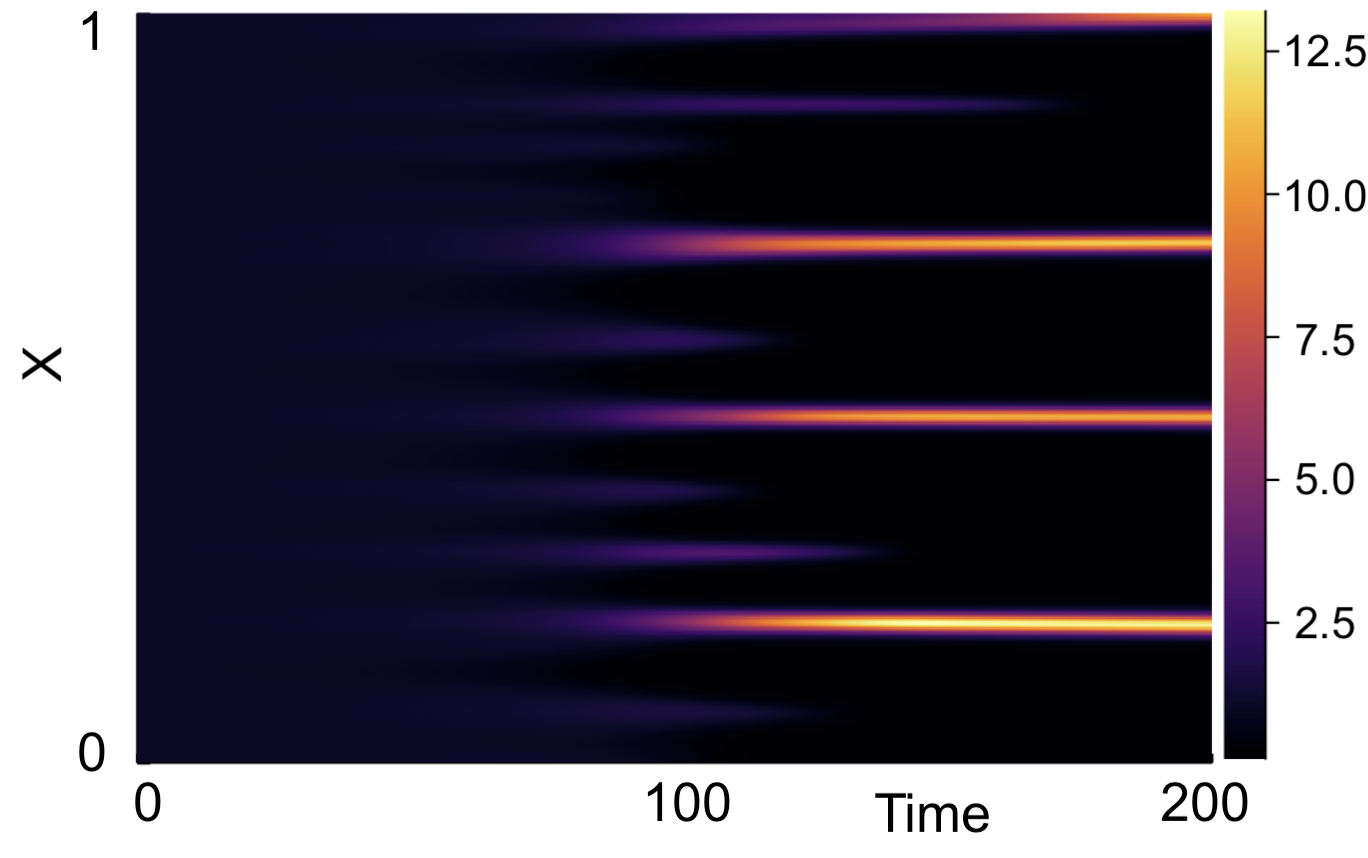}
        \caption{Numerical simulation of fixed delay case with $\tau=2.$}
        \label{}
    \end{subfigure}
    \hfill
    \begin{subfigure}[t]{0.45\textwidth}
        \centering
        \includegraphics[width=7cm,height=5cm]{fixt2.png}
        \caption{Numerical simulation with skewed distribution of $\rho=-10$. Distribution parameters are $\mu=2.70(3 s.f.)$ and $\omega=0.891(3 s.f.)$.}
        \label{}
    \end{subfigure}
    \hfill
    \begin{subfigure}[t]{0.45\textwidth}
        \centering
        \includegraphics[width=7cm,height=5cm]{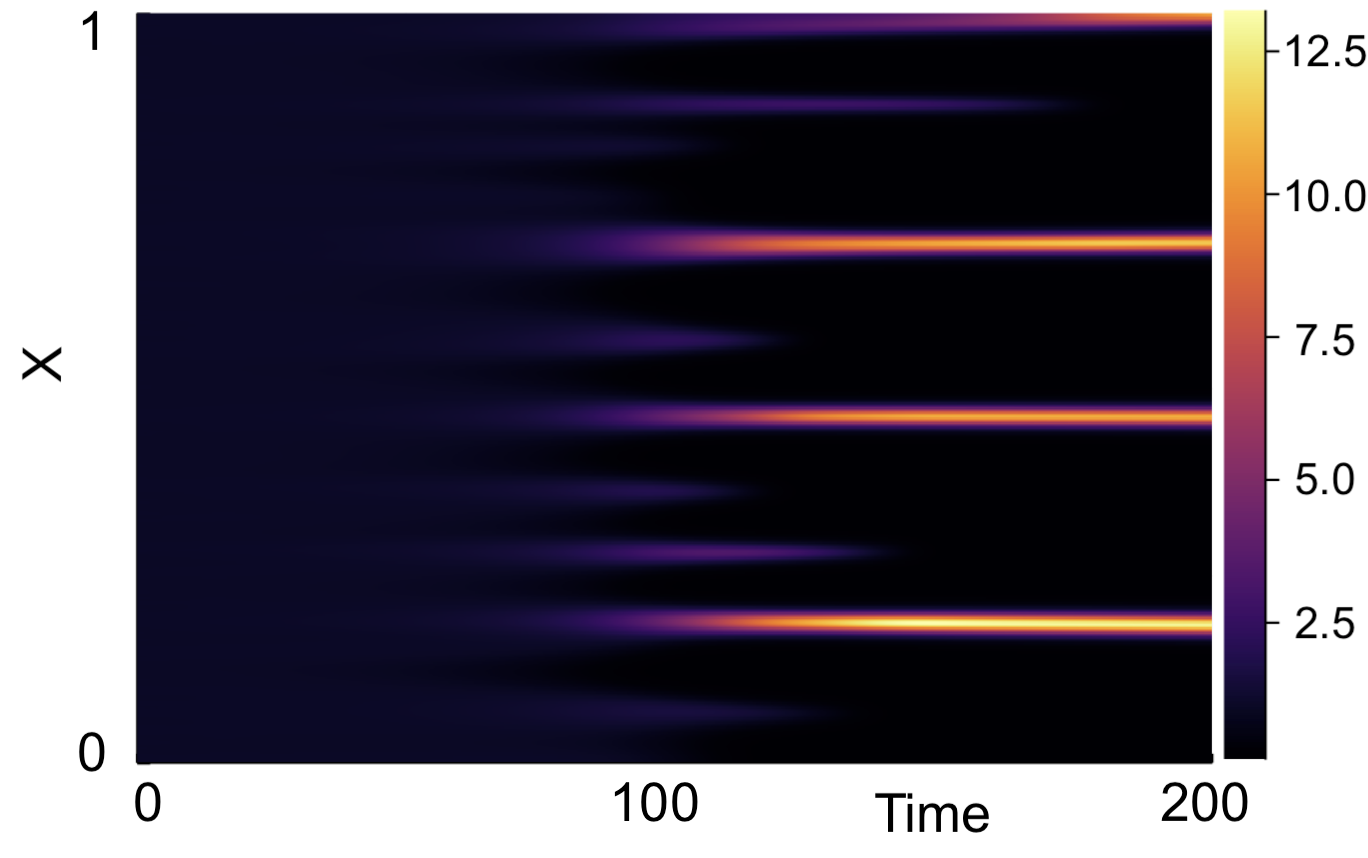}
        \caption{Numerical simulation with skewed distribution of $\rho=10$. Distribution parameters are $\mu=1.59(3 s.f.)$ and $\omega=0.525(3 s.f.)$.}
        \label{}
    \end{subfigure}
    \caption{Numerical rsesults for $(a,b)=(0.1,0.9)$ with $\rho=-10,10$ and $\tau=2$. Parameters $\epsilon^2=0.001$ and $L^2=9/2$. Initial conditions given by \eqref{firstic} and boundary conditions by \eqref{neumannbc}.}
    \label{fig:linskew2}
\end{figure}

% tau = 4
\begin{figure}[H]
    \centering
    \begin{subfigure}[t]{0.45\textwidth}
        \centering
        \includegraphics[width=7cm,height=5cm]{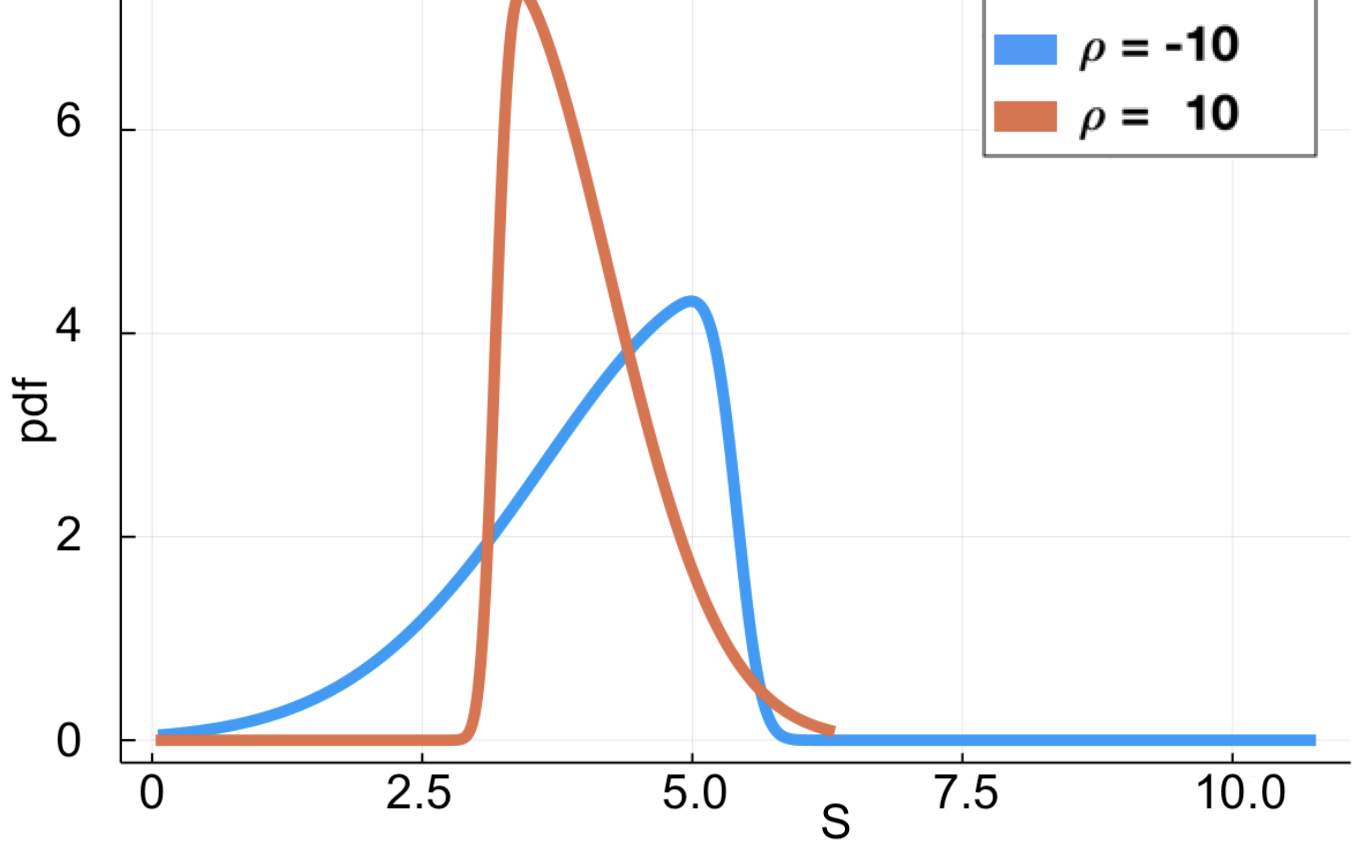}
        \caption{pdfs of skewed truncated Gaussian distributions, with $\rho=-10,10$. Both pdfs have mean $\tau=4$.}
        \label{}
    \end{subfigure}
    \hfill
    \begin{subfigure}[t]{0.45\textwidth}
        \centering
        \includegraphics[width=7cm,height=5cm]{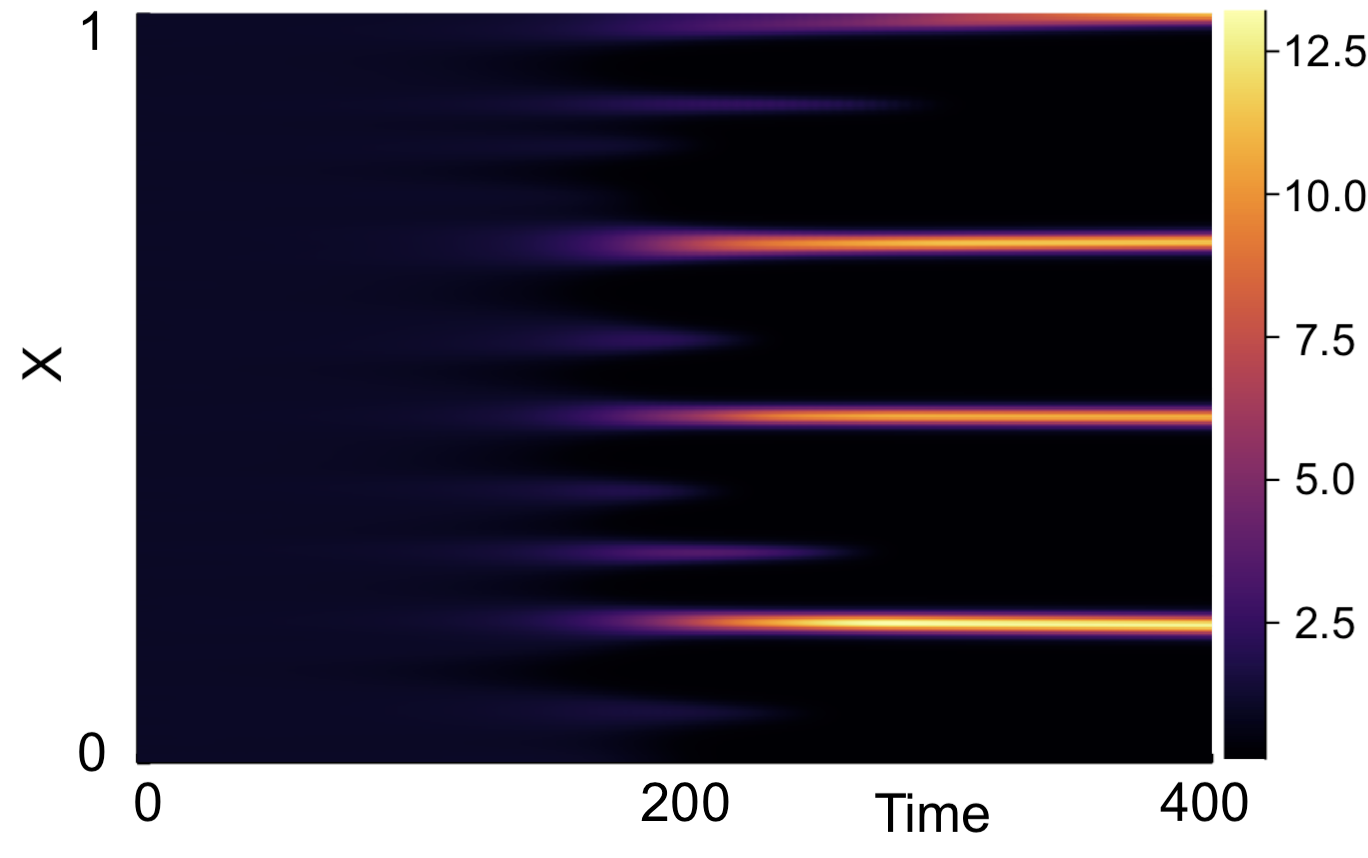}
        \caption{Numerical simulation of fixed delay case with $\tau=4.$}
        \label{}
    \end{subfigure}
    \hfill
    \begin{subfigure}[t]{0.45\textwidth}
        \centering
        \includegraphics[width=7cm,height=5cm]{fixt4.png}
        \caption{Numerical simulation with skewed distribution of $\rho=-10$. Distribution parameters are $\mu=5.41(3 s.f.)$ and $\omega=1.79(3 s.f.)$.}
        \label{}
    \end{subfigure}
    \hfill
    \begin{subfigure}[t]{0.45\textwidth}
        \centering
        \includegraphics[width=7cm,height=5cm]{fixt4.png}
        \caption{Numerical simulation with skewed distribution of $\rho=10$. Distribution parameters are $\mu=3.17(3 s.f.)$ and $\omega=1.05(3 s.f.)$.}
        \label{}
    \end{subfigure}
    \caption{Numerical rsesults for $(a,b)=(0.1,0.9)$ with $\rho=-10,10$ and $\tau=4$. Parameters $\epsilon^2=0.001$ and $L^2=9/2$. Initial conditions given by \eqref{firstic} and boundary conditions by \eqref{neumannbc}.}
    \label{fig:linskew4}
\end{figure}

% tau = 8
\begin{figure}[H]
    \centering
    \begin{subfigure}[t]{0.45\textwidth}
        \centering
        \includegraphics[width=7cm,height=5cm]{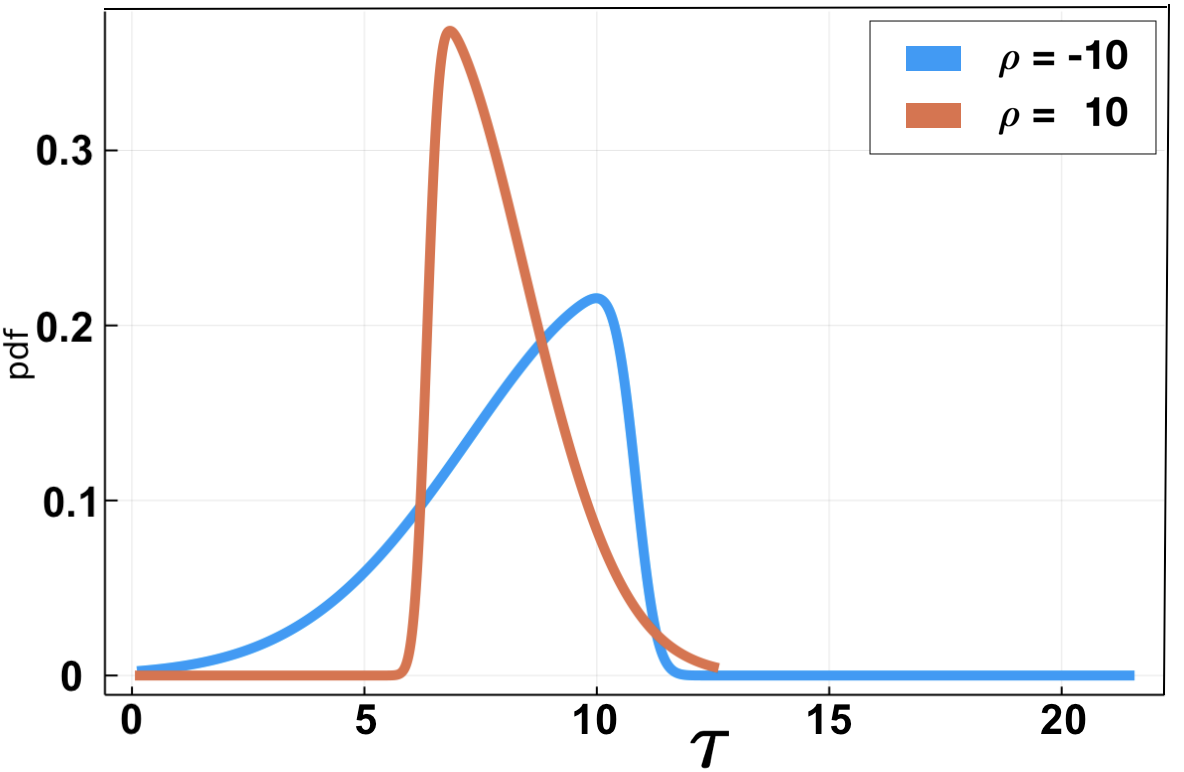}
        \caption{pdfs of skewed truncated Gaussian distributions, with $\rho=-10,10$. Both pdfs have mean $\tau=8$.}
        \label{}
    \end{subfigure}
    \hfill
    \begin{subfigure}[t]{0.45\textwidth}
        \centering
        \includegraphics[width=7cm,height=5cm]{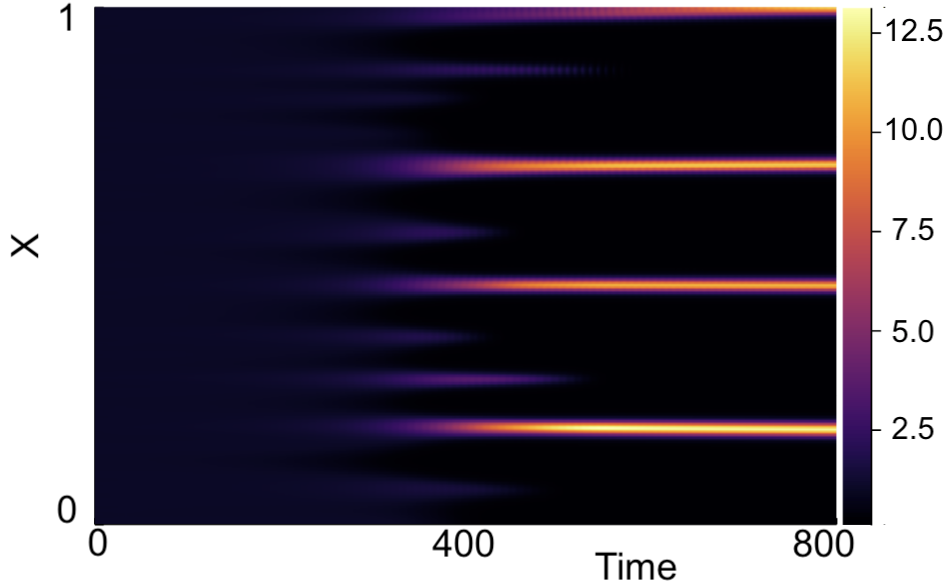}
        \caption{Numerical simulation of fixed delay case with $\tau=8$.}
        \label{}
    \end{subfigure}
    \hfill
    \begin{subfigure}[t]{0.45\textwidth}
        \centering
        \includegraphics[width=7cm,height=5cm]{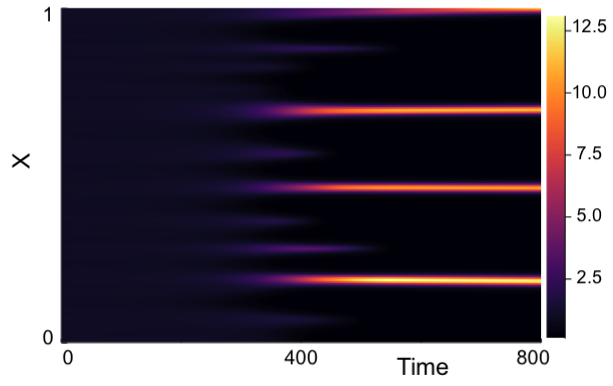}
        \caption{Numerical simulation with skewed distribution of $\rho=-10$. Distribution parameters are $\mu=10.8(3 s.f.)$ and $\omega=3.58(3 s.f.)$.}
        \label{}
    \end{subfigure}
    \hfill
    \begin{subfigure}[t]{0.45\textwidth}
        \centering
        \includegraphics[width=7cm,height=5cm]{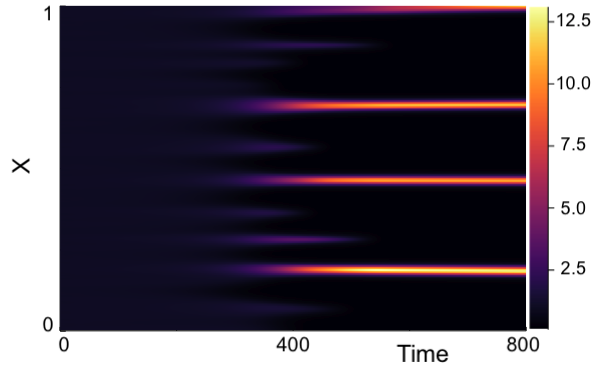}
        \caption{Numerical simulation with skewed distribution of $\rho=10$. Distribution parameters are $\mu=6.34(3 s.f.)$ and $\omega=2.09(3 s.f.)$.}
        \label{}
    \end{subfigure}
    \caption{Numerical rsesults for $(a,b)=(0.1,0.9)$ with $\rho=-10,10$ and $\tau=8$. Parameters $\epsilon^2=0.001$ and $L^2=9/2$. Initial conditions given by \eqref{firstic} and boundary conditions by \eqref{neumannbc}.}
    \label{fig:linskew8}
\end{figure}

%next line adds the Bibliography to the contents page
\addcontentsline{toc}{chapter}{Bibliography}
%uncomment next line to change bibliography name to references

\printbibliography
     %use a bibtex bibliography file refs.bib
%bibliographystyle{plain}  %use the plain bibliography style

\end{document}